\documentclass[cup6b]{cupbook}

\usepackage{amsmath}
\usepackage{amssymb}
\usepackage{graphicx}
\usepackage{rotating}
\usepackage{natbib}
\usepackage{color}

\begin{document}

\def\vs{\vskip 8pt} \def\vss{\vskip 6pt} \def\vsss{\vskip 2pt}
\parskip = 0pt 

\def\makeheadline{\vbox to 0pt{\vskip-30pt\line{\vbox to8.5pt{}\the
                               \headline}\vss}\nointerlineskip}
\def\toppageno{\headline={\hss\tenrm\folio\hss}}
\def\footnoterule{\kern-3pt \hrule width \hsize \kern 2.6pt \vskip 3pt}

\pretolerance=15000  \tolerance=15000
\def\ts{\thinspace}  \def\cl{\centerline}
\def\ni{\noindent}   \def\nnnhi{\noindent \hangindent=10pt}
                     \def\nnhi{\noindent \hangindent=17pt}
                     \def\ihi{\indent \hangindent=22pt}
                     \def\iihi{\parindent=22pt \indent \hangindent=22pt}
\def\h{\hfill}       \def\bk{\kern -0.3em}  \def\b{\kern -0.1em}
\def\r0{$\rho_0$}    \def\rc{$r_c$}
\def\0{\phantom{0}}  \def\1{\phantom{1}}  \def\d{\phantom{.}}
\def\etal{{\it et~al.\ }}  \def\huge{$\phantom{0000000000000000000000000}$}
\def\eg{{\it e.{\ts}g.}}
\def\gapprox{$_>\atop{^\sim}$}  \def\lapprox{$_<\atop{^\sim}$}
\def\ltapprox{\hbox{$<\mkern-19mu\lower4pt\hbox{$\sim$}$}}
\def\gtapprox{\hbox{$>\mkern-19mu\lower4pt\hbox{$\sim$}$}}

\def\mltapprox{\raise2pt\hbox{$<\mkern-19mu\lower5pt\hbox{$\sim$}$}}

\newdimen\sa  \def\sd{\sa=.1em  \ifmmode $\rlap{.}$''$\kern -\sa$
                                \else \rlap{.}$''$\kern -\sa\fi}
              \def\dgd{\sa=.1em \ifmmode $\rlap{.}$^\circ$\kern -\sa$
                                \else \rlap{.}$^\circ$\kern -\sa\fi}
\newdimen\sb  \def\md{\sa=.06em \ifmmode $\rlap{.}$'$\kern -\sa$
                                \else \rlap{.}$'$\kern -\sa\fi}
\def\s{\ifmmode ^{\prime\prime} \else $^{\prime\prime}$ \fi}
\def\min{\ifmmode ^{\prime} \else $^{\prime}$ \fi}

\def\eg{{e.{\thinspace}g.\ }}
\def\ie{{\it i.{\thinspace}e. }}

\def\omit#1{\empty}
\def\o#1{\empty}


\def\thechapter{}


\author[John Kormendy]{John Kormendy\\Department of Astronomy, University of Texas at Austin,\\
        2515 Speedway, Stop C1400,
        Austin, Texas 78712-1205, USA,\\
        kormendy@astro.as.utexas.edu;\\ \\
        Max-Planck-Institut f\"ur Extraterrestrische Physik,\\
        Giessenbachstrasse, D-85748 Garching-bei-M\"unchen, Germany;\\ \\
        Universit\"ats-Sternwarte, Scheinerstrasse 1, D-81679 M\"unchen, Germany}

\chapter{Secular Evolution in Disk Galaxies}

\abstract
Self-gravitating systems evolve toward the most tightly bound configuration that is 
reachable via the evolution processes that are available to them.  They do this by 
spreading -- the inner parts shrink while the outer parts expand -- provided that 
some physical process efficiently transports energy or angular momentum outward.  
The reason is that self-gravitating systems have negative specific heats.  As a result,
the evolution of stars, star clusters, protostellar and protoplanetary disks, 
black hole accretion disks and galaxy disks are fundamentally similar.~How 
evolution proceeds then depends~on~the evolution processes that are available to
each kind of self-gravitating system.  These processes and their consequences for
galaxy disks are the subjects of my lectures and of this Canary Islands Winter School. \vs

      I begin with a review of the formation, growth and death of bars.  Then
      \hbox{I review the slow (``secular'') rearrangement of energy, angular momentum,} 
\ni and mass that results from interactions between stars or gas clouds and
collective phenomena such as bars, oval disks, spiral structure and triaxial dark halos. 
The ``existence-proof'' phase of this work is largely over:~we~have a good heuristic understanding
of how nonaxisymmetric structures rearrange disk gas into outer rings, inner rings and stuff 
dumped onto the center.  The results of simulations correspond closely to the morphology of 
barred and oval galaxies.  Gas that is transported to small radii reaches high densities.
Observations confirm that many barred and oval galaxies have dense central concentrations 
of gas and star formation.  The result is to grow, on timescales of a few Gyr, dense central
components that are frequently mistaken for classical (elliptical-galaxy-like) bulges but 
that were grown slowly out of the disk (not made rapidly by major mergers).  The resulting
picture of secular galaxy evolution accounts for the richness observed in galaxy structure.

\includegraphics{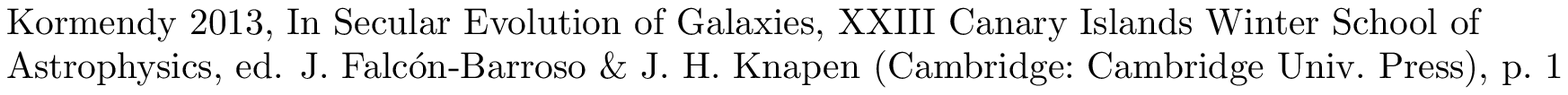}

      We can distinguish between classical and pseudo bulges because the latter  retain 
a ``memory'' of their disky origin.  That is, they have one or more characteristics of disks:
(1) flatter shapes than those of classical bulges, 
(2) correspondingly large ratios of ordered to random velocities, 
(3) small velocity dispersions $\sigma$ with respect to the Faber-Jackson correlation between 
    $\sigma$ and bulge luminosity, 
(4) spiral structure or nuclear bars in the ``bulge'' part of the light profile, 
(5) nearly exponential brightness profiles and 
(6) starbursts.  
So the cleanest examples of pseudobulges are recognizable.  However, pseudo and classical bulges can coexist in the same galaxy. 
\vskip 4pt

       I review two important implications of secular evolution: 
\vskip -17pt
\phantom{Invisible text to fix line spacing so that the abstract fits on 2 pages}
     \begin{enumerate}[(a)]\listsize
     \renewcommand{\theenumi}{(\alph{enumi})}
\item[(1)]{The existence of pseudobulges highlights a problem with our theory of galaxy 
formation by hierarchical clustering.  We cannot explain galaxies that are completely bulgeless.  
Galaxy mergers are expected to happen often enough so that every giant galaxy should have a classical bulge.  
But we observe that bulgleless giant galaxies are common in field environments.  We now realize that 
many dense centers of galaxies that we used to think are bulges were not made by mergers; they were 
grown out of disks.  So the challenge gets more difficult.  
This is the biggest problem faced by our theory of galaxy formation.} \vskip 4pt

\item[(2)]{Pseudobulges are observed to contain supermassive black holes (BHs), but they do not 
show the well known, tight correlations between BH mass and the mass and velocity dispersion of 
the host bulge.   This leads to the suggestion that there are two
fundamentally different BH feeding processes.  Rapid global inward gas transport in galaxy mergers leads 
to giant BHs that correlate with host ellipticals and classical bulges, whereas local and more stochastic 
feeding of small BHs in largely bulgeless galaxies evidently involves too little energy feedback to result 
in BH-host coevolution.  It is an important success of the secular evolution picture that morphological
differences can be used to divide bulges into two types that correlate differently with their BHs.} 
\end{enumerate}
\phantom{Invisible text to fix line spacing so that the abstract fits on 2 pages}
\vskip -17pt

       I review environmental secular evolution{\ts}--{\ts}the transformation of gas-rich, star-forming 
spiral and irregular galaxies into gas-poor, ``red and dead'' S0 and spheroidal (``Sph'') galaxies.
I show that Sph galaxies such as NGC\ts205 and Draco are not the low-luminosity end of the 
structural sequence (the ``fundamental plane'') of elliptical galaxies.  Instead, Sph galaxies have 
structural parameters like those of low-luminosity~S$+$Im~galaxies. Spheroidals are continuous
in their structural parameters~with~the disks of S0 galaxies. They are bulgeless S0s.
\hbox{S$+$Im\ts$\rightarrow${\ts}S0$+$Sph} 
transformation involves a variety of internal (supernova-driven baryon ejection) and environmental processes 
(e.{\ts}g., ram-pressure gas stripping, harassment, and starvation).

\vskip 4pt

       Finally, I summarize how hierarchical clustering and secular processes can be combined into a 
consistent and comprehensive picture of galaxy evolution.

\vfill\eject

\def\thechapter{1}

\section{Introduction}

      These lectures review the slow (``secular'') evolution~of~disk galaxies, both internally and environmentally driven.
As a heuristic introduction at a 2011 winter school, they emphasize a qualitative and intuitive understanding of physical processes. 
This provides a useful complement to Kormendy \& Kennicutt (2004), which is a more complete review of technical details and
the literature.  Since this is a school, my lectures will be as self-contained as possible.
There will therefore be some overlap with the above review~and with 
Kormendy (1981, 1982b, 1993b, 2008a,{\ts}b);
Kormendy \& Cornell (2004);
Kormendy \& Fisher (2005, 2008) and
Kormendy \& Bender (2012).

\subsection{Outline}

      The secular evolution of disk galaxies has deep similarities to the evolution of all other kinds of 
self-gravitating systems.  I begin by emphasizing these similarities.  In particular, the growth of pseudobulges
in galaxy disks is as fundamental as the growth of stars in protostellar disks, the growth of black holes in black hole 
accretion disks and the growth of proto-white-dwarf cores in red giant stars.  A big-picture understanding of these 
similarities is conceptually very important.  The associated physics allows us to understand what kinds of galaxies 
evolve secularly and what kinds do not.  This review discusses only disk galaxies; secular evolution of ellipticals 
is also important but is less thoroughly studied.

      Galaxy bars are important as ``engines'' that drive secular evolution, so I~provide a heuristic
introduction to how bars grow and how they die.  Then I review in some detail the evolution processes that are driven
by bars and by oval disks and the formation of the various kinds of structures that are built by these processes.
I particularly emphasize the growth and properties of pseudobulges.  Based on this, I summarize how we recognize 
pseudobulges and connect them up with our overall picture of galaxy formation. 

      Two consequences (among many) of secular evolution are particularly important.  I review the problem of 
understanding pure-disk galaxies.  These are galaxies that do not contain classical bulges.  We infer that they 
have not experienced a major merger since the first substantial star formation.  Many have barely experienced 
secular evolution.  We do not know how these galaxies are formed.  Second, I review evidence that classical bulges
coevolve with their supermassive black holes but pseudobulges do not. 

      Next, I discuss secular evolution that is environmentally driven.  Here, I
concentrate on the evidence that gas-rich, star forming spiral and irregular galaxies are transformed into gas-poor, ``red and
dead'' S0 and spheroidal galaxies.  I particular emphasize the properties of spheroidals -- that is, tiny dwarfs such as Fornax,
Draco and UMi and larger systems such as NGC~147, NGC~185 and NGC~205.  These are, in essence, bulgeless
S0 galaxies.  And I review the various transformation processes that may make these objects.

      Finally, I tie together our picture of galaxy formation by hierarchical clustering and galaxy merging 
(lectures by Isaac{\ts}Shlosman, Nick{\ts}Scoville and Daniela{\ts}Calzetti)
and the secular evolution that is the theme of this School. 

\subsection{Fast versus slow processes of galaxy evolution}

      Kormendy \& Kennicutt (2004) emphasize that the Universe is in transition from early times when galaxy
evolution was dominated by fast processes -- hierarchical clustering and galaxy merging -- to a future
when merging will largely be over and evolution will be dominated by slow processes (Fig.~1.1).

\vfill

\begin{figure}[h]



\includegraphics{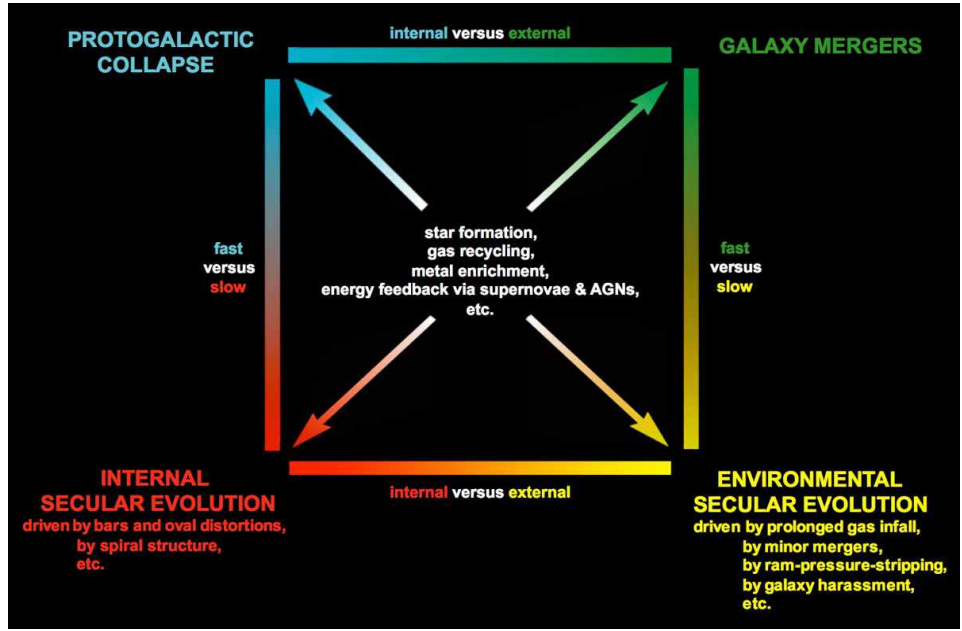}

\caption{\lineskip=0pt \lineskiplimit=0pt
Processes of galaxy evolution updated from Kormendy (1982b) and from Kormendy
\& Kennicutt (2004).  Processes are divided vertically into fast (top) and slow (bottom).  
Fast evolution happens on a free-fall (``dynamical'') timescale, $t_{\rm dyn} \sim (G \rho)^{-1/2}$, where
$\rho$ is the density of the object produced and $G$ is the gravitational constant.  Slow means many 
galaxy rotation periods.  Processes are divided horizontally into ones that happen internally 
in one galaxy (left) and ones that are driven by environmental effects (right).  The processes at center 
are aspects of all types~of~galaxy evolution.  My lectures are about slow processes, both internal 
(Sections 1.2{\ts\ts}--{\ts}1.6) and environmentally driven (Section 1.7).
}

\end{figure}

\eject

      We have a well developed picture of galaxy formation that mostly involves the processes in the upper-right 
corner of Fig.~1.1.  Quantum density fluctuations in non-baryonic, dynamically cold dark matter form immediately 
after the Big Bang and then get stretched by the expansion of the Universe.  Gravity drives hierarchical clustering
that causes these fluctuations to grow, separate out from the expansion of the Universe, collapse and form galaxy halos.
The baryons first go along for the ride and later cool inside the halos to form stars and visible galaxies.  Spiral galaxies 
form when halos quiescently accrete gas that dissipates~and~forms~disks.  Ellipticals form when galaxies collide and merge; 
then dynamical violence scrambles disks into ellipsoidal ellipticals.  It is a convincing story, developed in 
Toomre (1977a); 
White \& Rees (1978); 
Kauffmann \etal (1993); 
Steinmetz \& Navarro (2002, 2003), 
and many other papers.  Quoting Binney (2004),
``Cold Dark Matter theory has now reached the point at which it should be admitted as a Candidate Member to the Academy of 
Established Theories, so that it can sit alongside the established theories of Maxwell, Einstein, and Heisenberg.''

      Now we are making a transition to a time in the far future when the Universe will have expanded so much that most mergers that
ever will happen will already have happened.  Even now, major mergers -- defined as ones in which the less-massive progenitor is within a
factor of (say) \hbox{5\ts--\ts10} of the mass of the bigger progenitor -- are uncommon.  Minor mergers remain common now but will also get
less common in the future.  As this happens, more and more galaxies spend more and more of their time not undergoing fast and violent
evolution events.  And between such events -- that is, between galaxy collisions and mergers -- galaxies in isolation do not just sit 
and age their stars.  Instead, galaxies evolve on slow timescales; that is, timescales that are much longer than the crossing time
or the collapse time.  We call such slow processes ``secular''.  At present, both fast and slow processes are important.  It is easy
to find (especially in cluster environments) good examples of galaxies whose histories have been dominated by fast processes.  They are 
the ellipticals and the classical bulges of disk galaxies.  Elsewhere (particularly in field environments), it is easy to find galaxies 
whose evolution has almost entirely been secular.  Both of these types of galaxies are relatively easy to recognize.  But it is also
important to understand that both kinds of processes can be important in a single galaxy, and in particular, that a galaxy can contain
both a classical (merger-built) bulge and a pseudo (secularly built) bulge.  Recognizing this is difficult and indeed not always
possible.  We will spend some considerable effort on understanding how to differentiate classical and pseudo bulges.

       Beginning with the seminal paper of Toomre (1977a), most work on galaxy formation over the last 35 years has 
concentrated on hierarchical clustering.  The idea of secular evolution got its start at almost the same time; some of the earliest
papers on the subject are Kormendy (1979a, 1979b, 1981) and Combes \& Sanders (1981).  But research on this subject remained for many
years a series of largely isolated ``cottage industries'' that did not penetrate the astronomical folklore.  This changed very 
rapidly~in~the~last~decade.  In particular, Kormendy \& Kennicutt (2004) aimed to combine the cottage industries into a general
and well articulate paradigm on secular evolution that complements hierarchical clustering.  Now, this subject has become a major 
industry.  Whole meetings have been devoted to it.  This is the motivation that underlies the present Canary Islands Winter School. 

\subsection{A comment about the name ``pseudobulges''}

      As in Kormendy \& Kennicutt (2004), I use the name ``pseudobulge'' for all high-density, distinct central components of 
galaxies that are grown slowly out of disks and not rapidly by galaxy mergers.  They divide themselves into at least three 
subtypes that involve different formation processes.  Boxy bulges in edge-on galaxies are 
bars seen side-on (Combes \& Sanders 1981).  Bars are  disk phenomena.  I will call these ``boxy pseudobulges''.  
Second, dense central components are grown out of disks when nonaxisymmetries transport gas toward the center where it feeds 
starbursts.  Often, these are recognizably more disky than merger-built bulges.  But they are not guaranteed to be flat.  
Still, it is often useful to refer to them as ``disky pseudobulges''  when we need to differentiate them from boxy pseudobulges. 
Third, nuclear bars are a recognizably distinct subset of disky pseudobulges.  What problem am I trying to solve by calling 
all of these ``pseudobulges''?

      Astronomers have a long history of inventing awkward names for things.  For historical reasons, we have two different names 
for the products of major galaxy mergers.  If a merger remnant does not have a disk around it, we call it an ``elliptical''.  But if 
an elliptical subsequently accretes cold gas and grows a new disk (e.{\ts}g., Steinmetz \& Navarro 2002), then we call it a ``bulge''.  
In particular, I will call it a ``classical bulge''.  This is inconvenient when one wants to refer to all elliptical-galaxy-like 
merger remnants without prejudice as to whether they have associated disks.  Some authors call these ``spheroidal galaxies''.  
This is exceedingly misleading, because the same name is used for dwarf galaxies such as Draco and Fornax that are essentially 
bulgeless S0s (see Section 1.7 here).  Another
alternative is to call them ``hot stellar systems''.  This is misleading, because some ellipticals have lower velocity
dispersions than some pseudobulges and indeed also than some lenses (which certainly are disk components).  I could {\it define\/}
the term ``ellipsoidals'' to mean both classical bulges and ellipticals.  But this name has no constituency.  So I will explicitly 
say ``bulges and ellipticals'' when I need to refer to both at once. 

      I do not want the same problem to happen with pseudobulges.  

\section{Self-gravitating systems evolve by spreading}

      This is the most fundamental section of this review.  I want to give~you~a heuristic, intuitive understanding
of why secular evolution happens in disk galaxies.  And I want to show you how the evolution of essentially all self-gravitating
systems is fundamentally similar.  The theme of this Winter School is the disk-galaxy version of that more general evolution.

      I made this point in Kormendy \& Fisher (2008) and in Kormendy (2008a).  However, it is important enough -- 
especially at a school -- that I repeat it here in detail.  The following is a slightly expanded and paraphrased
version of the discussion in Kormendy (2008a).

      The general principle that drives the evolution of self-gravitating systems is this: {\it it is energetically favorable 
to spread}.  That is, as energy or angular momentum is transported outward, the inner parts shrink and grow
denser, while the outer parts expand, grow more diffuse, and, in some cases, escape the system.  How to see this depends 
on whether the system is dominated by rotation or by random motions.

\subsection{If dynamical support is by random motions}

\def\ts{\thinspace}

      For systems that are dominated by velocity dispersion, the argument is
given by Lynden-Bell \& Wood (1968) and by Binney \& Tremaine (1987).  The essential
point is that {\it the specific heat of a self-gravitating system is negative.}
Consider an equilibrium system of $N$ particles of mass $m$, radius~$r$ and 
three-dimensional velocity dispersion~$v$. The virial theorem tells us that 
2{\ts}KE + PE = 0, where the kinetic energy is KE = $Nmv^2/2$ and the potential energy 
is PE = $-G(Nm)^2/r$. \hbox{The total energy $E \equiv$ KE $+$ PE = $-$KE is negative.} 
This is what it means to be a bound system.  But the temperature of the system is
measured by $v^2$: $mv^2/2 = 3kT/2$.  So the specific heat 
$C \equiv dE/dT \propto d(-Nmv^2/2)/d(v^2)$ is also negative.  In the above, $G$ is
the gravitational constant and $k$ is Boltzmann's~constant. 

      The system is supported by heat, so evolution is by heat transport.
If the center of the system gets hotter than the periphery, then heat tends to flow 
outward.  The inner parts shrink and get still hotter.  This promotes further
heat flow. The outer parts receive heat; they expand and cool.  Whether the
system evolves on an interesting timescale depends on whether there is an
efficient heat-transport mechanism.  For example, many globular clusters evolve
quickly by two-body relaxation and undergo core collapse.  Giant 
elliptical galaxies -- which otherwise would evolve similarly -- cannot do so
because their relaxation times are much longer than the age of the Universe.

\subsection{If dynamical support is by rotation}

      Tremaine (1989) provides a transparent summary of an argument due to
Lynden-Bell \& Kalnajs (1972) and to Lynden-Bell \& Pringle (1974).  A disk is
supported by rotation, so evolution is by angular momentum transport.  The
``goal'' is to minimize the total energy at fixed total angular momentum.  
A rotationally supported ring at radius $r$ in a fixed potential $\Phi(r)$ has
specific energy $E(r)$ and specific angular momentum $L(r)$ given by
$$
E(r) = {r\over2}{d\Phi \over dr} + \Phi~~{\rm and}~~
L(r) = \biggl(r^3 \ts{d\Phi \over dr}\biggr)^{1/2}~. \eqno{(1.1)}
$$
Then $dE/dL = \Omega(r)$, where $\Omega = (r^{-1} d\Phi/dr)^{1/2}$ is the
angular speed of rotation.  Disks spread when a unit mass at radius $r_2$
moves outward by gaining angular momentum $dL$ from a unit mass at 
radius $r_1 < r_2$. This is energetically favorable: the change in energy, 
$$
dE = dE_1 + dE_2 = \biggl[- \biggl({dE \over dL}\biggr)_1 
                          + \biggl({dE \over dL}\biggr)_2\ts\biggr]dL 
                 = \ts[-\Omega(r_1) + \Omega(r_2)]\ts dL\ts, \eqno{(1.2)}
$$
is negative because $\Omega(r)$ usually decreases outward.  ``Thus disk
spreading leads to a lower energy~state.  In general, disk spreading,
outward angular momentum flow, and energy dissipation accompany one another
in astrophysical disks'' (Tremaine 1989). 

\subsection{Spreading in various kinds of self-gravitating systems}

      The consequences are very general.  All of the evolution summarized in Fig.~1.2
happens because of the above physics.

      Globular clusters, open clusters and the compact nuclear star clusters in galaxies
are supported by random motions.  Absent any gas infall, they 
spread in three dimensions by outward energy transport.  The mechanism is
two-body relaxation, and the consequences are core collapse and the evaporation
of the outer parts.


\begin{figure*}

\vspace{3.6 truein}


\includegraphics{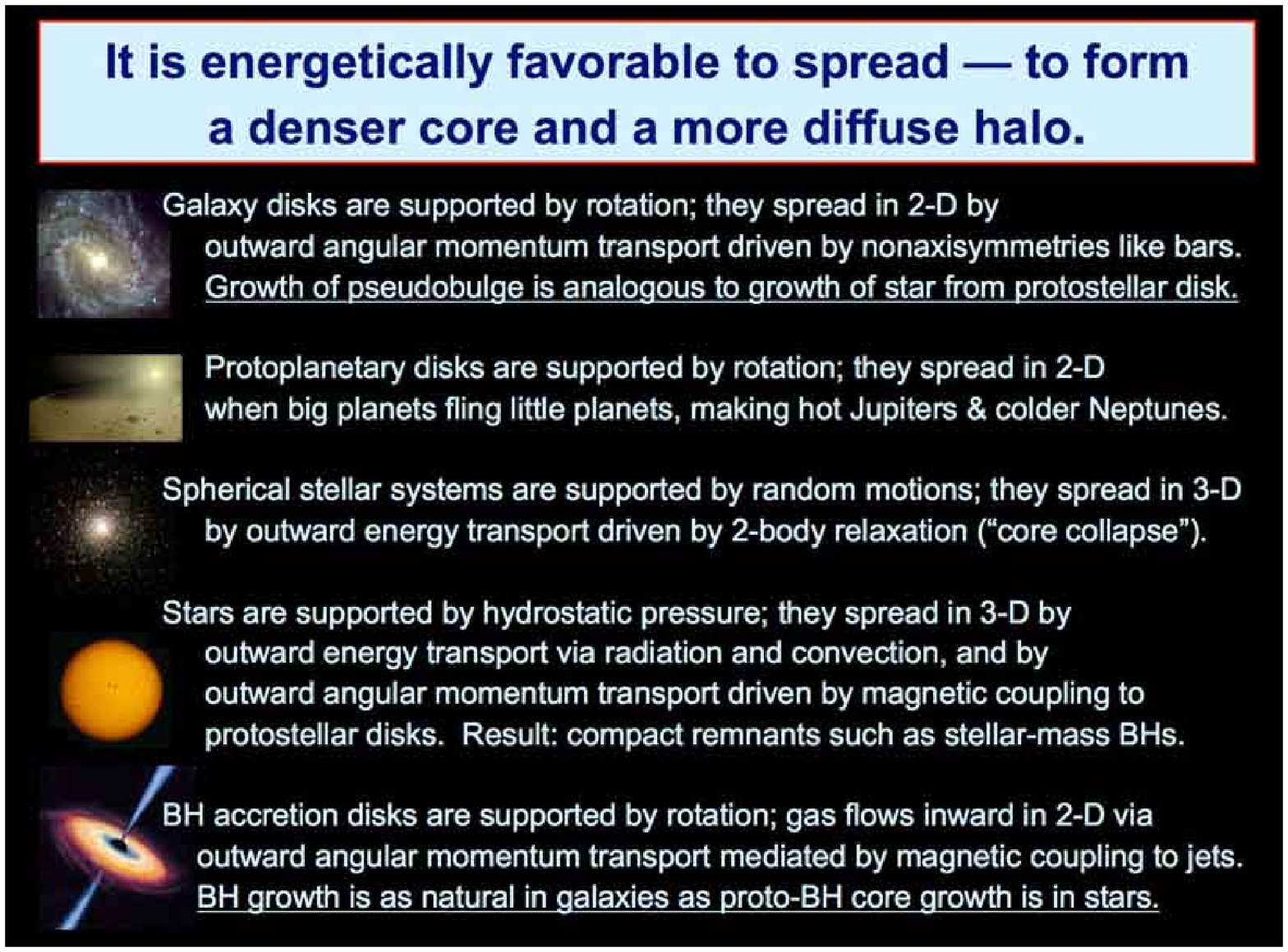}

\caption{\lineskip=0pt \lineskiplimit=0pt
Powerpoint slide that emphasizes the underlying similarity in evolutionary processes 
that shape various kinds of self-gravitating systems.  The secular evolution of galaxy
disks (top row) is the subject of my lectures.
}

\end{figure*}

      Stars are spherical systems supported by pressure.  They spread in three
dimensions by outward energy transport.  The mechanisms are radiation or
convection mediated by opacity.  Punctuated by phases of stability when nuclear
reactions replace the energy that is lost, stellar evolution consists of a
series of core contractions and envelope expansions.  One result is red
(super)giants containing cores that will become white dwarfs, neutron stars,
or stellar mass black holes.

      Protoplanetary disks are supported by rotation; they spread in two dimensions by outward angular momentum transport.  
The tendency toward energy equipartition between low-mass and high-mass objects has the consequence that big planets sink 
by flinging smaller planets outward.  If the evolution results from the interaction of a big planet with a collective
phenomenon such as a spiral density wave in gas or rubble, then the effect is the same.  The results are hot Jupiters and 
colder Neptunes or rubble.

      Protostars are spherical systems coupled to circumstellar disks by
magnetic fields that wind up because of differential rotation. This drives
jets that look one-dimensional but that really are three-dimensional; they
carry away angular momentum  and allow the inner circumstellar disk to shrink 
and accrete onto the star (Shu \etal 1994, 1995). 

      An accretion disk around a black hole is supported by rotation, so it evolves by angular 
momentum transport.   The evolution happens because of magnetic coupling between various parts 
of the accretion disk.  The details are complicated, but the net effect is that some material flows outward
as part of a jet and other material is accreted onto the black hole.  Note again that pictures of this process --
whether observations of radio jets or artists' conceptions of the accretion process -- almost always show 
narrow jets.  But, as in the case of protostars, outward angular momentum transfer must be involved in order
for some of the material to accrete onto the black hole.  

      Galactic disks are supported by rotation.  They spread in two dimensions by outward angular momentum transport.  
Efficient engines are provided by bars and by oval disks.  Like all of the above, \hbox{the evolution
is secular -- slow} compared to the collapse time or crossing time of the disk.  {\it The growth of a pseudobulge 
in a galactic disk is as natural as the growth of a star in a protostellar disk and as the growth of a black hole 
in a black hole accretion disk.}  Only the processes that transport angular momentum are different.

      This secular evolution is the subject of my paper and this Winter School.

\begin{figure*}

\vspace{5.2truein}





\includegraphics{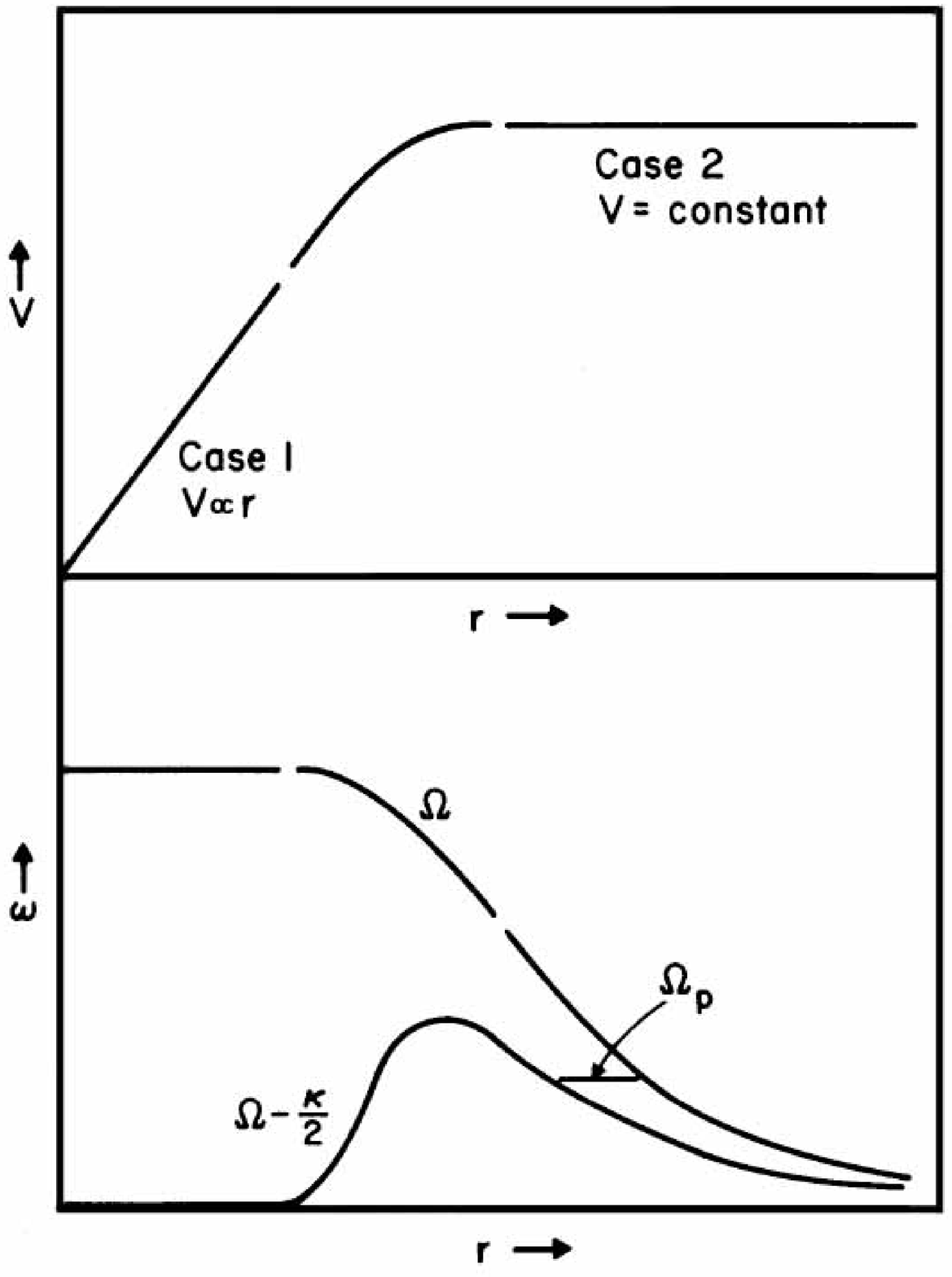}

\includegraphics{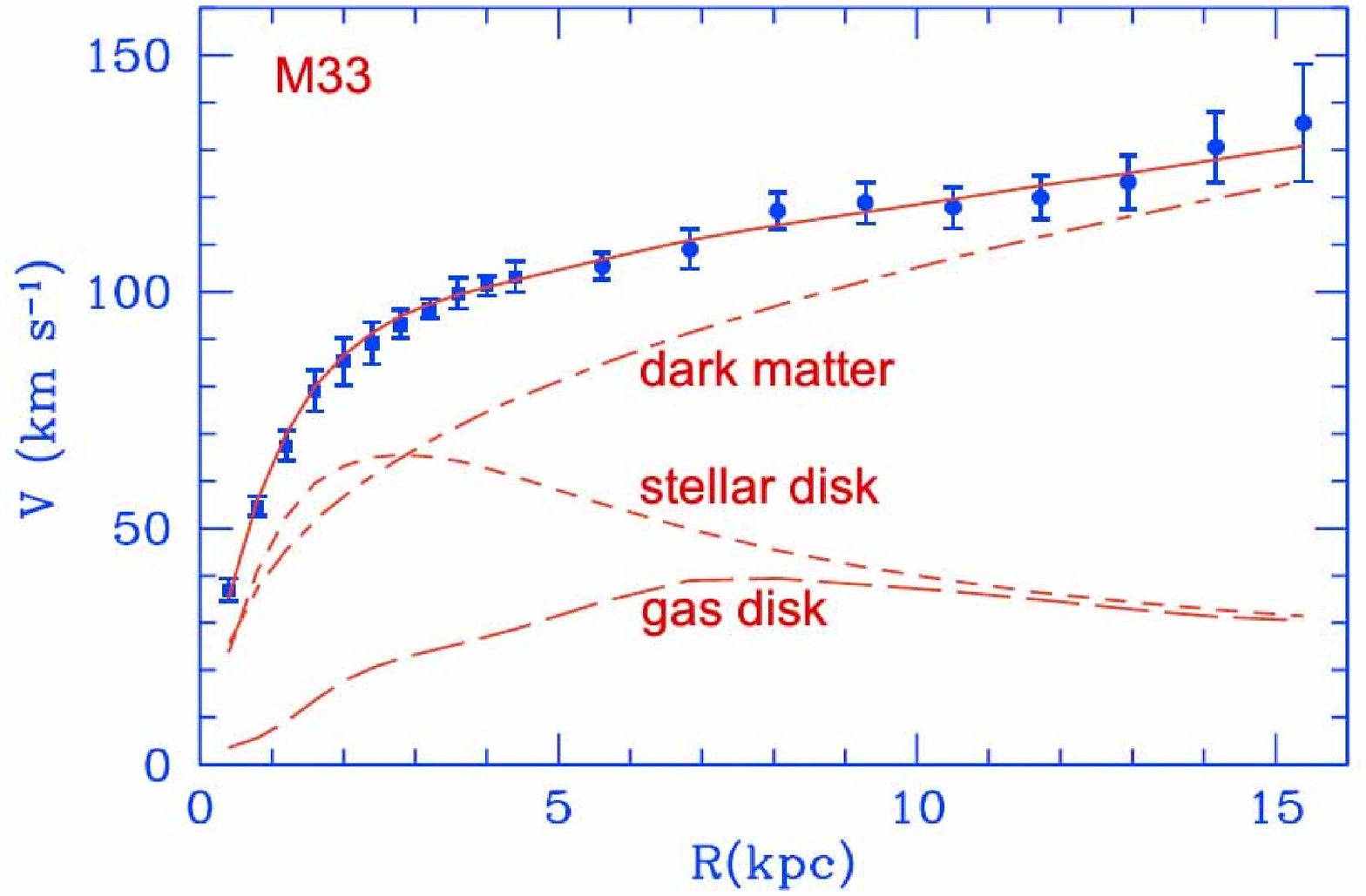}

\includegraphics{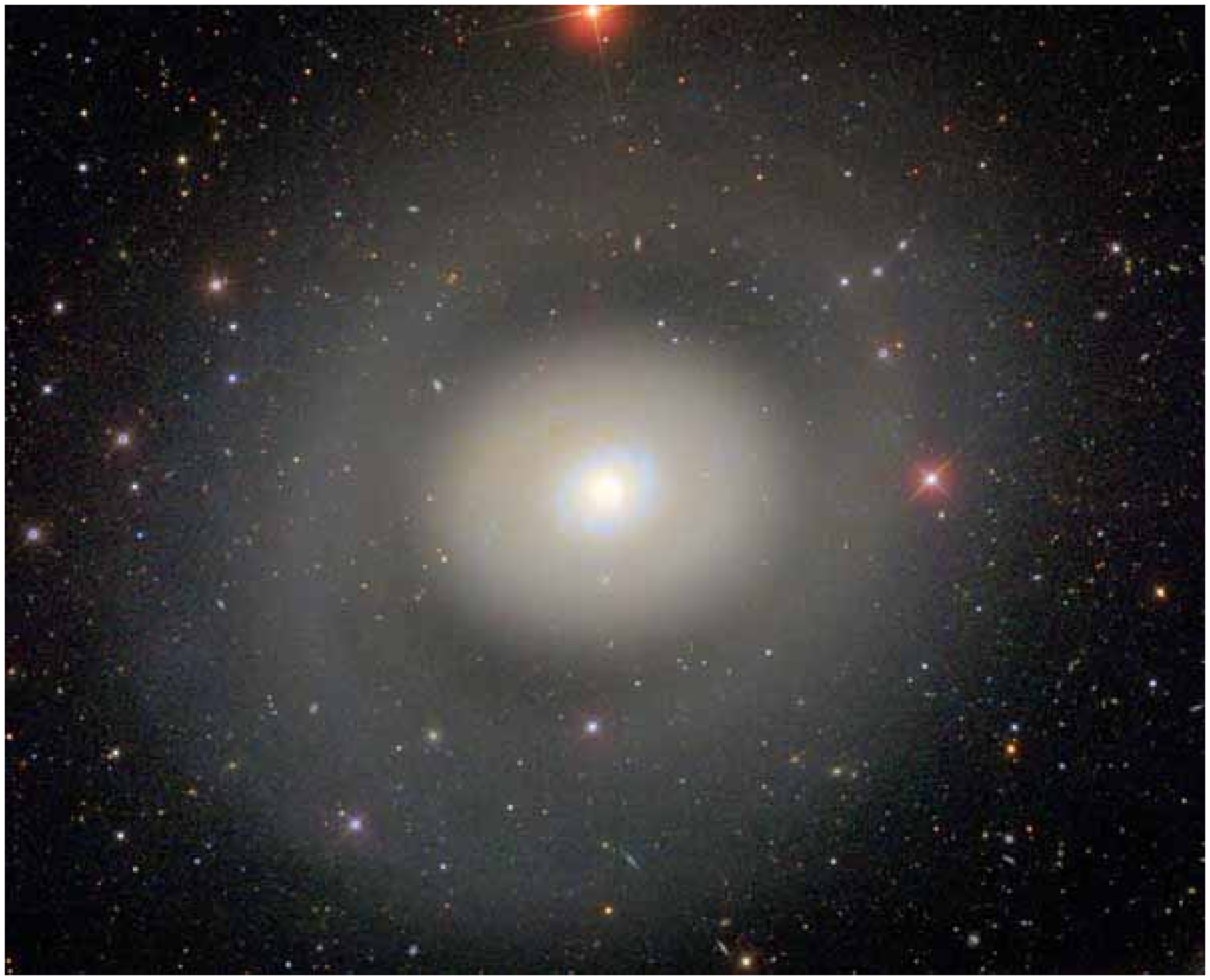}

\includegraphics{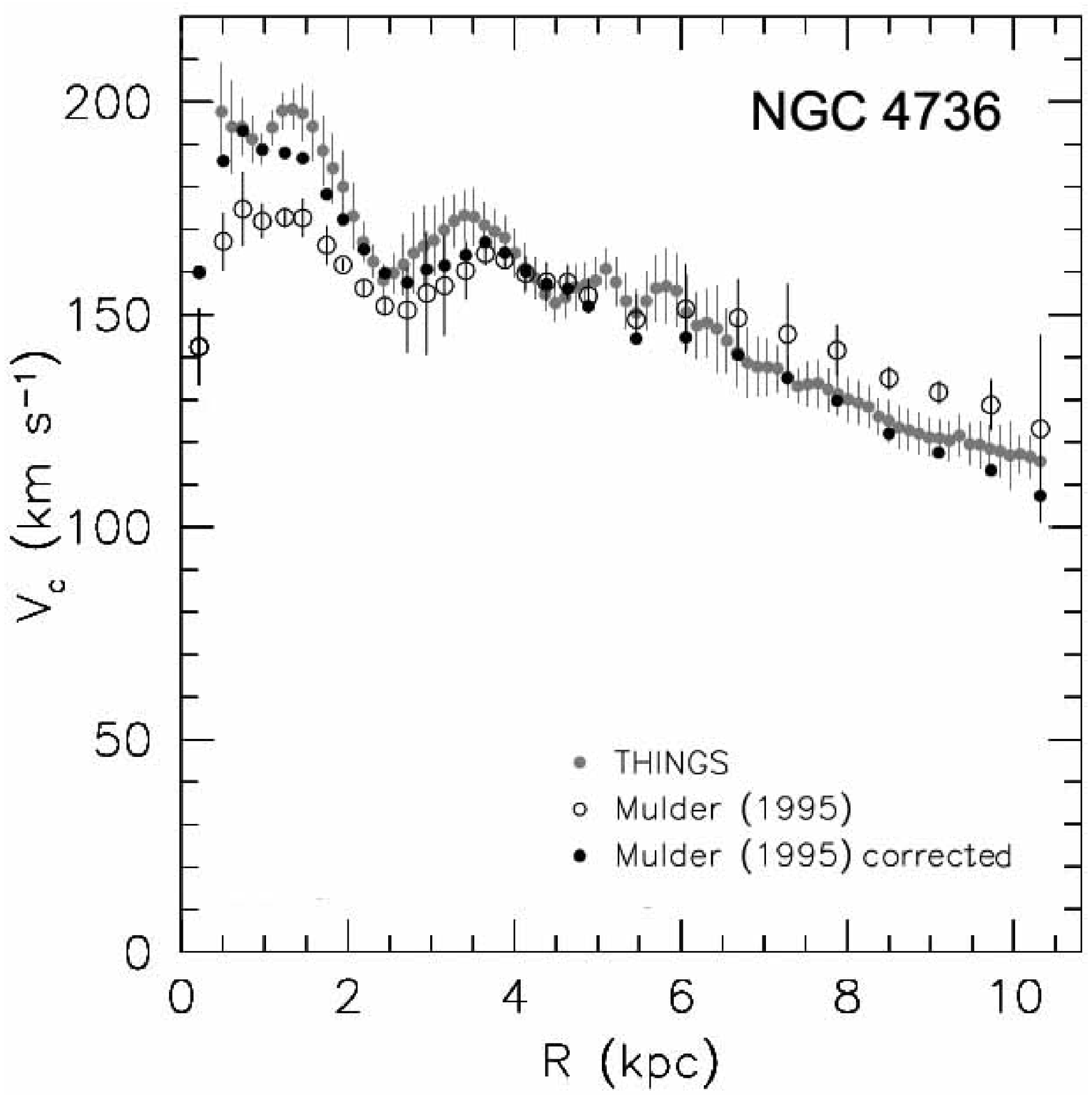}

\caption{\lineskip=0pt \lineskiplimit=0pt
Typical examples of spiral galaxies that do (NGC 4736, bottom) and do not (M{\ts}33, top) have mass distributions that
are conducive to secular evolution.  The top-left plots (Kormendy \& Norman 1979) show a generic rotation curve~$V(r)$
and associated angular velocity curves $\Omega(r) \equiv V(r)/r$ and $\Omega -\kappa/2$, where $\kappa$ is
the epicyclic frequency.  Where $V \propto r$, $\Omega =$ constant and secular evolution is not energetically favorable.
Even where $V(r)$ is turning downward from $V \propto r$ toward $V =$ constant, $\Omega$ decreases outward so slowly
that secular evolution is disfavored.  This is the situation throughout a galaxy like M{\ts}33 (top-right rotation
curve and rotation curve decomposition, from Corbelli \& Salucci 2000).  M{\ts}33 has neither a classical bulge nor a
significant pseudobulge.  In contrast, NGC 4736 (image at bottom-left, from {\tt http://www.wikisky.org}) 
has a rotation curve that decreases outward with radius, as derived by de Blok \etal (2008; bottom-right panel here).
Disk spreading is energetically very favorable.  Moreover,~NGC~4736~is a prototypical strongly oval galaxy,
so it has an engine for secular evolution.  Not surprisingly, it is our best ``poster child'' for secular evolution --
an unbarred galaxy with a pseudobulge identified by five classification criteria (Section 1.5.3).
}

\end{figure*}

\subsection{What kinds of galaxy disks evolve secularly?}

      For secular evolution to happen, a galaxy needs two things.  It needs a driving mechanism such as a bar, a 
globally oval disk (Section 1.3.3), or global spiral structure.  And disk spreading must be energetically favorable.  
Equation (1.2) tells us that it is more energetically favorable to transport angular momentum outward the more the angular
rotation velocity $\Omega(r)$ decreases outward.  This lets us divide galaxies into three types, as follows.

      Galaxies that lack driving agents should not show strong secular evolution.  Examples are unbarred galaxies
with classical bulges and weak or no global spiral structure (e.{\ts}g., M{\ts}31, NGC 2841 and NGC 4594, the 
Sombrero Galaxy).  We will see that these objects generally do not show centrally concentrated molecular gas and star
formation.  We take this as an sign that secular evolution largely is not happening.

      For galaxies with engines, Fig.~1.3 shows how the tendency to evolve depends on the mass distribution as 
revealed by the rotation curve $V(r)$.  Where $V(r)$ increases proportional to $r$ (or even rather more slowly),
the angular rotation velocity $\Omega(r)$ is constant (or very slowly decreasing outward).  This means that it is
not energetically favorable to evolve.   This is the situation in an Scd galaxy like M{\ts}33, where we 
find no pseudobulge.

      The Corbelli \& Salucci (2000) decomposition of the M{\ts}33 rotation curve into contributions due to the stellar disk,
the gas disk and the dark matter halo reveals an additional reason why secular evolution is disfavored.  The contribution of 
the dark matter to $V(r)$ is everywhere comparable to or larger than that of the visible matter.  But dark halos are
dynamically much hotter than disks.  They do not participate in disk dynamics.  The observation that the self-gravity of the 
disk is ``diluted'' by the gravity of the much stiffer halo is a further sign that secular evolution is disfavored.  
Kormendy \& McClure (1993) show that M{\ts}33 contains a nuclear star cluster (like a large globular cluster) but no 
significant evidence for a pseudobulge.

      In contrast, if a galaxy has a flat rotation curve, then $\Omega \propto 1/r$ and secular evolution is energetically
favorable (top-left panels of Fig.~1.3).  Even more extreme are galaxies in which the rotation curve decreases outward.
De Blok \etal (2008) conclude that the rotation curve of NGC 4736 decreases outward (Fig.~1.3, bottom-right).\footnote{This
result depends on the assumption that the gas is on circular orbits everywhere in the galaxy and that asymmetries in the velocity
field and in the galaxy image (Fig.~1.3, bottom-left) are due to warps. These assumptions are not valid.  NGC 4736 has
an oval disk (Bosma \etal 1977; Kormendy 1982b, 1993b; Kormendy \& Kennicutt 2004; Fig.~1.8 here).  A suitable analysis of the velocity 
field based on plausible axial ratios for the inner disk and outer ring has not been made.  However, it is likely that the true rotation
curve decreases outward more slowly than Fig.~1.3 suggests.  Nevertheless, disk secular evolution is more energetically favorable in 
NGC 4736 than it is in most galaxies.}  Under these circumstances, secular evolution should be especially important.  In fact, NGC 4736 
turns out to be our best ``poster child'' for secular evolution in an unbarred galaxy~(see~also~Fig.\ts1.6, Section 1.3.3., Fig.~1.8,
Fig.~1.28, Fig.~1.31, Fig.~1.35., Fig.~1.38, Table\ts1.3).

      It will turn out that angular momentum redistribution and therefore radial mass transport happens mostly to gas.  The
need for gas disfavors early-type galaxies and favors late-type galaxies.
  
      Putting all this together, we can already foresee what the observational review in the rest of this paper will 
reveal: {\it Secular evolution in galaxy~disks is, in the current Universe, happening mostly in intermediate-late-type
(e.{\ts}g.,~Sbc) galaxies.  These are the galaxies in which pseudobulges turn out to be most prominent.  Secular evolution 
is too slow to be important in the latest-type galaxies, because the mass distribution is too ``fluffy'', and so it is not
energetically favorable to transport angular momentum outward.  And secular processes no longer transport much gas in S0 and Sa 
galaxies, because they no longer contain much gas.  Nevertheless, (1) purely stellar secular processes are expected to happen in these 
galaxies, and (2) secular evolution is believed to have been important in the past, because many S0 galaxies are observed
to contain disky pseudobulges.}

      I turn next to a discussion of our motivation and where it will lead us.

\section{The study of galaxy evolution -- structural components}

      An important goal of extragalactic astronomy is to understand the structure of galaxies as encoded, e.{\ts}g., in the 
classification schemes of Hubble (1936), de Vaucouleurs (1959) and Sandage (1961, 1975).  An updated version of Hubble classification
is shown in Fig.~1.4, and a perpendicular cut through de Vaucouleurs' classification volume at stage Sb is shown in Fig.~1.5.
For each kind of galaxy, we want to understand the present structure and the formation and evolution processes which created
that structure.

      Our picture of galaxy formation by hierarchical clustering and our School's subject of secular evolution are
both important in creating the structures in Figs.~1.4 and 1.5.  Ellipticals and classical (elliptical-galaxy-like) bulges 
of disk galaxies form by hierarchical clustering and major mergers.~I review this in Section 1.8, and Isaac Shlosman 
discusses it in his lectures.  Bulges are usually classical in S0\ts--{\ts}Sb galaxies.~Secularly~built
pseudobulges are not distinguished in Hubble classification, but they are already important in some S0s,
and they become the usual central component at type Sbc and in later types (Kormendy \& Kennicutt 2004).  Unlike these permanent components,
spiral density waves are temporary features of disks that mostly last only as long as they have driving engines (Toomre 1977b, 1981).

\vfill

\setcounter{figure}{3}

\begin{figure}[h]




\includegraphics{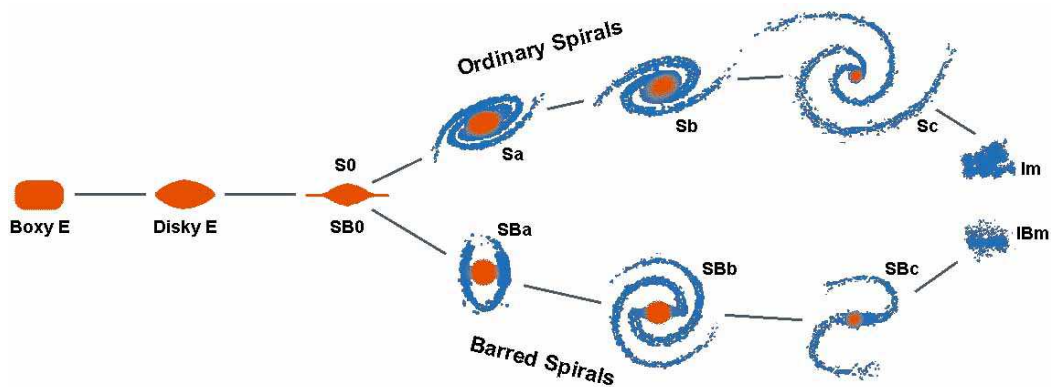}

\caption{Revised Hubble classification (Kormendy \& Bender 1996) recognizing the difference (the ``E{\ts}--{\ts}E dichotomy'') 
between boxy, slowly rotating ellipticals that have cores and disky-distorted, more rapidly rotating and coreless ellipticals that are 
continuous in their properties with S0s.  Ellipticals are understood to form via rapid processes; i.{\ts}e., by major mergers 
and violent relaxation.  These are reviewed in Section 1.8, where I discuss galaxy formation by hierarchical clustering.  
S0 and spheroidal galaxies (not included in Hubble types) 
form~--~I~suggest~--~at least mostly by environmental secular evolution (Section 1.7).  Spiral galaxies evolve partly by rapid processes 
(mergers make classical bulges, shown orange~in~the~figure).  However, the evolution of many spirals has been dominated by 
secular processes, including the formation of pseudobulges (also shown in orange, because Hubble classes do not distinguish between 
classical and pseudo bulges). 
}

\end{figure}

\cl{\null}
\vskip -28pt
\eject

\subsection{Inner rings ``(r)'' and outer rings ``(R)''}

      Figure 1.4 omits several regular features in galaxy structure that we need to understand.  
The most important of these are encoded in de Vaucouleurs' (1959) more detailed classification,
illustrated here in Fig.~1.5.  Rotate the ``tuning fork'' diagram in Fig.~1.4 about the E\ts--{\ts}S0 axis.  
In the plane of the page, de Vaucouleurs retains the distinction between unbarred (``SA'') and barred (``SB'') galaxies. 
The new, third dimension above and below the page is used for the distinction between galaxies that do or 
do not contain inner rings.  Fig.~1.5 is a perpendicular cut through this classification~volume.

\vfill

\begin{figure}[hb]



\includegraphics{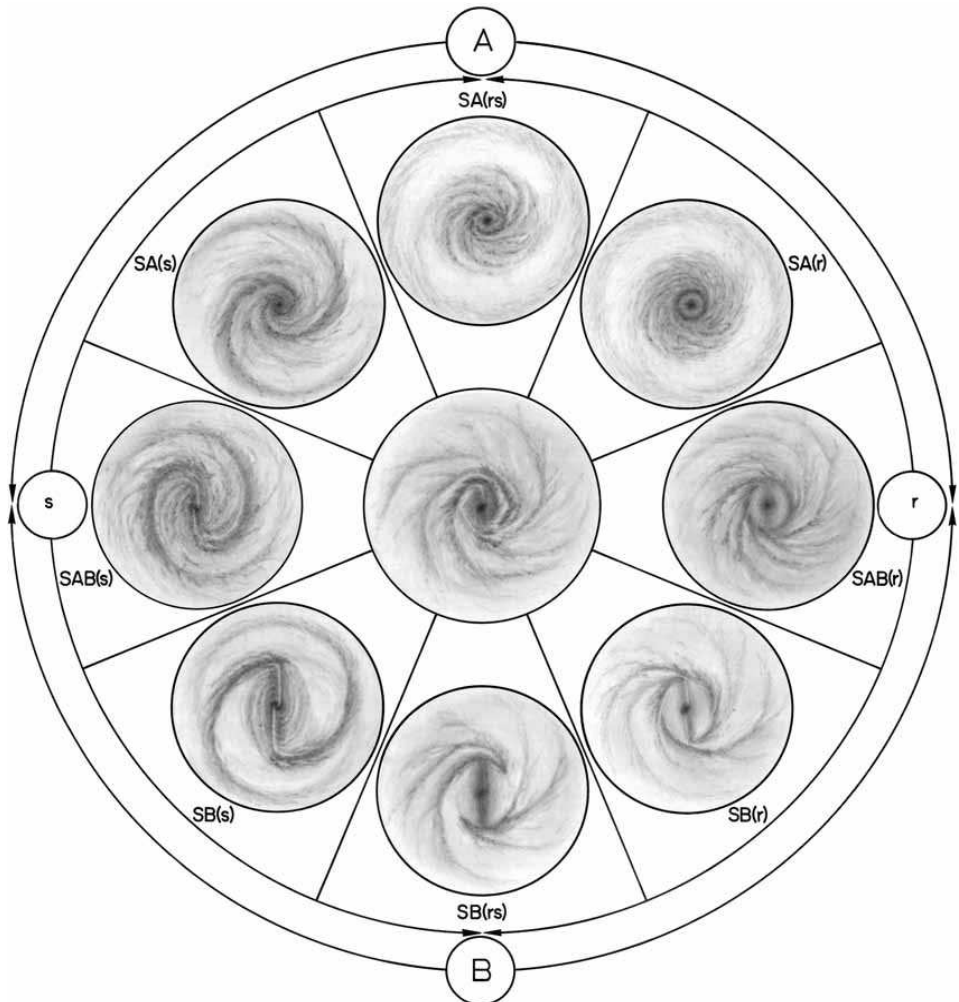}

\caption{Perpendicular cut at type Sb through the galaxy classification volume as defined by
de Vaucouleurs (1959).
I am grateful to Ron Buta for this version from the {\it de Vaucouleurs Atlas of Galaxies} (Buta \etal 2007).
}

\end{figure}

\eject

      Figure\ts1.6 (top) illustrates the distinction (bottom half of Fig.~1.5) between
SB(s) galaxies in which two global spiral arms begin at the ends of the bar and SB(r) galaxies
in which there is a ring of stars and gas surrounding the ends of the bar and the spiral arms start
somewhere on this ring.  In Figures 1.5 and 1.6, there is a prominent, nearly straight
dust lane on the rotationally leading side of the bar in SB(s) galaxies.  Notwithstanding Fig.~1.5,
there is generally no such dust lane in SB(r)~bars~(Sandage~1975).  Note in NGC 2523 that the inner
ring is patchy and blue and contains young stars, like the outer disk but unlike the relatively ``red
and dead'' bar and (pseudo)bulge.  This is a common property of inner rings.

\vfill

\begin{figure}[hb]


\includegraphics{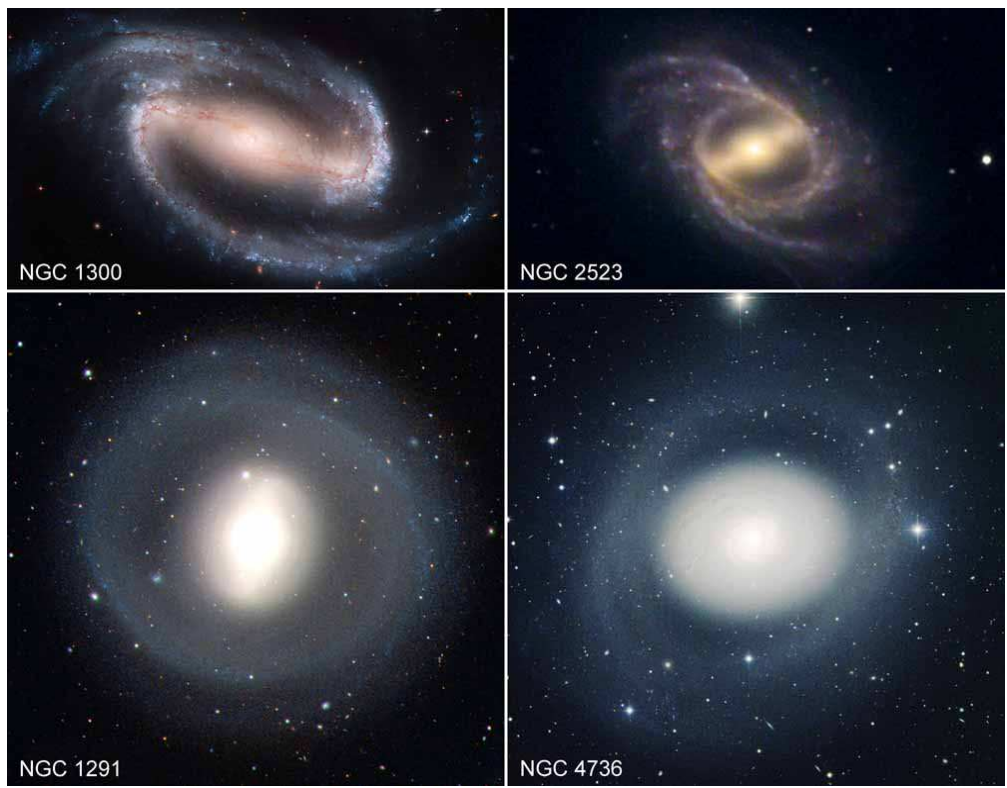}

\caption{(top) Contrast SB(s) and SB(r) structure: (left) SB(s)bc galaxy NGC~1300 has two global spiral 
arms that begin at the ends of the bar.  (right) The SB(r)bc galaxy NGC 2523 has a somewhat more
multi-armed but still global spiral pattern with arms that begin from the inner ring, not necessarily at the end of
the bar.
(bottom) Galaxies with outer rings: (left) the (R)SB(lens)0/a galaxy NGC 1291 and (right) the (R)SA(lens)ab galaxy 
NGC 4736.  The latter is a prototypical oval galaxy.  It contains a pseudobulge that is identified by no less than
five classification criteria from Section 1.5.3.
NGC 2523 is a Hubble Heritage image; NGC 2523 is from Adam Block (NOAO/AURA/NSF); NGC 1291 and NGC 4736 are from Kormendy 
\& Kennicutt (2004).
}

\end{figure}

\cl{\null}
\vskip -28pt
\eject

      Figure 1.6 (bottom) shows another feature that is recognized by de Vaucouleurs and Sandage although 
it is not given a separate classification~bin in diagrams like Fig.~1.5.  Both barred and unbarred galaxies 
can contain outer rings ``(R)'' at approximately twice the radius of the inner disk.  

      The properties of inner and outer rings are discussed (for example) in
Kormendy (1979b, 1981, 1982b),
Buta (1990, 1995, 2011, 2012), 
Buta \& Crocker (1991),
Buta \& Combes (1996) and
Buta \etal (2007).
I particularly use the following.
     \begin{enumerate}[(a)]\listsize
     \renewcommand{\theenumi}{(\alph{enumi})}
\item{Inner rings essentially always encircle the end of the bar.}
\item{Outer rings typically have radii $\sim 1.2$ times the radius of the bar.
      There is no overlap in the distribution of radii of inner and outer rings.}
\item{Both inner and outer rings typically contain H{\ts}{\sc i} gas and star formation.}
\item{Inner rings typically have intrinsic axial ratios of $\sim 0.85 \pm 0.10$ with the long
      axis parallel to the bar (Buta 1995).}
\item{Outer rings typically have intrinsic axial ratios of $\sim 0.87 \pm 0.14$.  In most galaxies,
       the long axis is perpendicular to the bar (as seen face-on), but a few outer rings are 
       elongated parallel to the bar (Kormendy 1979b; Buta 1995).}
\item{Figure 1.5 contains one mistake.  There is no continuity between the inner rings in unbarred and barred
      galaxies.  Instead, SA inner rings are similar to nuclear star-forming rings in SB galaxies (see
      Fig.~1.31 in Section 1.5.1).
      Nuclear rings are always much smaller than inner rings (Comer\'on \etal 2010).}
     \end{enumerate}
All of the above structural properties are natural products of the secular evolution that I review
in these lectures. 

\subsection{Lens Components ``(lens)''}

      Lenses are the final morphological component that we need to recognize observationally and to understand within our 
picture of galaxy evolution.  There is much confusion in the literature about lenses.  It is unfortunate that de Vaucouleurs 
chose to use the name ``lenticular'' for S0 galaxies.  I empathize: a bulge-dominated, edge-on S0 such as NGC~3115~or~NGC~5866 
is indeed shaped rather like an edge-on glass lens.  But this is due to the combination of a thick bulge and a thin disk; 
it does not happen because these S0s contain lens components.  In fact, most unbarred S0s do not 
contain lens components, whereas some Sa and Sb galaxies -- particularly barred ones -- do contain lens components.  We need 
to overcome the confusion that results from use of the misleading name ``lenticular galaxy''.  In all of my papers including 
this one, I avoid use of the name ``lenticular galaxy'' and always call these objects ``S0 galaxies''.  And I try to identify
lens components strictly using the following definition.

\vs

\begin{figure*}

\vskip 2.4truein



\includegraphics{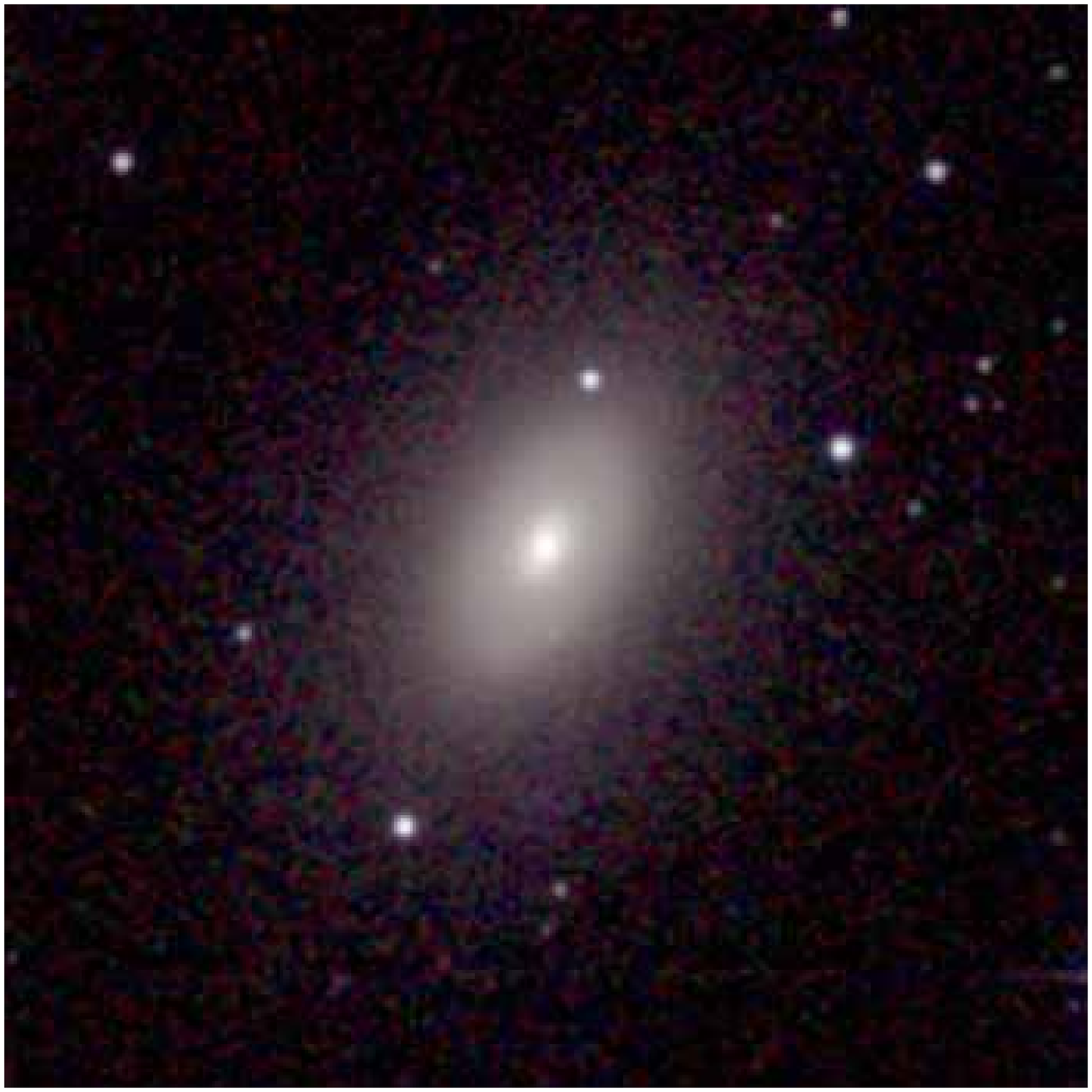}

\includegraphics{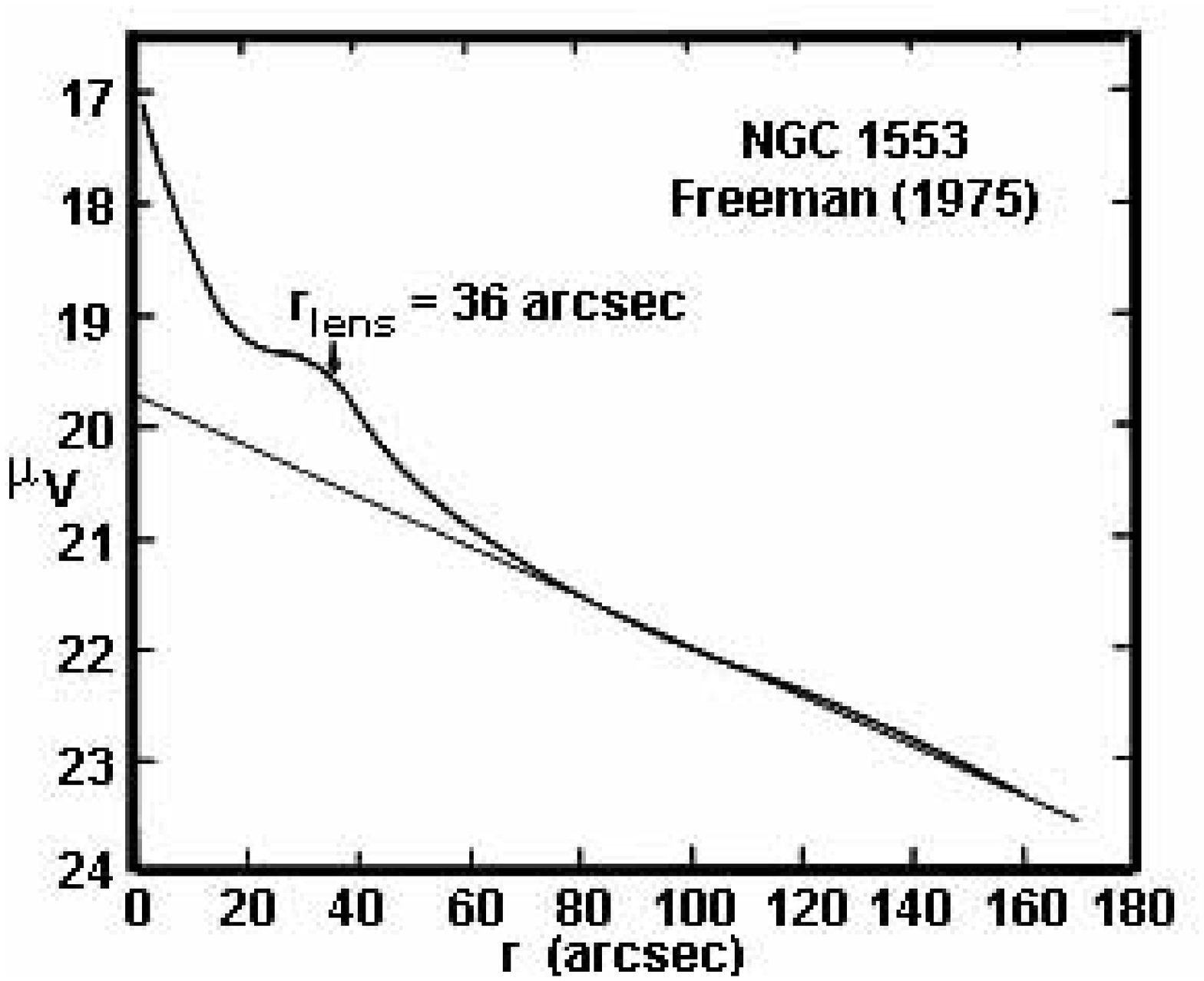}

\caption{(left) Color $JHK$ image of the SA(lens)0 galaxy NGC 1553 from the 2MASS Large Galaxy Atlas (Jarrett \etal 2003;
image from NED).  The lens is the elliptical shelf in the brightness distribution interior to the outer, exponential
disk.  It is best seen in the $V$-band, major-axis brightness profile shown in the right panel (Freeman 1975).
The straight line is an exponential fit to the disk profile.
}

\end{figure*}

      A lens component is an elliptical ``shelf'' in the brightness distribution of a disk galaxy.  That is, it has a shallow 
brightness gradient interior~to~a sharp outer edge.  Typical face-on lenses have axial ratios of $\sim$\ts0.85\ts--\ts0.9. 
Lenses occur most commonly in barred galaxies; then the bar almost always fills the lens in one dimension.  NGC 1291 in
Fig.~1.6 is an (R)SB(lens)0/a example of this configuration.  Lenses can also occur in unbarred galaxies; the prototypical 
example is the SA(lens)0 galaxy NGC 1553 (Fig.~1.7).  It is important to note that there is no overlap in the intrinsic
ellipticity distributions of bars and lenses.~Bars always have much smaller (pole-on) axial ratios of \lapprox \ts$1/4$.
So bars and lenses are clearly different, even though both occur only in disk galaxies.  A lens and an inner ring of the
same size can occur together (e.{\ts}g., NGC 3081).  The properties of lenses
are established in
Kormendy (1979b) and in
Buta (2011, 2012);
Buta \etal (1996, 2007).

      I will argue in Sections 1.4.3.4{\ts}--{\ts}1.4.3.6 that lenses are defunct bars.  That is, following 
Kormendy (1979b, 1981, 1982b), I suggest that bars evolve away into lenses when the secular evolution that
they drive increases the central mass concentration so much that the elongated bar orbits can no longer 
precess together at the same bar pattern speed $\Omega_p$.

      I use the notation SA(lens) and SB(lens), respectively, for unbarred and barred galaxies with lenses.
Buta \etal (2007) uses the letter ``el''; I do not do this only because it is difficult to distinguish
``el'' from the number~1.

\vfill\eject

      It will turn out that lenses, outer rings, inner rings and pseudobulges are~like~the~more 
familiar classical bulges, bars and disks in the sense that they are distinct components in the structure of galaxies that 
retain their essential character over many galaxy rotations.  They are different from spiral arms, which are
density waves in the disk that last -- more or less -- only as long as the ``engine'' that 
drives them.  Still,
there are many ways in which the above components can be combined into a galaxy.  This
results in de Vaucouleurs classifications that may be as complicated as (R)SB(r)bc
or (R)SB(lens)0/a (NGC 1291 in Fig.~1.6).  This complication makes many people
uneasy.  However, I emphasize: {\it Every letter in such a complicated galaxy classification
now has a physical explanation within our picture of secular galaxy evolution.}  It is a
resounding testament to the educated ``eyes'' of classifiers such as Allan Sandage and
Gerard de Vaucouleurs that they could pick out regularities in galaxy structure that 
ultimately were understandable in terms of galaxy physics.

\subsection{Oval Disks}

      Fundamental to our understanding of the engines that drive secular evolution is the observation that many 
unbarred galaxies nevertheless are not globally axisymmetric.  The inner disks of these galaxies typically
have intrinsic axial ratios of $\sim$\ts0.85.  Bars are more elongated (typical axial ratio $\sim$\ts0.2),
but a much smaller fraction (typically \lapprox \ts1/3) of the disk participates.  Observations indicate that
oval disks are approximately as effective in driving secular evolution as are bars.

      Oval galaxies can be recognized independently by photometric criteria
(Kormendy \& Norman 1979;
Kormendy 1982b) 
and by kinematic criteria
(Bosma 1981a, b).  They are illustrated schematically and with observations of prototypical galaxies in
Fig.~1.8.  The photometric criteria are:
     \begin{enumerate}[(a)]\listsize
     \renewcommand{\theenumi}{(\alph{enumi})}
\item{The disk consists of two nested ovals, each with a shallow brightness gradient interior to a sharp outer edge.}
\item{The outer oval has a much lower surface brightness than the inner oval.  It can consist of a featureless disk
      (e.{\ts}g., in an S0 galaxy) or spiral arms that make a pseudo ($\equiv$ not quite closed) outer ring or a
      complete outer ring.}
\item{The nested ovals generally have different axial ratios and position angles, so they must be oval if
      they are coplanar.  But the flatness of edge-on galaxies at these fairly high surface brightnesses shows that
      these disks are oval.}
     \end{enumerate}
NGC 4736 and NGC 4151 are outer ring and outer spiral arm versions of oval galaxies (Fig.~1.8).  I use the notation
SA(oval) or SB(oval) -- exactly analogous to SA(lens) or SB(lens) -- for galaxies with oval disks.

\vfill\eject

\centerline{\null}

\vfill

\begin{figure}[ht]

\vskip 1truein




\includegraphics{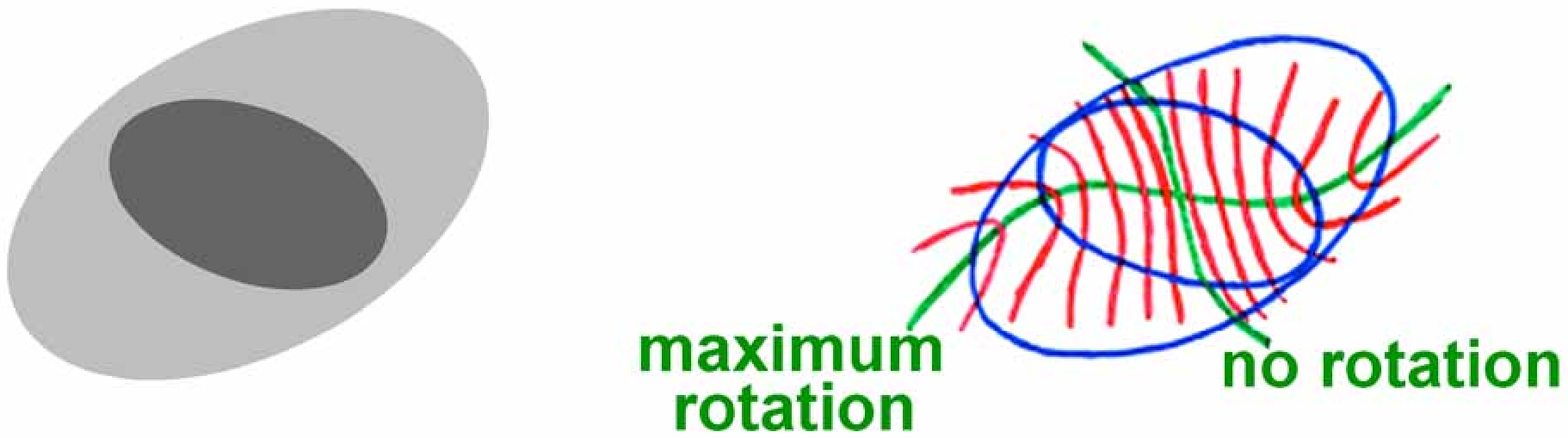}

\includegraphics{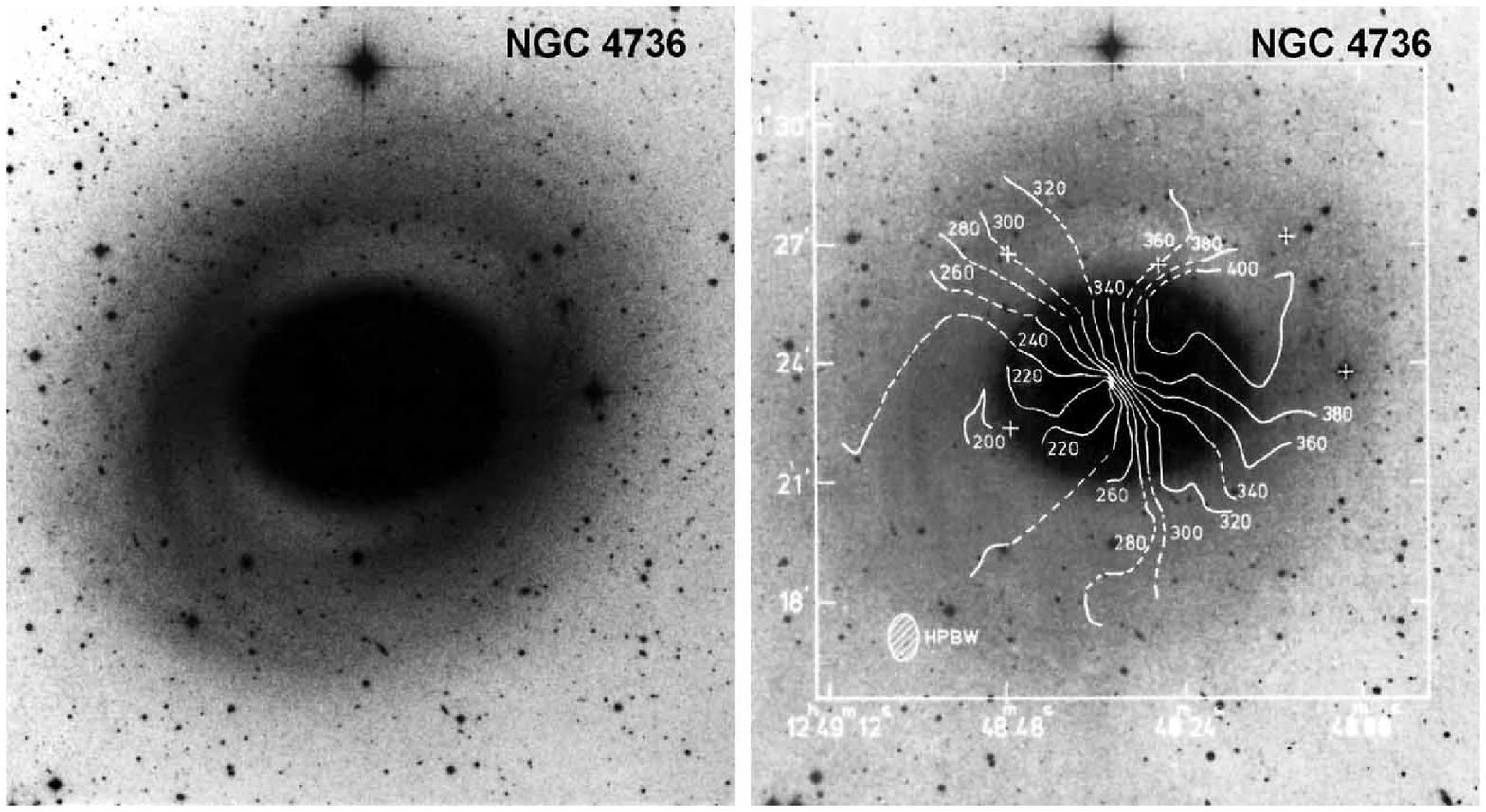}

\includegraphics{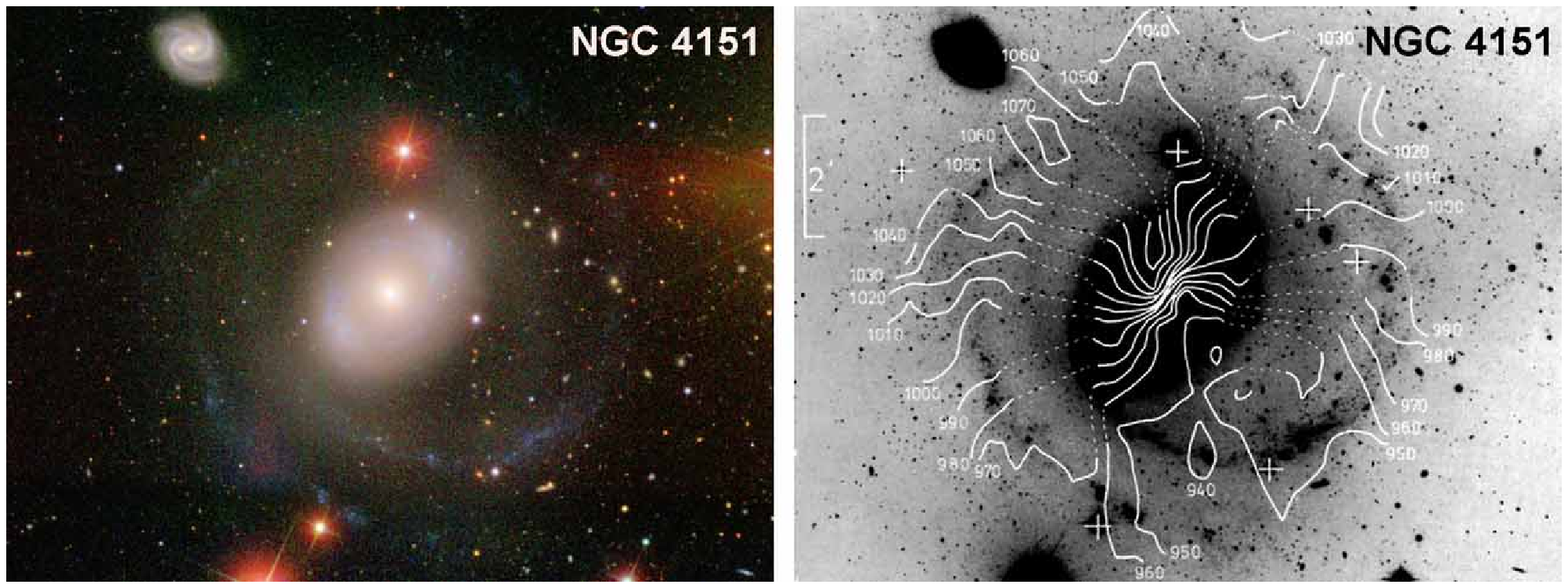}

\caption{Criteria for recognizing unbarred oval galaxies shown schematically (top) and with observations
of NGC 4736 and NGC 4151 (adapted from Fig.~9 in Kormendy \& Kennicutt 2004).  The image of NGC 4736 is from
my IIIa-J plate taken with the 1.9 m Palomar Schmidt telescope.  The image of NGC 4151 is from 
{\tt http://www.wikisky.org}.
The H{\ts}{\sc i} velocity contours of NGC 4736 and NGC 4151 are from
Bosma \etal (1977a)
and from
Bosma \etal (1977b), respectively.
}

\end{figure}

\eject

The kinematic criteria for recognizing ovals are that the velocity field is symmetric and regular, but:
     \begin{enumerate}[(a)]\listsize
     \renewcommand{\theenumi}{(\alph{enumi})}
\item{the kinematic major axis twists with radius;}
\item{the optical and kinematic major axes are different; and}
\item{the kinematic major and minor axes are not perpendicular.}
     \end{enumerate}
Here observation (a) is equally consistent with a warp.  But (b) and (c) point uniquely to an
oval velocity field that is seen at a skew orientation.  Given this conclusion, observation (a) then means that the
intrinsic orientations of the outer and inner ovals are different.  Usually, the outer oval is elongated perpendicular 
to the inner one, as seen face-on.  My favorite poster child for secular evolution, NGC 4736, shows all of these
effects.~The~classical~Seyfert\ts1 galaxy NGC 4151 shows them even more clearly (Fig.~1.8).

      Figure 1.9 shows four more oval galaxies.  I use it to emphasize four points.
\vskip -18pt
\phantom{Invisible text to improve line spacing.}
     \begin{enumerate}[(a)]\listsize
     \renewcommand{\theenumi}{(\alph{enumi})}
\item[(1)]{Oval galaxies are 
very recognizable and very common.~It is curious that they have not better penetrated the astronomical folklore.
Kormendy \& Kennicutt (2004) show that they are important engines for secular evolution.}
\end{enumerate}
\vfill

\begin{figure}[hb]


 \includegraphics{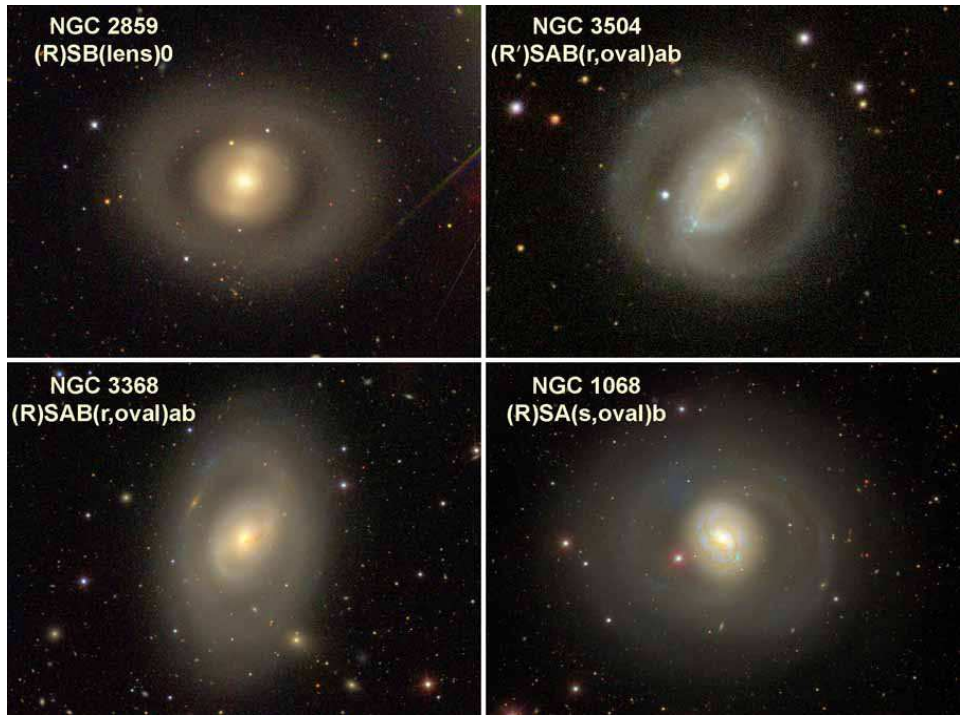}

\caption{Sequence of outer-ring galaxies that are structurally similar except that a bar embedded in the lens or 
inner oval structure ranges from strong at upper-left to invisible at lower-right.  All images are from 
{\tt http://www.wikisky.org}.
}

\end{figure}

\eject

     \begin{enumerate}[(a)]\listsize
     \renewcommand{\theenumi}{(\alph{enumi})}
\item[(2)]{There is a complete structural continuity in outer ring galaxies from (R)SB objects that have strong bars 
to (R)SA objects that are completely barless.\footnote{The (R)SA(oval)ab galaxy NGC 4736 
(Figs.~1.3, 
1.6, 
1.8,
1.28,
1.31, 
1.35, 
1.38)
contains a nuclear bar but no main bar.  The nuclear bar provides a ``connection'' to barred galaxies that is relevant
in Section 1.5.2.8.  But it is much too small to affect the overall evolution of the galaxy.  NGC 4736 is therefore
completely unbarred for the purposes of the present discussion.  Similarly, NGC 1068 (Fig.~1.9) also has a nuclear 
bar but no main bar.  I suggest that the connection between these two nuclear bars and their host oval disks is not
an accident -- that, in fact, oval disks are the late-type analogs of lenses and thus were once barred (see point 3).  
The hint is that some nuclear bars can survive even after the main bar has evolved away.}  NGC 2859 is one of the best 
examples of an (R)SB(lens)0 galaxy in which the bar fills the lens in one dimension.  Another excellent example is 
NGC 3945
(Kormendy 1979b; 
Fig.~1.17 and
Fig.~1.24 here).  NGC 3504 and NGC 3368 are two examples among many of galaxies that have
well defined oval disks and extremely weak bars.  I have typed them as (r,oval) and (s,oval) because there is a
bright blue rim of star formation around the outside of the inner oval, exactly as in many (r,lens) galaxies.
I am not aware of any measurements of the bar strength that take into account the oval structure.  But the bars
that are revealed by the 2MASS survey at $K^\prime$ band must be only a few percent of the disk light.
NGC 1068 is classified as unbarred by Sandage (1961) and by Buta \etal (2007); there is a hint of a
bar in the 2MASS image (Jarrett \etal 2003).  NGC 4736 is a purely unbarred, oval galaxy.$\dagger$}

\item[(3)]{Albert Bosma has advocated for many years that (lens) and (oval) components are the early-Hubble-type 
and later-Hubble-type versions of the same kind of structure.  For many years, I was unsure about whether or
not to believe this.  I am now convinced that Albert is exactly correct.  The reasons are discussed in
Sections 1.4.3.4{\ts}--{\ts}1.4.3.6.
Here, I juxtapose (R)SB(lens) and (R)SAB(oval) galaxies in Fig.~1.9 to emphasize their similarity.  I will 
continue to use this notation throughout this review, but I want to make it clear that \hbox{\it I believe that (lens) 
and (oval) structures are fundamentally the same.}}

\item[(4)]{There also is a continuity from complete outer rings (R) such as the one in NGC 2859 through spiral arms that
are distorted until they almost closed to form what is called a ``pseudoring'' (R$^\prime$) as in NGC 3504 through
to spiral arms that are less -- but  similarly -- distorted as in NGC 4151 (Fig.~1.8 here) and NGC 1300 
(compare Fig.~1.6 here with Fig.~13 in Kormendy 1979b).  See de Vaucouleurs (1959); de Vaucouleurs \etal (1991:~RC3)
and Buta (2011, 2012) for further discussion.}
\end{enumerate}
\phantom{gronk}
\vskip -13pt

      All of these continuities prove to have important physical interpretations within our developing picture
of disk-galaxy secular evolution.  Most of them are reviewed in the following sections.  The stellar dynamics of 
(R)S(r) galaxies are discussed in Lia Athanassoula's (2012) third lecture.

      Further examples of the above structures are shown in Ron Buta's (2012) 
lectures and in the {\it de Vaucouleurs Atlas of Galaxies\/} (Buta \etal 2007).

\vfill\eject

\subsection{Classical versus physical morphology of galaxies}

      At the start of a new science, it is common to classify the objects under study into ``natural groups'' 
(Morgan 1951) whose members share observed characteristics that are judged to be important.  The success of the 
taxonomy depends on how well the natural groups -- the classification bins -- prove to correlate~with~physics.  
The galaxy classification scheme of Hubble (1936) and Sandage (1961) as augmented by de Vaucouleurs (1959) won
a Darwinian struggle between taxonomic systems because it succeeded best in ordering galaxies by properties that 
helped us to understand galaxy formation and evolution.  Ron Buta (2012) reviews this subject in his lectures
at this School. 

      It is important -- and fundamental to the aims of this paper -- to contrast morphological galaxy classification as 
practiced in the early days of extragalactic research with what becomes necessary as the subject matures.  Sandage and Bedke (1994) 
emphasize the importance early on of not being led astray by preconceptions: 
``The extreme empiricist claims that no whiff of theory may be allowed into the {\it initial} classification procedures, decisions, 
and actions.''   Nevertheless, every classifier chooses which features to consider as important and which to view as secondary. 
Hubble succeeded because he chose properties that became important to our understanding.  Sandage recognizes this: 
 ``Hubble correctly guessed that the presence or absence of a disk, the openness of the spiral-arm pattern, and the degree of 
resolution of the arms into stars, would be highly relevant."  Hubble based his classification on choices
made with future interpretation in mind. \vsss

      In contrast (Kormendy 2004a): ``At the level of detail that we nowadays try to understand, the time has passed when we can 
make effective progress by defining morphological bins with no guidance from a theory.  [Breaking] down the wall between morphology 
and interpretation successfully has always been a sign of the maturity of a subject.  For example, without guidance from a theory, 
how would one ever conceive of the complicated measurements required to see solar oscillations or use them to study the interior 
structure of the Sun?  In the same way, we need the guidance of a theory to make sense of the bewildering variety of phenomena 
associated with galaxies and to recognize what is fundamental and what is not.  This motivates a `physical morphology' [Kormendy 1979a,{\ts}b,
1981, 1982b; Kormendy \& Kennicutt 2004], not as a replacement for classical morphology -- which remains vitally important~-- but as 
a step beyond it.  Physical morphology is an iteration in detail that is analogous to de Vaucouleurs's iteration beyond the 
Hubble tuning fork diagram.''

      One purpose of this review is to bring this process up to date.  Our aim is to engineer the best one-to-one correspondence that 
we currently can between recognized types of galaxies or galactic building blocks and physical processes of galaxy formation. 

      It is reasonable to expect that an improved understanding of galaxies will show that an initial classification
missed some physics.  Also, some features of galaxies could not, early on, be observed well enough to be included 
in the classification.   Two substantial additions and one small tweak to classical morphology are needed here.  The small
tweak is to recognize~that nuclear star-forming rings are distinct from and often occur together with inner rings (r)
in barred galaxies.  This was point (f) in Section 1.3.1.  The first major correction is to recognize and differentiate between 
real and counterfeit ellipticals (Section 1.7).   The second is to recognize and differentiate between real and counterfeit bulges.
The latter are  ``pseudobulges'' (Section 1.5). 

      An essential aspect of physical morphology is to treat galaxies as being composed of a small number of building blocks,
the distinct components in the mass distribution.  The problem that this solves and the benefits that it provides were described 
as long ago as Kormendy (1979a): 

      ``[Classical morphology] assigns a classification cell to each list of galaxy characteristics.  As we look at galaxies 
more and more closely, the list of observed characteristics becomes alarmingly large.  Thus the revised morphological types of 
de Vaucouleurs already require `about one hundred cells' (de Vaucouleurs 1963) and still do not recognize features such as lenses. 
It is difficult to attach dynamical significance to the bewildering variety of forms that the system describes.''

       In physical morphology, ``we aim to identify as distinct components groups of stars or gas whose structure, dynamics, and 
origin can profitably be thought of as distinct from the dynamics and origin of the rest of the galaxy.  \dots~~The large number 
of cells in classical morphology is now thought of as the many ways that components and their secondary behavior [such as spiral
structure] can be combined to make a galaxy.  The strength of this approach is twofold.  It breaks up the complicated problem of
galaxy structure into smaller and more manageable pieces to which it is easier to attach dynamical interpretations.  Secondly,
{\it investigations of correlations and interactions between components are very efficient in suggesting previously unrecognized
dynamical processes} [emphasis added].  Thus the component approach provides an efficient framework for studies of galaxy dynamics.''

      Throughout this review, my aim is to tweak the distinctions between galaxy components (and hence their names) until they 
are both correct descriptions and uniquely tied to physical processes of galaxy formation.

\section{A Heuristic introduction to bars and spiral structure}

      Bars are the most important drivers for secular evolution in galactic disks; we will see them again and again throughout 
this School.  For this reason, I now want to give you a heuristic introduction to bar dynamics and evolution. Lia Athanassoula 
(2012) will talk about this subject in more detail.  Here, I concentrate on a simple version of the essential
physics that you need to understand in order to have an intuitive feel for how bars form, grow and die.  Bars are density
waves in the disk.  That is, the bar pattern rotates more-or-less rigidly at a single pattern angular velocity $\Omega_p$,
whereas the angular velocity $\Omega(r)$ of material in the disk varies with radius.  This is even more true of spiral structure:
over most of a global spiral pattern, the stars and gas have $\Omega > \Omega_p$; they catch up to the spiral arms from behind, move 
through them in the rotation direction (lingering in the arms to give them their density enhancement) and then move on in the
rotation direction toward the next arm.  Since bars and spirals have some similarities, I briefly discuss spiral structure, too. 

\subsection{Orbital resonances in a galactic disk}

      Orbital resonances are the key to understanding bars and spirals.  The main resonances -- inner Lindblad resonance, corotation,
and outer Lindblad resonance -- will recur in many of the lectures at this school.  I therefore need to introduce them in some detail.

      The general orbit of a star in a galactic disk is an unclosed rosette, because the potential is not Keplerian 
(that is, the galaxy mass is distributed in radius and not all in one central point as in, for example, the Solar System).

      Figure 1.10 shows disk orbits as seen in a frame of reference that rotates clockwise (orange arrow) at the
pattern speed $\Omega_p$ of some coherent structure.  In this figure, I show a spiral arm (red), because spirals almost 
always trail in the rotation direction (that is, the arm is convex in the direction of rotation).  Thus it is easy to remember 
the rotation direction at a glance.~I work in the epicyclic approximation (Mihalas \& Routly 1968; Binney \& 
Tremaine 1987) in which radial excursions are small.  Then, at the radius where the stars and the pattern corotate, i.{\ts}e., 
where $\Omega_p = \Omega$ (green), a stellar orbit is a small elliptical epicycle around the mean radius.  
The motion around the ellipse is counterclockwise, because the forward velocity is higher when the star is closer to the center 
than average.  In the epicyclic approximation, the motion around the ellipse is simple harmonic.  Corotation is the most
important resonance in the galaxy, because the mean position of the star with respect to the global pattern never changes
as long as $\Omega_p$ is fixed.

\begin{figure*}

\vskip 5.5truein


 \includegraphics{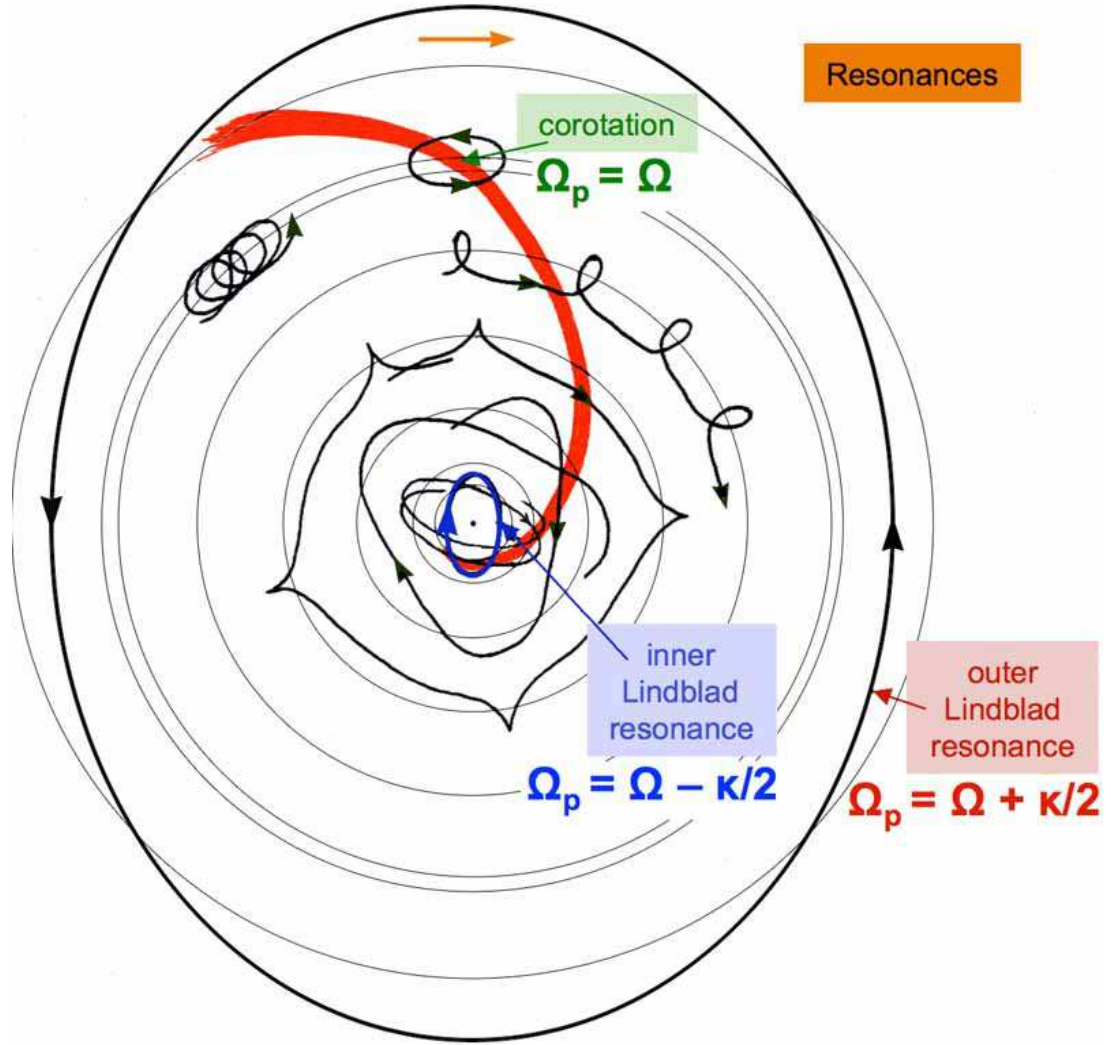}

\caption{Schematic illustration of galactic orbits as seen in a frame of reference that rotates clockwise
(orange arrow) at the pattern angular velocity $\Omega_p$ of some coherent structure such as a spiral
arm (red).  The pattern stays fixed as we view it, whereas stars drift with respect to the pattern 
forward (clockwise) inside corotation and backward (counterclockwise) outside corotation.
At the corotation resonance (green: $\Omega_p = \Omega$), a star with a small component of random
velocity added to $\Omega$ executes simple harmonic motion counterclockwise around a closed~ellipse (the
``epicyclic approximation'').  At the inner Lindblad resonance (blue: $\Omega_p = \Omega - \kappa/2$, where
$\kappa$ is the epicyclic frequency), the stellar orbit is closed and the star executes two radial
excursions for every revolution forward around the center.  At the outer Lindblad resonance (red: $\Omega_p = \Omega
+ \kappa/2$), the orbit is closed and the star executes two radial excursions for every revolution backward around
the center in the rotating frame.  At other radii, several of which are shown, the stellar orbits are unclosed.
}

\end{figure*}

\eject

      We will need the ``epicyclic frequency'' $\kappa$ of the small radial and azimuthal excursions around 
the mean motion,
$$
\kappa^2 = {{2V}\over{r}}\biggl({V \over r} + {{dV}\over{dr}}\biggr)~, \eqno{(1.3)}
$$
where $V(r)$ is the rotation velocity as a function of radius $r$.  

      If we move slightly inward from corotation toward the galactic center, then a star
oscillates around almost-closed ellipses while drifting forward with respect to the global pattern (leftmost
orbit illustrated in Fig.~1.10).

      Contining toward smaller radii (next three orbits inward in Fig.~1.10), the forward drift gets
faster until the backward loop disappears and we end up with an unclosed rosette orbit that involves somewhat
more than two radial excursions for every revolution forward in the rotation direction around the center.  It is 
clear that, as we move farther inward and the forward drift rate continues to increase, the orbit again becomes 
almost closed, now with almost exactly two radial excursions for every revolution (this is the second-innermost
orbit shown in the figure).

      Continuing toward still-smaller $r$, we arrive at ``inner Lindblad resonance'' (ILR) where the (blue) orbit 
is closed and the star executes exactly two radial excursions while it drifts forward with respect to the global
pattern by one complete revolution around the center.  This happens where $\Omega_p = \Omega - \kappa/2$.
ILR is again an important resonance, because the star repeatedly has the {\it same phase} in its radial oscillation
when it has the {\it same position} with respect to the global pattern.  So the star interacts more strongly
with the pattern near ILR than it does elsewhere in the disk.  Moreover, ILR gives us a first-order understanding of 
bar dynamics, as discussed in the next section.

      Starting at corotation and moving outward, stars drift backward with respect to the global pattern
as seen in our rotating frame.~At~some~large~$r$, the orbit is closed as the star makes
two radial oscillations for every revolution backward around the center (the orbit is not illustrated).  This is ``outer Lindblad resonance''
(OLR), and it happens where $\Omega_p = \Omega + \kappa/2$.

      The generic rotation curve $V(r)$ and frequencies $\Omega(r)$ and $\Omega(r) - \kappa(r)/2$ are
shown in Fig.~1.3.  Where $V \propto r$, $\kappa = 2\ts\Omega$ and $\Omega - \kappa/2 = 0$.  Where \hbox{$V =$ constant,}
$\kappa = \sqrt{2}\ts\Omega$ and $\Omega - \kappa/2 = (1 - 1/\sqrt{2})\ts\Omega$ decreases proportional to $\Omega$.
In between, $\Omega - \kappa/2$ first rises and then falls with increasing $r$ as the rotation curve turns downward from
$V \propto r$ to $V \simeq$ constant.  {\it The radial range over which $\Omega - \kappa/2$ varies little with radius
turns out to be crucial to the formation of bars and spiral arms.  The evolving height and radial extent of the $\Omega - \kappa/2$
maximum proves to control the fate of bars.  And the natural pattern speed for bars proves to be $\Omega_p \simeq
\Omega - \kappa/2$.}  As follows:

\subsection{Bars and spirals as almost-kinematic density waves}

      The essential insight into the dynamics of bars and global spirals is due to Lindblad (1956) and Kalnajs (1973; 
see Toomre 1977b for a lucid review).  It is illustrated in Fig.~1.11.  Especially for outward-rising rotation curves (M{\ts}33)
and even for nearly constant or falling rotation curves (M{\ts}81), the angular precession rate $\Omega - \kappa/2$ of
ILR orbits is small and nearly constant over much of the galaxy.~\hbox{Suppose that $\Omega$\ts$-$\ts$\kappa/2$ is
{\it exactly\/} constant~with~$r$.}  Then:~{\it If closed ILR orbits are arranged to produce a trailing spiral enhancement in density 
(top left) or a straight bar (top right), the orbits will precess together at exactly the same $\Omega - \kappa/2$ and so will
preserve the spiral or bar shape.  These are purely kinematic density waves.}

\begin{figure}[hb]

\vskip 3.7truein


 \includegraphics{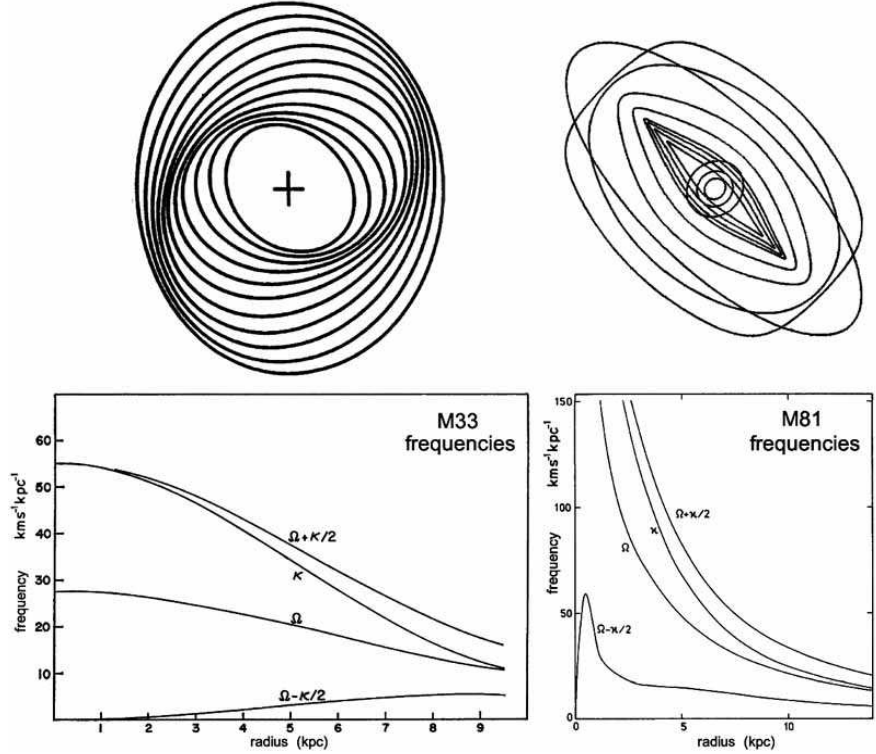}

\caption{(bottom) Frequencies $\Omega$, $\kappa$ and 
$\Omega \pm \kappa/2$ for almost-circular~orbits~in~M{\ts}33 (Shu \etal 1971) and M{\ts}81
(Visser 1980).  Both galaxies have large radial ranges in which $\Omega - \kappa/2 \simeq$
constant and ILR orbits precess almost together.
(top left) From Kalnajs (1973), elliptical ILR orbits aligned to produce a spiral density wave. 
\hbox{(top right)} From Englmaier \& Gerhard (1997), examples of principal orbit families for a bar 
oriented at $-45^{\circ}$.  The elongated orbits parallel to the bar are the $x_1$ family out of
which the bar is constructed.  If $\Omega - \kappa/2 =$ constant with radius over the extent 
of the $x_1$ orbits, they precess together and preserve the bar shape. 
Interior to ILR, the $x_2$ orbits are perpendicular to the bar. (The almost-square orbit has four 
radial oscillations for each revolution; it is a 4:1 ultraharmonic resonance orbit.)
}

\end{figure}

\eject

      In practice, $\Omega - \kappa/2$ is not quite constant.  To make real bars~or 
spirals, {\it it is the job of self-gravity to keep the orbits precessing exactly together.}

      This discussion is a heuristic argument for another important conclusion. {\it The natural
pattern speed of a bar or global spiral is $\Omega_p \simeq \Omega - \kappa/2$.}  Note in Fig.~1.11 that this
is much slower than $\Omega$ over most of the galaxy.  That is, stars and gas revolve around the galactic center
more quickly than the pattern rotates; they catch up to the pattern from behind, participate in it for a time --
lingering in the pattern and thereby enhancing its density (Toomre 1981) -- and then continue onward in the forward
rotation direction.  For bars, corotation happens 
near the end of the bar or slightly beyond it (see Sellwood \& Wilkinson 1993 for a review).

      Some barred galaxies that have substantial (pseudo)bulges also have nuclear bars $\sim$ one-tenth of the 
radius of the main bar (Section 1.5.2.8).  When the galaxy mass distribution is very centrally concentrated, $\Omega - \kappa/2$ 
has~a high maximum at small radii (frequency curves for M{\ts}81 in Fig.~1.11).  
{\it The natural pattern speed of a nuclear bar is roughly equal to the local value of $\Omega - \kappa/2$
and is therefore much higher than the pattern speed of the main~bar.}  This is seen in numerical simulations
(Debattista \& Shen 2007; Shen \& Debattista 2009) and in real galaxies (Corsini \etal 2003).
 
      The $x_1$ family of closed bar orbits shown in Fig.~1.11 and the similar, non-periodic orbits that are
trapped around them by the bar's self-gravity involve large perturbations from circular orbits.  Strong bars are nonlinear.
Then the approximations that we have made are not accurate.  In particular, the true pattern speed is rather larger than 
the local value of $\Omega - \kappa/2$ (e.{\ts}g., Shen \& Debattista 2009). 
Nevertheless, the epicyclic approximation still captures the essence of the physics (Ceverino \& Klypin 2007).

      In particular, for any realistic $\Omega_p \sim \Omega - \kappa/2$, there must be an inner Lindblad resonance 
in a centrally concentrated galaxy such as M{\ts}81, but there cannot be an ILR in a pure-disk galaxy such as M{\ts}33.
Inside ILR, the closed ``$x_2$ orbits'' are elongated perpendicular to the bar (Fig.~1.11).  So they cannot 
be used to construct the bar and to maintain its self-gravity.  We will see in Section 1.5 that the secular evolution of a barred galaxy 
is in part to transport gas toward the center and thereby to build a pseudobulge.  {\it As 
secular evolution increases the central concentration of a barred galaxy, the inner maximum in $\Omega - \kappa/2$ gets 
higher and wider~in~radius.  Therefore it gets more difficult for self-gravity to keep the decreasing number of $x_1$ orbits 
precessing together.~Also, the number of (damaging) $x_2$ orbits increases.~The bar gets weaker.  Thus, bars naturally commit suicide 
by the secular evolution that they drive.  This happens preferentially in early-type galaxies, not in galaxies such as M{\ts}33.}  
All of these heuristic predictions prove to be accurate.

\vfill\eject

\subsection{The growth, structural evolution and death of bars}

      This section is an overview of the life histories of bars -- from rapid growth through secular evolution in ``middle age''
through eventual death -- at the level that we need in this Winter School.  I concentrate in this section on the bars themselves;
the effects of bars on the other components in their galaxies will be the subject of later sections.  I will not cover the rich dynamics 
of bars in their nonlinear phases.  For this work, the best review is Sellwood \& Wilkinson (1993).~Contact between the complicated 
nonlinear dynamics that is reviewed there and the simplified story that I tell~here~is~not~always~good.  That's where the frontier 
in this subject lies.  Here, I will summarize the most important bar evolution processes based on a series of seminal papers. 
Sellwood \& Wilkinson (1993, hereafter SW93) provide more references, and Athanassoula (2012) reviews -- and, indeed, 
advances -- some of these subjects in her lectures at this School.  

      My orientation is a little different from that of SW93.  They take -- as much as possible,
in this complicated subject -- a formal approach rooted in the rigorous mathematics of galaxy dynamics.  I take an observer's
approach.  That is, I try to benefit as much as possible from the insight gained from analytical and $n$-body studies,
but I also put equal emphasis~on~letting~the observations suggest (or, in some cases, prove) which processes are at work.
Sometimes observations hint that a process (e.{\ts}g., evolution of bars~to~lenses) takes place even when the theory of 
such a process is not well formulated.  But asking the right question is the efficient road to progress.  So this observational 
guidance is very useful.  One key is to remember which ideas are hypotheses and which are proven.  And it is especially 
important not to overinterpret the temptingly rich array of possibilities that are inherent in this subject.  I will try to be clear 
about the confidence with which various ideas are suggested.  And I will emphasize what I believe are the most important 
remaining problems.

\subsubsection{Bar instability}

      Massive, cold, rotating disks are famously unstable to the formation of bars (e.{\ts}g., 
Ostriker \& Peebles 1973;
Toomre 1977b, 1981;
SW93).  
Many $n$-body studies start with a stellar disk that is highly unstable and then follow the evolution
of the bar that results.  Most results that are derived in this way are probably realistic, modulo (important!)~the neglect 
of gas and dark matter.  However, it is important to realize that bar evolution almost certainly does not start this way.  
It is unrealistic to think that a disk grows axisymmetrically to a high mass, meanwhile using up its gas to make stars, and 
only then discovers (``Oh, my God!'') that it is bar-unstable.

      In a paper entitled ``Most Real Bars are Not Made by the Bar Instability'', Sellwood (2000) expresses a related worry.
Most barred galaxies have enough central mass concentration so {\it the bar essentially must have an ILR.}  That~is,
$\Omega - \kappa/2$ is similar to the curve for M{\ts{81} in Fig.~1.11.  Our heuristic arguments in \S\ts1.4.2 then
make plausible the result that also emerges from more detailed studies (e.{\ts}g., Toomre 1981), namely that a bar instability 
never gets started when the resulting bar would have a strong ILR.  Sellwood acknowledges and dismisses the possibility that the 
disk might not originally have allowed an ILR, i.{\ts}e., that the central concentration was manufactured by the bar only after it formed. 
One piece of supporting evidence was a conclusion (Abraham \etal 1999) that strongly barred galaxies were rare at $z > 0.5$;
this is now known to be incorrect (Jogee \etal 2004).  More immediate was the concern that building up the central concentration
weakens~the~bar.  I will suggest below that bar suicide happens, and it happens at a measurable (pseudo)bulge-to-disk
ratio.  But I, too, have worried about why strong bars can coexist with strong ILRs.~The evidence will suggest that, when the central
mass grows slowly {\it while the bar has a strong angular momentum~sink,} a nonlinear bar can -- for a while -- continue to grow
even when it already has an ILR.  But the more important point for the present section is this: {\it present\/} disk conditions
do not favor the growth of bars by a simple instability. \vs

      Possible implications:

     \begin{enumerate}[(a)]\listsize
     \renewcommand{\theenumi}{(\alph{enumi})}
\item{Bar formation may be a threshold process that begins and then proceeds slowly and perhaps episodically
      (Sellwood 2000) as the disk grows massive enough with respect to its dark halo (see also Mihos \etal 1997).}
\item{The formation of the stars that now make up the bars in early-type galaxies happened long
      ago in progenitor galaxies about which we know very little.~~E.{\ts}g., the
      boxy bulge $\equiv$ almost-end-on bar of our Galaxy is made of stars that are very old and enhanced
      in $\alpha$ elements (see Renzini 1999 for a review).  
      So its stars formed over a period of \lapprox1~Gyr.
      Gas fractions were much higher then than they are now (Genzel~et~al.~2006).  
      Gas may be centrally important to bar formation.  Observations of high-redshift disks that are nevertheless
      younger than our Galaxy's bulge stars show surprises such as $\sim 10^8$-$M_\odot$ clumps
      and high velocity dispersions (Genzel \etal 2006; F\"orster~Schreiber~et~al.~2009).  The clumps are 
      believed to form by violent instabilities, and they eventually merge to form at least some
      classical bulges (Elmegreen \etal 2008; Section 1.8.1 here).  None of this sounds like the typical initial conditions
      assumed in $n$-body simulations of bar formation.  We have no reason to be confident that we
      know how to start those simulations with realistic initial conditions.}
\item{Contrariwise, a bar may form long after star formation makes the stellar disk.  One way is thought
      to be through tidal tickling by neighboring galaxies (e.{\ts}g.,
Noguchi 1987, 1988; 1996, which also discusses gas infall;
Gerin \etal 1990;
Barnes \etal 1991;
Mihos \etal 1997).
It is difficult to test this (for example) by looking for an excess of barred galaxies in dense environments such as the Coma cluster.  
The reason is that any extra tendency to form bars where there are many galaxy encounters must compete with and may lose out to the 
stabilizing effects of disk heating (Marinova \etal 2012; see Kormendy \& Bender 2012 for evidence that disk heating happens even in 
the Virgo cluster).}
\item{Occam's razor is a dangerous weapon.~Answers are not guaranteed~to~be~simple.  All the above -- i.{\ts}e.,~a series of episodic,
      encounter-driven growth spurts that punctuate a steady, slower growth by outward angular momentum transport as envisaged
      by Sellwood (2000) -- may happen.  {\it It is important to understand that global spiral structure connected with a bar or inner
      ring is a signature of outward angular momentum transport and therefore a sign that the bar is growing stronger.}}
     \end{enumerate}

      In the next sections, I review the life histories of bars and the evolution that they drive in their host galaxies,
keeping the above issues in mind.  As needed, I emphasize where simulations do not yet include important~physics.
Limitations include 
(1) that many simulations do not include gas, and gas physics is variously but always importantly simplified, 
(2) that continued infall of gas and small galaxies is not included, 
(3) that feedback from hot young stars, from supernovae and from active galactic nuclei is generally not included and may be important, 
    and especially 
(4) that simulated galaxies are usually not as multicomponent as real galaxies.  
Two-body relaxation is uninterestingly slow in galaxy disks.
Evolution is driven by the interactions of stars and gas with collective phenomena. 
Physical morphology (\S\ts1.3.4) invites us to investigate the interactions of different nonaxisymmetric components in galaxies.  
This is where the action is.

\subsubsection{$N$-body simulations:\\Rapid bar growth via instability\\followed by secular growth via angular momentum transport}

      Among the many published $n$-body studies of bar formation (SW93), I focus on Sparke \& Sellwood (1987), because they analyze 
their simulation in ways that tell us what we need to know.  The initial condition~is~an~axisymmetric galaxy with a bulge-to-total 
ratio of 0.3.~The bulge is a rigid Plummer\ts(1911) sphere with size $\sim$\ts$1/5$ that of the disk.  A rigid potential is a 
limitation:~the bulge gives the galaxy an ILR, and it is interesting to see the (surprising lack of) consequences, but the bar and
bulge~cannot~interact.  The disk is a Kuz'min (1956)\ts--{\ts}Toomre (1963) model, density $\propto 1/(a^2 + r^2)^{3/2}$, where $r$ 
is radius and $a$ measures the scale length of the density falloff.  With a Toomre (1964) {\it axisymmetric\/} stability parameter of 
$Q = 1$, the initial disk was very unstable.  The disk contained 50,{\kern 0.4pt}000 particles whose gravity was softened on a length 
scale of $0.1{\ts}a$.  Other limitations are that the model contains no gas and no ``live'' dark matter halo of simulated particles; 
both of these limit the extent to which the bar sees a sink for angular momentum that allows it to grow.  More detailed simulations 
can now be run, but this subject does not get the attention that it deserves, and in particular, analyses are not usually as physically
motivated and compelling as the one in Sparke \& Sellwood (1987).  There is much to be learned from this simulation, as follows.

      Figure 1.12 shows the evolution of the disk density, and Fig.~1.13 shows the growth in amplitude of the bar and
the evolution of the pattern speed of the bar and its associated spiral structure.  Time is measured in natural units 
(discussed in Sparke \& Sellwood 1987).  Beginning immediately and extending to time 50, the disk rapidly grows a two-armed
instability that is bar-shaped near the center and a global, two-armed~spiral~farther~out.  The bar rapidly grows stronger:~more stars 
participate, the axial ratio of the bar decreases, the bar grows longer and its pattern speed decreases.  
Figure 1.13 dramatically reveals the important result that {\it the pattern speed of the bar is larger than the pattern speed 
of the spiral at all times.}  This is consistent with our heuristic argument: the bulge creates a near-central maximum in
$\Omega - \kappa/2$, and the pattern speeds of the bar at relatively small radii and the spiral at relatively larger radii are
comparable to \hbox{but slightly larger than the local value~of~$\Omega - \kappa/2$.}~(Note:~in Fig.~1.13, frequency\ts$=$\ts$2\Omega_p$,
because the pattern is bisymmetric.)  {\it Because $\Omega_p$ is smaller for 
the spiral than for the bar, the spiral continually shears away from the bar.~It is no surprise that spiral arms do not
necessarily start at the end of the bar, especially in galaxies such as NGC 2523, which have well developed inner rings and which
therefore are dynamically mature.}

      Figure 1.13 makes it clear that the rapid, instability-driven initial growth of the bar is followed by a distinct later
phase of secular growth, when the bar amplitude increases and $\Omega_p$ decreases more slowly than in the early, rapid phase.
In this collisionless model, the outer disk has been heated so much that the spiral structure is weak during the secular phase 
(Fig.~1.12).  In real galaxies, there are two reasons why the later growth is more dramatic.  First, galaxies contain gas; 
(1) its response is stronger than that of the stars and, in fact, dissipative, and (2) gas also keeps the stellar disk cooler 
and more responsive, because the stars that form from it are formed with low velocity dispersions.  Second, the dark halo in a
real galaxy provides an additional sink for the disk angular momentum.  

\vfill\eject

      The bar ends roughly at corotation.  Therefore, stars stream through the bar density wave in the forward direction.
Analysis of the orbital structure of the late-stage model shows that -- as expected -- the bar is made almost exclusively of 
orbits trapped around the $x_1$ family of closed orbits that align with the bar.  Despite the presence of an IRL,, there are
almost no orbits analogous to the $x_2$ family (the model has only 50,{\kern 0.4pt}000 particles, and not all orbits were checked.)
Additional details of the orbit structure are discussed in Sparke \& Sellwood (1987).

\begin{figure}[hb]

\vskip 5.3truein


 \includegraphics{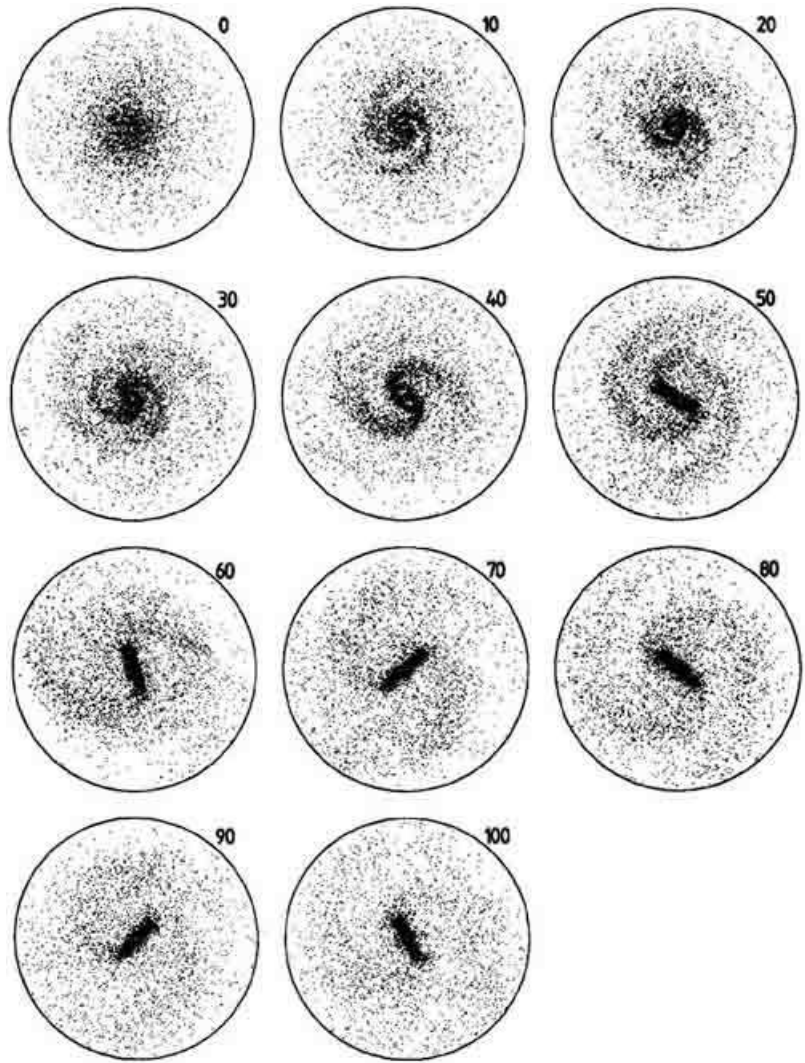}

\caption{Bar formation in the $n$-body simulation of Sparke \& Sellwood (1987, Fig.~1).  Times are in natural
units that are discussed in their paper.  Only one particle in ten is included, and the bulge component is not shown.
}

\end{figure}

\eject

\cl{\null} \vfill

\begin{figure}[hb]


\vskip 10pt


 \includegraphics{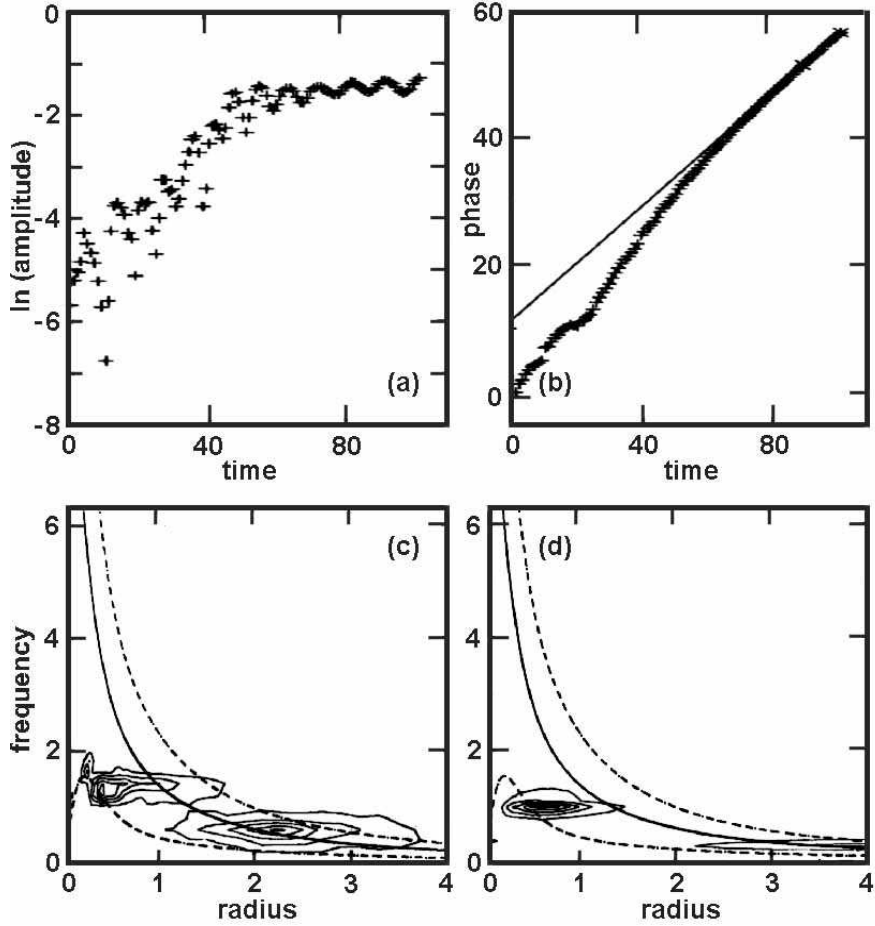}

\caption{Evolution of the bar in the $n$-body model of Sparke \& Sellwood (1987).  Panels (a) and (b) show the
amplitude and phase of the bar as a function of time.  A straight line is fitted to the phases at late times.
It shows that the pattern speed decreases quickly during the early, rapid growth of the bar and then more slowly
during the later, secular growth.  The bottom panels show contours of the square root of the power in the bar
or spiral perturbation as a function of frequency $=$ $2\Omega_p$ and radius.  Panel (c) is for the early part of the rapid growth
(up to time 31) and panel (d) is for the slow growth phase at times 40 to 103.  A very important conclusion is that
the bar has a higher pattern speed than the spiral structure.  Also, note that the pattern speeds of both the bar
and the spiral arms decrease with time, as shown for the bar in panel (b).  The smooth curves show $2 \Omega$
(solid curve) and $2 \Omega \pm \kappa$ (dashed curves) for the initial, axisymmetric model.  They apply only
approximately during the later, nonlinear phases.  Nevertheless, it is clear that both the bar and the spiral
have pattern speeds that are similar to but slightly larger than the local values of $\Omega - \kappa/2$.  Also
note that the bar develops despite the fact that it has an ILR.  These are Figs.~2 and 3 from Sparke \& Sellwood (1987).
Similar results are illustrated in Bournaud \& Combes (2002).
}

\end{figure}

\eject

\begin{figure*}

\vskip 2.8truein


 \includegraphics{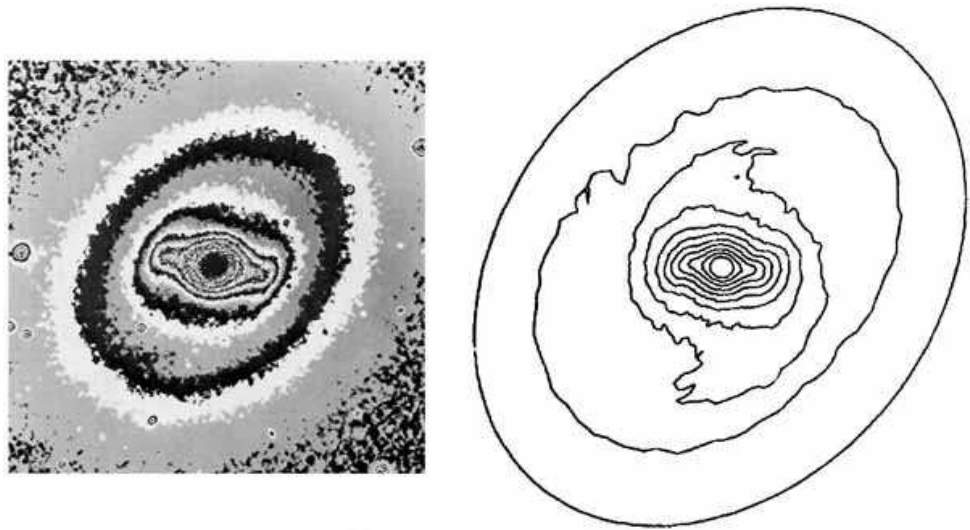}

\caption{Comparison of the $n$-body bar model of Sparke \& Sellwood (1987: isodensity contours at right) with 
the SB0 galaxy NGC 936 (Kormendy 1983: isophote map at left).  This is Fig.~6 from Sparke \& Sellwood (1987).
}

\end{figure*}

      Comparison to real SB0 galaxies shows that the $n$-body bar model of Sparke \& Sellwood (1987) is realistic 
in many respects.  It has the shallow density gradient and sharp outer edge of real bars in early-type galaxies.  
It has the rectangular shape observed, e.{\ts}g., in NGC 936 (Fig.~1.14).  And it shows qualitatively similar
non-circular streaming motions.  At the end of the simulation, the disk is too hot to allow
much spiral structure; therefore, the bar has no ``live'' component with which it can effectively interact, and
as a result, it is very robust.  The situation in NGC 936 is similar: the Toomre (1964) stability parameter in
the outer disk is $Q \simeq 3$ to 4, so essentially no small-scale structure is possible (Kormendy 1984b).  Not 
surprisingly, the purely stellar-dynamical $n$-body model of Sparke \& Sellwood (1987) is a realistic simulation 
of an SB0 galaxy but not of a gas-rich barred spiral.

      Shortcomings of the model are easily linked to limitations in the physics:

    \begin{enumerate}[(a)]\listsize
     \renewcommand{\theenumi}{(\alph{enumi})}
\item{The model is two-dimensional and therefore cannot include orbital
      complexity (including chaos) that is fundamentally three-dimensional.  In particular, the bar cannot
      buckle and hence thicken in the axial direction (Section 1.4.3.3).}
\item{The lack of gas means that the model can simulate SB0 but not spiral galaxies.}
\item{The bar cannot grow very much at late times because the disk gets too hot.  The same problem is well known
      in the context of understanding spiral structure.  Quoting Toomre (1990): ``The gravitational forces involved
      in [spiral structure] have the unwelcome side effect that they tend to increase the mean epicyclic motions
      of any stars \dots~which take part.  Typically, in just a few revolutions, these would-be participants become
      simply too `hot' to contribute appreciably to any more fun.  In other words, the vigor of these apparent
      `spiral instabilities' was quickly recognized to doom any hopes for their longevity -- unless one adds some
      form of cooling \dots~such as might be natural in a gas but surely not in a disk of stars.''  Toomre concluded that
      ``It really seems    
      as if we typically need at least to double the known amounts of reasonably active mass in the disks of Sc galaxies --
      beyond the gas amounts that they are known to possess~-- in order that such galaxies appear about as few-armed
      as they often are.  {\it And from where can they have obtained \dots~such relatively cool additional disk material
      if not from recent infall\ts}'' (emphasis added).  So another limitation of the Sparke \& Sellwood
      model -- and of other $n$-body simulations~-- is the lack of continued cosmological infall of
      cold gas (Combes 2008a; Bournaud \& Combes 2002).}
\item{Lacking gas, the bar cannot commit suicide by transporting gas to the center and building a pseudobulge that makes it more
      difficult for $x_1$ orbits to precess together.  I believe that $n$-body studies overestimate the robustness of bars
      when they give them no opportunity to increase the central mass concentration.}
\item{Concurrently, the bar cannot continue to grow stronger if the model does not include a ``live halo'' of particles
      that can act as an angular momentum sink.} 
\end{enumerate}

\vskip -5pt
      These comments are not meant to be critical of Sparke~\&~Sellwood~(1987).  Sellwood \& Carlberg (1984) recognize
the importance of gas infall to the maintenance of spiral structure.  Sparke \& Sellwood (1987)
is a clean study with the machinery that was available at the time.  It focuses on fundamentals that we need to understand.  
Many $n$-body studies both before and since have led to similar 
conclusion and have further extended our understanding of bar evolution (e.{\ts}g., Athanassoula 2003, 2005, 2012).

      My remarks are meant to highlight how much this subject presents opportunities to students now.~There is a danger
that we may miss important physics because we do not yet observe the early stages of bar formation.  This is 
an opportunity for observers of high-$z$ galaxies.  Modulo this uncertainty:

      {\it I suggest that bars grow secularly via an ongoing competition between outward angular momentum transport that
strengthens the bar and the buildup of pseudobulges that weaken the bar.}  This competition allows some barred galaxies 
to have both strong bars and high-mass (pseudo)bulges that create (otherwise destructive) ILRs near the center.

      Both the disk and the dark halo are angular momentum sinks.  For the halo, this conclusion is based mostly on 
$n$-body simulations (e.{\ts}g.,
Sellwood 1980, 2006, 2008;
Athanassoula 2003, 2005;
Athanassoula \& Misiriotis 2002),
but note that Valenzuela \& Klypin (2003) conclude that the effect is often overestimated. 
For the disk, global spiral structure is the visible sign that the process is ongoing.  

      Cold gas is important to the evolution for many reasons.  It makes the disk more responsive, both via
its small velocity dispersion and by making new stars that keep the disk cold.  Radial redistribution of 
gas is one product of angular momentum transport.  And especially important is the conclusion reviewed in the next
sections that gas driven inward to the center builds substantial pseudobulges in many galaxies.
The natural frequencies at which ILR orbits want to precess therefore become less constant with radius.
When this effect wins the above competition, the bar is weakened or destroyed.  Continued cosmological infall
of cold gas is central to all of these processes.

      I argue in the rest of this review that many aspects of the above story are well understood but
that many opportunities remain to be explored.

\subsubsection{Vertical buckling of bars and the formation of box-shaped bulges}

      The vertical thickening of bars into ``box-shaped bulges'' is reviewed from a theoretical perspective in SW93
and from an observational perspective in Kormendy \& Kennicutt (2004).  My review here is brief.

      $N$-body simulations show that strong bars thicken in the axial direction until they look peanut-shaped when 
viewed side-on and box-shaped when viewed at the most common orientations between side-on and end-on.  The first
paper to show this -- Combes \& Sanders (1981) -- was one of the earliest papers on galaxy secular evolution.  Two
distinct processes are responsible.

      Figure 1.15 shows the evolution of an $n$-body bar that, seen side-on, turns into a ``box-shaped bulge''
via a buckling instability in the axial direction (Raha \etal 1991).  The instability happens after the bar
is well formed, and \phantom{0000000000}

\vfill

\begin{figure}[hb]

\vskip 1truein


 \includegraphics{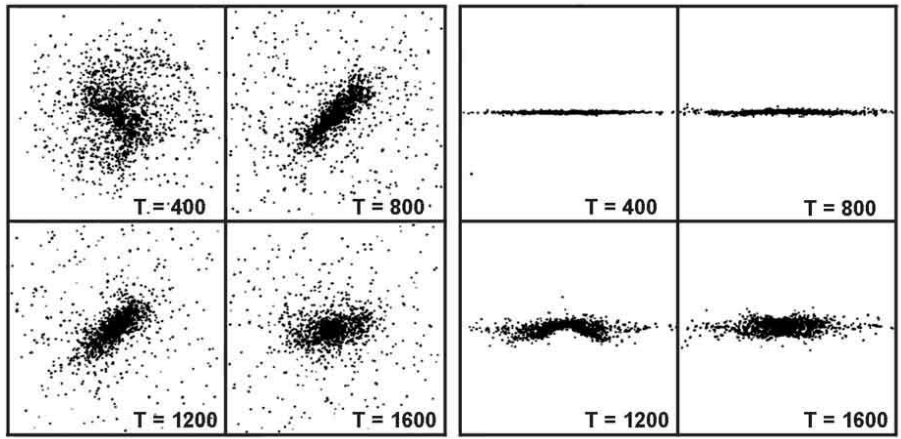}

\caption{Face-on (left) and edge-on (right) views of the vertical buckling of an $n$-body bar
and the formation of a ``box-shaped bulge'' (Figs.~1 and 2 from Raha \etal 1991).  The rotation
period at the end of the bar is $\Delta T \simeq 200$ time units.
}

\end{figure}

\eject

\noindent its growth is about as rapid as that of the original bar instability.  Two bar rotations later,
the bar has weakend considerably, grown more centrally concentrated, and \hbox{become box- or peanut-shaped as seen
side-on.}  \hbox{The resemblance to box-shaped bulges in edge-on galaxies (Section\ts1.5.2.9),} already noted by Combes \&
Sanders (1981) is compelling.  Many other \hbox{$n$-body} studies reach similar results (e.{\ts}g.,
Athanassoula \& Misiriotis 2002;
Athanassoula 2003, 2005;
Athanassoula \etal 2005;
Shen \etal 2010).

      A second mechanism by which bars may thicken in the axial direction is via resonances between the rotation
of the bar and the vertical oscillations of disk stars (e.{\ts}g.,
Pfenniger 1984, 1985;
Combes \etal 1990;
Pfenniger \& Norman 1990, 1991;
Pfenniger \& Friedli 1991;
Hasan \etal 1993).
When the vertical oscillation of stars is in resonance with the azimuthal rotation of a bar, each star repeatedly 
encounters the bar at a similar phase~in~its~orbit.  This pumps up the vertical velocity dispersion of disk stars 
and gives the center a boxy appearance when viewed edge-on.  Unlike bar buckling, resonant heating is 
inherently a secular process.

      The relative importance of these two processes in making the observed ``box-shaped bulges'' is not known.
However, the most important conclusion appears robust regardless of the answer: when we see a boxy bulge in an edge-on
galaxy, it is part of the bar and not a merger remnant.  We therefore call these features ``box-shaped pseudobulges''.
I return to them in Section 1.5, when I discuss the evolution that bars drive in their galaxies.

\subsubsection{Bar suicide and the origin of lens components.\\I.~$N$-body simulations}

      Early $n$-body simulations suggested that bars are very robust (see SW93).  The reason is
now clear: those simulations contained no gas and so could not engineer big changes in the
central concentration of the mass distribution.  More recent simulations which allow the galaxy to grow a central mass 
-- either ``by fiat'' or more naturally by rearranging the disk gas -- show that {\it bars tend 
to commit suicide by the secular evolution that they drive.}  In particular, they force disk
gas to fall toward the center, where it builds up a high-central-concentration pseudobulge, and this causes
the bar to weaken.  Pseudobulge growth is covered in later sections.  The death of bars and their evolution --
I suggest -- into lens components is discussed here.

      The earliest simulations of this process were motivated by the observation that most galaxies contain central 
supermassive black holes (Kormendy \& Richstone 1995).   The earliest paper that I know, Norman \& Hasan (1990),
already got essentially the modern answer:~``It is estimated~that~a~black~hole with mass equal to 17\ts\% the total mass
is required to destroy the [$x_1$] family of orbits and hence the bar.''  No galaxy is known to have such a big black hole
(Kormendy \& Ho 2013), but we now know that a centrally concentrated pseudobulge -- although less effective than a 
point mass -- has a similar effect.  Hasan \& Norman (1990) is a detailed followup.  Pfenniger \& Norman (1990) 
concentrated on pseudobulge building by a combination of inward radial gas flow and resonant vertical heating, and
they reached similar conclusions about bars destruction.  Subsequent discussion has centered on the question of 
how much central mass is needed to affect a particular degree of bar weakening.  Apparent disagreements between 
papers mainly result from the fact that fluffier central mass concentrations are less damaging to bars. 

      Why central mass concentrations are destructive is easy to see using Fig.~1.11 and was summarized in Section 1.4.2.  As the 
central mass increases, the central maximum in $\Omega - \kappa/2$ becomes higher and broader in radius (see the panel for 
M{\ts}81).  This shrinks the radius range in which $\Omega$\ts$-$\ts$\kappa/2$ varies slowly with radius and in which~it~is 
possible to have $\Omega_p$\ts$\sim$\ts$\Omega$\ts$-$\ts$\kappa/2$.  To put it differently, the radius range between ILR and 
corotation in which $x_1$ orbits align with the bar gets smaller.  Recall that $x_2$ orbits interior to ILR align perpendicular
to the bar.  So, as the central concentration increases, there are fewer bar-supporting $x_1$ orbits, and they
have a harder time all precessing at the same rate $\Omega_p \sim \Omega - \kappa/2$.  As the bar potential weakens,
self-gravity becomes less able to keep all the orbits precessing together.  Once an orbit ``escapes'' from its alignment
with the bar, it phase-mixes azimuthally in a short time.  Even if there is no big change in orbital eccentricities,
the long, thin shelf in surface brightness that makes up the bar will, \hbox{I suggest (Kormendy 1979b), tend to phase-mix}
azimuthally into a lower-surface-brightness and more axially symmetric shelf in surface brightness that has the same sharp 
outer edge that the bar had.  This is precisely a description of a lens component.  It is part of the reason why I suggest
that, as strong bars weaken, the stars that escape from them grow a lens.  I suggest further that we see galaxies at
all phases of this evolution: some have bars and no lenses; many have bars embedded in lenses of the same major-axis size,
and some have lenses with no bars.

      The rest of this section reviews further papers on bar suicide and observational evidence that lenses are defunct bars.

      Combes (2008b,\ts2010,\ts2011) gives  brief theoretical \hbox{reviews of bar-to-lens} evolution.~She also
reviews an additional important aspect of bar dissolution.  When gas loses angular momentum and flows
toward the center, the stellar bar receives that angular momentum that is lost by the gas.  This makes the orbits 
rounder and the bar weaker.  Papers that emphasize this effect include Bournaud \& Combes (2002) and Bournaud \etal (2005).


\vfill\eject

\cl{\null}\vfill

\begin{figure}[hb]

\vskip 5.truein


 \includegraphics{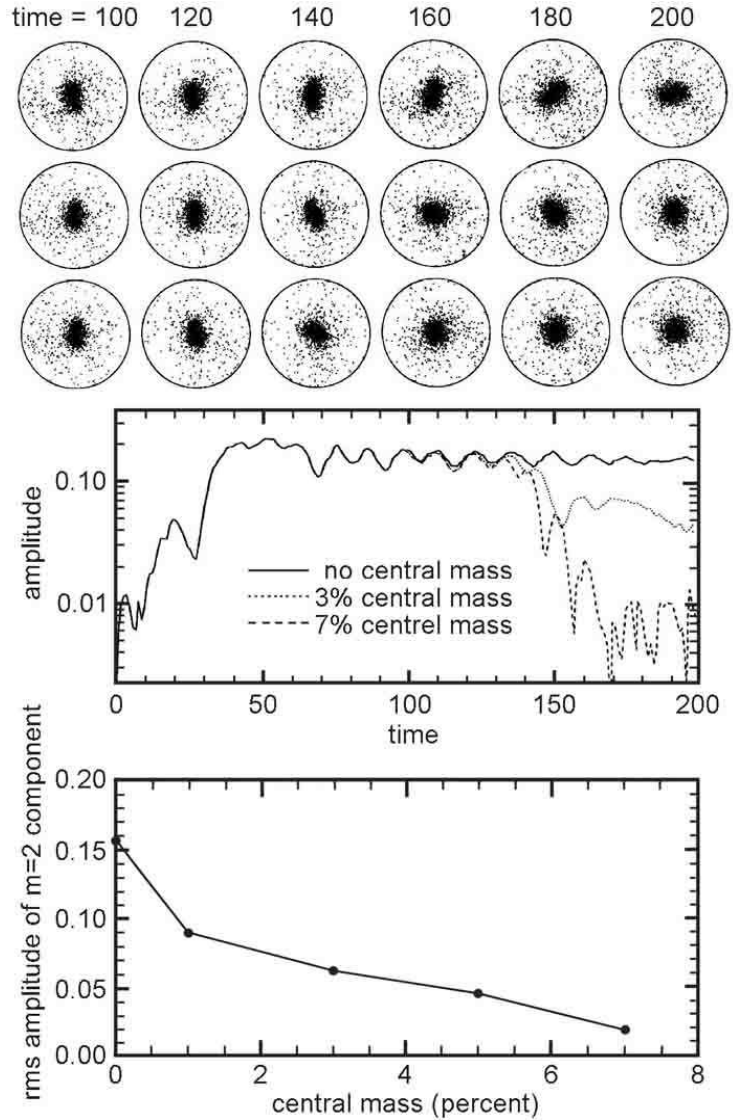}

\caption{Death of an $n$-body bar as the central mass concentration is increased (adapted from Figs.~1, 2 and 3 of
Norman \etal 1996).  An initial axisymmetric disk was evolved to time 100, by which time a stable bar had developed (leftmost
configuration shown in each of the three rows of particle distributions~at~the~top).  The top three rows show the time 
evolution as a point mass is grown at the center.  Its mass shrinks from bulge size to zero size between $t = 100$ to $t = 150$.
At the end, it makes up 0\ts\%, 3\ts\% and 7\ts\% of the mass of the disk (top row to bottom~row).  In the absence of a
central point mass, the bar is stable (top).  The bar is weakened progressively more as the central mass increases. 
The bar amplitude is shown in the middle panel.  The bottom panel shows the $t = 200$ amplitude as a 
function of the final central point pass.  Note that, whereas higher point masses more thoroughly destroy the bar, its 
amplitude is non-zero even for a 7\ts\% central point mass.
}

\end{figure}

\cl{\null}\vskip -27pt

\eject

\cleardoublepage

      My lectures included an example from Norman \etal (1996) of an $n$-body simulation of the weakening of a bar by a 
growing central mass.  This is illustrated in Fig.~1.16.  A Kuz'min -- Toomre disk that contains 75\ts\% of the mass is 
the only component modeled with (50,{\kern 0.4 pt}000) ``live'' particles.  The rest of the mass is in a rigid bulge.  This 
is modeled as the sum of two Plummer spheres, initially of the same size; the aim was to let one of them shrink in radius
once the bar was well established.  The disk was initially given a Toomre stability parameter of $Q = 1$ ensuring that a bar 
would~form.  This already guarantees that the initial bar will coexist with a bulge whose $B/T$ ratio
is typical of observed galaxies.  Figure 1.16 shows that the inner part of the disk formed a bar by time 50; if nothing
further was done to the model (top row of particle distributions and ``no central mass'' amplitude line in the middle panel),
the bar remained stable in amplitude to time 200.  

      Between time 100 and time 150 (about four bar tumbling periods), the lower-mass ``bulge'' Plummer sphere was shrunk
by a factor of 50 in radius.  Note that this turns it effectively into a point mass, not a pseudobulge with a realistic
scale length.  No mass was added, so the overall equilibrium of the disk was essentially unaffected.  But the amplitude of 
the bar decreased as the central point mass was shrunk.~Even a 3\ts\% central point mass reduced the bar amplitude substantially.
A 7\ts\% point mass killed it almost completely.

      Norman \etal (1996) explored the mechanism whereby the bar is destroyed in some detail, both in the above, 
two-dimensional simulation and in three-dimensional simulations that give similar results.  In essence, moving some 
mass to the center shrinks the phase space (e.{\ts}g., radial range) of the $x_1$ orbits.  As ILR moves
outward, some $x_2$ orbits appear, and these do not support the bar.  Some orbits become chaotic.  At some point (time
$\sim$\ts$140 \pm 5$ in Fig.~1.16), the bar coherence quickly breaks down and the disk becomes more axisymmetric.  Note
that the bar is not destroyed completely.

      Norman and collaborators acknowledge that these are ``highly idealized calculations intended to study just one aspect 
of the rich secular evolution of barred galaxies.  The contraction of a rigid mass component is, of course, highly artificial,''
and more to the point, the pseudobulges that weaken real bars have length scales $\sim 1/10$ those of their associated disks;
they are not point masses.  With Norman \etal (1996), we emphasize that the model is constructed so a relatively large
bulge coexists comfortably~with~the~bar.  Figure 1.16 should be taken as a ``proof of concept'' of bar suicide, not
as a definitive measurement of how much central mass concentration is required to achieve it.  I use Norman \etal (1996)
and not more detailed simulations as my example here because my purpose is primarily pedagogical,~and~I~don't want the
essential theme to get lost in the details of orbit analysis.

      The real world is substantially more complicated in ways that make it difficult to predict via simulations how much 
central mass concentration a bar can tolerate.  No available simulation realistically follows the competition between (1) the 
bar growth that results from allowing the disk and dark halo to be angular momentum sinks, with the disk kept cold and continually 
replenished with cosmologically infalling gas, and (2) the secular evolution that builds up the central mass concentration and
that thereby fights the growth of the bar.  While the pseudobulge grows both through gas infall and star formation and
(e.{\ts}g., Norman \etal 1996) by the redistribution of stars, resonances such as (but not limited to) ILR sweep through the central 
regions and continually keep a fresh supply of material responsive to resonant interactions with the bar.  

      So it is no surprise that some papers suggest that central mass concentrations as small as a few percent of the disk mass 
are enough to destroy bars
(e.{\ts}g., Berentzen \etal 1998;
Sellwood \& Moore 1999;
Hozumi \& Hernquist 1999, 2005;
Bournaud \etal 2002, 2005;
Hozumi~2012)
whereas others (e.{\ts}g.,
Friedli \& Benz 1993; 
Norman \etal 1996;
Athanassoula \etal 2005)
find that larger masses are necessary.  These apparent disagreements are largely resolved by a fundamental conclusion
due to Shen \& Sellwood (2004): ``{\it For a given [central] mass, dense objects cause the greatest reduction in bar amplitude,
while significantly more diffuse objects have a lesser effect\/}'' (emphasis added).
The culprit in bar suicide is certainly not the central supermassive black hole (Shen \& Sellwood 2004) or even a
nuclear star cluster, since these typically have masses of $\sim$ 0.2\ts\% of the mass of a classical bulge and a smaller 
fraction of the mass of a pseudobulge or disk (see Kormendy \& Ho 2013 for a review).  

      The results of Shen \& Sellwood (2004) suggest that pseudobulges mainly are responsible for destroying bars.  For a ``soft''
central mass whose size is comparable to that of a pseudobulge, Shen and Sellwood find that a 5\ts\% central mass decreases the bar strength 
to $\sim$\ts60\ts\% of its undamaged value, whereas a 10\ts\% mass decreases the bar strength to $\sim$\ts1/3 of its undamaged value.  
Orbital analysis reveals that the bar dies because the phase space of $x_1$ orbits shrinks, and many $x_1$ orbits become chaotic as the 
bar decays.  In most of these models, the dark matter halo was a rigid potential and therefore not an angular momentum sink.  Test runs 
with a live halo show that the bar is then slightly harder to destroy.  Similarly, the disk is dissipationless and does not accrete cold 
gas from the cosmological web, so it is not as good an angular momentum sink as the disk in a real spiral galaxy.  Nevertheless, these results
provide a realistic picture of what it takes to destroy a bar, at least as judged by a comparison with real galaxies (Section 1.4.3.5).

\eject

\vfill\eject

\cl{\null}

\vfill

\begin{figure}[hb]


 \includegraphics{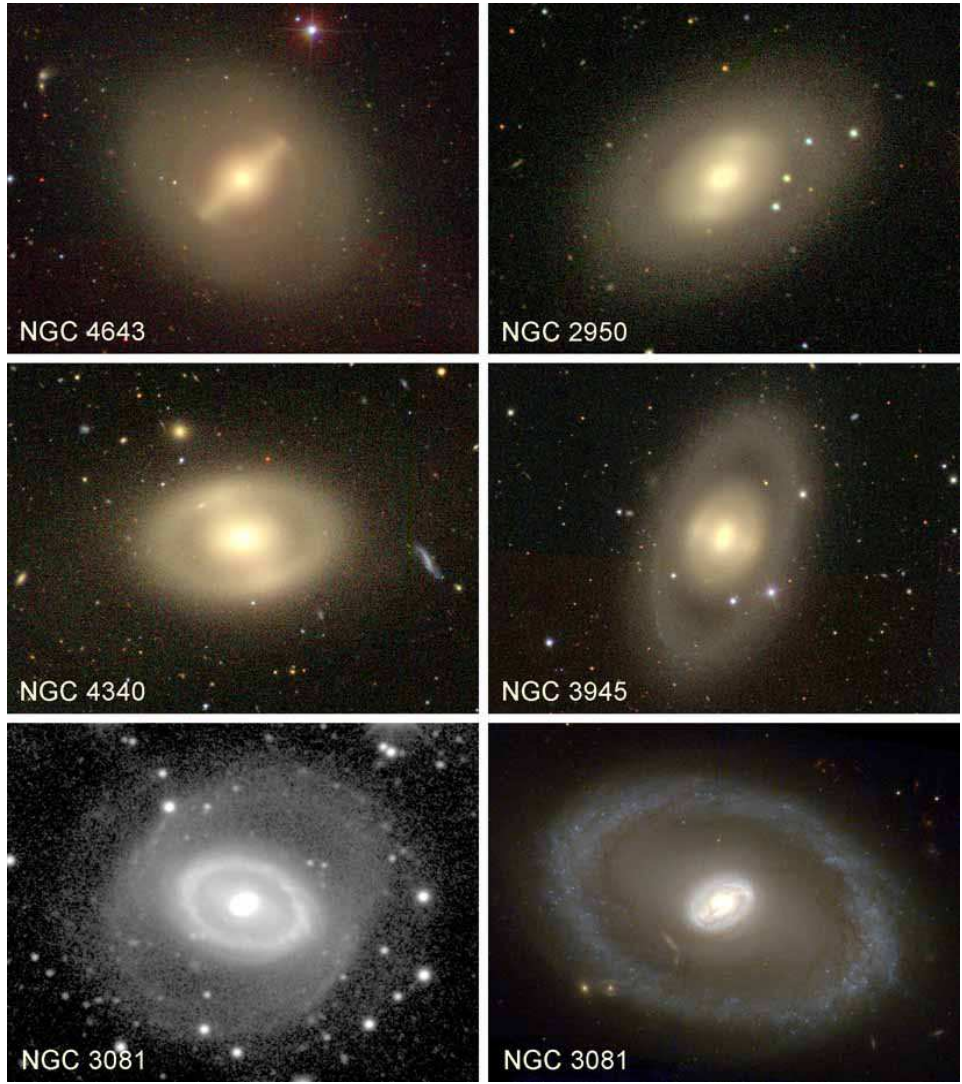}

\caption{Typical galaxies in which a bar fills a lens component in one dimension.  This figure shows a sequence from a strong bar 
with essentially no lens (NGC 4643) through galaxies that have both bars and lenses (NGC 2950, NGC 4340, NGC 3945) to NGC 3081, 
which has an exceedingly weak bar embedded in a bright lens with a very bright, star-forming inner ring around the rim of the lens.~The 
{\it HST\/} image~at right is an enlargement of the inner ring, lens, bar and pseudobulge.~In NGC 3081, the pseudobulge replicates the 
main B(r,lens) structure with a nuclear~bar.  All images are from {\tt http://www.wikisky.org} except those of NGC 3801, which
are from the {\it de Vaucouleurs Atlas of Galaxies} (Buta \etal 2007: left image) and {\it HST} (Buta \etal 2004: right image).  
Similar galaxies are shown in earlier figures.
In Fig.~1.6, NGC 1291 is an (R)SB(lens)0/a galaxy similar to NGC 3945 here.
In Fig.~1.9, NGC 2859 is an (R)SB(lens)0 galaxy similar to NGC 3945.  The other galaxies in that figure are
later-Hubble-type versions of the sequence shown here.  
}

\end{figure}

\cl{\null}
\vskip -28pt

\eject

\subsubsection{Bar suicide and the origin of lens components. II. Observations}

      Long before we knew of a process that could destroy bars, Kormendy~(1979b, 1981, 1982b) suggested that bars weaken
with time and evolve into lenses.  Observations on which this suggestion was based include the following.

    \begin{enumerate}[(a)]\listsize
    \renewcommand{\theenumi}{(\alph{enumi})}
\item{Lens components are very common in barred galaxies; in the sample of 121 SB galaxies studied by Kormendy (1979b),
      54\ts\% of SB0$^-$ to SBa galaxies have lenses.  The relative light contribution of the bar
      and lens varies widely from ${\rm bar}/\kern 0.5pt{\rm lens} \gg 1$ to ${\rm bar}/\kern 0.5pt{\rm lens} \ll 1$. Figure 1.17 shows
      such a sequence.  I illustrate many B(lens) structures in these lectures to emphasize how common they are.}
\item{Lens components occur but are rare in unbarred S0\ts--{\ts}Sa galaxies.  Examples are shown in
      Figs.~1.7 and 1.18.  It is likely that the inner ``oval disks'' in later-type (typically Sb) galaxies
      are lenses, too.  If B $\rightarrow$ lens evolution occurs, this suggests that it goes nearly to completion
      (i.{\ts}e., ${\rm bar}/{\rm total}$ light $\rightarrow$ 0)
      rarely in early-type galaxies and more commonly in mid-Hubble-type galaxies.}
\item{Lens components and bars in early-type galaxies have similar radial brightness profiles; i.{\ts}e., shallow
      surface brightness gradients interior to a sharp outer edge.  The brightness profiles of the lenses measured in
      Kormendy \& Bender (2012) are well fitted by S\'ersic (1968) functions with $n \simeq 0.5$ (i.{\ts}e., Gaussians).}
\item{When bars and lenses occur together, the bar almost always fills the lens in one dimension 
     (Kormendy 1979b, 
      Buta \etal 2007;
      Buta 2011, 2012;
      Fig.~1.17 and references there).  
      Points (c) and (d) imply that, to make a lens, it is sufficient to azimuthally phase-mix bar orbits.
      Points (a) and (b) hint that such a process occurs in many barred galaxies but that bars are completely destroyed
      only rarely.  This is consistent with Section 1.4.3.4: moderate ``soft'' mass concentrations weaken bars but do not
      necessarily destroy them.}
\end{enumerate}

\vfill

\vskip 1.3truein

\begin{figure}[hb]


 \includegraphics{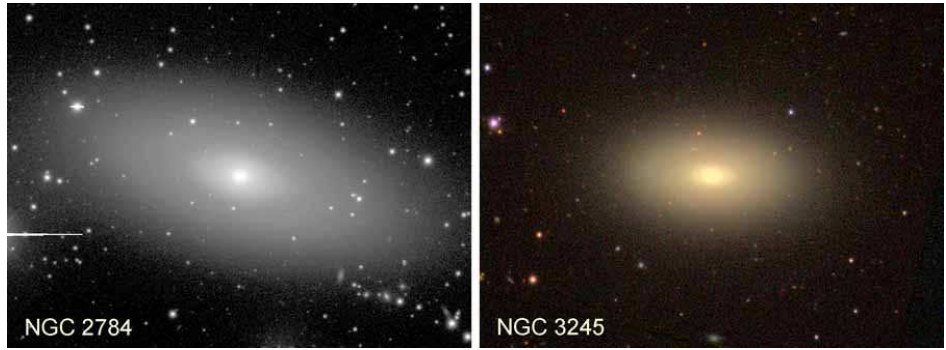}

\caption{{\kern -1.5pt}The prototypical unbarred S0 galaxy with a lens is NGC\ts1553~(Fig.~1.7).  This figure shows two more examples.
NGC 2784 is from the {\it de Vaucouleurs Atlas} and NGC 3245 (rotated so north is at left) is from {\tt http://www.wikisky.org}.  The lenses
are the elliptical shelves in surface brightness just outside the (pseudo)bulges.
Several earlier figures show barless, late-type analogs which probably also are lenses 
(NGC 4736 in Figs.~1.3, 1.6 and 1.8; NGC 4151 in Fig.~1.8).
}

\end{figure}

\cl{\null}
\vskip -28pt

\cl{\null}
\vskip -30pt

    \begin{enumerate}[e]\listsize
    \renewcommand{\theenumi}{(\alph{enumi})}
\item[(e){\kern -5pt}]{~Figure 1.19 shows a compelling observation that supports B $\rightarrow$ lens evolution.  The top panels show 
rotation velocity $V$ and velocity dispersion~$\sigma$ along the major axes of the barred S0(lens) galaxy
NGC\ts3945 (Fig.~1.17) and the unbarred S0(lens) galaxy NGC 1553 (Fig.~1.7).  The NGC 3945 bar is oriented along 
the minor axis, so the major-axis observations in Fig.~1.19 really measure the lens.  The bottom panel shows $V/\sigma$ for these
two galaxies plus NGC 2784 (Fig.~1.18) and two more SB0 galaxies.  I conclude: (1) {\it The inner part of each lens is hotter than
its corresponding pseudobulge.  That is, $\sigma$ is higher and $V$ is smaller in the lens than in the pseudobulge.}
(2) {\it This behavior and the $V/\sigma$ radial profiles are the same in barred and unbarred lenses.}
(3) {\it And $V/\sigma$ in these lenses is similar to $V/\sigma$ along the ridge line of the strong bar
         in NGC 936} (Kormendy 1983).}
\end{enumerate}

\vfill

\vskip 1.3truein

\begin{figure}[hb]


 \includegraphics{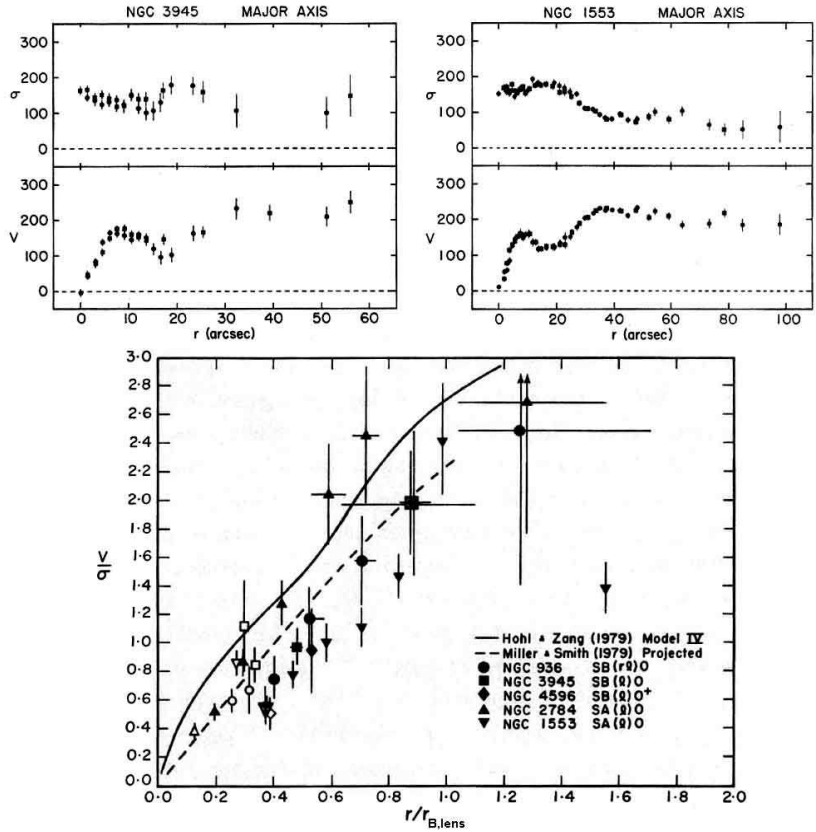}

\caption{(top) Major-axis rotation velocity $V$ and velocity dispersion $\sigma$ data in NGC 3945 [SB(lens)0] and in NGG 1553  
[SA(lens)0].  The lens radius is 51$^{\prime\prime}$ in NGC 3945 and 36$^{\prime\prime}$ in NGC 1553.  The instrumental dispersion 
is $\sim$\ts100 km s$^{-1}$ for NGC 3945 and $\sim$\ts65 km s$^{-1}$ for NGC 1553.  (bottom) Local ratio of 
rotation velocity to azimuthal velocity dispersion in lenses of barred and unbarred galaxies. Radii are normalized to the 
radius of the lens for galaxies and to the corotation radius for the n-body models.  Horizontal ``error bars'' show the 
range in radii over which data were averaged. Open symbols refer to measurements contaminated by bulge light.  This figure is 
taken from Kormendy (1981, 1982b, 1984a).
}

\end{figure}

\cl{\null}
\vskip -28pt

\eject

    \begin{enumerate}[e]\listsize
    \renewcommand{\theenumi}{(\alph{enumi})}
\item[\phantom{(e)}]{Thus the relative importance of ordered motions (rotation) and random motions is similar in lenses and bars,
                     both observed and $n$-body-simulated.  In particular, at about 40\ts\% of the radius $r_{\rm lens}$ of the lens,
                     $V/\sigma \sim 0.8$.
                     Also, $\sigma$ is the line-of-sight velocity dispersion, so the total velocity dispersion is likely
                     to be at least a factor of $\sqrt{2}$ larger.  This means that the ratio of rotational kinetic energy
                     to random kinetic energy is $\sim 1/3$ in the inner parts of both bars and lenses.  That is, stellar orbits
                     are similarly far from circular.  Lenses are parts of disks (Freeman~1975), and yet they are
                     hotter than their associated (pseudo)bulges.  The close similarity of $V^2/\sigma^2$ in bars and lenses
                     (cf.~Bosma \etal 2010) means that it is relatively easy for bars to evolve into lenses -- it is enough if $x_1$ orbits
                     escape the persuasion to precess together that is supplied by bar self-gravity; it is not necessary to 
                     change their energies and angular momenta drastically.
              }
\item[(f){\kern -5pt}]{~The fact that we see arbitrarily weak bars embedded in strong lenses is a powerful argument that these bars were once
           stronger (Kormendy 2004b).  NGC 3081 in Fig.~1.17 is the best example.  It contains an exceedingly faint bar, 
           but almost all of the inner light is in the lens -- which is {\it much\/} rounder~--~and~the pseudobulge. {\it  Can we really 
           imagine that one or two percent of the stars~in~a~disk discover that they are bar-unstable, that they make a nice, 
           highly elongated bar, and that most of the disk meanwhile stays just slightly oval?  This is not what happens in $n$-body
           simulations of bar formation.  There, essentially all of the disk participates in the instability, and most disk stars that
           live interior to the final radius of the bar participate in the bar.}  That's why $n$-body bars buckle in the axial direction --
           they are self-gravitating structures.  Lenses, on the other hand, appear to be flat (e.{\ts}g., Kormendy 1982b and
           Kormendy \& Bender 2012).  Thus the observation (Figs.~1.17 and 1.18) that there is a complete continuity from
           ${\rm bar}/{\rm lens} \gg 1$ to ${\rm bar}/{\rm lens} \ll 1$ to ${\rm bar}/{\rm lens} = 0$, together with the above results,
           strongly suggests that bars evolve into lenses.  
           The fact that barless lenses are rare at early Hubble types implies that any evolution is relatively slow or that it stops
           when a galaxy gets transformed into an S0.  The observation of barless ovals that appear to be equivalent to lenses in galaxies 
           such as NGC 4736 and NGC 1068 implies that the evolution goes to completion more readily in galaxies that contain gas.} 
\end{enumerate}

\phantom{More invisible text to fix line spacing}
\vskip -20pt

      All of these suggestions are comfortably consistent with -- and arguably could have been predicted from -- our picture that bars drive 
secular evolution that builds up the central density by forming pseudobulges that are harmful to the survival of the bar.

      I do not want to argue too forcefully for this picture.  I especially do not want to suggest that 
bar $\rightarrow$ lens evolution is the only thing~that~happens.  But the most economical suggestion that is consistent with 
observations and simulations is that bars evolve away as they increase the central concentration of the galaxy.~One result is lens 
components that~have~the same radii, shallow brightness gradients, and sharp outer edges as their parent bars.  Some tests of this idea
are suggested in Section 1.4.3.7.

\vfill

\subsubsection{Bar suicide and the origin of lens components.\\III.~What pseudobulge mass is required to destroy bars?}

\vskip -5pt

      We get a preliminary idea of what $PB/T$ ratio divides SA and SB galaxies by examining values for intermediate-Hubble-type
oval galaxies that are illustrated in this paper.  Two more such galaxies are shown in Fig.~1.20.
Table 1.1 shows~the~results.  The sample is small.
Also, $PB/T$ ratios of SAB galaxies vary widely, perhaps because the central concentrations of pseudobulges
vary widely.  Nevertheless, we derive a clean result, as follows.

      {\it Table 1 suggests that realistic pseudobulges with $PB/T \simeq 0.34 \pm 0.02$ are massive enough to destroy bars 
essentially completely and to convert them into lenses.  This is a preliminary result, reported here for the first time.} 

\cl{\null} \vskip -60pt \cl{\null}

 \begin{table}[h]
  \caption{Pseudobulge-to-total luminosity ratios for \phantom{0000000}barred~and~unbarred~oval~galaxies}
    \begin{tabular}{c|lcl}
     \hline \hline
     {Galaxy} & {~~~~Type}                    &      {$PB/T$} & Source\\
     \hline
     NGC 2273 & \phantom{(R$^{\prime}$)}SBa   &  $0.19 \pm 0.02$             & Kormendy \& Bender (2013) \\
     NGC 3393 & (R$^{\prime}$)SBab            &  $0.27 \pm 0.06$             & Kormendy \& Bender (2013) \\
     \hline
     NGC 3081 & \phantom{(R$^{\prime}$)}SAB0/a&  $0.11\phantom{\pm 0.000}$   & Laurikainen \etal (2010)  \\
     NGC 3368 & \phantom{(R$^{\prime}$)}SABab &  $0.25\phantom{\pm 0.000}$   & Kormendy \& Bender (2013) \\
     NGC 4151 & (R$^{\prime}$)SABab &  \phantom{0}$0.327\phantom{\pm 0.000}$ & Gadotti (2008) \\
     \hline
     UGC 3789 & (R)SAab                       &  $0.32 \pm 0.03$             & Kormendy \& Bender (2013) \\
     NGC 1068 & (R)SAb                        &  $0.35 \pm 0.05$             & Kormendy \& Bender (2013) \\
     NGC 4736 & (R)SAab                       &  $0.36 \pm 0.01$             & Kormendy \& Bender (2013) \\
     \hline \hline
    \end{tabular}
  \label{sample-table}
\end{table}

\vskip 1.truein

\begin{figure}[hb]


 \includegraphics{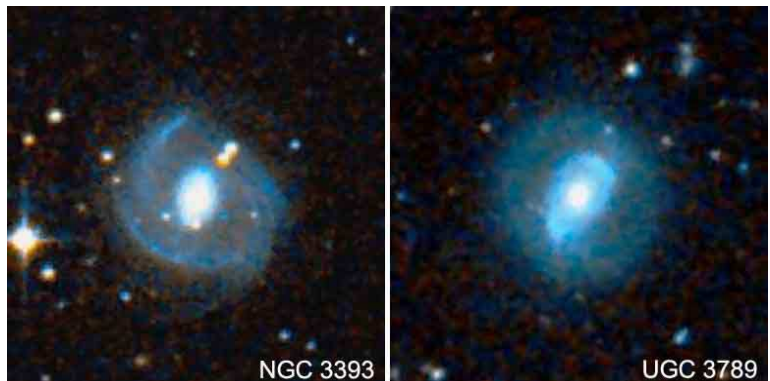}

\caption{Oval galaxies that bracket the $PB/T$ value which divides galaxies that do and do not have bars.
The barred galaxy NGC 3393 has $PB/T = 0.27 \pm 0.06$;
the unbarred galaxy UGC 3789 has $PB/T = 0.32 \pm 0.03$. 
This figure from Kormendy \& Ho (2013) is based on Digital Sky Survey images from {\tt http://www.wikisky.org}.
}

\end{figure}

\cl{\null}
\vskip -28pt
\eject

\subsubsection{Remaining puzzles and suggestions for further work}

      Our understanding of the lives and deaths of bars is now reasonably well developed.  A few particularly
obvious problems and suggestions about how to address them are outlined below.  

    \begin{enumerate}[a]\listsize
    \renewcommand{\theenumi}{(\alph{enumi})}
\item{How do bars form?  Despite the long history of seeing violent instabilities when $n$-body systems
      are given small Toomre $Q$ by {\it fiat\/}, and despite convincing hints that tidal tickling helps,
      we understand bar formation less well than we think.  We need to investigate bar growth both observationally
      by observing distant galaxies and by using simulations. Closely related and very important are (b)~and~(c).} \vskip 2pt
\item{We need to study bar evolution via the competition between angular momentum sinks
      (the outer disk and dark halo), which make the bar grow stronger, and the increasing importance 
      of ILR as the bar moves material inward, which makes the bar grow weaker.  This is a complicated
      modeling job.  But the great richness of possible interactions between nonaxisymmetric components is lost when
      the relevant components either are not included or are turned into fixed potentials.  I do not criticize 
      published papers for trying to distill clean subsets of the physics that we want to study.  But there are
      effects that we cannot find if component interactions are switched off.  In particular,
      it is likely that a bar can coexist with a more massive (pseudo)bulge if it grew in the context of the above
      competition.} \vskip 2pt
\item{Including gas is critically important in studying bar evolution.  Gas dissipation is essential for the inward
      flow of the material that builds pseudobulges.  Whether recycled, not yet used up, or accreted from outside, gas
      helps to keep the disk cold enough to be responsive.  Without it, $n$-body disks get so hot that
      spiral structure switches off and the disk gets less effective at absorbing angular momentum.  Cosmological
      accretion of cold gas may be essential~for~realistic~bar~evolution.}   \vskip 2pt
\item{On the observational front, we need to measure the luminosity and mass functions of elliptical galaxies, classical bulges,
      pseudobulges and disks as a function of environmental density.  Early observations demonstrate that the answers are 
      different in clusters and in the field (Sandage \etal {\kern -1pt}1985a; Kormendy{\ts}{\it et{\ts}al.\/}{\ts}2010).  
      But we need more quantitative measures of the relative importance of secular evolution and hierarchical clustering 
      and merging as a function of environment.} \vskip 2pt
\item{The idea that bars evolve into lenses needs to be tested further.  Note how many 
      relevant observations and theoretical hypotheses date from the 1980s.  This is an opportunity
      for today's students.  We need absorption-line kinematic measurements of bars and lenses.  We need 
      photometry and component decomposition to see, e.{\ts}g.,~whether the bar/lens luminosity
      ratio tends to be smaller as the (pseudo)bulge is more massive or more centrally concentrated.  Caution:~many 
      decomposition codes use the assumption
      that each component has a flattening that is independent of radius.  This is certainly wrong and commonly leads to
      unphysical conclusions (e.{\ts}g., about numbers of components).  And we need mass models to determine halo properties.
      There is much work to be done.}
\end{enumerate}

\vfill\eject

\section{Secular evolution and the growth of pseudobulges}

      This section reviews the processes of secular evolution in disk galaxies and the structures that they produce.  
I begin with bar-driven evolution but also discuss why evolution happens in more than just barred galaxies.   Much 
emphasis will be on the high-density central parts of galaxies that in the past were confused with merger-built bulges.  
As explained in Section 1.1.3, I call such components ``pseudobulges'' if -- to the best of our knowledge -- they were not 
made by mergers but rather were grown out of the disk.

      The essential qualitative results of secular evolution were known more than 30 years ago at the time of my Saas-Fee
lectures (Kormendy 1982b).  From that review, Table 1.2 here lists the major structural components in galaxies and what we then
thought we knew about their origin.~All of the suggested formation mechanisms are strongly supported by more recent~work and 
form the main subjects of these lectures. The only change that I have made in reproducing Table 1.2 is to remove a suggestion 
that the transition~from \phantom{000000000}

\vfill

\vskip 2.truein

\begin{figure}[hb]


 \includegraphics{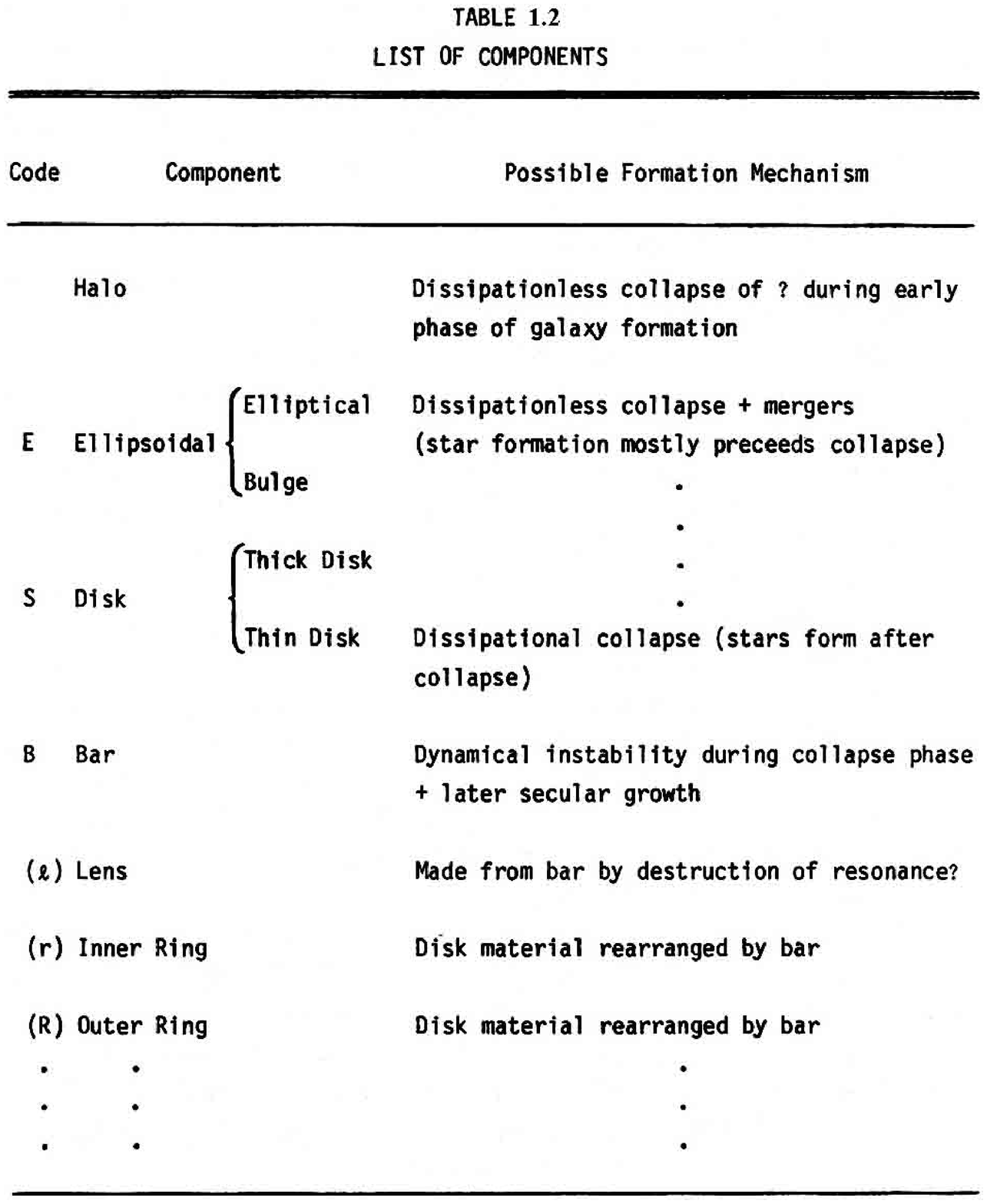}


\end{figure}

\cl{\null}
\vskip -28pt
\eject

\noindent  ellipticals and bulges to thick and thin disks is continuous.  The only change in emphasis that I would make 
           in reformulating the table now would be to emphasize major mergers as the origin of classical bulges and ellipticals 
           and note that dissipation and starbursts occur naturally in wet mergers.

      To bring this story up to date, let's start by revisiting the definition of a classical bulge.  The clearest 
statement that captures the intention~of morphologists such as Sandage and de Vaucouleurs is by Alvio Renzini (1999): 
``A bulge is nothing more nor less than an elliptical galaxy that happens to live in the middle of a disk.''  There is a world 
of information hidden in this beguilingly simple defintion:

      First, what is an elliptical?  Formally, this subject is beyond~the~scope of these lectures.~We understand that ellipticals 
form by major~galaxy~mergers and not by secular processes.  Still, we cannot understand pseudobulges without knowing something about
classical bulges and ellipticals.  Moreover, some lecturers at this school (notably Isaac Shlosman and Nick Scoville) focus 
on our standard picture of the evolution of structure by hierarchical clustering and galaxy mergers.  I try to connect this story with 
our picture of secular evolution in Section\ts1.8.  Some properties of elliptical galaxies and classical bulges are compared in
Section 1.7.  Here, I list the properties that we need in the present section.  Classical bulges and elliptical galaxies

    \begin{enumerate}[a]\listsize
    \renewcommand{\theenumi}{(\alph{enumi})}
\item{have smooth, nearly elliptical isophotes;}
\item{have S\'ersic (1968) function $\log{I(r)} \propto r^{-1/n}$ brightness profiles that fall steeply toward larger radii $r$
      with 2 \lapprox\ts$n$ \lapprox \ts4 for most bulges (e.{\ts}g., Fisher \& Drory 2008);}
\item{satisfy the same ``fundamental plane'' correlations (Djorgovski \& Davis 1987; Faber \etal 1987; Djorgovski \etal 1988; 
      Bender \etal 1992) between the ``effective radius'' $r_e$ that contains half of the total light, the ``effective surface
      brightness'' $\mu_e$ at $r_e$ and the total absolute magnitude $M_V$ (Fisher \&{\ts}Drory~2008; Kormendy \& Bender 2012) 
      as illustrated here in Fig.~1.68 and}
\item{generally rotate approximately as rapidly as oblate spheroidal stellar systems that have isotropic velocity dispersions
      and that are flattened mainly by rotation (e.{\ts}g., Kormendy \& Illingworth 1982).}
\end{enumerate}
\noindent Point (c) is the most important, because some of (a), (b) and (d) are shared by other kinds of ellipsoidal stellar 
systems called ``spheroidals'' (Section\ts1.7).

      I adopt the Renzini definition, because it embodies the idea that I want to emphasize,
namely that the definition of a component should be made in terms of its formation physics as well as its observed properties.

\begin{figure*}

\vskip 1.44truein


 \includegraphics{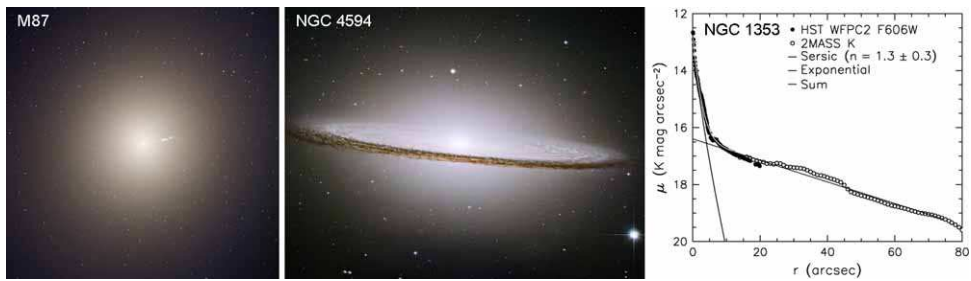}

\caption{Illustrating the surrogate definitions of a bulge: M{\ts}87 is a giant elliptical,
and NGC 4594 is a classical-bulge-dominated Sa galaxy seen almost edge-on.  These are Hubble Heritage
images.  The right panel shows the surface brightness profile along the major axis of the Sb galaxy
NGC 1353; we use it again in Fig.~1.33.  It is decomposed into an outer exponential disk profile and 
an inner (it will turn out: pseudobulge) profile with a S\'ersic index of $n = 1.3 \pm 0.3$.}

\end{figure*}

      The problem with our definition is that it is difficult to apply, especially to galaxies that are not seen edge-on.
Therefore it is common to use one of two surrogate definitions as illustrated in Fig.~1.21.  When a galaxy disk is seen
nearly edge-on (e.{\ts}g., NGC 4594, at center), then the part of the galaxy that is like an elliptical (e.{\ts}g., M{\ts}87, at left) 
is the high-surface-brightness, centrally concentrated component that is thicker than the disk.  When the galaxy is seen nearly face-on, 
we cannot see such a central thickening.  But we know from galaxies like NGC 4594 that the brightness profile of the bulge is 
much steeper than that of the disk and extends, near the center, to much higher surface brightnesses than those of the disk.  
NGC 1353 in Fig.\ts1.21 shows a clear separation between the high-surface-brightness central (pseudo)bulge and the outer 
exponential disk.  It also illustrates how we decompose the surface brightness profile of a face-on galaxy into bulge and disk parts 
that add up to the observed profile.  Bulge-disk decomposition is discussed in Section 1.7.4.1.  Here, I note the surrogate definition 
of a bulge that is used when a galaxy is seen nearly face-on: it is the central part of the galaxy that is defined by the extra light 
above the inward extrapolation of the outer, exponential or S\'ersic disk profile.  That is, the bulge-dominated part of NGC 1353 
is the part at $r$ \lapprox \ts10$^{\prime\prime}$.  It happens to be a pseudobulge rather than a classical bulge, but this does 
not affect the surrogate definition.

      The good news is that classifying bulges by using the surrogate definitions is easy.  The bad news is that we don't know what
they are physically.  That is, there is no guarantee that the surrogate definitions always identify central components that originate
via the same formation physics.  In fact, I will try to convince you in this section that the surrogate definitions often
identify high-surface-brightness central components that are nothing like ellipticals, especially in late-type galaxies.

\vfill\eject

      Therefore, I adopt a different approach that is more physically motivated but also more hazardous (Kormendy \& Kennicutt 2004).
I define classical bulges to be elliptical galaxies that happen to live in the middle of a disk.  For this to make
sense physically, we need to check that there exist central components of disk galaxies that have essentially all of the properties
of comparable-luminosity elliptical galaxies. This check has been made; it is discussed in Sections 1.7.4.4 and 1.8.1.
The intention is that classical bulges, like ellipticals, form via major galaxy mergers.  In contrast, I define a pseudobulge as the
dense central part of a galaxy -- either a thick component that bulges above and below an edge-on disk or the extra light at small
radii above the inward extrapolation of the outer disk profile -- that was constructed from the disk by (mostly) secular processes. 
For this definition to make sense, we need to see convincing evidence that secular disk evolution happens and that it
makes pseudobulges that we know how to identify.  We have such evidence; the classification criteria are listed in Section 1.5.3. 
There turn out to be three generic kinds of pseudobulges (Fig.~1.22).

      These definitions involve different practical problems.  We always know what we are talking about physically.  But it may be 
difficult to apply the classification criteria.  I prefer this problem to the ones above that are inherent in a descriptive classification. 
My job in the rest of this section is to present the case for a physically motivated definition of pseudobulges.

\vfill

\vskip 2.truein

\begin{figure}[hb]


 \includegraphics{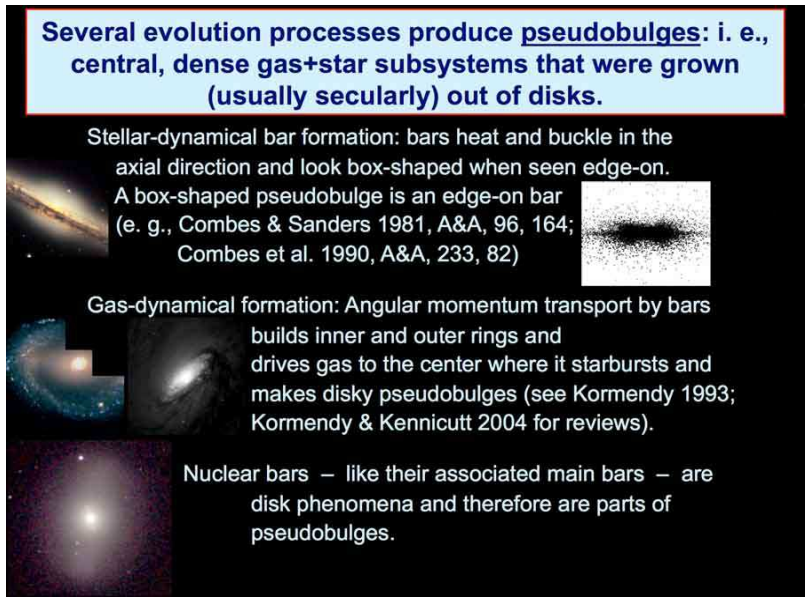}

\caption{Powerpoint slide illustrating three kinds of pseudobulges. Section 1.5.2.9 discusses boxy pseudobulges
as vertically buckled bars.  Construction of disky pseudobulges is discussed here.  For nuclear bars, see
Section 1.5.2.8.}

\end{figure}

\cl{\null}
\vskip -28pt
\eject

\def\etal{{\it et al.\ }}

\subsection{The response of gas to a rotating bar:\\Construction of outer rings, inner rings and pseudobulges}

      Figure 1.23 shows the fundamental results of bar-driven secular evolution.  The disk spreads (Section 1.2).
Gas at large radii gains angular momentum; it moves to still larger radii and is shepherded into an outer ring near OLR. 
This is identified with the outer rings in (R)SB galaxies (Fig.~1.24).  Gas at small radii gives up some
of its angular momentum and falls to the center.  High-density gas likes to make stars; we see starbursts 
in \hbox{Figs.~1.29\ts\ts--\ts1.31}.  We identify the result as a pseudobulge.  In between, gas is focused into 
a ring around the end of the bar, the inner ring seen in SB(r) galaxies~(Fig.\ts1.24).

\vfill

\begin{figure}[hb]


 \includegraphics{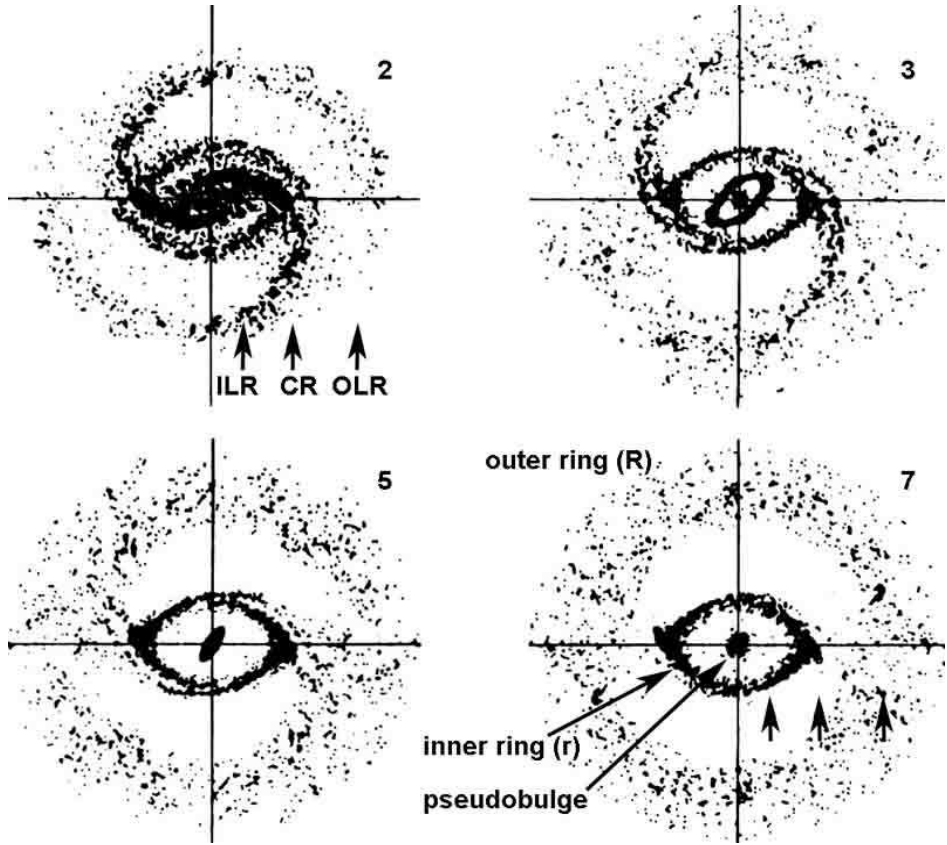}

\caption{Gas particle distributions in a ``sticky particle'' simulation of the response
of cold gas to a rotating bar.  The bar is not shown but, in each panel, is  horizontal and has a diameter equal to 
that of the inner ring in the last panel.  The panels show the gas response after 2, 3, 5 and 7 bar rotations 
(numbers at upper-right in each panel).  The arrows in the first and last panels indicate the position of the inner
Lindblad resonance (ILR), the corotation resonance (CR) and the outer Lindblad resonance (OLR).  From Simkin \etal (1980).}

\end{figure}

\eject

      Figure\ts1.24 shows the close correspondence between the features produced in the Simkin \etal (1980)
simulation and the morphology of barred galaxies.  More sophisticated hydrodynamic calculations clarify
the physics (Fig.~1.26), but it is remarkable that just letting gas particles stick together when they
collide is enough to reach the most basic conclusions.

\vfill

\begin{figure}[hb]

\vspace{4.45 truein}


 \includegraphics{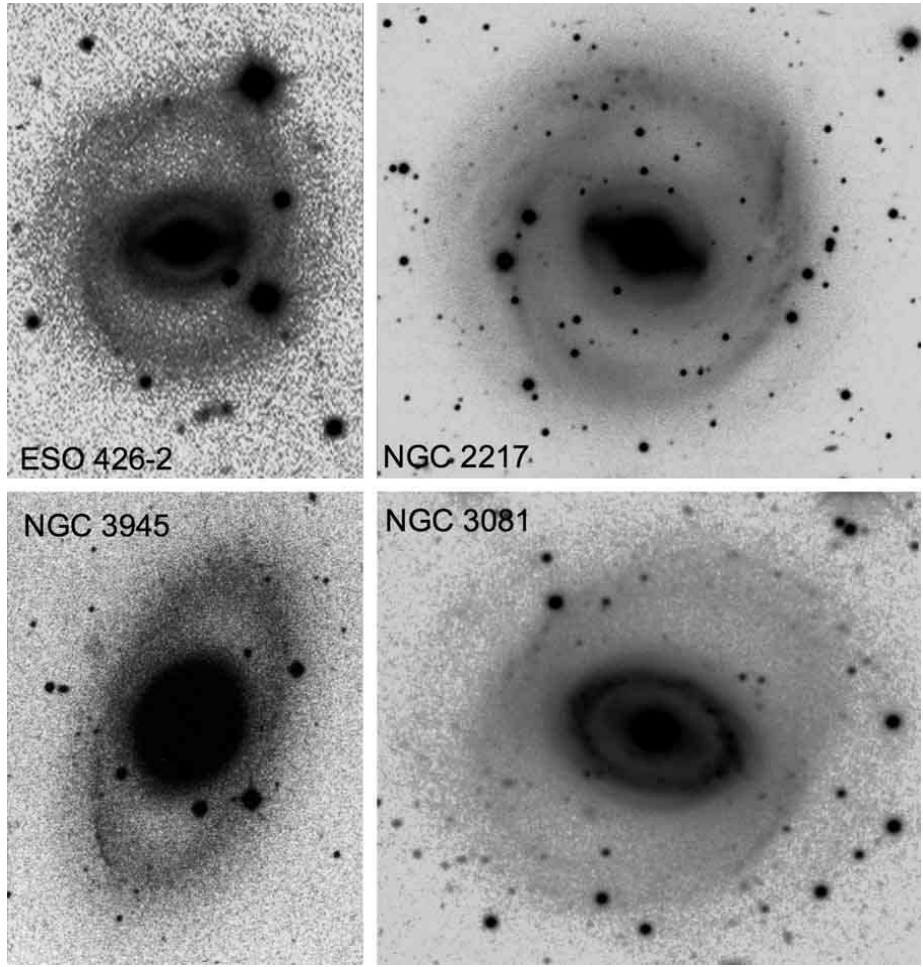}

\caption{Examples of galaxies which show the morphological features that were produced in the sticky particle
         simulation of Simkin \etal (1980).  NGC 3945 is from a photographic plate that I took with the Mount
         Wilson 100-inch telescope; the others are from the {\it de Vaucouleurs Atlas of Galaxies} (Buta \etal 2007).
         All four galaxies have complete or almost complete outer rings (R).  ESO 426-2 and NGC 3081 have both
         an outer ring and an inner ring (r); such galaxies are rare.  NGC 2217 contains a lens component that
         is filled by the bar in one dimension.  NGC 3945 and NGC 3081 are also illustrated in Fig.~1.17.
         Additional galaxies with similar morphology are shown in Figs.~1.3, 1.6, 1.8, 1.9, 1.17, 1.36, 1.37 and 1.45.}

\end{figure}

\eject

      Figure 1.25 explores the difference between SB(r) and SB(s) galaxies.  Sanders \& Tubbs (1980) investigated how the
response of gas to a bar varies with bar pattern speed and strength.    Fast bars have small corotation radii; e.{\ts}g., 
$r_{\rm cor}$/(disk scale length) = 0.7 in the bottom-left model panel.  They drive SB(s) structure like that in NGC 1300.
Slower bars end near corotation ($r_{\rm cor}/a = 1.1$) and drive SB(r) structure like that in NGC 2523.  Since bars slow 
down as they evolve (Fig.~1.13) and since it takes time to collect gas into an inner ring,
we conclude that SB(s) galaxies are dynamically young and that SB(r) galaxies are dynamically mature.

\vfill

\begin{figure}[hb]


\vfill 


 \includegraphics{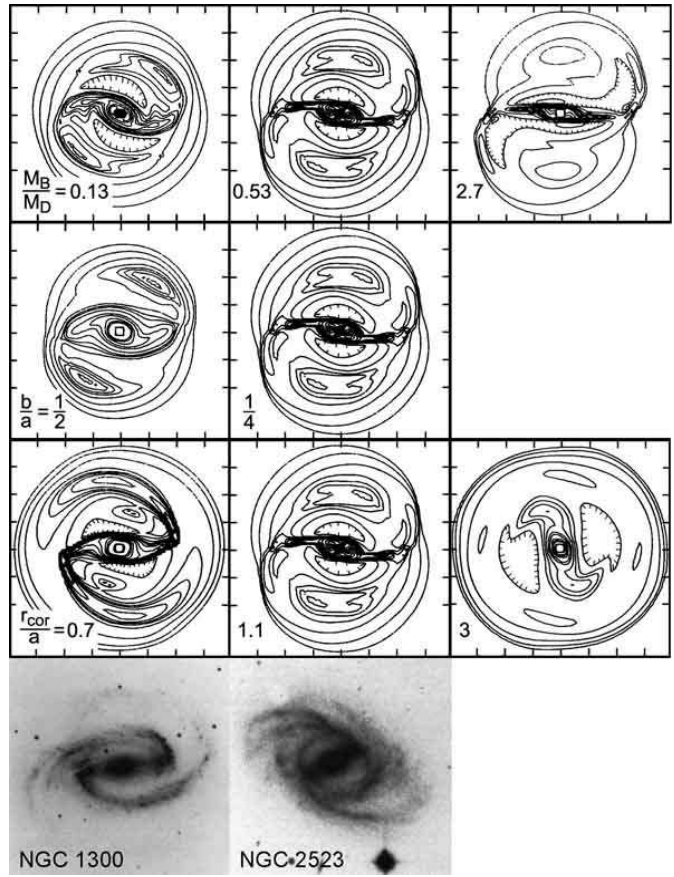}

\caption{Contours of steady-state gas density in response to a bar (adapted 
from Sanders \& Tubbs 1980, who also show intermediate cases).  The bar is horizontal and has a length equal to four 
axis tick marks.  The top row explores the effect of varying the ratio $M_B/M_D$ of bar mass to disk mass.
The second row varies the bar axial ratio $b/a$.  The third row varies the bar pattern speed, parametrized by the 
ratio $r_{\rm cor}/a$ of the corotation radius to the disk scale length.  The middle column is the same standard model 
in each row; it resembles the SB(r) galaxy NGC 2523.  The left panels resemble the SB(s) galaxy NGC 1300.
The right panels take the parameters to unrealistic extremes; they do not resemble real galaxies.
}

\end{figure}

\cl{\null} \vskip -30pt

\eject

      Consistent with this conclusion is the observation (Sandage 1975) that dust lanes on the rotationally
leading sides of bars are common in SB(s) galaxies but rare in SB(r) galaxies (cf.~Figs.~1.6, 1.24, 1.25).~The 
gas between the inner ring and the pseudobulge has largely been depleted in SB(r) galaxies (see
Kormendy \& Barentine 2010:~Fig.~1.46 here and Barentine \& Kormendy 2012 for examples).

      Sanders \& Tubbs (1980) also investigated the effects of varying the strength of the imposed bar potential,
measured both by the ratio of bar mass to disk mass $M_B/M_D$ (top row of simulation results in Fig.~1.25) and by the
axial ratio $b/a$ of the bar (middle row of simulation results).  Very weak bars ($M_B/M_D = 0.13$) or bars that
are not very elongated ($b/a = 1/2$, more like an oval disk than like a bar) produced weak SB(s) structure that
does not resemble well developed SB(s) galaxies such as NGC 1300.  Parameters that are more realistic in matching
strongly barred galaxies ($M_B/M_D = 0.53$, $b/a = 1.4$, and $r_{\rm cor}/a = 1.1$; i.{\ts}e., the standard model
shown in the middle column of model results) provided the best match to an SB(r) galaxy.  This model shows a linear 
maximum in gas density that is parallel to the bar and offset from it in the forward rotation direction like the straight 
dust lanes in SB(s) galaxies.  But the bar potential and gas content are imposed by the initial conditions; they are 
not self-consistently evolved from more axisymmetric initial conditions.  This accounts for the (rarely observed) 
coexistence of the offset gas density maximum in the bar and the (r) structure.  Finally, note that, 
when the bar is much stronger than bars in real galaxies ($M_B/M_D = 2.7$) or much slower than in real galaxies 
($r_{\rm cor}/a = 3$), the gas response fails to resemble that in observed galaxies.

      Better simulations generally confirm and expand on the above conclusions (e.{\ts}g.,
Salo \etal 1999;
Rautiainen \& Salo 2000;
Rautiainen \etal 2005;
Treuthardt \etal 2009;
see Kormendy \& Kennicutt 2004 for further review). 
In particular, detailed hydrodynamic models of the gas response in the region of the bar confirm the offset density 
maximum seen in the Sanders \& Tubbs (1980) model and greatly clarify the nature of bar-driven secular evolution.  

      The most illuminating simulations are by Athanassoula (1992), shown here in Fig.~1.26.  She explored 
the response of inviscid gas to an imposed bar, concentrating on the production of gas shocks and their relation 
to dust lanes.  If the mass distribution is sufficiently centrally concentrated to result in an ILR, then she
found that straight gas shocks~-- which we identify with dust lanes -- are produced that are offset in the forward 
(rotation) direction from the ridge line of the bar.  The ``$x_2$ orbits'' that align perpendicular to the bar inside 
ILR push the offset to be largest near the center.  This is seen in Fig.~1.26 and Fig.~1.27 as well as in Fig.~1.5 and Fig.~1.6.

\vfill\eject

\cl{\null}

\vfill

\begin{figure}[hb]


 \includegraphics{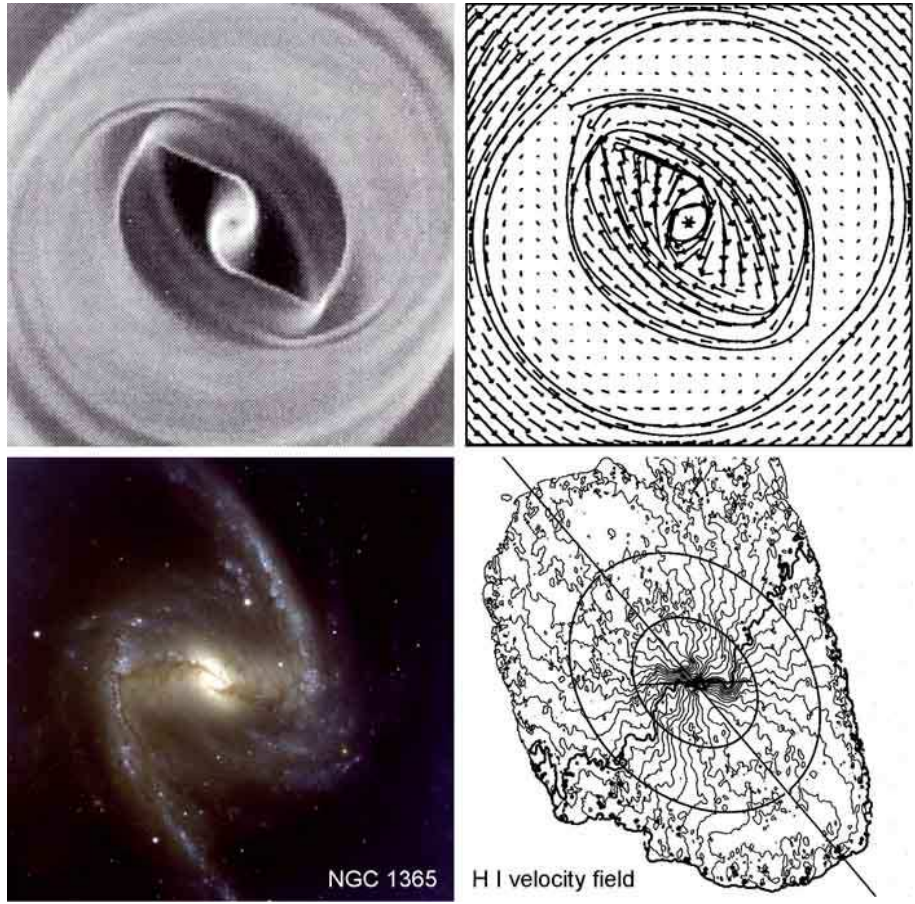}

\caption{Comparison of the gas response to a bar (Athanassoula 1992 model 1) with NGC 1365.
The galaxy image is from the VLT and is reproduced courtesy of ESO.  In the models, the bar 
potential is oriented at $-45^{\circ}$ to the vertical, parallel to the lowest-density (dark)
part of the gas density grayscale distribution at upper-left.  The bar axial ratio is 0.4 and its 
length is approximately half of the box diagonal.  The top-right panel shows the velocity field;
arrow lengths are proportional to flow velocities.  Discontinuities in gas velocity indicate 
shocks at the narrow ridge lines of high gas density in the upper-left panel.   High gas 
densities are identified with dust lanes in real galaxies.  The model correctly reproduces the 
observations that dust lanes are offset in the forward rotation direction from the ridge line of
the bar, that they are offset by larger amounts nearer the center, and that very near the center, 
they curve and become nearly azimuthal.  The bottom-right panel shows the H{\ts}{\sc i} velocity 
field of NGC 1365 from Lindblad \etal (1996).  The contour interval is 20 km s$^{-1}$.  The
velocity contours crowd strongly in the dust lanes shown at lower-left (the scales of the two 
panels are slightly different).  This supports the interpretation that the dust lanes 
are signatures of shocks in the gas velocity field.  Shocks are signs that the gas 
loses energy.  It must fall toward the center.  In fact, NGC 1365 has high gas densities and
active star formation in its bright center (Lindblad 1999; Curran \etal 2001a, b).  
Adapted from Fig.~7 of Kormendy \& Kennicutt (2004).}

\end{figure}

\cl{\null}
\vskip -28pt
\eject

      The consequence of shocks is that they inevitably imply that gas flows toward the center.  
Because the shocks are nearly radial, gas impacts them at a steep angle.  Therefore the velocity that is
lost in the shock is mainly rotational.  This robs the gas of energy and forces it to fall toward the center.

      Important confirmation of these results is provided by the observation that H{\ts}{\sc i}
gas velocity contours crowd closely just where we observe dust lanes in optical images (e.{\ts}g., NGC 1365
in Fig.~1.26 and NGC 1530 in Fig.~1.27).  The H{\ts}{\sc i} observations do not have high enough spatial resolution
to prove that a shock must be present, but the rapid change in the velocity field coincident with the dust
lane strongly supports the above evolution picture.  

      It is easy to get a heuristic understanding of the gas velocity field shown in Fig.~1.26.
Gravitational torques produced by the bar accelerate the gas as it approaches the bar.  As a result,
it climbs to larger radii as it crosses the ridge line of the bar.  Then, as it leaves the bar, the bar
potential minimum is behind the gas and decelerates it.  Incoming gas overshoots a little before it plows 
into the departing gas, so the shocks are nearly radial but offset from the ridge line of the bar in the 
rotationally forward direction.

\vfill

\begin{figure}[hb]


 \includegraphics{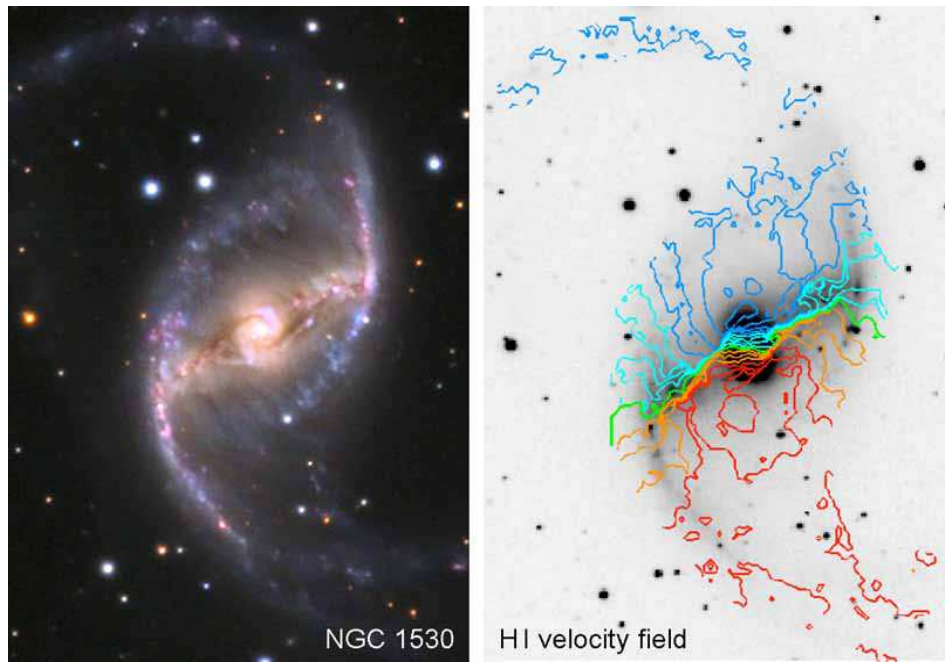}

\caption{NGC 1530 is another excellent example of the crowding of H{\ts}{\sc i} velocity contours at the
         position of the dust lane in an SBb galaxy.  The image is from
         {\tt http://www.caelumobservatory.com/gallery/n1530.shtml} courtesy of Adam Block.
         The velocity field is from Regan \etal (1997) courtesy of Michael Regan.}

\end{figure}

\eject

      Do observations show central concentrations of gas in barred galaxies and in other galaxies (e.{\ts}g.,
oval disks) in which the above secular evolution is expected to happen?  And do they {\it not\/} show such
central gas concentrations in the {\it absence\/} of engines for secular evolution?  Figure 1.28 provides six
examples of the general result that the answer is ``yes''.

\vfill

\vskip 2.truein

\begin{figure}[hb]


 \includegraphics{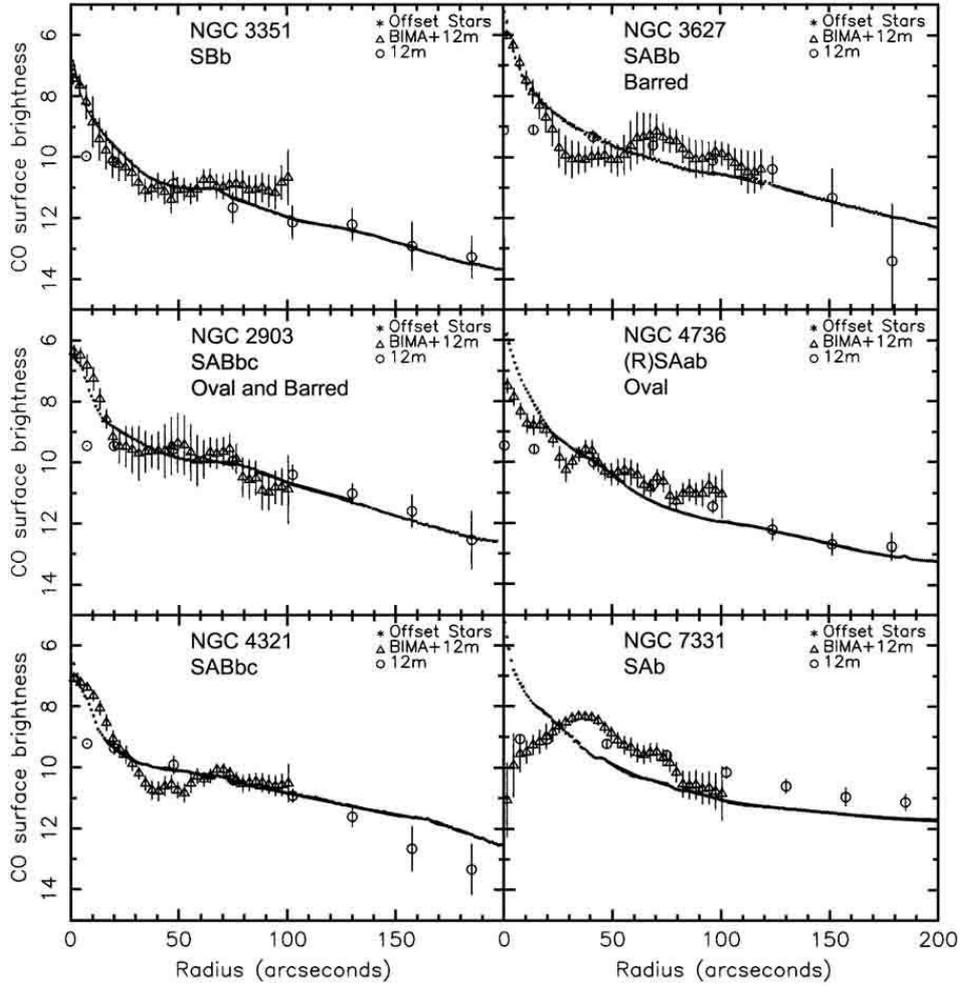}

\caption{Radial profiles of CO gas and stellar $K$-band surface brightness from the BIMA SONG (Fig.~20 from Kormendy \&
         Kennicutt 2004, adapted from Regan \etal 2001).  CO surface brightness is in magnitudes of Jy km s$^{-1}$ arcsec$^{-2}$ 
         with zeropoint at 1000 Jy km s$^{-1}$ arcsec$^{-2}$.  The stellar surface brightness profiles have been shifted 
         vertically to the CO profiles. Morphological types are from the RC3 (de Vaucouleurs \etal 1991) and oval disks are 
         identified in Kormendy  \& Kennicutt (2004).   NGC 4736 is a prototypical pseudobulge also illustrated  in 
         Figs.~1.3, 1.6 and 1.8.  
         All galaxies in this figure except NGC 7331 have structures that are expected to cause gas to flow toward the center.  
         NGC 7331 is included to show the very different CO profile in a galaxy with a probable classical bulge (cf.~Fig.~1.29). 
         }

\end{figure}

\cl{\null}
\vskip -28pt
\eject

     What happens to the infalling gas?  Concentrated into a small volume, it gets very dense.
Crunching gas likes to make stars.  The Schmidt~(1959)~-- Kennicutt (1989, 1998a, 1998b) law that star formation
rates increase faster than linearly with gas density makes this explicit (Fig.~1.32 here).  Confirming our
expectations, observations point to enhanced star formation, often in spectacular starbursts, near the centers of barred
and oval galaxies.  In particular, {\it Spitzer Space Telescope\/} mid-infrared observations in the 24 $\mu$m bandpass 
are sensitive to warm dust that reradiates light from hot young stars (Fig.~1.29).  Survey results by Fisher (2006) and 
Fisher \etal (2009) show ubiquitous central star formation in medium- and late-type barred and oval galaxies but not
in galaxies with classical bulges (Fig.~1.29).

\cl{\null}

\vfill

\vskip 2.truein

\begin{figure}[hb]


 \includegraphics{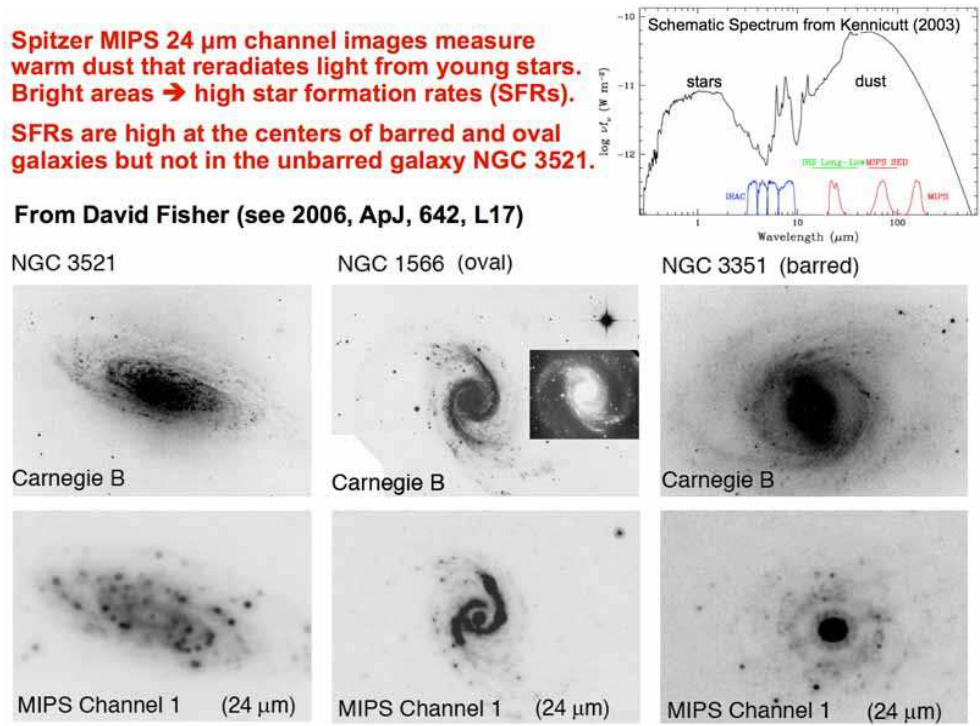}

\caption{Powerpoint slide showing typical high central star formation rates (SFRs) in galaxies
         that contain driving agents for secular evolution (bars and oval disks) but not in galaxies
         that contain no driving agent (NGC 3521).~A schematic galaxy spectrum (top right, from
         Kennicutt 2003) shows the separate, roughly black body spectra of stellar populations in galaxies
         (``stars'') and of warm dust (``dust'').  The {\it Spitzer Space Telescope} MIPS 24 $\mu$m bandpass
         is sensitive to dust that reradiates light from young stars. Thus SFRs are high where galaxies are bright at
         24 $\mu$m.  The upper row of images from the {\it Carnegie Atlas of Galaxies}
         (Sandage \& Bedke 1994) show typical galaxies in $B$ band.  The bottom row of {\it Spitzer\/} 
         24 $\mu$m images show high central SFRs in the pseudobulges of NGC 1566 and NGC 3351 but not in
         the classical bulge of NGC 3521.  This figure appears courtesy of David B.~Fisher.}

\end{figure}

\cl{\null}
\vskip -28pt
\eject

      More detailed views of this star formation are shown in Figs.~1.30~and~1.31.  Figure 1.30 shows NGC 5236, a particularly close 
and well known example.  The dust lanes on the rotationally trailing sides of the spiral arms are believed to have the same cause
as the ones on the leading sides~of~the~bar: they trace shocks where the gas enters the spiral arms.  Star formation is triggered
there by gas compression.  Since $\Omega > \Omega_p$, the gas moves forward, beyond the shock, during the time that it takes
stars to form.  As a result, the ridge line of bright young stars and H{\ts}{\sc ii} regions that ``are strung out like pearls along the arms''
(Baade 1963) is offset forward of the dust lanes.  Our picture of how spiral density waves stimulate star formation and hence are traced by 
young stars is discussed in
Roberts (1969);
Dixon (1971);
Shu \etal (1973) and
Roberts \etal (1975) and is reviewed in
Toomre (1977b).

      Returning to bar-driven inward gas transport and its consequences, the central regions of NGC 5236 are dominated by 
intense star formation 
(Harris \etal 2001;
Knapen \etal 2010).  
Similarly, the whole of the central region of NGC 1365 (Fig.~1.26) is undergoing a starburst 
(Kristen \etal 1997;
Galliano \etal 2008;
Elmegreen \etal 2009)
A less intense but still prototypical example is NGC 1300 (Fig.~1.6; Knapen \etal 2006; Comer\'on \etal 2010).

Spiral galaxies in which there is no significant ILR and in which the spiral structure extends to the center
can also have central starbursts (for example, NGC 4321,
Knapen et al.~1995a, b; 
Sakamoto et al.~1995).

\vfill

\begin{figure}[hb]


 \includegraphics{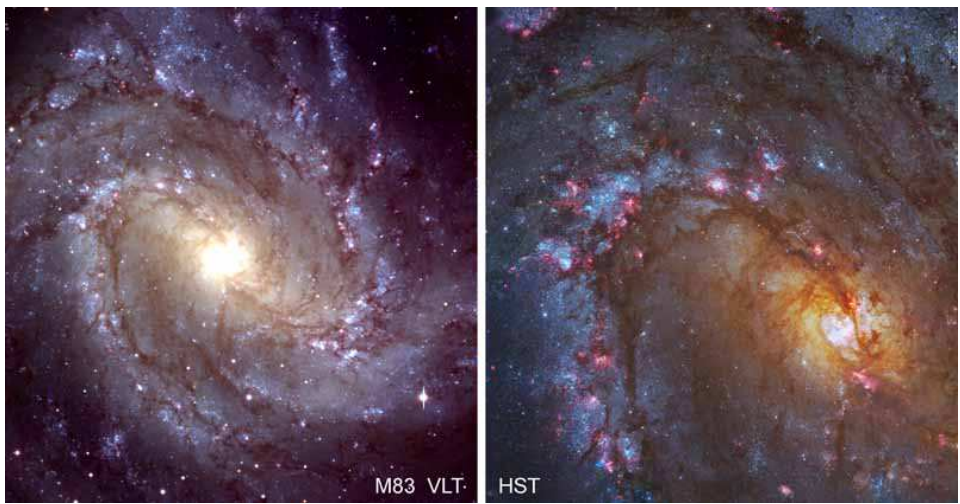}

\caption{M{\ts}83 = NGC 5236 is a prototypical face-on SAB(s)c galaxy with dust lanes on the trailing sides of the
         spiral arms and the leading sides of the bar.  The center is undergoing
         a spectacular starburst.  The left image is from
         the ESO VLT ({\tt http://www.eso.org/public/images/eso9949a/} with color balance tweaked 
         to better match the enlargement at right, which is a Hubble Heritage image). 
         \omit{({\tt http://heritage.stsci.edu/2009/29/index.html}).} 
         }

\end{figure}

\eject

      Nuclear starburst rings are particularly compelling examples of the star formation that -- we suggest -- results 
from inward transport~of~disk~gas.  Figure 1.31 shows examples. Three of these galaxies are barred, but NGC~4736 
is in my prototypical unbarred, oval galaxy (Figs.~1.3, 1.6, 1.8).  It emphasizes again that numerical simulations 
and observations both imply that oval galaxies evolve secularly in the same way that barred galaxies do.

\vfill

\begin{figure}[hb]

\vspace{4.45 truein}


 \includegraphics{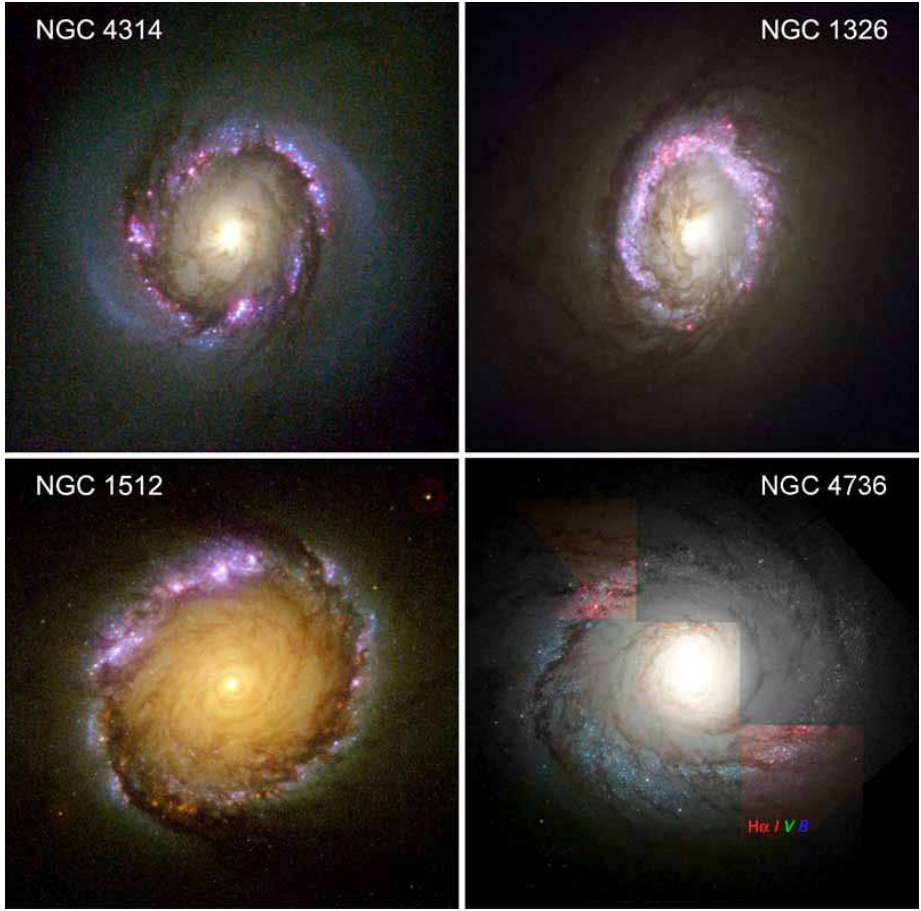}

\caption{Nuclear star-forming rings in barred and oval galaxies (update of Fig.~8 in Kormendy \& Kennicutt 2004). 
         Sources: NGC 4314 -- Benedict \etal (2002);
                  NGC 1326 -- Buta \etal (2000) and Zolt Levay (STScI);
                  NGC 1512 -- Maoz \etal (2001); 
                  NGC 4736 -- Zolt Levay (STScI).  
         The NGC 4736 panel was made from nonoverlapping images in the bandpasses indicated by the colors used in
         these RGB renditions.~All colors are available in only two places around the ring, but they make it clear 
         that this star-forming ring in a prototypical oval galaxy (see Figs.~1.3, 1.6 and 1.8)
         is closely similar to the other nuclear rings, which occur in barred galaxies.  Note that these
         nuclear rings are distinct from and always smaller than the inner rings that encircle bars (see
         point (f) in Section 1.3.1).
         }

\end{figure}

\eject

      The star formation discussed in this section is not associated with galaxy mergers and instead is closely
connected with probable engines for secular evolution and with observed features (such as radial dust lanes
in bars) which suggest that such evolution is in progress.  We conclude with some confidence that it is building
pseudobulges.  Are the gas supplies and star formation rates (SFRs) consistent with reasonable growth times for
pseudobulges of the observed masses?  

      Kormendy \& Kennicutt (2004) address this for nuclear star-forming rings (Fig.~1.32).  The rings have larger
SFR densities $\Sigma_{\rm SFR}$ and total gas densities $\Sigma_{\rm gas}$ than do spiral galaxies in general, 
defining a Schmidt-Kennicutt relation similar to the well known  $\Sigma_{\rm SFR} \propto \Sigma_{\rm gas}^{1.4}$.
Absent continued gas infall, the \phantom{00000000000000000}

\vfill

\begin{figure}[hb]


 \includegraphics{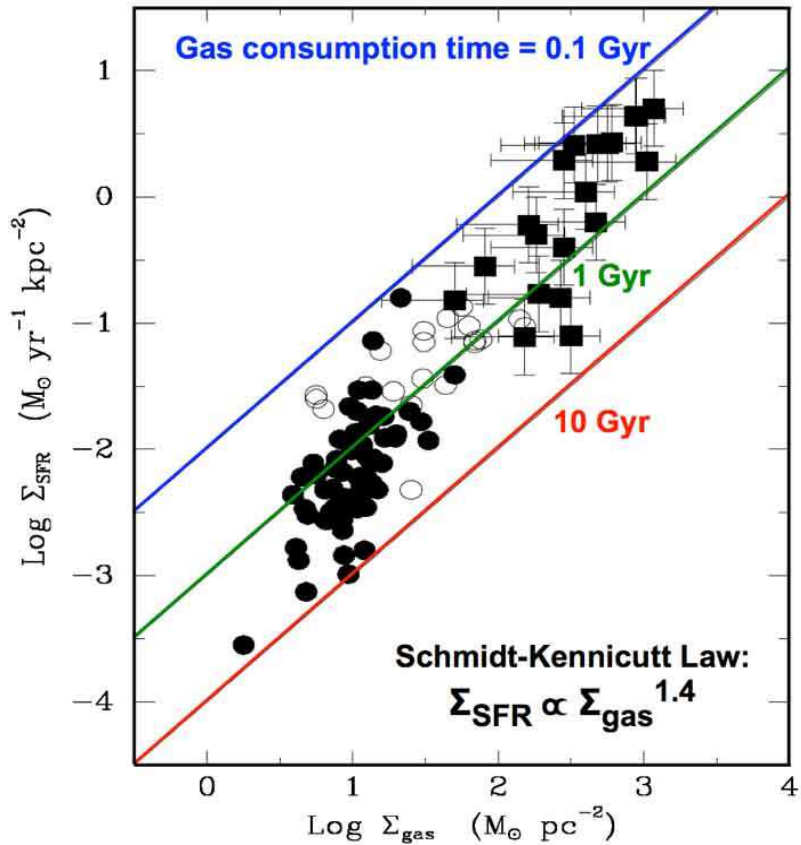}

\caption{Correlation between SFR surface density and   total gas surface density for circumnuclear star-forming rings
  (filled squares) compared to disk-averaged values for spiral galaxies (filled circles) and the 
   centers of these galaxies (open circles).  The nuclear ring data are compiled in Kormendy \& Kennicutt (2004, from
   which this figure is taken); the other data are from Kennicutt (1998b).  Diagonal lines show
   gas consumption timescales if no additional gas is supplied to the central region.
         }

\end{figure}

\eject

\noindent currently available gas would be consumed in $\sim$\ts0.2\ts--\ts2 Gyr.  However, I have argued that the
central gas is continually replenished.  Thus the observed SFRs of 
0.1\ts--\ts10 $M_\odot$ yr$^{-1}$ are consistent with the formation of pseudobulges with masses $\sim$\ts$10^8$\ts--\ts$10^{10}$
$M_\odot$ over several Gyr.  These values are reasonable.  

      Fisher \etal (2009) and Fisher \& Drory (2010) use the {\it Spitzer Space Telescope\/} to study SFRs and star formation
histories in classical and pseudo bulges in detail.  They conclude (2009) that ``All bulges are found to be forming 
stars irrespective of bulge type (pseudobulge or classical bulge).   [However,] classical bulges have the lowest specific SFR 
[SFR per unit stellar mass], implying growth times that are longer than a Hubble time, thus the present-day SFR does not 
likely play a major role in the evolution of classical bulges.  [In contrast,] at present-day SFRs, the median pseudobulge 
could have grown the present-day stellar mass in 8 Gyr.  In almost all galaxies in our sample, the specific SFR of the bulge 
is higher than that of the outer disk.
This suggests that almost all galaxies are increasing their $B/T$ through internal star formation.  The SFRs in pseudobulges 
correlate with their structure.  More massive pseudobulges have higher SFR density, this is consistent with the stellar 
mass being formed by moderate, extended star formation. Bulges in late-type galaxies have similar SFRs as pseudobulges in 
intermediate-type galaxies and are similar in radial size. However, they [have lower masses]; thus, they have much shorter 
growth times, $\sim$\ts2 Gyr. Our results are consistent with a scenario in which bulge growth via internal star formation 
is a natural and near ubiquitous phenomenon in disk galaxies.''

      I want to emphasize the contrast between star formation in major mergers and star formation during secular evolution 
(Kormendy \& Kennicutt 2004).  Merger starbursts last \lapprox ~{\kern -1pt}a few hundred million years.  Most classical
bulges and ellipticals are seen long after the assembly events that constructed them.  Therefore, they are mostly seen
to contain old stars.  In contrast, secular
 evolution is inherently long-term.  Except in S0 pseudobulges, we usually see star formation in action.  
{\it If star formation is ubiquitous, it must be secular.}  
As a result, ongoing star formation that is not observed to be associated with morphological indicators of a merger in progress 
(such as tidal tails) is the first pseudobulge classification criterion listed in Section 1.5.3.

      {\it We have a detailed picture of internal secular evolution in galaxy disks.  Central molecular gas concentrations 
and starbursts are closely associated with bars and oval disks that act as evolution engines.  They are not generally 
found in classical bulges.  They are also closely associated with dust lanes and H{\ts}I velocity crowding in bars that are
signatures of secular evolution.  These correlations argue in favor of internal secular evolution and against the 
idea that these features could be produced by large numbers of minor mergers.}

\subsection{The observed properties of pseudobulges}

      Quoting Kormendy \& Kennicutt (2004): ``How can we tell whether a (pseudo)bulge is like an elliptical or 
whether it formed secularly?  The answer and the theme of this section is that pseudobulges retain enough
memory of their disky origin so that the best examples are easily recognizable.''   As we move from ``proof of concept'' 
into work on large samples of galaxies, we will have to face the difficulty that classification gets difficult when --
as we must expect -- both a classical bulge and a subsequently grown pseudobulge are present.  In such cases, bulge-pseudobulge
decomposition is necessary (e.{\ts}g., Erwin \etal 2003). In this 
section, I discuss the properties of more prototypical, pure pseudobulges to show how they differ from classical bulges
and ellipticals.  Section 1.5.3 then lists the classification criteria, numbered according to the subsections in the following
discussion.

\subsubsection{Pseudobulges of spiral galaxies show ongoing star formation}

      As noted in the previous section, ongoing {\it central\/} star formation in relatively normal spiral galaxies that show 
no signs of a merger in progress is a strong pseudobulge indicator.  The work by Fisher \etal (2009, 2010) shows that such
star formation is generally present in morphologically classified pseudo but not classical bulges.  If star formation
is ubiquitous, it must be secular.  

\subsubsection{Pseudobulges are flatter than classical bulges}

This is a two-part criterion.  First, when the galaxy is highly inclined and the ellipticity profile 
$\epsilon(r)$ tells us the relative flattening of various components, pseudobulges are often (not always)~seen to be flatter 
than classical bulges.  Commonly a part of them is as flat as the associated outer disks.  Second, 
pseudobulges in spiral (but not S0) galaxies usually show spiral structure all the way to the center of the galaxy.  Classical bulges, 
like ellipticals, are essentially never flatter than $\epsilon = 0.6$ .  They cannot show spiral structure.
This classification criterion can be applied even to face-on galaxies.

      The connection between pseudobulge flatness and secular evolution has been made since the earliest papers on this subject.
Kormendy (1993b) describes the prototypical pseudobulge in NGC 4736 like this: ``The central brightness profile
is an $r^{1/4}$ law that reaches the high central brightness characteristic of a bulge (Boroson 1981).  However, the $r^{1/4}$-law
component shows a nuclear bar and spiral structure to within a few arcsec of the center.  Bars are disk phenomena.  More
importantly, it is not possible to make spiral structure in a bulge.  Thus the morphology already shows that the $r^{1/4}$-law
profile belongs to a disk.  This is shown more quantitatively by [rapid rotation]'', which is discussed here in Section 1.5.2.3.

      The importance of spiral structure is further emphasized by Courteau \etal (1996): ``Many of these [late-type] galaxies 
[in their sample] show spiral structure continuing into the central regions. \dots~We invoke secular dynamical evolution and
gas inflow via angular momentum transfer and viscous transport'' as the interpretation of the disky central structure.

      Spectacular examples of disky pseudobulges often with spiral structure emerge from {\it HST} imaging surveys.  
The best known of these is by Carollo and collaborators
(Carollo \etal 1997, 1998;
Carollo \& Stiavelli 1998;
Carollo 1999;
Carollo \etal 1999, 2002;
Seigar \etal 2002).
They refer to bulges that are structurally unlike ellipticals as ``irregular bulges'' and  note that:
``The widespread presence of star formation in the irregular bulges support scenarios in which a fraction of bulges 
form relatively late, in dissipative accretion events driven by the disk.''  Figures 1.33 and 1.34 show examples.

\cl{\null}

\vfill

\vskip 2.truein

\begin{figure}[hb]


 \includegraphics{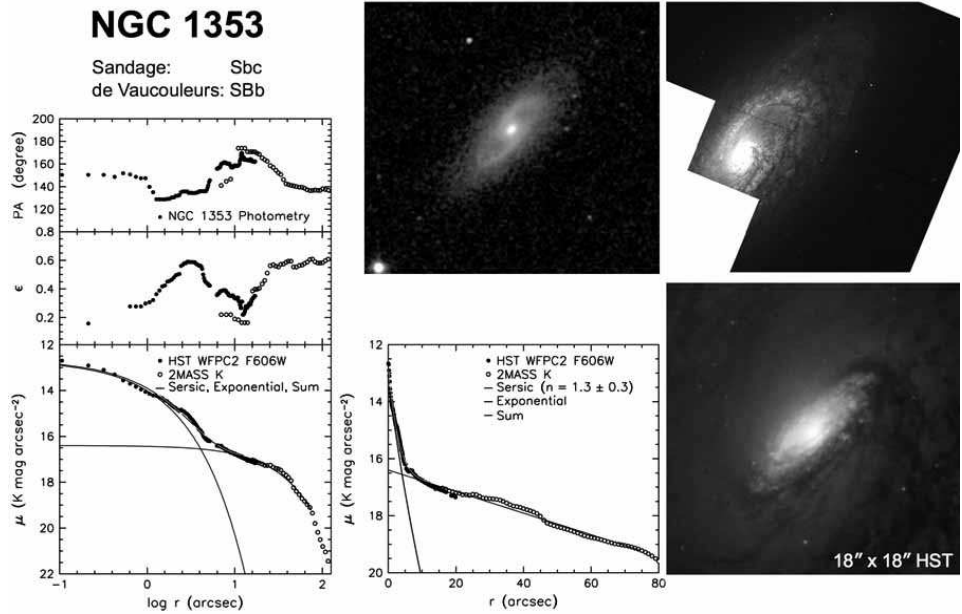}

\caption{NGC 1353 pseudobulge: the images show (left to right and top to bottom)
the 2MASS $JHK$ image with a field of view of 4\md4 $\times$ 4\md4 (Jarrett \etal 2003),
an {\it HST\/} WFPC2 F606W image (Carollo \etal 1998), and 
an 18$^{\prime\prime}$ $\times$ 18$^{\prime\prime}$ zoom of the F606W image.
The plots show surface photometry with the {\it HST\/} profile shifted to the $K$ band; 
$\mu$ is surface brightness, $\epsilon$ is ellipticity and PA is position~angle. 
A bulge-disk decomposition of the major-axis profile into a S\'ersic function plus an exponential disk
(curves) shows that a ``bulge'' dominates at radii $r$ \lapprox \ts10$^{\prime\prime}$.  
This component is identified as a pseudobulge 
(1) because it has the same apparent flattening as the disk (compare the 18$^{\prime\prime}$ and large-field views),
(2) because it shows small-scale spiral structure that can only be sustained in a disk,
(3) because patchiness indicates the presence of dust and star formation and
(4) because its S\'ersic index $n = 1.3 \pm 0.3$ is less than 2.  From Kormendy \& Kennicutt (2004) which shows
more examples.
         }

\end{figure}

\cl{\null}
\vskip -28pt
\eject

      Figure 1.34 shows {\it HST\/} images of the central parts of Sa\ts--{\ts}Sbc galaxies.  All show flat shapes, spiral structure, of
patchy star formation.  These are the central regions that, by the surrogate definitions, would be identified as the
galaxies' bulges.  It seems safe to conclude that no one who saw these images would define bulges as mini-ellipticals that live
at the centers of disks.

\vfill

\vskip 2.truein

\begin{figure}[hb]


 \includegraphics{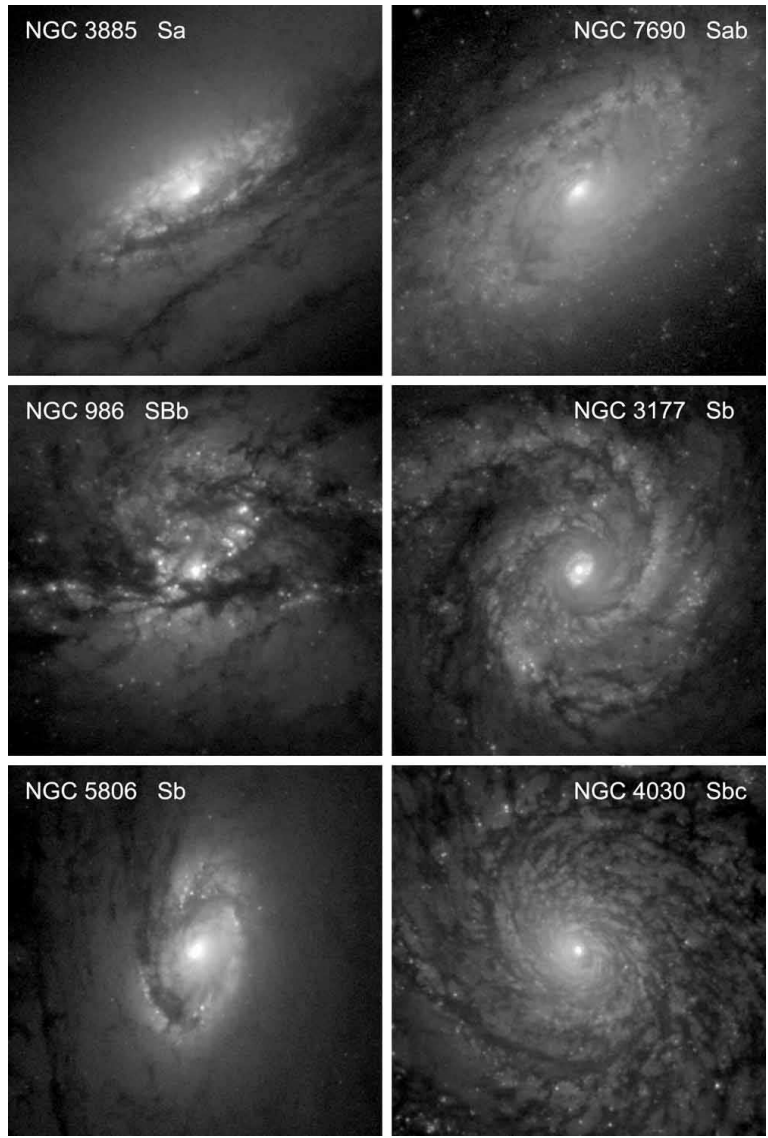}

\caption{Sa -- Sbc galaxies whose ``bulges'' have disk-like properties.  Each panel shows an 18$^{\prime\prime}$ $\times$
18$^{\prime\prime}$ region around the galaxy center extracted from {\it HST\/} WFPC2 F606W images taken and kindly provided 
by Carollo \etal (1998).  North is up and east is at left.  Intensity is proportional to the logarithm of the galaxy surface 
brightness.  From Kormendy \& Kennicutt (2004).
         }

\end{figure}

\cl{\null}
\vskip -28pt
\eject

\subsubsection{Pseudobulges are more rotation-dominated than classical bulges}

      Pseudobulges were first recognized because rotation is more important with \hbox{respect to random motions than it is in classical 
bulges.  Via~the~\hbox{$V_{\rm max}/\sigma$\ts--\ts$\epsilon$}} diagram (Fig.\ts1.35) for galaxies such as NGC\ts3945 and NGC\ts4736,
Kormendy (1982a,{\ts}b) concluded that some ``SB bulges are more disklike than SA bulges'' and: ``{\it a significant 
fraction of the bulge in many SB galaxies may consist of disk gas which has been transported to the center by the bar.~As the gas accumulates, 
it forms stars and builds up a centrally concentrated stellar distribution which is photometrically like a bulge but dynamically like a disk.}'' 
Again, the fundamental ideas about secular evolution have been with us for a long time.  These results are brought up to date with more 
recent long-slit spectroscopy in Fig.\ts1.35.  The result is that $V_{\max}/\sigma$ is larger at a given $\epsilon$ in pseudobulges 
(most filled symbols) than in classical bulges (open symbols) or in ellipticals (crosses).

\vfill

\vskip 2.truein

\begin{figure}[hb]


 \includegraphics{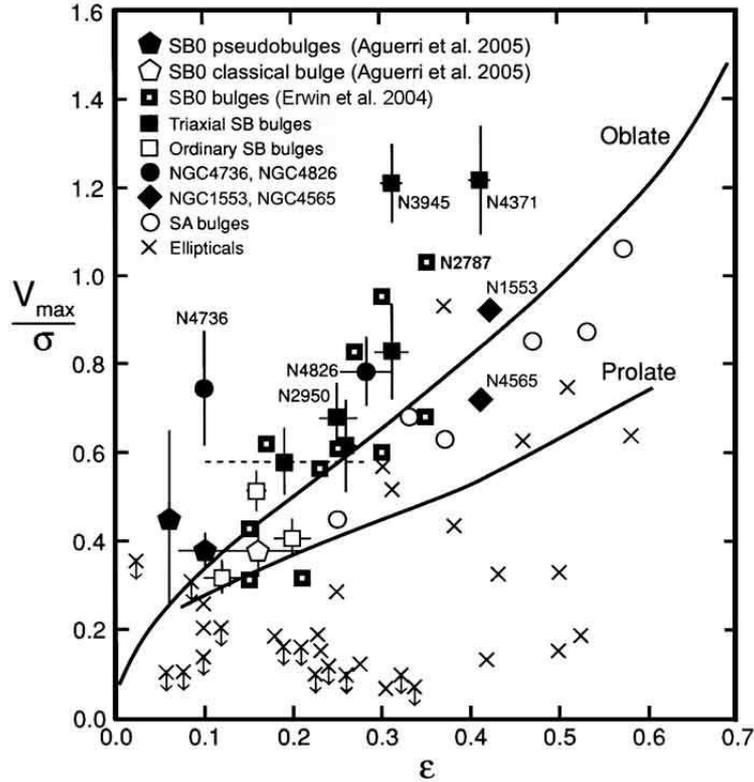}

\caption{Relative importance of rotation and random velocity as a function 
of ellipticity $\epsilon$ = (1\thinspace$-${\thinspace}axial ratio) for various kinds of
stellar systems.  Here $V_{\rm max}/\sigma$ is the ratio of the maximum rotation
velocity to the mean velocity dispersion interior to $r_e$.  The 
``Oblate'' line describes oblate-spheroidal systems that have
isotropic velocity dispersions and that are flattened by rotation; 
it is a consequence of the tensor virial theorem (Binney \& Tremaine 1987).  
From Kormendy \& Fisher (2008).
         }

\end{figure}

\cl{\null}
\vskip -28pt
\eject

      The disky dynamics of pseudobulges are beautifully illustrated by the SAURON team's integral-field
spectroscopy.  This result is particularly important, so Figs.~1.36 and 1.37 show three examples in detail.
In Fig.~1.36, the surface brightness profile tells us the part of NGC 4274 that the surrogate definitions
would identify as the bulge.  It dominates at $r$ \lapprox \ts10$^{\prime\prime}$.  The two-dimensional
kinematic and line-strength maps then show that, compared to the rest of the inner parts of the galaxy,
the component at $r$ \lapprox \ts\ts10$^{\prime\prime}$ rotates more rapidly, has a lower velocity dispersion,
and has stronger H{\ts}$\beta$ absorption lines.  The latter result means that the central disky structure
is made of younger stars than the rest of the galaxy.  As Falc\'on-Barroso \etal (2006) and Peletier \etal 
(2007a, b; 2008) note, all this is very consistent with our picture of bar-driven secular evolution.

      Figure 1.37 shows closely similar results for the SB galaxies NGC~3623~and NGC 5689.  It is
remarkable how clearly the central component is a rapidly rotating, cold disk in all three
galaxies. The younger ages of the pseudobulges also supports our evolution picture.  Note again that
all three galaxies are barred.

\vfill

\vskip 2.truein

\begin{figure}[hb]



 \includegraphics{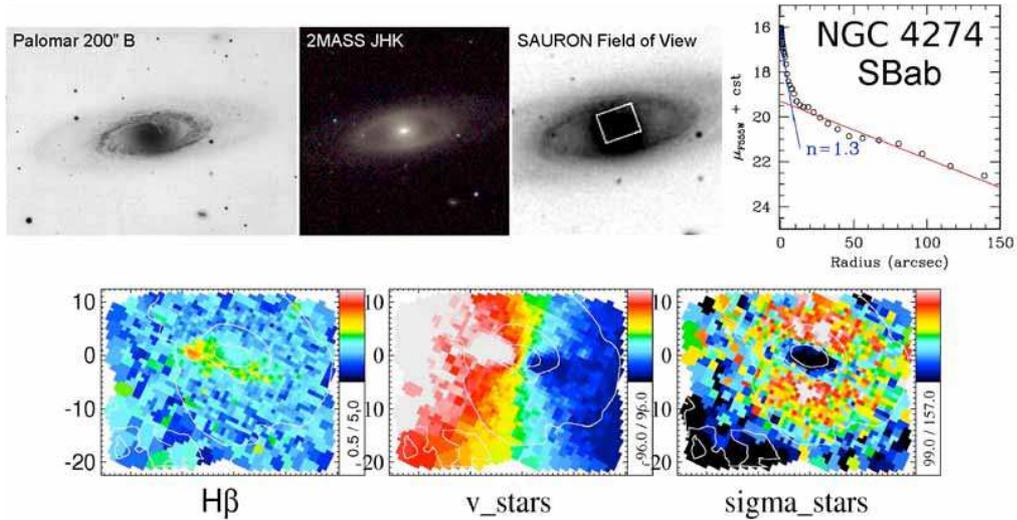}

\caption{SAURON integral-field spectroscopy of the disky pseudobulge of the Sa
galaxy NGC 4274.  The galaxy is barred, but the bar is foreshortened, because it is
oriented nearly along the minor axis.  It fills an inner ring, as is normal in
SB(r) galaxies (Kormendy 1979b).~The brightness profile (upper-right)
is decomposed into a S\'ersic (1968) function plus an exponential disk.  The
S\'ersic function has $n = 1.3$, i.{\thinspace}e., $n < 2$, as in other pseudobulges
(Section 1.5.2.6).  The pseudobulge dominates the light at radii $r$~\lapprox 
\thinspace10$^{\prime\prime}$.  The kinematic maps (Falc\'on-Barroso et al.~2006)
show that this light comes from a disky component that is more rapidly rotating (center), 
lower in velocity dispersion (right), and stronger in H$\beta$ line strength (left,
from Peletier et al.~2007a) and hence younger than the rest of the inner galaxy.
From Kormendy \& Fisher (2008) as adapted from Peletier \etal (2007b, 2008).
         }

\end{figure}

\cl{\null}
\vskip -28pt
\eject

\cl{\null}

\begin{figure}[ht]

\vskip 4.79truein



 \includegraphics{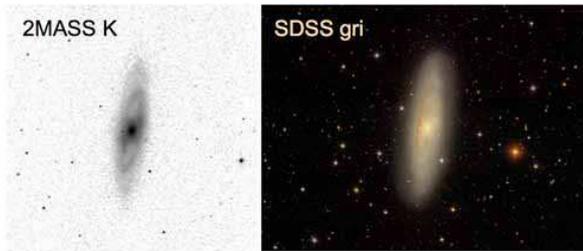}

 \includegraphics{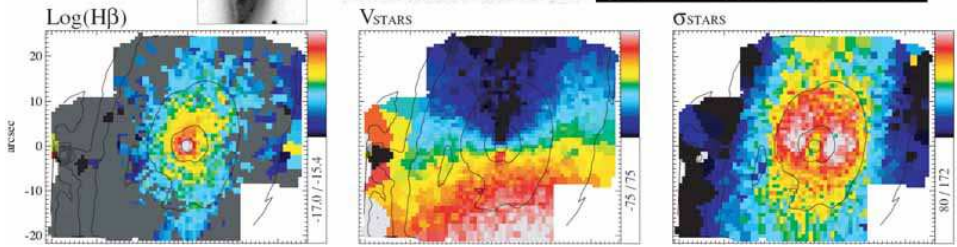}

\caption{SAURON integral-field spectroscopy of the disky pseudobulges of the SB0
galaxy NGC 5689 and the SABa galaxy NGC 3623 (adapted from Falc\'on-Barroso \etal 2006).
Again, both bars are oriented nearly along the minor axis.  As in NGC 4274, the bar
in NGC 3623 fills an inner ring.  Also as in NGC 4274, the kinematic maps show that 
both galaxies contain pseudobulges that are prominently disky and more rapidly rotating (center), 
lower in velocity dispersion (right), and stronger in H$\beta$ line strength (left) and 
hence younger than the rest of the inner galaxy.  It is important to emphasize that
pseudobulges generally show more than one (here: three) of the classification criteria
listed in Section 1.5.3).
         }

\end{figure}

\cl{\null}
\vskip -40pt
\cl{\null}

\subsubsection{Pseudobulges have small velocity dispersions for their luminosities}

      Pseudobulges have smaller $\sigma$ than do classical bulges of the same $M_B$ 
(e.{\ts}g., Kormendy \& Illingworth 1983; 
Kormendy \& Kennicutt 2004;
Gadotti \& Kauffmann 2009; 
cf.~Falc\'on-Barroso \etal 2006;
Peletier \etal 2007a).~This could partly be due to low $M/L_B$ ratios resulting from young stars.

\clearpage

\subsubsection{Almost all pseudobulges have bulge-to-total ratios $PB/T$ \lapprox \ts1/3.\\
               Bulges with $B/T$ \gapprox \ts1.2 are classical.}

      Fisher\ts\&{\ts}Drory (2008) study the properties of \hbox{(pseudo)bulges~in~79, S0\ts--{\ts}Sc} galaxies to clarify our 
classification criteria.  They show that, in general, different classification criteria agree in distinguishing pseudobulges 
from classical bulges.  To do this work, they need first to classify bulges using purely morphological criteria (Fig.~1.38).  
This has the disadvantage that morphology alone cannot always identify pseudobulges in S0 galaxies.
But it lets them study a large sample, and it allows a fair test of (for example) the already strong hint (see
Kormendy \& Kennicutt 2004) that the S\'ersic indices of pseudobulges are smaller than those of classical bulges.

\vfill

\begin{figure}[hb]


 \includegraphics{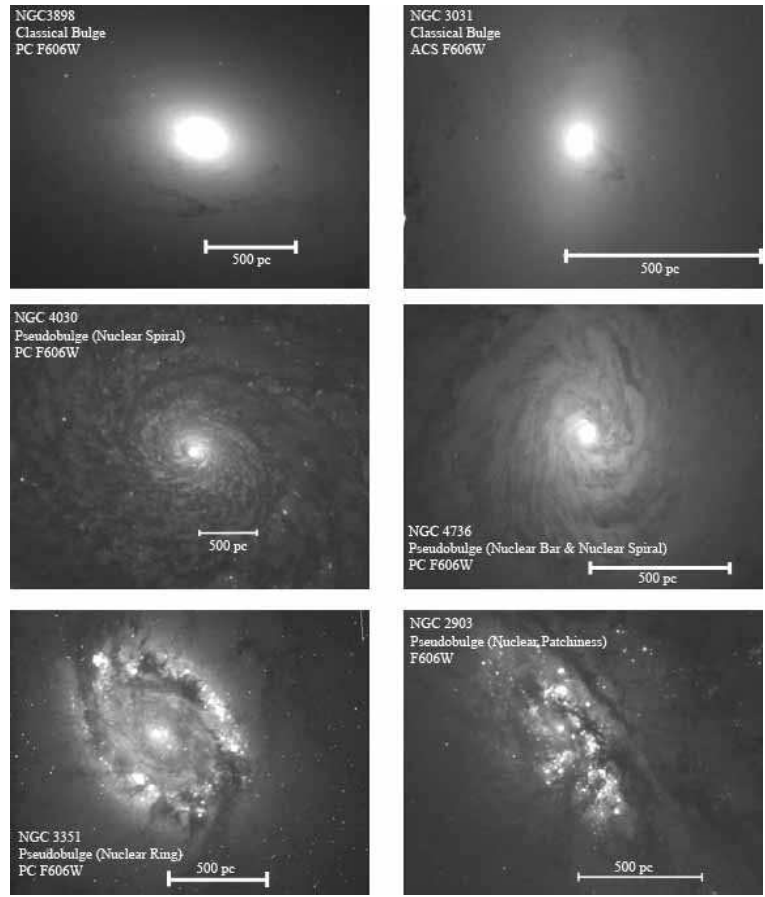} 

\caption{Morphological features used by Fisher \& Drory (2008; 2010) to classify bulges as classical or pseudo. 
Classical bulges have smooth, nearly featureless and nearly elliptical isophotes as do elliptical galaxies
(NGC 3998, NGC~3031~=~M{\ts}81).  Pseudobulges show one or more of the following:
nuclear bars, 
spiral structure, or 
patchy star formation (bottom four galaxies).
From Fisher \& Drory (2010).
         }

\end{figure}

\cl{\null}
\vskip -28pt
\eject

      Fisher \& Drory (2008) measure surface brightness profiles of their galaxies using {\it Hubble Space Telescope\/} 
({\it HST\/}) archival images, their own large-field images from the McDonald Observatory 0.9 m telescope and images 
from archives such as the Sloan Digital Sky Survey (SDSS).  Combining data from many sources provides accurate 
profiles over wide dynamic~ranges.  They decompose the brightness profiles into disk components with exponential 
profiles and (pseudo)bulge components with S\'ersic (1968) function profiles.  Figure 1.39 is an example.  

      This work further confirms the pseudobulge properties discussed in previous sections and provides
several new results:

      About half of early-type galaxies contain pseudobulges; almost all Sbc galaxies contain pseudobulges, and 
as far as we know, no Sc or later-type galaxy has a classical bulge (see Fig.~1.40).  This is in excellent agreement
with results discussed in Kormendy \& Kennicutt (2004).

      Pseudobulges are often as flat as their associated disks; classical bulges are thicker
than their associated disks.  NGC 3031\ts={\ts}M{\ts}81 and NGC 3169 in Fig.~1.39 are examples.

\vfill

\begin{figure}[hb]


 \includegraphics{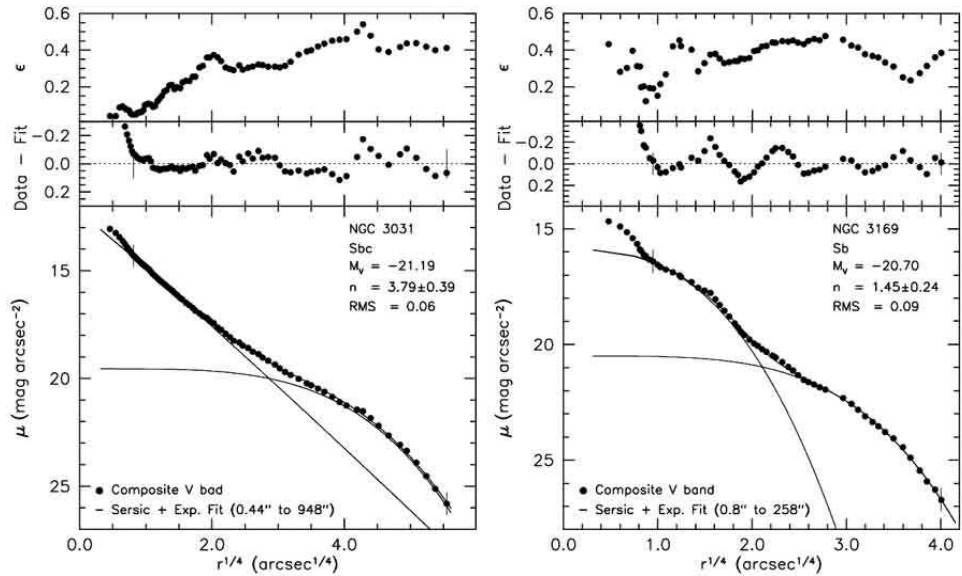} 

\caption{Photometric decompositions of a galaxy with a classical
bulge (M81,{\ts}left) and one with a pseudobulge (NGC 3169).  Classification is 
carried out as in Fig.~1.38.  In the bottom panels, the filled circles are 
the major-axis profiles and the solid curves are the S\'ersic function (pseudo)bulge, 
the exponential disk and their sum, which is almost invisible atop the data points.
The fit is made between the vertical dashes; note that both galaxies contain
nuclear star clusters (``nuclei'') in addition to their bulges.
The S\'ersic index and the fit RMS are given in the key.  The middle panel shows the
deviations of the fit from the data in more detail.  The top panel shows the
total ellipticity profile.   From Fisher \& Drory (2008).
         }

\end{figure}

\cl{\null}
\vskip -28pt
\eject

\cl{\null}

\begin{figure*}

\vskip 3.3truein


 \includegraphics{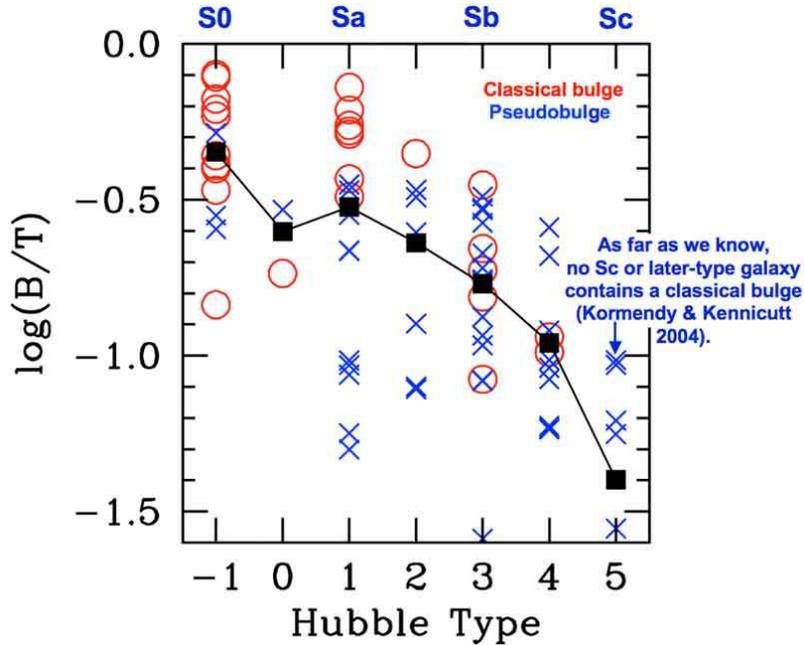} 

\caption{Logarithm of (pseudo)bulge-to-total luminosity ratio $B/T$ versus
Hubble type.  Bulges are classified as in Fig.~1.38.  Black squares show how the average $B/T$
for pseudobulges and classical bulges together correlates with Hubble type as 
expected from its use as a Hubble classification criterion (Sandage 1961). Note: some
pseudobulges have log $B/T$\ts$<$\ts$-1.6$.  From Fisher \& Drory (2008).
         }

\end{figure*}

\cl{\null}
\vskip -28pt

     Pseudobulge-to-total luminosity ratios are almost always $PB/T$ \lapprox \ts0.35 (Fig.~1.40),
consistent with the expectation that one cannot secularly convert {\it almost all\/} of a disk into a bulge 
(Kormendy \& Kennicutt 2004).  Small bulge-to-total luminosity ratios $B/T$ do not guarantee that a 
bulge is pseudo, but large $B/T$ \gapprox \ts0.5 guarantee that it is classical.

\subsubsection{Most pseudobulges have S\'ersic $n < 2$.  Classical bulges have $n \geq 2$.}

      Figures 1.33 and 1.36 show that the pseudobulges in NGC 1353 and NGC 4274 have S\'ersic indices
$n = 1.3 \pm 0.3$; i.{\ts}e., almost-exponential brightness profiles.  The pseudobulge in NGC 3169
(Fig.~1.39) has $n = 1.45 \pm 0.24$, which is still  close to exponential.  In contrast, the classical
bulge of M{\ts}81 has $n = 3.79 \pm 0.39$.  This difference turns out to be a general result.

      Kormendy \& Kennicutt (2004) review the history of this result.~The~idea that bulges in
late-type galaxies have nearly exponential (not $r^{1/4}$-law) brightness profiles dates back to 
work by Andredakis \& Sanders (1994) and Andredakis \etal (1995).  At first, no connection was made 
to pseudobulges; the result was just that $n$ is smaller in bulges of later-Hubble-type galaxies.

\eject

      Then Courteau \etal (1996) carried out bulge-disk decompositions for $>$\ts300 galaxies
and concluded that $>$ 2/3 of them -- especially at late Hubble types -- are best fitted by 
double exponentials, one for the bulge and one for the disk.   As a diagnostic of formation processes, 
Courteau \etal (1996) went on to examine the ratio $h_b/h_d$ of the scale lengths  of the inner and outer
exponentials. They found that $h_b/h_d = 0.08 \pm 0.05$ and concluded that: ``Our measurements of exponential 
stellar density profiles [in bulges] as well as a restricted range of [bulge-to-disk] scale lengths
provide strong observational support for secular evolution models.
Self-consistent numerical simulations of disk galaxies evolve toward a double
exponential profile with a typical ratio between bulge and disk scale lengths
near 0.1 (D.~Friedli 1995, private communication) in excellent agreement with
our measured values''.

     Fisher \& Drory (2008) find closely similar results.  The effective radii $r_e$ of pseudobulges and the 
scale lengths $h_d$ of their associated disks are coupled, whereas the same is not true for classical
bulges.  Specifically, the mean ratio is ${<}r_e/h_d{>} = 0.21 \pm 0.10$ for 53 pseudobulges
but ${<}r_e/h_d{>} = 0.45 \pm 0.28$ for 26 classical bulges.  Since $r_e = 1.678{\ts}h_b$ for an 
exponential, the above result for pseudobulges corresponds approximately to 
${<}h_b/h_d{>} = 0.13 \pm 0.06$ (the conversion is approximate because the pseudobulges
are not exactly exponentials).  This confirms the results by Courteau \etal (1996) and 
similar results by MacArthur \etal (2003).  Courteau and MacArthur did not calculate $h_b/h_d$
separately for classical and pseudo bulges and therefore found a correlation with Hubble type 
rather than different distributions of $h_b/h_d$ for the two types of bulges. But Courteau and MacArthur, 
like Fisher and Drory, interpreted the connection between bulge and disk properties as a sign
that secular evolution and pseudobulge formation become more important at later Hubble
types. 

      As a result of this and other work (see especially Weinzirl \etal 2009), the connection between pseudobulges and 
small S\'ersic indices is by now well established.  Classical bulges, on the other hand, have S\'ersic indices like 
those of elliptical galaxies, which have $n$ \gapprox \ts2 (e.{\ts}g., Kormendy \etal 2009, hereafter KFCB).  Moreover,
 the S\'ersic indices $n \simeq 2$ to 3 of ellipticals are well understood as natural results of their formation by major 
galaxy mergers (e.{\ts}g., Hopkins \etal 2009a, c; see Kormendy \etal 2009 for a review).  So the connection between
the $n$ \gapprox \ts\ts2 S\'ersic indices of classical bulges with their formation by major mergers is also well established.

      Therefore, even though we do not fully understand how disk secular evolution determines the resulting
pseudobulge S\'ersic indices, we see strong enough correlations between $n$ and other pseudobulge properties
so that we can use the S\'eersic index as a classification criterion (Figs.~1.41, 1.42). 

\vfill\eject

\cl{\null}

\begin{figure*}

\vskip 4.38truein



 \includegraphics{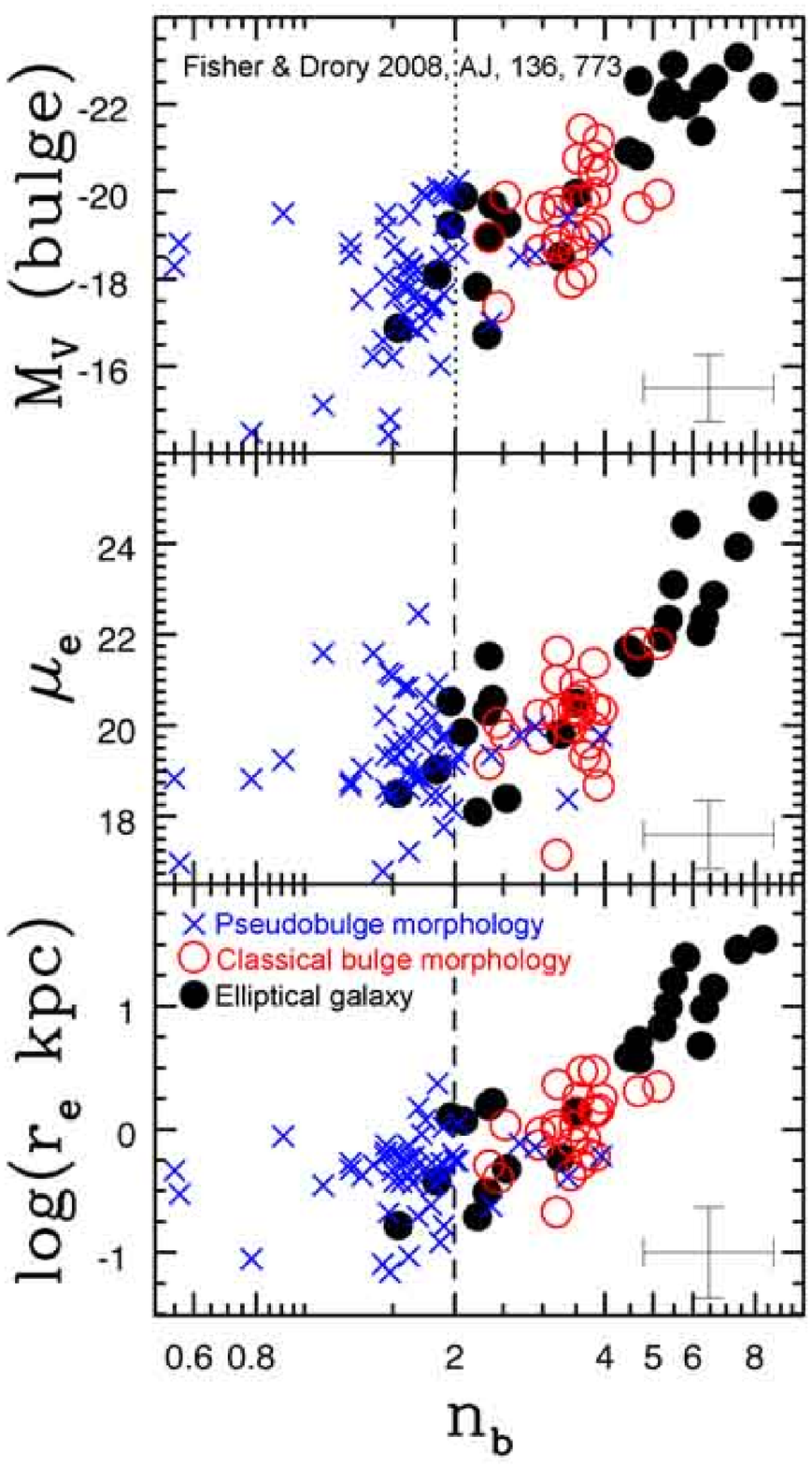} 

 \includegraphics{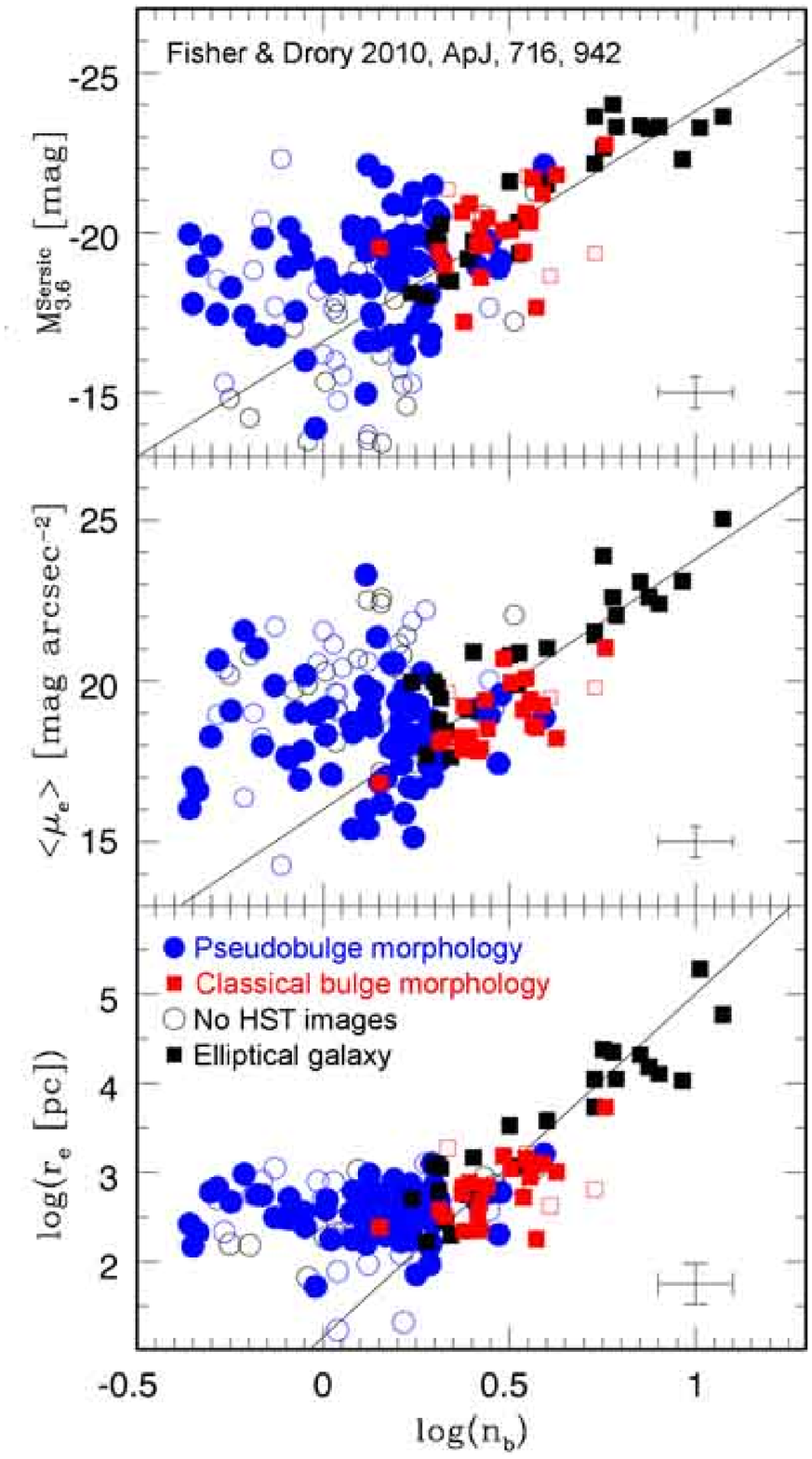} 

\caption{Correlations with (pseudo)bulge S\'ersic index $n_b$ of (top to bottom) bulge absolute magnitude,
         effective brightness at (left) or averaged interior to (right) the half-light radius $r_e$ and effective
         radius $r_e$.  The $V$-band date in the left plots are from Fisher \& Drory (2008); the 3.6 $\mu$m-band 
         {\it Spitzer Space Telescope} data in the right plots are from Fisher \& Drory (2010).  In all panels, 
         the central components are classified morphologically as in Fig.~1.38.
         }

\end{figure*}

\cl{\null}
\vskip -35pt

      Therefore classical bulges have S\'ersic $n \geq 2$, whereas pseudobulges usually
have $n < 2$.  Note in Fig.~1.41 and in the earlier discussion of NGC 4736 that some pseudobulges 
do have S\'ersic indices as big as $n \simeq 4$.  Therefore it is always important to apply as many 
classification criteria as possible.

\subsubsection{Fundamental plane parameter correlations}

      Figure 1.41 provides an introduction to the correlations between effective radius $r_e$, effective brightness
$\mu_e$ at $r_e$, velocity dispersion $\sigma$ and bulge absolute magnitude $M_V$ (the ``fundamental plane''). 
Classical bulges satisfy the fundamental plane correlations of elliptical galaxies.
Some pseudobulges do so, too.  But many have larger $r_e$ and fainter $\mu_e$ than do classical bulges.

\vfill\eject

\cl{\null}

      Figure 1.68 updates the observation that classical bulges and ellipticals have the same correlations.
Simulations of major galaxy mergers reproduce these correlations (e.{\ts}g.,
Robertson \etal 2006;
Hopkins \etal 2008, 2009b).

      Earlier versions that emphasize pseudobulges are shown in Figs.~1.42 and 1.43.  Many pseudobulges
satisfy the correlations for bulges and ellipticals, but in general, they show substantially larger scatter.

\vfill

\begin{figure}[hb]


 \includegraphics{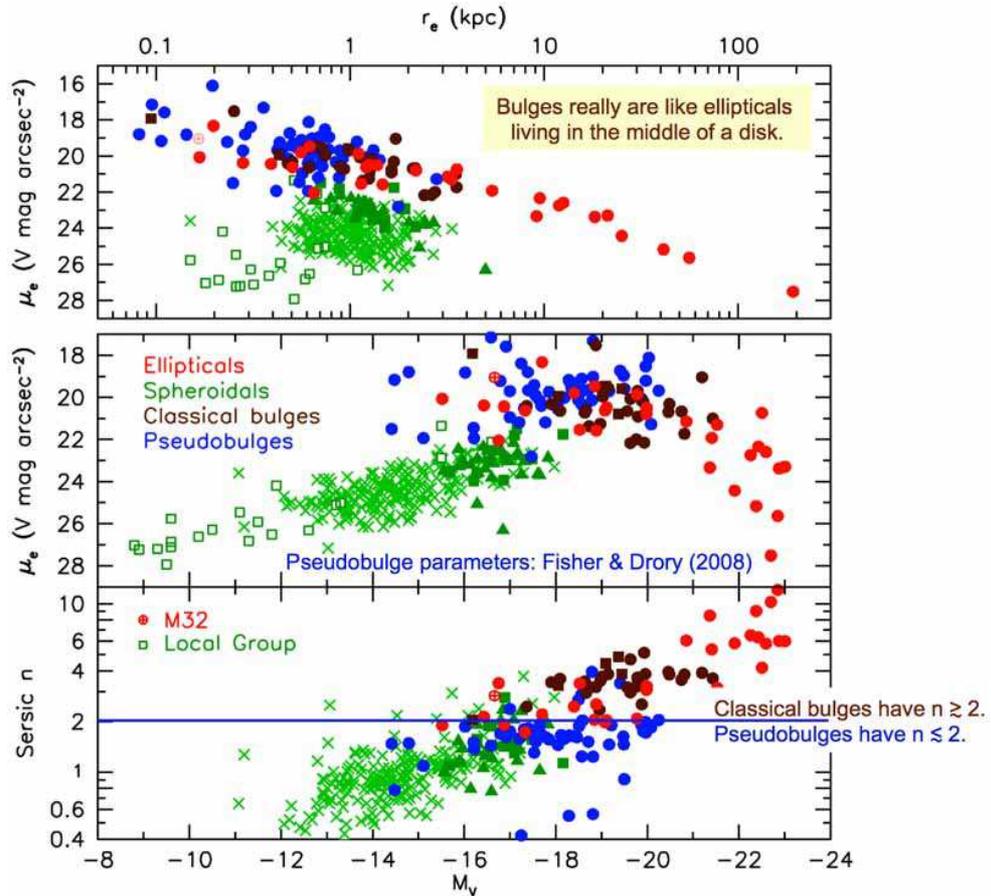} 

\caption{Structural parameter correlations for pseudobulges (blue), classical bulges (brown), ellipticals (red), 
         and spheroidal galaxies (green).  Pseudobulge data and most bulge points are from Fisher \& Drory (2008). 
         The ellipticals, five bulge points and the green squares are from Kormendy \etal (2009: KFCB). Green 
         triangles show all spheroidals from Ferrarese \etal (2006) that are not in KFCB. Crosses show all 
         spheroidals from Gavazzi \etal (2005) that are not in KFCB or in Ferrarese \etal (2006). Open squares 
         are Local Group spheroidals (Mateo 1998; McConnachie \& Irwin 2006). The bottom panels show major-axis
         S\'ersic index $n$ and effective surface brightness $\mu_e$ versus galaxy or bulge absolute magnitude.
         The top panel shows $\mu_e$ versus effective radius $r_e$ (the Kormendy 1977b relation, which shows the
         fundamental plane almost edge-on).   From Kormendy \& Fisher (2008).
         }

\end{figure}

\cl{\null}
\vskip -28pt
\eject

      Figure 1.43 shows the $\mu_e$\ts--\ts$r_e$ correlation from Carollo (1999).  I emphasize this version because 
it shows more extreme pseudobulges together with spheroidal galaxies, galactic nuclear star clusters and globular clusters.  
Spheroidals form a sequence perpendicular to the correlation for ellipticals (Figs.~1.42, 1.61).  Globular clusters are 
different from both ellipticals and spheroidals (Kormendy 1985, 1987).  The comparison that is important here is the one 
between bulges plus ellipticals, pseudobulges and galactic nuclei.

      Figures 1.42 and 1.43 show that some pseudobulges satisfy the parameter correlations for classical bulges and 
ellipticals, but many deviate by having brighter $\mu_e$ (Kormendy \& Bender 2012) or fainter $\mu_e$ (see also
Falc\'on-Barroso \etal 2002;
Kormendy \& Fisher 2008;
Gadotti 2009).  
All available data suggest that {\it pseudobulges fade out by becoming low in surface brightness, not by becoming like 
nuclear star clusters.  Nuclear star clusters are not faint versions of pseudobulges.  Indeed, tiny pseudobulges and 
normal nuclei coexist in Scd galaxies like M{\ts}101 and NGC 6946 (Kormendy \etal 2010).}

      The fundamental plane is only secondarily useful for bulge classification.  Many pseudobulges satisfy 
the correlations for classical bulges, so use of the correlations as the only classification method (Gadotti 2009;
Gadotti \& Kauffmann 2009) is not feasible.  Extreme pseudobulges can be identified because their parameters deviate from the E
correlations in Fig.~1.42, but these objects can usually be robustly classified using other criteria anyway.

\vfill

\begin{figure}[hb]


 \includegraphics{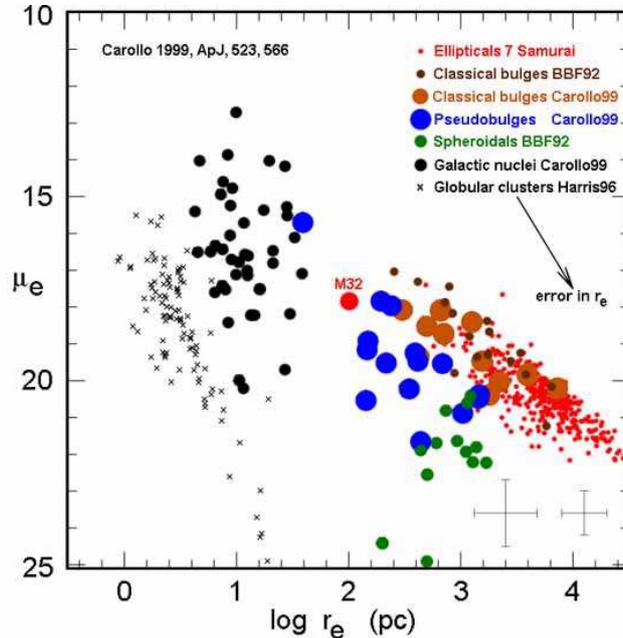} 

\caption{Effective surface brightness versus effective radius for various kinds of stellar 
         systems (Carollo 1999; this version is from Kormendy \& Fisher 2008).
         }

\end{figure}

\cl{\null}
\vskip -28pt
\eject

\subsubsection{Nuclear bars}

      Figure 1.44 shows examples of nuclear bars at the centers of barred and oval galaxies.  Nuclear bars are well known 
phenomena (de Vaucouleurs 1959; Sandage 1961; Kormendy 1981, 1982b; Buta \etal 2007; Buta~2011,~2012).  They are always 
much smaller than the main bars in which they are embedded.~Observations are consistent with the suggestion that their physics 
is essentially the same as that of main bars.  How they form is not known in detail, but it is reasonable to expect that the 
growth of stellar disky pseudobulges out of inflowing cold gas can result in a bar instability.  

      Pattern speeds $\Omega_p$ of nuclear bars are well predicted by Section 1.4.2.  They live at small radii where stellar 
densities are high and where $\Omega - \kappa/2$~is~large.  So $\Omega_p \sim \Omega - \kappa/2$ should be larger 
for nuclear bars than it is for main bars.  This is seen in $n$-body models
(Debattista \& Shen 2007; 
Shen \& Debattista 2009).
Direct measurement of $\Omega_p$ is difficult, but Corsini \etal (2009) use the Tremaine \& Weinberg (1984) method on
NGC 2950 and conclude that the main and nuclear bars have different pattern speeds.
For present purposes, different $\Omega_p$ is sufficiently well established by the observation that nuclear bars
have random orientations with respect to their main bars (Fig.~1.44).  

      Bars are disk phenomena.~A nuclear bar therefore identifies a pseudobulge.

\vfill

\begin{figure}[hb]


 \includegraphics{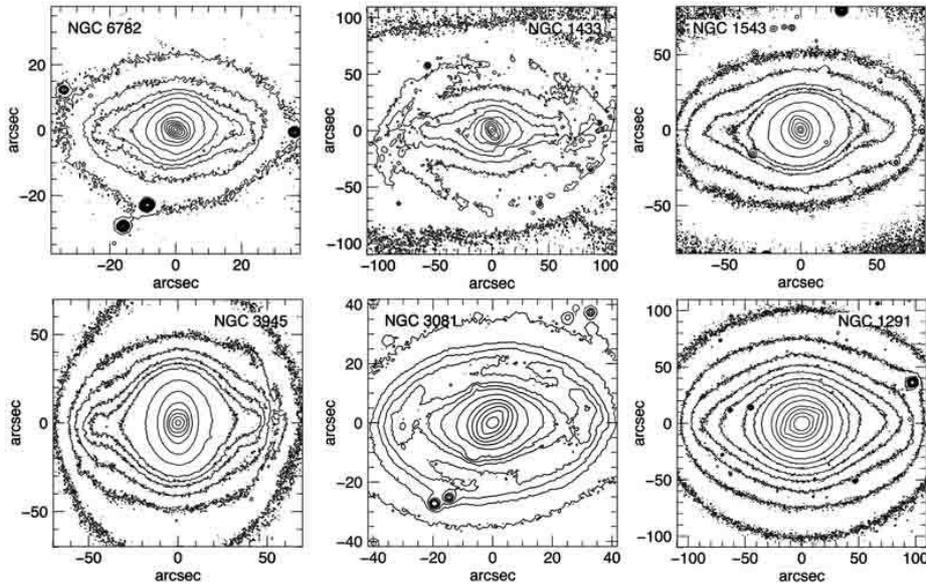} 

\caption{Bars within bars.  The main bar is rotated until it is horizontal.  Contour levels are close together in intensity 
at large radii and widely spaced in intensity in the nuclear bars.  NGC 3081 and NGC 1433 have inner rings.  
NGC 1291 is also shown in Fig.~1.6.
NGC 3081 and NGC 3945 are shown in Figs.~1.17 and 1.24.
The images are courtesy Ron Buta; this figure is from Kormendy \& Kennicutt (2004).
The nuclear bar in NGC 2950 (see above discussion) is illustrated in Fig.~1.17.
         }

\end{figure}

\cl{\null}
\vskip -28pt
\eject

\subsubsection{Boxy pseudobulges are edge-on bars.}

      Section 1.4.3.3 reviews how bars buckle and thicken in the vertical 
direction and consequently look like ``box-shaped bulges'' when seen edge-on.  These are parts of
disks, so I call them ``boxy pseudobulges''.  The one in our Galaxy is particularly clearcut, because we are
close enough to it so that, from our perspective, the near side looks taller than the far side (Fig.~1.45).

\vfill

\begin{figure}[hb]


 \includegraphics{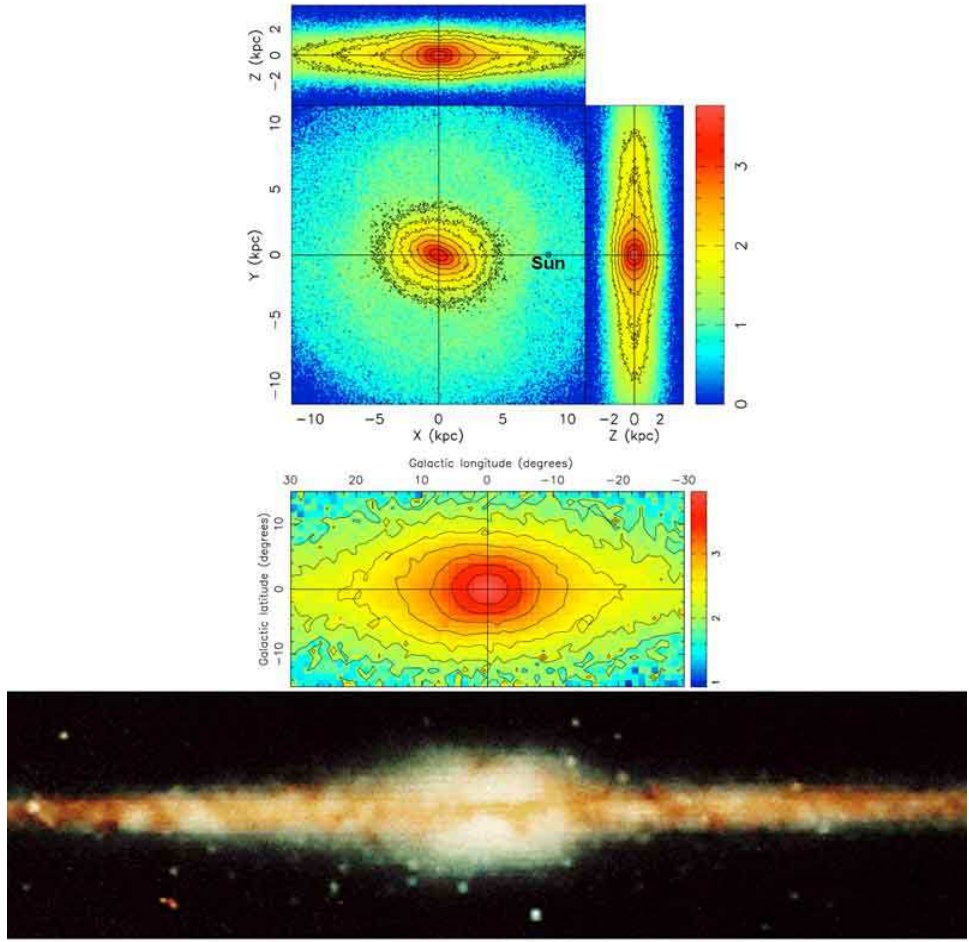} 

\caption{The boxy pseudobulge = almost-end-on bar of our Galaxy in (bottom: from~{\tt
http://www.nasaimages.org/luna/servlet/detail/nasaNAS$\sim$5$\sim$5$\sim$24223}
         {\tt $\sim$127634:COBE-s-View-of-the-Milky-Way})
         a COBE infrared view 
         and (top) an $n$-body model from Shen \etal (2010).  The top three model panels show the face-on
         and side-on views as projected and seen from far away.  Seen almost side-on (top), the bar looks
         peanut-shaped.  Seen almost end-on from the direction of our Sun (right-hand view), it looks boxy 
         when seen from far away.~Viewed instead~from within our
         Galaxy at the position of the Sun (bottom model panel), the near side of the bar is significantly closer 
         than the far side.  Therefore the near side looks taller than the far side,
         exactly as in the COBE image (Blitz \& Spergel 1991).
         }

\end{figure}

\cl{\null}
\vskip -28pt
\eject

      Observations which further show that boxy bulges are edge-on bars are reviewed in Kormendy
\& Kennicutt (2004).  The well known example of NGC 4565 (Fig.~1.46, from Kormendy \& Barentine 2010) turns out 
to be an SB(r) galaxy with a second, tiny pseudobulge at its center that is distinct from the boxy bar (Fig.~1.47).  

\vfill

\begin{figure}[hb]


 \includegraphics{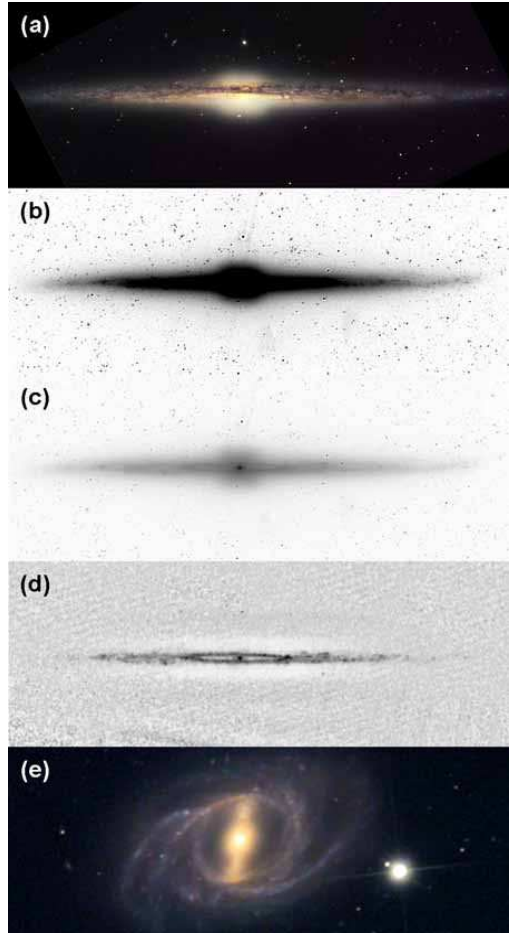} 

\caption{(a) Optical and (b,{\ts}c) {\it Spitzer} 3.6 $\mu$m negative images of NGC 4565 shown at 
different stretches to emphasize (b) the boxy bar and (c) an inner ring and pseudobulge.  The newly 
detected pseudobulge is the central bright dot.  (d) {\it Spitzer} IRAC 8 $\mu$m negative image 
showing PAH emission and therefore star formation from the inner ring and outer disk.  Because the inner ring is 
dark inside at 8\ts$\mu$m, we conclude that the dark inside seen at 3.6 $\mu$m is not caused by dust absorption.  
Rather, the ring really is dark inside.  Therefore NGC 4565 is an SB(r)b galaxy, i.{\ts}e., an almost-edge-on  
analog of NGC 2523 (bottom panel).  The NGC 2523 image has been scaled so the inner ring has the same apparent 
radius as in NGC 4565 and rotated to the apparent bar position angle inferred for NGC 4565.  We suggest that 
NGC 2523, if oriented as in the bottom panel and inclined still more until we observed it almost edge-on, would 
show the features seen in the NGC 4565 images. 
         }

\end{figure}

\cl{\null}
\vskip -28pt
\eject

      Finding the tiny pseudobulge hidden inside the boxy bar of NGC 4565 solves a long-standing puzzle and
cements an important implication for the statistics of $(P)B/T$ luminosity ratios.  Compare NGC 4565 with any 
more face-on barred galaxy, such as NGC 2523 in Fig.~1.46.  Face-on galaxies show a (pseudo)bulge, a bar,
and an outer disk; i.{\ts}e., three or more components.  The edge-on galaxy NGC 4565
shows only a ``boxy bulge'' and a disk.  As long as we thought that boxy distortions were minor features of bulges,
this was no problem -- many unbarred Sb galaxies have just a bulge and a disk.  But if the box
in NGC 4565 is an edge-on bar, then the galaxy contains a bar and a disk; i.{\ts}e., only two components.~This 
is not seen in face-on galaxies except at very late Hubble types.  Where is the ``bulge'' in NGC 4565?

      Figures 1.46 and 1.47 show the answer.  The minor-axis profile of the boxy structure is exponential.  
Inside this structure,
there is a clearly distinct, tiny central component that has $n = 1.33 \pm 0.12$ and that therefore is a second,
``disky'' pseudobulge.  The important implication is that $PB/T = 0.06 \pm 0.01$ is {\it much smaller\/} than
the value $B/T \simeq 0.4$ (Simien \& de Vaucouleurs 1986) for the box.  Closely similar results are found for
the edge-on ``boxy bulge'' galaxy NGC 5746 (Barentine \& Kormendy 2012).  Not counting boxy bulges because 
they are bars, {\it $(P)B/T$ ratios are much smaller than we have thought for essentially all edge-on galaxies with 
boxy bulges.  In particular, the {\it classical\/}-bulge-to-total ratio in our Galaxy is $\sim 0$} (Shen \etal 2010).

\vfill

\begin{figure}[hb]

 \includegraphics{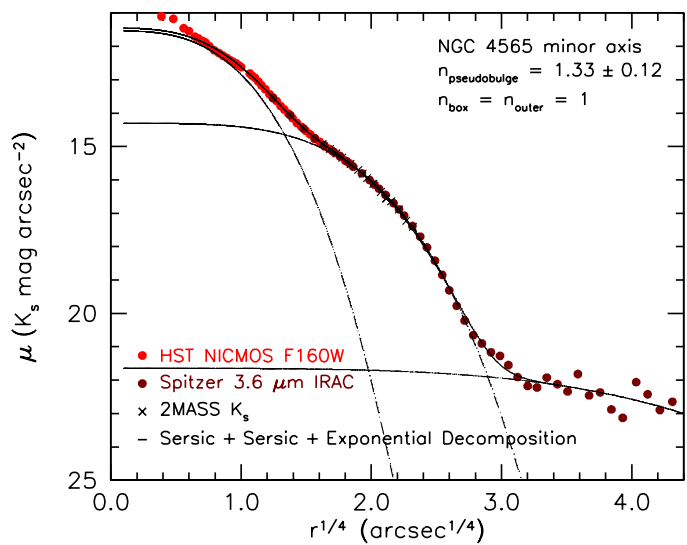} 

\caption{Minor-axis profile of NGC 4565 from Kormendy \& Barentine (2010).
The dashed lines show a decomposition of the profile into components in order of increasing radius: 
a Seyfert nucleus (Ho \etal 1997) or nuclear star cluster that is not included in the fit, the 
pseudobulge (S\'{e}rsic), box-shaped bar (S\'{e}rsic) and outer halo (exponential).  The solid line is the sum of the components.
         }

\end{figure}

\cl{\null}
\vskip -28pt
\eject

\subsection{Pseudobulges classification criteria}

      The bulge-pseudobulge classification criteria updated from Kormendy \& Kennicutt (2004) are listed here for convenience.  
They are identified by the number $m$ in the Section 1.5.2.$m$ in which they were discussed.

    \begin{enumerate}[e]\listsize
    \renewcommand{\theenumi}{(\alph{enumi})}

\item[(1){\kern -3pt}]{If the center of the galaxy is dominated by Population I material (young stars, gas and dust), 
                       but there is no sign of a merger in progress, then the bulge is at least {\it mostly\/} pseudo.  }

\item[(2){\kern -3pt}]{Pseudobulges often have disky morphology; e.{\ts}g., their apparent flattening 
                       is similar to that of the outer disk, or they contain spiral structure all the way in to
                       the center of the galaxy.  Classical bulges are much rounder than their disks
                       unless the galaxy is almost face-on, and they cannot have spiral structure.}

\item[(3){\kern -3pt}]{Pseudobulges are more rotation-dominated than classical bulges 
                       in the \hbox{$V_{\rm max}/\sigma$\ts--\ts$\epsilon$} diagram. 
                       Integral-field spectroscopy often shows that the central surface brightness excess over the inward 
                       extrapolation of the outer disk profile is a flat central component that rotates rapidly and that has a 
                       small velocity dispersion.}

\item[(4){\kern -3pt}]{Many pseudobulges are low-$\sigma$ outliers in the Faber-Jackson (1976) correlation between (pseudo)bulge 
                       luminosity and velocity dispersion.  A similar signature is that $\sigma$ {\it decreases} from the disk into 
                       the pseudobulge.}

\item[(5){\kern -3pt}]{Small bulge-to-total luminosity ratios do not guarantee that a bulge
                       is pseudo,~but almost all pseudobulges have $PB/T$ \lapprox \ts0.35.~If $B/T$ \gapprox \ts0.5, the bulge is classical.}

\item[(6){\kern -3pt}]{Most pseudobulges have S\'ersic index $n < 2$, whereas almost all classical 
                       bulges have $n \geq 2$.  The processes that determine the small S\'ersic indices of 
                       pseudobulges are not understood, but the correlation of small $n$ with other
                       pseudobulge indicators is so good that this has become a convenient classification
                       criterion.  Note, however, that some pseudobulges do have S\'ersic indices as big as 4.}

\item[(7){\kern -3pt}]{Classical bulges fit the fundamental plane correlations for elliptical
                       galaxies.  Some pseudobulges do, too, and then these correlations are not helpful for
                       classification.  But more extreme pseudobulges are fluffier than classical bulges;
                       they have larger $r_e$ and fainter surface brightnesses $\mu_e$ at $r_e$.  These
                       can easily be identified using fundamental plane correlations.}

\item[(8){\kern -3pt}]{In face-on galaxies, the presence of a nuclear bar shows that a pseudobulge dominates the central light.  
                       Bars are disk phenomena.  Triaxiality in giant ellipticals involves completely different physics -- 
                       slow (not rapid) rotation and box (not $x_1$ tube) orbits.  }

\item[(9){\kern -3pt}]{In edge-on galaxies, boxy bulges are edge-on bars; seeing one is sufficient to identify a pseudobulge.  
                       The boxy-nonrotating-core side of the ``E{\ts}--{\ts}E dichotomy'' of elliptical galaxies into two kinds 
                       (Kormendy \etal 2009) cannot be confused with boxy, edge-on bars because boxy ellipticals -- even if they 
                       occur in disk galaxies (and we do not know of an example) -- are so luminous that we would measure
                       $B/T > 0.5$.  Then point (5) would tell us that this ``bulge'' is classical.}

     \end{enumerate}

      It is important always to apply as many classification criteria~as~possible.

\vfill\eject

\subsection{Secular evolution and hierarchical clustering}

      We now have a well articulated paradigm of secular evolution in galaxy disks that complements our
standard picture of hierarchical clustering.  There is no competition between these two galaxy formation pictures --
both are valid, and their relative importance depends on cosmological lookback time and on environment.  I have emphasized 
that the Universe is in transition from early times when the rapid processes of hierarchical clustering
were most important in controlling galaxy evolution to future times when galaxy merging will largely have finished and
slow, internal processes will dominate.  

      In the present Universe, the mass in bulges (including ellipticals) and that in disks are roughtly equal
(Schechter \& Dressler 1987;
Driver~{\it et al.}~2007;
Gadotti 2009;
Tasca \& White 2011).
Uncertainties are large, with estimated ratios of the mass in bulges to that in disks as large as
$\sim$\ts2 (Fukugita \etal 1998).  One reason is that the relative numbers of disks and merger remnants
is a strong function of environment:~disks predominate in the field, whereas most giant ellipticals live in galaxy clusters (Section\ts1.6.1).

      The ratio of mass in classical bulges to that in pseudobulges has not yet been determined for large and unbiased 
galaxy samples.  It is reasonable to expect that the {\it ratio of masses\/} is not large.  

      In contrast, the {\it ratio of numbers\/} of pseudobulges to numbers of classical bulges could easily be \gapprox 1.
Kormendy \& Kennicutt (2004) estimated that most Sa galaxies contain classical bulges, that in Sb galaxies, classical and
pseudo bulges are comparably common and that most Sbc galaxies contain pseudobulges.  Recent work suggests that there are
more pseudobulges at early Hubble types than Kormendy \& Kennicutt thought
(Fisher \& Drory 2008;
Weinzirl \etal 2009).
S0 galaxies contain pseudobulges more often than Sa galaxies, consistent with parallel-sequence galaxy classifications in which 
S0s form a sequence in $(P)B/T$ that parallels the sequence of spirals (Section 1.7).  At late Hubble types, Sc\ts--{\ts}Im 
galaxies appear never to contain classical bulges.~Many Scd\ts--{\ts}Sm galaxies do not contain pseudobulges, either;
M{\ts}33 is an example, and the main reason is that $\Omega(r)$ depends little enough on radius so that it is not energetically 
profitable to transport angular momentum outward (Section 1.2.4).

      I conclude with a point of perspective from Kormendy \& Kennicutt (2004).  In the early 1970s, when I was a 
graduate student, Hubble classification was in active use, but we also knew about many regular features in galaxy disks, 
such as lens components, nuclear, inner and outer rings, nuclear bars and boxy bulges, that we did not understand and that 
often were not even included in the classification.  We also knew many peculiar galaxies, no two of which look alike; they did 
not fit comfortably into Hubble classification, and we did not understand them, either.  Now both the peculiar galaxies and the 
structural regularities are becoming well understood within two paradigms of galaxy evolution that got their start in the late 1970s. 
The peculiar galaxies were once normal but now are undergoing tidal interactions or are galaxy mergers in progress (Section 1.8.1).
And structures such as rings and lenses that are seen in many galaxies are products of the secular evolution of relatively isolated 
galaxy disks.  Between collisions, galaxies do not just sit quietly and age their stellar populations.  Galaxies represent snapshots 
of moments in time during dynamical evolution that goes on today and that will contine to go on for many billions of years to come. 

\section{Astrophysical implications of pseudobulges}

      The general implications of pseudobulges for galaxy formation are the main subject of this Winter School.~Here, I focus on two 
additional astrophysical implications that came -- at least to me -- as a surprise.  First is the challenge that classical-bulge-less 
galaxies (even ones that contain pseudobulges) present for our picture of galaxy formation by hierarchical clustering and merging.
Second is the lack of any tight correlation between the masses of supermassive black holes and the properties of pseudobulges.

\omit{Some parts of these lecture writeups also overlap my past reviews of this subject (Kormendy 1979a, 1979b, 1981, 1982, 1993, 2008; 
      Kormendy \& Kennicutt 2004; Kormendy \& Cornell 2004; Kormendy \& Fisher 2005, 2008).} 

\subsection{A challenge for our theory of galaxy formation\\by hierarchical clustering and merging:\\Why are there so many pure-disk galaxies?}

      Look at any movie of a numerical simulation of hierarchical clustering in action.  Your overwhelmingly strong impression
will be that the lives of dark matter halos are violent.  They continually collide with and accrete smaller halos, 
which -- by and large -- approach from random directions.  And virtually no halo grows large\footnote{In this review, as in 
Kormendy \etal (2010), I will adopt the sufficient and practical definition that a ``large'' galaxy is one in which the circular orbit rotation 
velocities of massless test particles at large radii are $V_{\rm circ} \geq 150$ km s$^{-1}$.} without undergoing at least a few major
mergers between progenitors of comparable mass.

      {\it Given this merger violence, how can there be so many bulgeless galaxies?  The puzzle has two parts.
How does hierarchical clustering prevent the formation of classical bulges that are the scrambled-up remnants of
the progenitor stars that predate the merger?  And how does the merger assembly of galaxy halos prevent the 
destruction of large but very flat disks, at least some of which are made in part of old stars.  Bulgeless disks are
not rare.}

\vfill\eject

      Figure\ts1.48 shows the purest examples of this problem, the iconic \hbox{late-type} galaxies whose edge-on 
orientations make it clear that they have no classical bulges and no signs of pseudobulges.  Such~galaxies~are~common 
(Karachentsev \etal 1993; Kautsch \etal 2006).  UGC 7321 is studied~by Matthews \etal (1999a); Matthews (2000); 
Banerjee \etal (2010).  IC\ts5249 is studied by van der Kruit \etal (2001).  Matthews \etal (1999b), Kautsch (2009) and 
van der Kruit \& Freeman (2011) review superthin galaxies.


      It is a challenge to explain these galaxies.  It helps that they~are~not~large: Hyperleda lists rotation 
velocities of $V_{\rm circ} = 95$, 79, 97 and 92 km s$^{-1}$ for UGC 7321, IC 2233, IC 5249 and UGC 711, respectively.
This is smaller than $V_{\rm circ}$\ts=\ts$135 \pm 10$ km s$^{-1}$ in M{\ts}33 (Corbelli\ts\&{\ts}Salucci 2000; Corbelli 2003). \phantom{00000}

\vfill

\begin{figure}[hb]


 \includegraphics{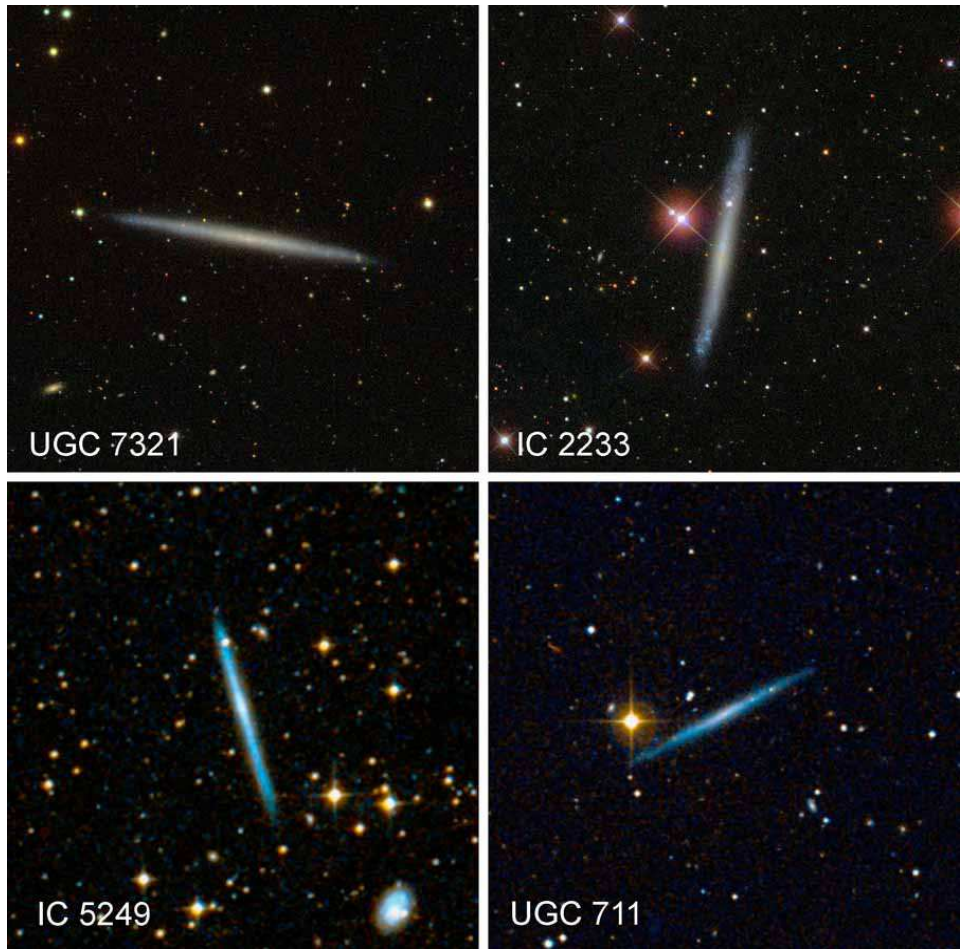}

\caption{Edge-on, completely bulgeless, pure-disk galaxies.  All images are from {\tt http://www.wikisky.org}; the top
         galaxies are from the SDSS and the bottom ones are from DSS.  The DSS images have a bluer color balance 
         than the SDSS.
         }

\end{figure}

\cl{\null}
\vskip -28pt
\eject

\noindent Explaining bulgeless disks is least difficult for dwarf galaxies.  They suffer fewer mergers and tend to accrete 
gas in cold streams or as gas-rich dwarfs
(Maller \etal 2006; 
Dekel \& Birnboim 2006; 
Koda \etal 2009; 
Brooks \etal 2009; 
Hopkins \etal 2009d, 2010). 
Energy feedback from supernovae counteracts gravity most effectively in dwarf galaxies
(Dekel \& Silk 1986; 
Robertson \etal 2004; 
D'Onghia \etal 2006;
Dutton 2009; 
Governato \etal 2010). 
Attempts to explain pure disks have come closest to success in explaining dwarf galaxies
(Robertson \etal  2004; 
Governato \etal 2010). 

      Kormendy \etal (2010) conclude that the highest-mass, pure-disk galaxies are the ones that most constrain our formation picture.
They inventory all giant galaxies ($V_{\rm circ} \geq 150$ km s$^{-1}$) at distances $D \leq 8$ Mpc within which we can resolve small
enough radii to find or exclude even the smallest~bulges.  Table 1.3 documents the $B/T$ and $PB/T$ luminosity ratios for these
galaxies.  Giant, bulgeless galaxies are not rare.  Figure 1.49 shows the most extreme galaxies in which $B/T = 0$ rigorously.
Kormendy \etal (2010) emphasize \phantom{00000}

\vfill

\begin{figure}[hb]


 \includegraphics{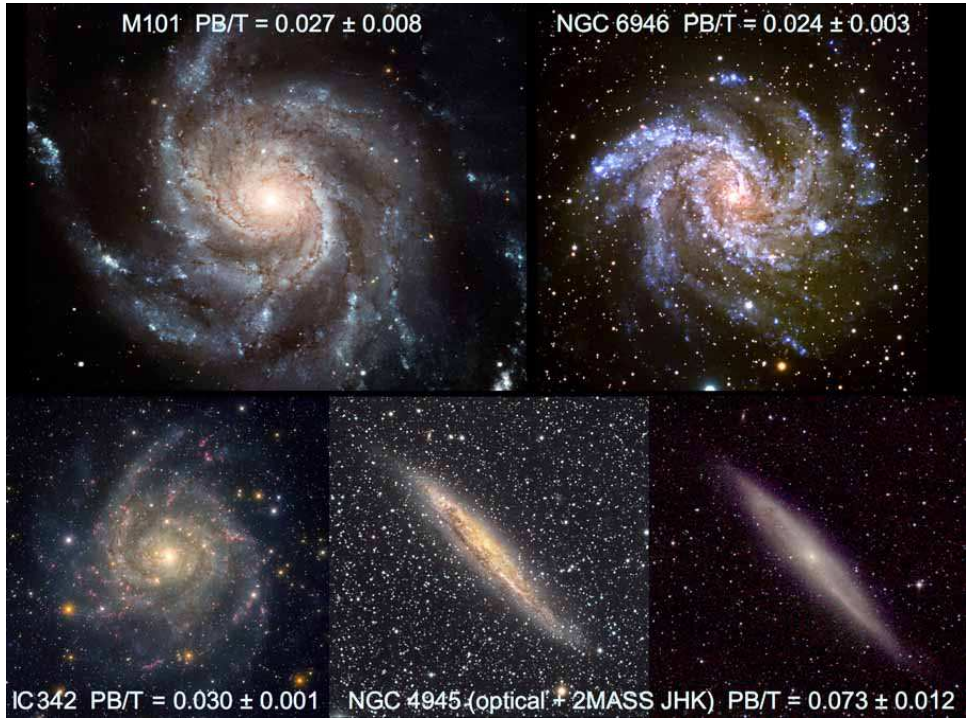}

\caption{Face-on, completely bulgeless, pure-disk galaxies.  All four galaxies have bulge-to-total luminosity ratios
         of $B/T = 0$.  They have the smallest pseudobulges in the local sample of giant galaxies (outer rotation
         velocities~$V_{\rm circ}$\ts$>$\ts150~km~s$^{-1}$; for these galaxies,~174--210 km s$^{-1}$)  in Table 1.3.~The  
         pseudobulge-to-total luminosity ratios $PB/T$ are given in the figure.  Unless otherwise noted, the images are from 
         {\tt http://www.wikisky.org} or Kormendy \etal (2010).
         }

\end{figure}

\cl{\null}
\vskip -28pt
\eject

\noindent that {\it ``we do not have the freedom to postulate classical bulges which have arbitrary properties (such as low surface brightnesses)
that make them easy to hide.  Classical bulges and ellipticals satisfy well-defined fundamental plane correlations} (Fig.~1.68).
{\it Objects that satisfy these correlations cannot be hidden in the above galaxies. 
So $B/T = 0$ in 4/19 of the giant galaxies in our sample.''}  Seven more galaxies contain pseudobulges; since we believe that these are grown
secularly out of disks and not made via mergers, these are pure-disk galaxies from the hierarchical clustering point of view.  Four more
galaxies contain classical bulges smaller than any that are made in hierarchical clustering simulations.  Only M{\ts}31 and M{\ts}81 have
classical bulges with $B/T \simeq 1/3$, and only two more galaxies are ellipticals with $B/T = 1$   

      Fisher \& Drory (2011) derive similar statistics in the $D \leq 11$ Mpc volume.

\vfill

\includegraphics{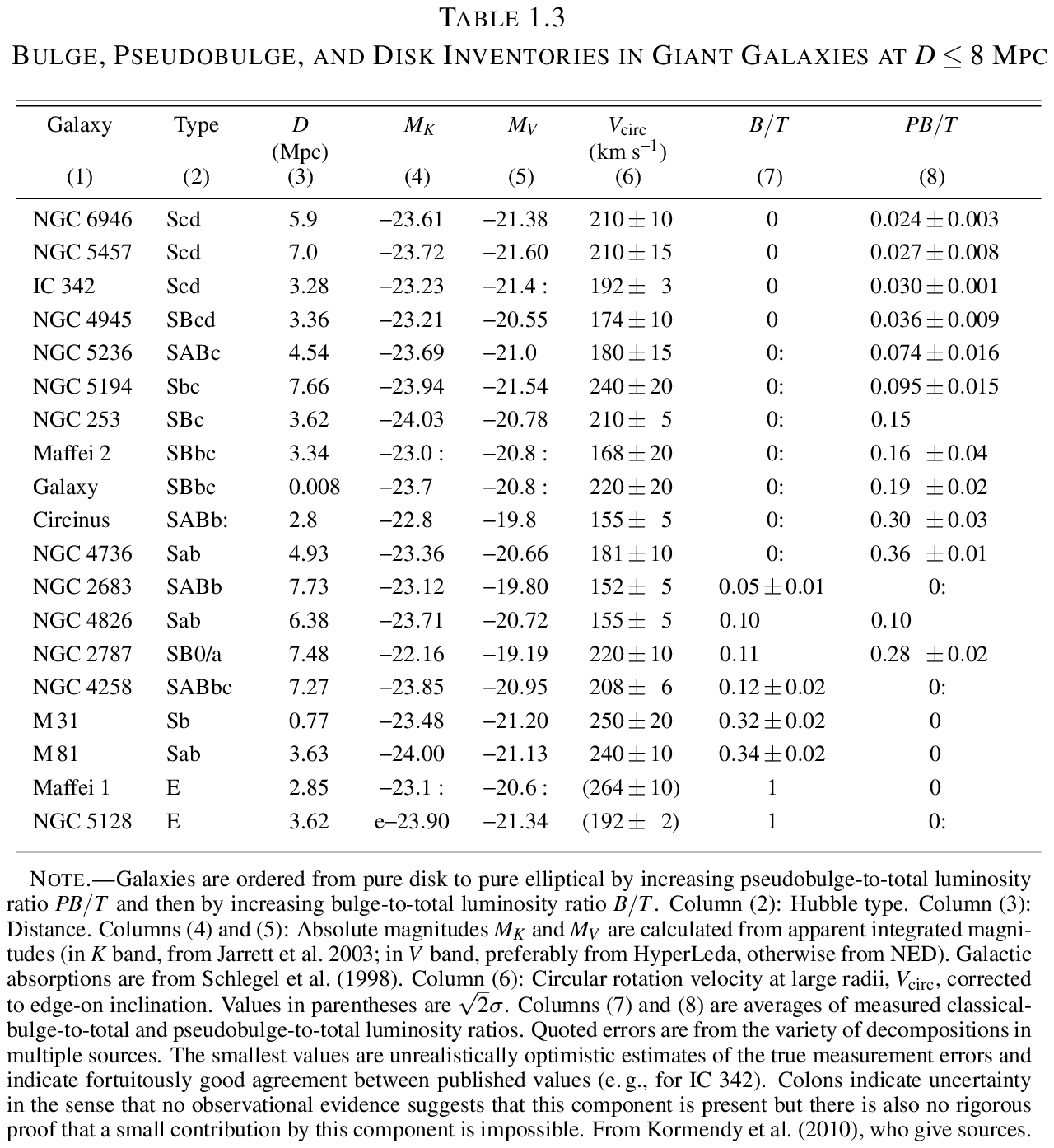}

\cl{\null}
\vskip -28pt
\eject

      Kormendy \etal (2010) conclude that {\it giant bulgeless galaxies do not form the rare tail of a distribution of formation histories 
that include~a~few fortuitously mergerless galaxies.  In the field, the problem of forming giant, pure-disk galaxies by 
hierarchical clustering is acute.  In contrast, in the Virgo cluster, \gapprox \ts2/3 of the stellar mass is in merger
remnants.  Therefore the problem of explaining pure-disk galaxies is a strong function of environment.}

      This is a sign that AGN feedback, the physics popularly used to address the problem, is not the answer.  The effectiveness of
energy feedback depends on galaxy mass.  In contrast, galaxies tell us that environment is the controlling factor.  Giant, 
pure-disk galaxies (Fig.~1.49) are more massive than small ellipticals.  And the thin disk of our Galaxy{\ts}--{\ts}which,
given its boxy bar, is a giant pure-disk galaxy{\ts}--{\ts}contains stars as old as 10 Gyr (Oswalt \etal 1996; Winget \& Kepler 2008).  So I 
suggest that the solution to the pure-disk problem is not to use energy feedback to delay disk star formation in order to give the halo
time to grow without forming a classical bulge.  I believe that a viable solution must use the environmental dependence of the
pure-disk galaxy problem in an essential way (e.{\ts}g., Peebles \& Nusser 2010).

\vskip -24.1pt
\cl{\null}

\subsection{\hbox{Supermassive black holes do not correlate with pseudobulges}}

       Kormendy \& Ho (2013) review a modest revolution that is in progress in studies of supermassive black holes
(BHs) in galaxy centers.  For~more~than a decade, observed BH demographics have suggested a simple picture
in which BH masses $M_\bullet$ show a single correlation each with many properties of their host galaxies (Fig.\ts1.50).
Most influential was the discovery of a tight correlation between $M_\bullet$ and the velocity dispersion $\sigma$ 
of the host bulge~at radii where stars mainly feel each other and~not~the~BH
(Ferrarese \& Merritt 2000;
Gebhardt \etal 2000; 
Tremaine \etal 2002;
G\"ultekin \etal 2009).  
Correlations are also observed between $M_\bullet$ and bulge luminosity 
(Kormendy 1993a; 
Kormendy \& Richstone 1995; 
Magorrian \etal 1998),
bulge mass 
(Dressler 1989;
McLure \& Dunlop 2002; 
Marconi \& Hunt 2003;
H\"aring \& Rix 2004;); 
core parameters of elliptical galaxies
(Milosavljevi\'c \etal 2002;
Ravindranath \etal 2002;
Graham 2004; 
Ferrarese \etal 2006;
Merritt 2006;
Lauer \etal 2007; 
Kormendy \& Bender 2009),
and globular cluster content
(Burkert \& Tremaine 2010;
Harris \& Harris 2011).
These have led to the belief that BHs and bulges coevolve and regulate each other's growth (e.{\ts}g.,
Silk \& Rees 1998;
Richstone \etal 1998;
Granato \etal 2004;
Di Matteo \etal 2005;
Springel \etal 2005;
Hopkins \etal 2006;
Somerville \etal 2008).

     This simple picture is now evolving into a richer and more plausible story in which BHs correlate
differently with different kinds of galaxy components.    BHs do not correlate at all with galaxy disks 
(Kormendy \& Gebhardt 2001; 

\vfill\eject

\noindent Kormendy \etal 2011; Kormendy \& Ho 2013), although some pure-disk galaxies contain BHs (see
Ho 2008 for a review).
And despite contrary~views,
(Ferrarese 2002;
Baes \etal 2003;
Volonteri \etal 2011),
it is clear that BHs do not correlate tightly enough with dark matter halos to imply any special relationship
between them beyond the fact that dark matter controls most of the gravity that makes hierarchical clustering happen
(Ho 2007;
Kormendy \& Bender 2011;
Kormendy \& Ho 2013).  So BHs coevolve only with bulges.

      What about pseudobulges?  They are closely connected with disks, but some contain BHs. 
The best example is our Galaxy (Genzel \etal 2010).

      Hu (2008) finds, for a small sample, that BHs in pseudobulges have smaller $M_\bullet$ than BHs in classical bulges
and ellipticals of the same $\sigma$.  Graham (2008) reports the possibly related result that barred galaxies also deviate from the 
$M_\bullet$\ts--\ts$\sigma$ relation in having small $M_\bullet$, but interpretation is complicated by the fact that some 
of his barred galaxies contain classical bulges (e.{\ts}g., NGC 1023, NGC 4258), some contain pseudobulges (e.{\ts}g., 
NGC 3384, our Galaxy) and some contain both (NGC 2787).  More definitively, results similar to Hu's are found by
Nowak \etal (2010) and by
Greene \etal (2010).

      Kormendy \etal (2011) and Kormendy \& Ho (2013) now show for larger samples that pseudobulges correlate little enough 
with $M_\bullet$ so coevolution is not implied (Fig.~1.50).  This simplifies the problem of coevolution by focusing
our attention on galaxy mergers.  It is a substantial success of the secular evolution picture that a morphological
classification of bulges separates them into two kinds that correlate differently with BHs.

\vfill

\cl{\null}

\vfill

\begin{figure}[hb]

\includegraphics{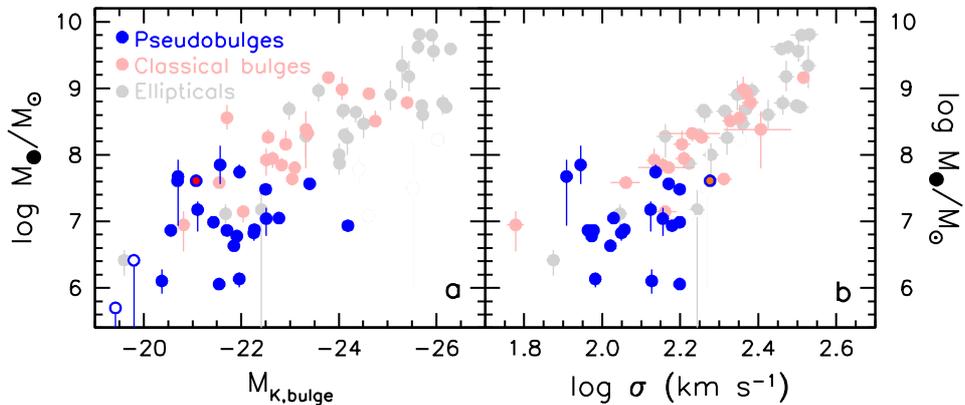}

\caption{Correlation of dynamically measured BH mass $M_\bullet$ with (left) \hbox{$K$-band} bulge absolute magnitude 
    and (right) velocity dispersion averaged inside $r_e$.
    Pseudobulges with dynamical BH detections are shown with blue filled circles and those with $M_\bullet$ upper limits 
    are shown with blue open circles.  
    NGC 2787 may have both a small classical and a large pseudo bulge (Erwin \etal 2003); its blue symbol has a red center.  
    Classical bulges and ellipticals are shown in ghostly light colors to 
    facilitate comparison. This is a preliminary figure from Kormendy \& Ho (2013).
         }

\end{figure}

\cl{\null}
\vskip -28pt
\eject

\section{Environmental secular evolution:\\The structure and formation of S0 and spheroidal galaxies}    
                                                                                                         
      Research on internal secular evolution is now a major industry,~but~work~on environmental secular evolution
still is a series of important but disconnected cottage industries.  We need to make the subject an integral 
part of our standard picture of galaxy evolution.  This section reviews 
environmental secular evolution, following Kormendy \& Bender (2012).

      Our theme is that dwarf spheroidal (dSph) galaxies such as Draco and UMi and higher-luminosity Sph galaxies such as
NGC 205 are transformed, `\hbox{`red and dead''} spiral and irregular galaxies and that many S0 galaxies similarly are transformed 
earlier-type spirals.  That is, Sph galaxies are bulgeless S0s.  The easiest way to introduce this theme is using Fig.~1.51.  
This is one of the best-known extragalactic images, but it is not widely realized that it includes an easy way to form a 
mental picture of the difference between elliptical and spheroidal galaxies.   In doing so, it speaks directly to the 
fundamental question: What is an elliptical galaxy (Fig.~1.52)?


\vfill

\begin{figure}[hb]


 \includegraphics{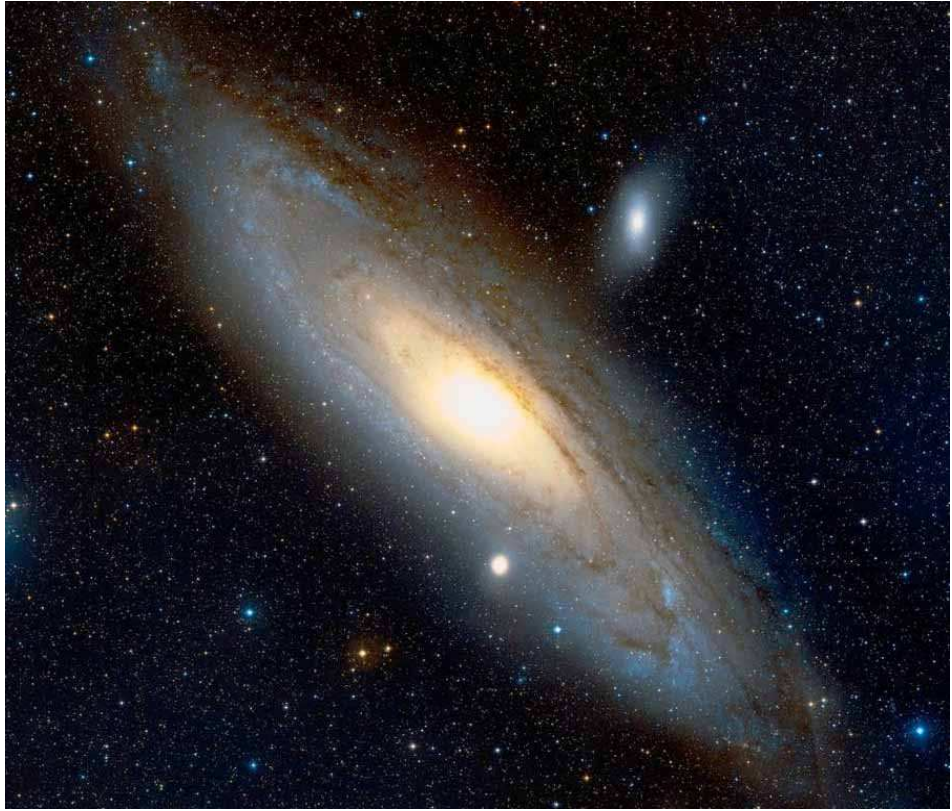}

\caption{M{\ts}31 (Sb, center), M{\ts}32 (E, lower companion) and NGC 205 (Sph, upper companion)
from the Digital Sky Survey via {\tt http://www.wikisky.org}.}  

\end{figure}

\cl{\null}
\vskip -28pt
\eject

\subsection{What is (and what is not) an elliptical galaxy?}

      In Fig.~1.51, M{\ts}31 is an Sb spiral with a classical bulge;~\hbox{$B/T = 0.34 \pm 0.03$} 
(Kormendy \etal 2010).  Absent the~disk, the 
bulge is indistinguishable from a smallish elliptical.  M{\ts}32~is~one~of~the tinest true ellipticals, with a 
$V$-band absolute magnitude of $M_V \simeq -16.7$ (Kormendy \etal 2009).  It is small and dense and
commonly called a ``compact elliptical'' (cE).  But compactness is not a disease; it is mandated by the
physics that makes the Fundamental Plane (Figs.~1.57, 1.68).  In fact, M{\ts}32 is an entirely 
normal example of a tiny elliptical (Kormendy~{\it et~al.}~2009).   In contrast, NGC 205 is the most 
luminous example in the Local Group of a galaxy that satisfies the morphological definition of an 
elliptical but  that differs quantitatively from ellipticals (as a result, it is typed ``E5 pec'').
\hbox{{\it NGC~205 has the same luminosity as M{\ts}32,}} $M_V \simeq -16.6$ (Mateo 1998).   It looks 
different because it is larger, lower in surface brightness and shallower in surface brightness gradient.  
Measured quantitatively, these differences put NGC 205 near the bright end of a sequence of elliptical-looking 
galaxies that is disjoint from{\ts}--{\ts}in fact, almost perpendicular to{\ts}--{\ts}the sequence of ellipticals 
and classical bulges in Figs.\ts1.43, 1.57, 1.68{\ts}--{\ts}1.69 and 1.71. {\it This means that NGC 205 is not an elliptical.}  

\vfill

\begin{figure}[hb]


 \includegraphics{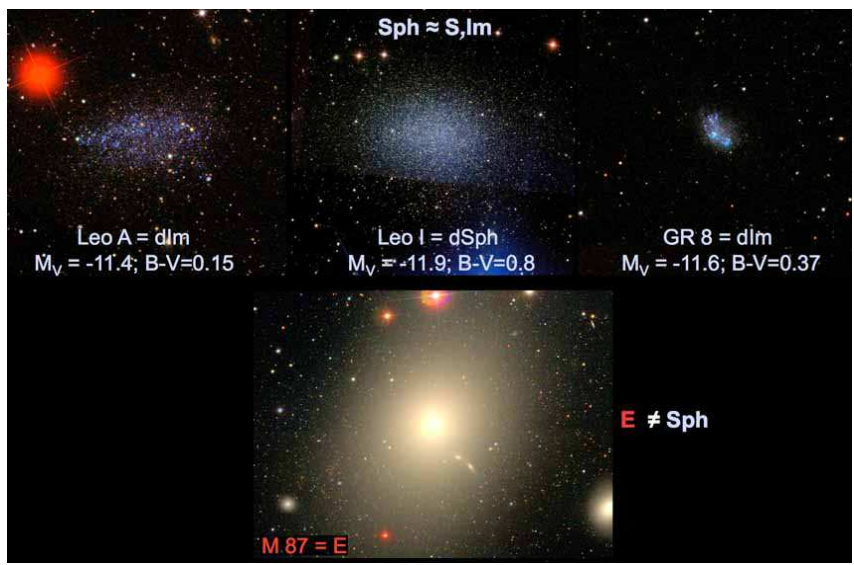}

\caption{What is an elliptical galaxy?  Morphologically, the dwarf galaxy Leo~I (top-middle panel)
         resembles the dwarf irregulars Leo{\ts}A and GR{\ts}8 in its low surface brightness and shallow 
         brightness gradient.~But it resembles the giant elliptical~M{\ts}87 in having elliptical isophotes
         and no cold gas.  Since only the latter characteristics morphologically define
         ellipticals, Leo I is often called a ``dwarf elliptical''.  However, 
         purely morphological criteria prove unable to distinguish objects that have
         different formation~histories.\ts~Leo I turns out to be related to dI galaxies,
         not to ellipticals.\ts~So I do not call it a dwarf elliptical; rather, I call it
         a dwarf spheroidal (dSph)~galaxy.
         }

\end{figure}

\cl{\null}
\vskip -28pt
\eject

\noindent Finding gas and young stars in it supports this conclusion.  We call such objects ``spheroidal galaxies'' (Sphs), 
adapting a name (``dwarf spheroidal'') that is in common use for smaller examples.  The fact that NGC 205 has surface 
brightnesses similar to those of the disk of M{\ts}31 is not an accident.
A variety of evidence leads to the conclusion that Sph galaxies are defunct late-type galaxies whereas classical
bulges and ellipticals are the remnants of major galaxy mergers.  This story is the subject of the present section.

      Recall (Section 1.3.4) how classical morphologists attach no interpretation to descriptions, whereas physical morphologists
try to construct a system in which classification bins uniquely separate objects that have different origin and evolution.  
I emphasized there that, even though classical morphologists try to avoid interpretation, they nevertheless makes choices about 
which features to view as important and which to view as secondary.  Figure\ts1.52 illustrates how this results in a problem 
for classifying elliptical galaxies.  In its {\it isophote shape\/}, Leo I resembles the elliptical galaxy M{\ts}87.  However, 
in its {\it surface brightness}, it resembles the irregular galaxies Leo{\ts}A and Gr\ts8.  Hubble classification is based mainly
on isophote shape, so it has been common to call galaxies like Leo I and NGC 205 ``dwarf ellipticals'' (e.{\ts}g., Sandage\ts1961).  
{\it But there has never been any guarantee that structural morphology identifies physically different kinds of objects.}
Figure 1.53 makes this point concrete.

\vfill

\begin{figure}[hb]


 \includegraphics{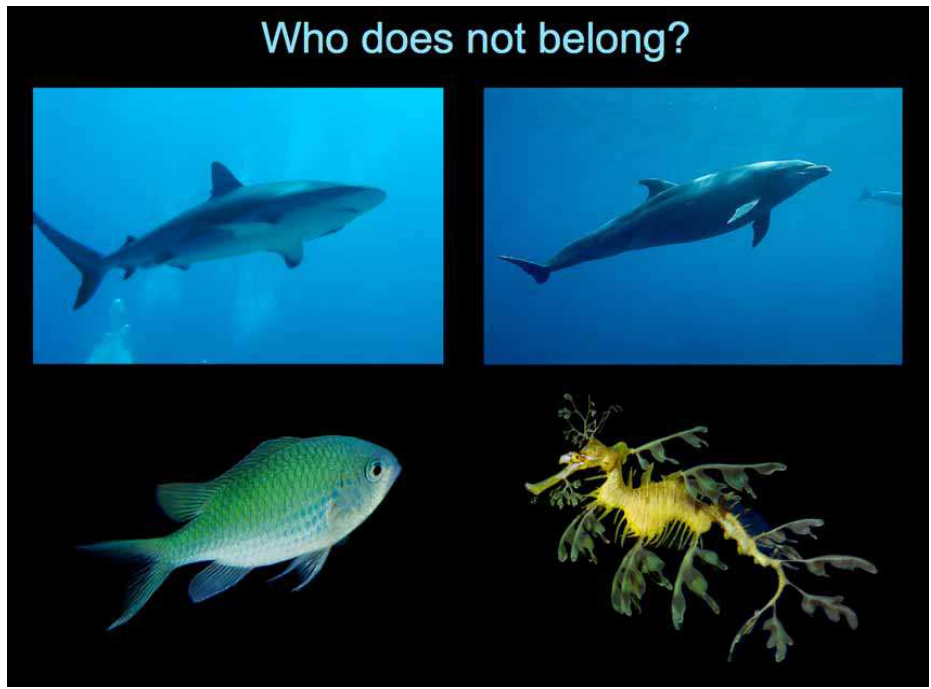}

\caption{The danger of classifying using only morphology.  Who does not belong?
         }

\end{figure}

\cl{\null}
\vskip -28pt
\eject

\cl{\null}

\begin{figure*}

\vskip 3.6truein


 \includegraphics{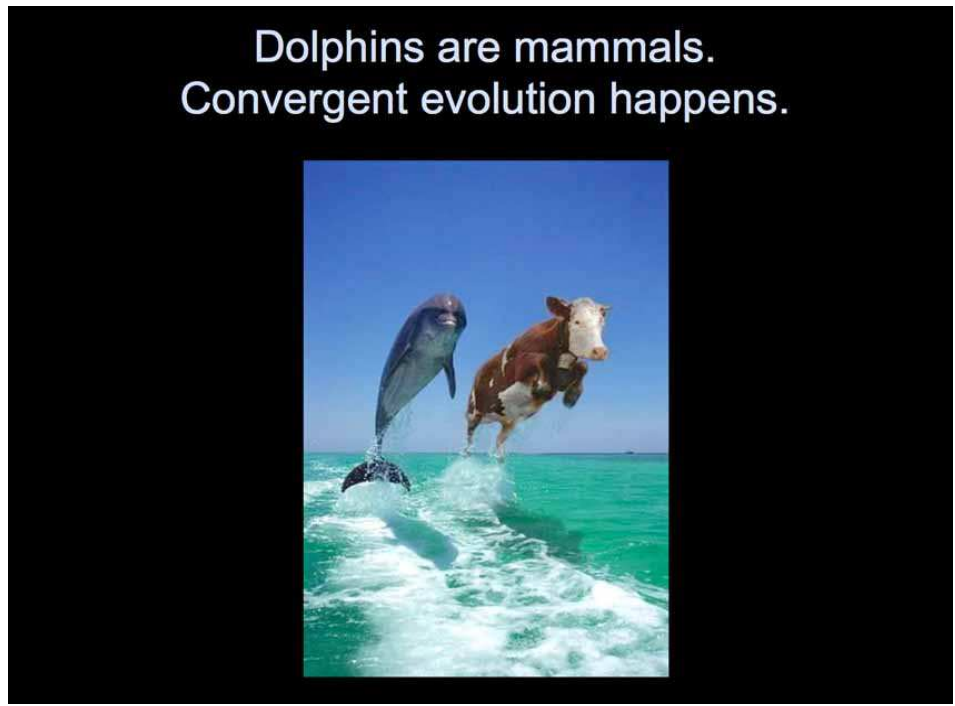}

\caption{Dolphins are Mammals.  Convergent evolution happens.  It happens to galaxies
         as well as to creatures on Earth, and elliptical and spheroidal galaxies prove
         to be examples.  They look morphologically similar but have different formation histories.
         I warmly thank Douglas Martin ({\tt http://www.dolphinandcow.com}) for permission to use this figure.
         }

\end{figure*}

\cl{\null}
\vskip -37pt

      Who does not belong in Fig.~1.53?  The answer is of course well known (Fig.~1.54).  Dolphins (Fig.~1.53, top right) are mammals, 
even though they are morphologically similar to sharks (Fig.~1.53, top left).  To make a living, both need to be well streamlined,
strong swimmers.  Convergent evolution made them that way.  In contrast, a leafy seadragon (Fig.~1.53, bottom right:
{\tt http://picasaweb.google.com/lh/photo/cEq5cwlB2\_cmufKXlOKJcg}) is a kind of seahorse whose main need is good camouflage
to avoid predators.  So, even though it is a fish, its morphology has evolved to be very different from that of a shark.
A ``Hubble classification'' of sea animals that was superficially based on visible structural characteristics could mistakenly 
combine sharks and dolphins into the same or closely related classification bins and could miss the more subtle (but more important)
differences that distinguish sharks and sea dragons from dolphins and cows.  Which parameters best distinguish the physical differences 
that are most important to us is not necessarily obvious without detailed study.

      Convergent evolution happens to galaxies, too. 

\vfill

\eject

\subsection{The E -- Sph dichotomy} 

      Why did we ever think that Leo{\ts}I and NGC\ts205 are ellipticals?~The~answer is that 
research on galaxies began with descriptive classical morphology (e.{\ts}g.,
Hubble 1936; 
de Vaucouleurs 1959;
Sandage 1961), 
and then the above galaxies satisfy the definition of an elliptical.  However, {\it we will see in
Fig.~1.59 that Sandage et al.\ts(1985b) had no trouble in distinguishing between E and dE galaxies of the
same luminosity.}  If this sounds surpassingly strange to you, you have the right reaction.
I will come back to this point below.

      Astronomers are conservative people -- this is often healthy -- and most people clung to the idea
that galaxies like Leo I and NGC 205 are ellipticals even after hints to the contrary started to appear.
Figure 1.55 shows an example.  Ellipticals (filled circles) have higher surface brightness at lower
galaxy luminosities, whereas ``dwarf ellipticals'' (open~circles~and~crosses) have lower surface
brightnesses at lower luminosities.  M{\ts}32 is consistent with the extrapolation of the E sequence.
However, at that time, we thought that M{\ts}32 is compact because it has been tidally 
truncated by M{\ts}31 (King 1962; Faber 1973).  Bingelli \etal (1984) therefore concluded that E and dE galaxies form
a continuous but not monotonic sequence in surface brightness as a function of luminosity.  Meanwhile,
one could wonder whether the two sequences in Fig.~1.55 already hint at different formation physics.

\vfill

\begin{figure}[hb]


 \includegraphics{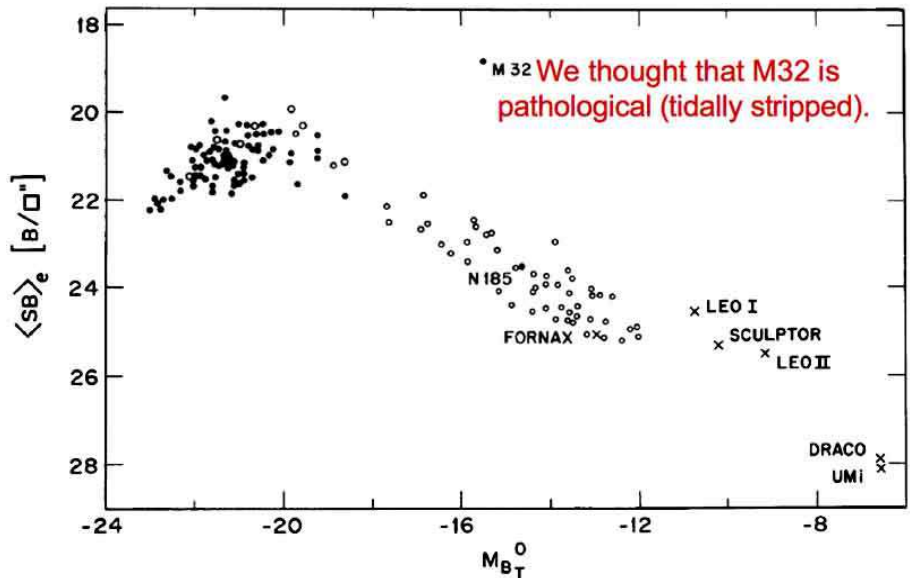}

\caption{Parameter correlations for elliptical and ``dwarf elliptical'' galaxies from
         Bingelli \etal (1984).  These authors suggested that giant and dwarf ellipticals
         form a continuous but not monotonic sequence in mean surface brightness as a function
         of absolute magnitude $M_{\rm BT}$ and that M{\ts}32 -- which deviates prominently
         from this sequence -- is pathological.
         }

\end{figure}

\cl{\null}
\vskip -30pt

\eject

\cl{\null}

\begin{figure*}

\vskip 2.0truein


 \includegraphics{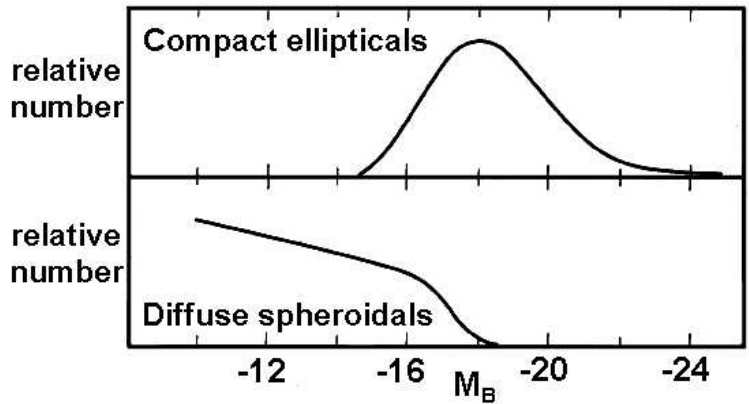}

\caption{Luminosity functions of (top) normal elliptical galaxies roughly from M{\ts}32 to M{\ts}87 and
         (bottom) spheroidal galaxies roughly from Draco and UMi to galaxies such as NGC{\ts}205 (Wirth
         \& Gallagher 1984).  At that time, ``spheroidals'' were commonly called ``dwarf ellipticals''.
         This figure then shows that the smallest non-dwarf ellipticals have lower luminosity than the biggest
         dwarf ellipticals.
         }

\end{figure*}

\cl{\null}
\vskip -45pt

Wirth \& Gallagher (1984) were the first to suggest that M{\ts}32-like compact ellipticals and not the more diffuse 
galaxies like Draco and Leo{\ts}I and NGC\ts205 form the faint end of the luminosity sequence of elliptical galaxies.  
This was based on a successful search for {\it relatively isolated\/} dwarf compact ellipticals which resemble
M{\ts}32.  The new compact ellipticals and the well known ones that are companions to larger galaxies were found 
to lie along the extrapolation to lower luminosity of the correlations for {\it normal\/} ellipticals
of parameters such as effective radius and velocity dispersion.  With respect to this family of normal ellipticals,
``the diffuse ellipticals are a distinct structural family of spheroids whose properties begin to diverge from those 
of the classical ellipticals at an absolute magnitude of $M_B \sim -18$.  At $M_V = -15$, these two families differ in
mean surface brightness by nearly two orders of magnitude.  The key point to note for this discussion is that, in
the range \hbox{$-18$ \lapprox \ts$M_B$ \lapprox \ts$-15$,} {\it both} structural classes of elliptical galaxies coexist''
(Wirth \& Gallagher 1984).  This implies that the luminosity functions of elliptical and spheroidal 
galaxies differ as shown in Fig.~1.56.

      The Wirth \& Gallagher (1984) paper was largely based on four newly found, free-flying compact ellipticals. 
The competing idea (Faber 1973) that compact ellipticals are tidally truncated was largely based on three galaxies, 
M{\ts}32, NGC 4486B and NGC 5846A; then the diffuse dwarfs would be the faint extension of the E sequence.
With both conclusions based on small numbers of galaxies, it was not clear which picture is correct.
The rest of this section reviews the now very strong evidence that Wirth \& Gallagher (1984) were presciently close
to correct in almost every detail, including Fig.~1.56.

\eject

      As a graduate student at Caltech in the early 1970s, I was brought up on the picture that ellipticals
form a continuous, non-monotonic sequence in their structural parameters from the brightest to the faintest galaxies~known.
Then, in the 1980s, I gained access to two important technical advances.  The first was CCD detectors that are linear in 
sensitivity over large dynamic ranges.  The second was the Canada-France-Hawaii telescope (CFHT), which had the best ``seeing''
then available on any optical telescope.  These allowed me to study the central structure of galaxies in unprecedented detail.  
The results revolutionized my picture of 
ellipticals.  They confirmed and extended Wirth \& Gallagher (1984), whose ideas I was not aware of until the end of my work. 
The story is instructive for students, so I describe it here in detail, abstracted from a popular article in {\it Stardate\/} 
magazine (Kormendy 2008b).

      My CFHT surface photometry showed an unexpected result (Fig.~1.57).  Ellipticals define the sequence of red points:~less 
luminous ones are smaller and higher in surface brightness from M{\ts}87 to M{\ts}32.~This much was expected; for bright galaxies, 
it is the correlation shown by the filled~circles~in~Fig.~1.55.  Importantly, the high-resolution CFHT photometry helps to fill 
in the gap between M{\ts}32 and the other ellipticals.\ts~This makes M{\ts}32 look less peculiar. \phantom{000000}  

\vfill

\begin{figure}[hb]


 \includegraphics{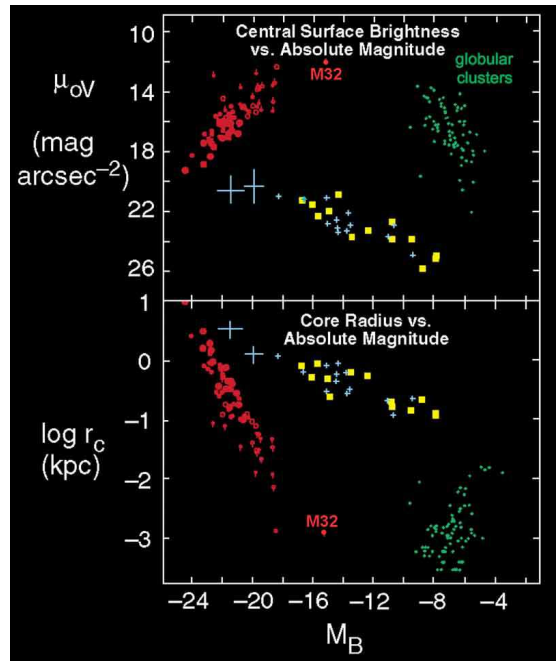}

\caption{Kormendy (1985, 1987) showed with much larger samples that E and Sph
         galaxies form disjoint sequences in parameter space (cf.~Wirth \& Gallagher 1984).
         Sphs (yellow) are not faint ellipticals (red).  Instead, their parameter
         correlations are almost identical to those of  dwarf spiral and irregular
         galaxies (blue).  This figure shows approximate central surface brightness
         and King (1966) core radius, both corrected as well as possible for PSF blurring,
         versus $B$-band absolute magnitude.
         }

\end{figure}

\cl{\null}
\vskip -45pt
\eject

\noindent The surprise was the behavior of the ``dwarf ellipticals'', shown in Fig.~1.57 by yellow points.  Using near-central
parameters rather than parameters measured within the effective radii $r_e$ as in Fig.~1.55, it is clear that dwarf ellipticals
do not satisfy the correlations for elliptical galaxies.  Less luminous dwarf ellipticals are lower -- not higher -- in surface brightness.  
A gap has appeared between ordinary and dwarf ellipticals.  Wirth \& Gallagher's (1984) conclusions are confirmed with a much larger sample.

      Size and density are diagnostic of galaxy formation, so I realized at this point that dwarf ellipticals are not ellipticals at all.  
As one point after another got plotted and intermediate cases failed to show up, my previous picture of elliptical galaxies fell apart.  
Kuhn (1970) captures exactly what happens in a scientist's mind when his understanding of a subject falls apart.  Quoting Kormendy (2008b):
``The first reaction was consternation.  What have I screwed up?  I checked my data reduction.  I considered whether my galaxy sample 
could be biased.  Nothing seemed wrong.  Better data just led in an unexpected direction.  I had to accept the new result: dwarf ellipticals 
are not ellipticals.  But then we should not call them ``dwarf ellipticals''.  The smallest such companions to our Milky Way had 
sometimes been called dwarf spheroidals.  So, to minimize the departure from tradition, I called all such objects ``spheroidals''. 
The biggest ones in Virgo are only as luminous as an average elliptical, but they are giant spheroidals.''

    ``If spheroidals are not ellipticals, what are they?  Kuhn describes what happens next.  Deprived of the guidance of any previous 
understanding of a subject, a scientist in the midst of a scientific revolution does not know what to do next.  In turmoil and in desperation, 
wild ideas get tried out, most of them wrong.  I plotted in my diagrams all the other kinds of stellar systems that I knew about.  
I plotted globular clusters of stars [green points], spiral galaxy disks [two large blue plus signs, each an average for several galaxies
from Freeman 1970],  and irregular galaxies [blue plus signs].  The globulars were unconnected with ellipticals and 
spheroidals.  But the irregulars and spirals were a surprise.  They showed exactly the same correlations as the spheroidals.  Aha!  A new 
picture was emerging.  Maybe spheroidals are related to spirals and irregulars.  They have almost the same structure.  They don't 
contain gas and young stars, which are common in spirals and irregulars.  And they have smoother structure.  But I realized that, if 
the gas were removed or converted into stars, dynamical evolution of the now-gasless spheroidal would smooth out its formerly patchy 
structure within a few galactic rotations. We knew that the dwarf spheroidal companions of the Milky Way had varied star formation 
histories.  A few contain only old stars, as ellipticals do, but most experienced several bursts of star formation, and the most recent 
burst was a few billion or even as little as a few hundred million years ago.  What are galaxies that have not yet had their last burst of
star formation and that therefore still must contain gas?  This is not a controversial question [Kormendy \& Bender 1994].  They are irregulars.  
I realized: if we looked at the Milky Way's dwarf spheroidals when the Universe was half of its present age, about half of them would 
still be irregulars. Irregulars have been turning into Sphs gradually over most of the history of the Universe.  In the Virgo cluster,
lots of processes can make this happen.  The most obvious is ram-pressure stripping: as an irregular galaxy falls into Virgo 
for the first time, it rams into the million-degree gas that fills the cluster, and its cold gas gets swept away.  It started to look 
like no accident that the irregulars in Virgo live around the outside of the cluster, while the center is inhabited by spheroidals
[Binggeli \etal 1987].''

     ``Within a few days, I had a new picture.~Spheroidals are defunct~spiral and irregular galaxies converted by their environment 
to look like ellipticals.  This helped our picture of galaxy formation, because we already knew that ellipticals form by galaxy mergers, 
whereas, quoting Tremaine (1981), `Dwarf elliptical satellite galaxies cannot form by mergers with other satellites since their relative 
velocities are too high.'  We were in trouble when we had to find a single formation process that could explain NGC 4472, one of 
the biggest galaxies in the nearby Universe, and dwarf spheroidals that are a million times less luminous and that look like fragile, 
gossamer clouds of stars [Fig.~1.58 here].  But they look like the smallest irregulars, minus gas and young stars [Fig.~1.52 here].  
So this problem was solved.  I~reported these results [at a workshop in Rehovot, Israel], and they were~well received.''
The result that E and Sph galaxies are different is called the E{\ts}--{\ts}Sph dichotomy.

\vfill

\begin{figure}[hb]


 \includegraphics{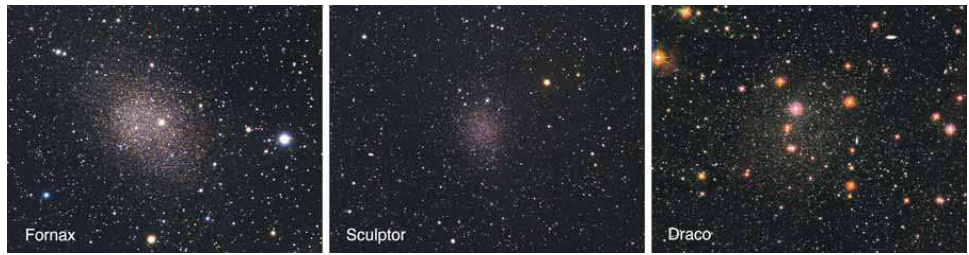}

\caption{Luminosity sequence of dSph satellites of our Galaxy.~Fornax,~Sculptor and Draco have absolute magnitudes of
         \hbox{$M_V = -13.2$, $-11.1$ and $-8.8$,} respectively (Mateo 1998), and correspondingly decreasing surface brightnesses 
         (see Fig.~1.57).  Draco is the cloud of faint stars in the right panel; the bright stars with the 
         instrumentally-produced red halos are foreground stars in our Galaxy.  Contrast M{\ts}87 in Fig.~1.52.  Could M{\ts}87
         and Draco really have similar formation histories, with different results only because 
         changing the mass tweaks the formation physics?  The results reviewed here imply that the answer is ``no''.
         We now believe that M{\ts}87 is a remnant of the dynamical violence of galaxy mergers, whereas Draco formed
         quescently as a dwarf irregular that lost its gas long ago.   From Kormendy (2008b).
         }

\end{figure}

\cl{\null}
\vskip -45pt
\eject

\subsection{Mixed reactions to the E -- Sph dichotomy}

      Scientific research is a quintessentially human enterprise, as reactions to the above result illustrate:

      The essential theoretical understanding of why Sph and S$+$Im galaxies have lower stellar densities at lower galaxy 
masses followed immediately.  Dekel \& Silk (1986) ``suggest that {\it both the dI's and the dE's} [here:~dSphs] 
{\it have lost most of their mass\/} in [supernova-driven] winds after the first burst of star formation, and that this process determined 
their final structural relations.  The dI's somehow managed to retain a small fraction of their original gas, while the dE's either have 
lost all of their gas at the first burst of star formation or passed through a dI stage before they lost the rest of the gas and turned dE.''
Our story here adds detail on dI\ts$\rightarrow${\ts}dSph transformation processes but otherwise is based on exactly the above picture.

      Reactions among observers have been more mixed.  The reasons are many and revealing and occasionally entertaining; they range from
innate conservatism to specific scientific arguments to turf wars.  I will concentrate on the part of this history that is most instructive 
for students.

      I already noted that many astronomers are conservative:~they~do~not~easily discard a picture that they believed in for many years.  
This is healthy~-- imagine what would happen if we chased, willy-nilly, after every outrageous idea that got proposed.  It
is prudent to treat new ideas with respect, but in a mature subject, it is uncommon for a long-held, well-supported picture to be completely
wrong.  The situation is more tricky when subjects are young and not yet well developed.  This proved to be such a case.  Nevertheless,
it is understandable that people who had long been involved in research on dwarf galaxies reacted to the above developments with some ambivalence.  
In particular, the group of Sandage, Binggeli, Tammann and Tarenghi wrote a series of papers on the Virgo cluster in the mid-1980s, some before
and some after the Wirth \& Gallagher (1984) and Kormendy (1985, 1987) papers.  Struggles with the new ideas were evident in some of the later
papers.  The nature of these struggles reveals how seeds of the new ideas could have been recognized in the older results.  I belabor this 
point because the conceptual blindness that results when we embrace a paradigm of how nature works always threatens our ability to see something
new.  {\it As you do your research, it is healthy to be careful and conservative but also prudent to ask yourself: Am I missing something because
of paradigm-induced conceptual blindness?} Kuhn (1970) provides a perceptive discussion of this subject.

      Figure 1.55 already illustrated how one hint -- the opposite slopes of the surface-brightness--luminosity correlations -- was contained 
in previous work.  

\eject

\cl{\null} 

\begin{figure*}

\vskip 2.8truein


 \includegraphics{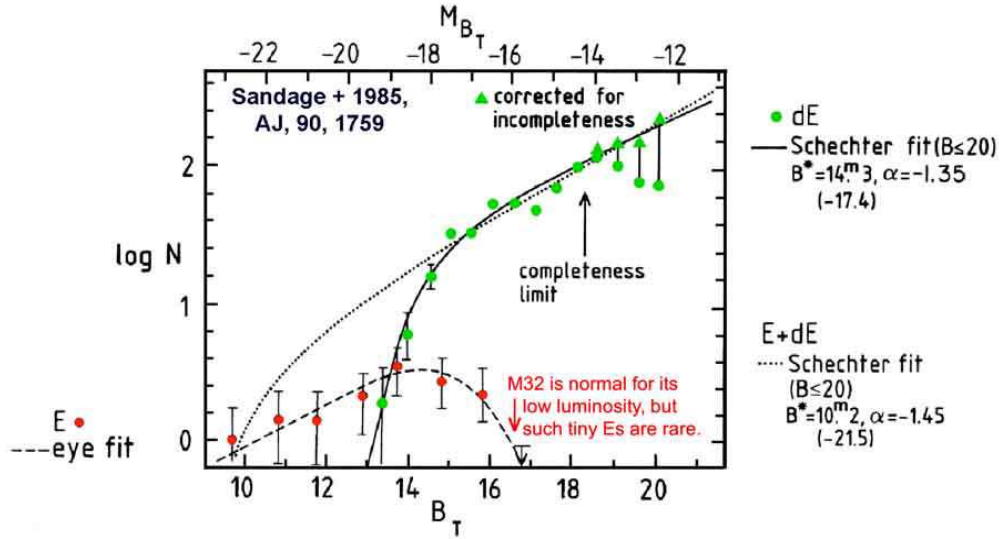}

\caption{Luminosity functions of elliptical and spheroidal galaxies in the Virgo cluster.
         This figure is adapted from Sandage \etal (1985b), who used the traditional name ``dE'' for
         spheroidals.  I have updated the Hubble constant from $H_0 = 50$ to 70 km s$^{-1}$ Mpc$^{-1}$. 
         Magnitudes are in $B$ band.  From Kormendy \& Bender (2012).
         }

\end{figure*}

\phantom{gronk}

\vskip -40pt

      Figure 1.59 is a better illustration (Sandage \etal 1985b).  It shows with data on Virgo galaxies the result that is shown 
schematically~in~Fig.~1.56.  {\it The luminosity function of ellipticals is bounded at high and low $L$}. M{\ts}32 has normal properties 
for its low luminosity, but such tiny ellipticals are rare.  In contrast, spheroidal galaxies (which Sandage \etal 1985b~call~``dEs'') 
never are very bright, but they get rapidly more common at lower luminosities until they are lost in the detection noise.
The steep faint-end slope of the luminosity function had been recognized 
for a long time (Zwicky 1942, 1951, 1957) and is built into the well known Schechter (1976) analytic luminosity function.
But, to the best of my knowledge, Sandage \etal (1985a, b) and Binggeli \etal (1988) were the first to measure luminosity functions separately
for different morphological types of galaxies and to show that only Sph galaxies have luminosity functions that continue to rise with 
decreasing luminosity to the detection limit of the data.  This is the solid result in Fig.~1.59.  Here is the incongruity:

      {\it Sandage et al.~(1985b) distinguish between elliptical and dwarf elliptical galaxies of the same luminosity.}
Quoting Kormendy \& Bender (2012): ``A dwarf version of a creature is one that, when mature, is smaller than the normal sizes of
non-dwarf versions of that creature. \dots~And yet, [Fig.~1.59] invites us to imagine that the smallest non-dwarf ellipticals are 
20 times less luminous than the brightest `dwarf ellipticals'.''

      Sandage and collaborators recognized and struggled with this incongruity.  Quoting Sandage \& Binggeli (1984):~``The distinction 
between E and dE types is made on morphological grounds alone, using surface brightness
as the criterion.  Normal E galaxies have a steep radial profle (generally following an $r^{1/4}$ law)
with high central brightness.~The~typical~dE~has~a {\it nearly flat} radial profile, following
either a King [1966] model with a small concentration index or equally well an exponential law.  
The morphological transition from E to dE is roughly at $M_B \simeq -18$, but there is overlap.''
Recognition of this difference dates back at least to Baade (1944):
``NGC\ts147 and NGC\ts185 are elliptical nebulae of very low luminosity.  In structure, they deviate
considerably from what is considered the typical E-type nebula.  In both objects the density gradient
is abnormally low.''
Binggeli \etal (1985) also recognized the quantitative similarity beween spheroidals~and~irregulars; 
their Virgo ``membership criteria applied are: (1) dE and Im members have low surface brightness. \dots''.
Soon afterward, Sandage \etal (1985b) admit that ``We are not
certain if this [E{\ts}--{\ts}dE dichotomy] is totally a tautology due merely to the arbitrary classification criteria
that separate E from dE types \dots~or if the faint cutoff in the [E luminosity function] has physical
meaning related to the properties of E and dE types.  In the first case, the problem would be only one
of definition.~In the~second, the fundamental difference in the forms of the luminosity functions of E and
dE types\ts\dots{\ts}would suggest that two separate physical families may, in fact, exist with {\it no\/} 
continuity between them (cf.~Kormendy 1985 for a similar conclusion).''  Revising a long-held picture can
be uncomfortable.

      Within a few more years, Binggeli \etal (1988) recognized that ``The distinction 
[between] Es and dEs must almost certainly mean that the two classes are of different origin [Kormendy 1985, Dekel \& Silk 1986].  
This is also supported by the fact that the luminosity functions of Virgo Es and dEs [are different].''  And Binggeli
\& Cameron (1991) concluded that ``there are no true intermediate types between E and dE.  The [E{\ts}--{\ts}dE] dichotomy 
is {\it model-independent\/}'' (emphasis in the original).

       But psychology did not lose its hold on people.  Binggeli changed~his~mind: in a section entitled ``The E--dE dichotomy
and how it disappears'', Jerjen \& Binggeli (1997) emphasize the observation that, in a plot of brightness profile S\'ersic
index {\it versus} $M_{BT}$, E and dE galaxies show a continuous correlation.  They conclude that compact ellipticals like M{\ts}32 and its
analogs in Virgo are ``special'' and that dEs form the extension of the ellipticals to low $L$.  However, this is not the only 
relevant correlation.  The observations which suggest the dichotomy had not disappeared.  And the fact that one can find parameters 
of galaxies that are insensitive to the differences between two types does not prove that the two types are the same.  Many
parameters are continuous between ellipticals and spheroidals.  E.{\ts}g., the content of heavy elements is not only a continuous 
function of luminosity for ellipticals and spheroidals, it is essentially the same continuous function for spirals and irregulars, too
(e.{\ts}g., Mateo 1998). If we looked only at element abundances, we would be blind to all structural differences encoded in Hubble types. 

      More recent criticisms of the E -- Sph dichotomy are reviewed in Kormendy \etal (2009) and in Kormendy \& Bender (2012).
The arguments involve technical details such as sample size and profile analysis techniques.  These are of less immediate
interest, and any discussion of them quickly gets long.  I~therefore refer readers to the above papers for our answers
to the criticisms.  A few are relevant here and will be discussed below.  But the best way to address uncertainty about
the E -- Sph dichotomy is to observe larger samples of galaxies and to address more general scientific questions, as follows.

\vskip -30pt

\phantom{000000000000}

\subsection{\hbox{Confirming the E--Sph dichotomy with large galaxy samples}}

      Kormendy \etal (2009: KFCB) extend the sample size of the parameter correlations in Fig.~1.57 by measuring
brightness profiles for all known ellipticals in the Virgo cluster and combining these with data on $\sim$\ts275 Sph galaxies. 
Examples are shown in Fig.~1.60.  Data from many sources were combined to construct composite profiles over large dynamic ranges.  
S\'ersic functions fit most of the galaxy light to remarkable precision:~over the fit ranges (vertical dashes in Fig.~1.60), 
the average RMS deviation = 0.040 mag arcsec$^{-2}$ for the whole KFCB sample.  Kormendy (2009) further added ellipticals
from Bender \etal (1992) and Sphs from Chiboucas \etal (2009).  The updated Fig.~1.57 is shown in Fig.~1.61.

\cl{\null}

\vfill

\begin{figure}[hb]

\includegraphics{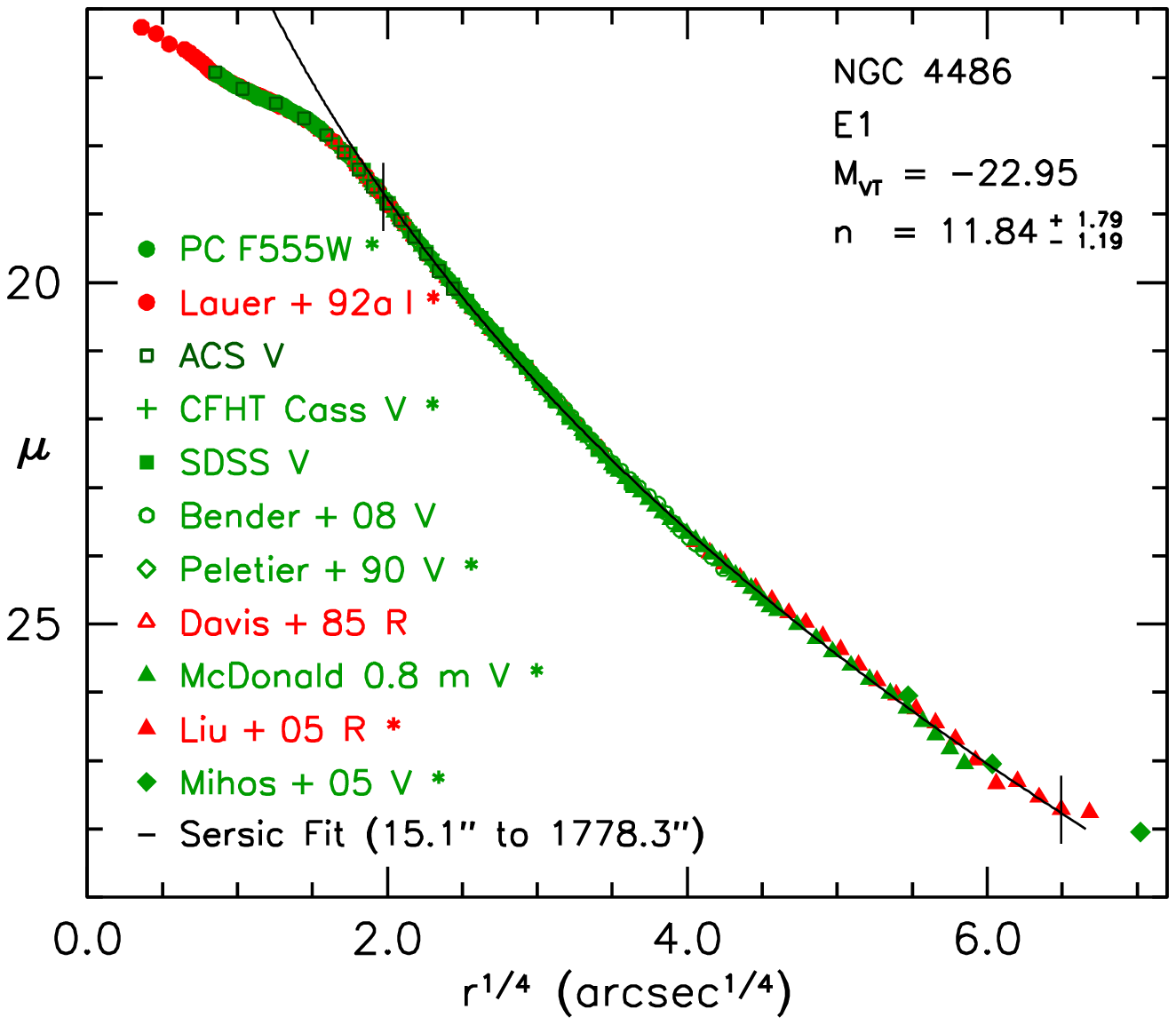}
\includegraphics{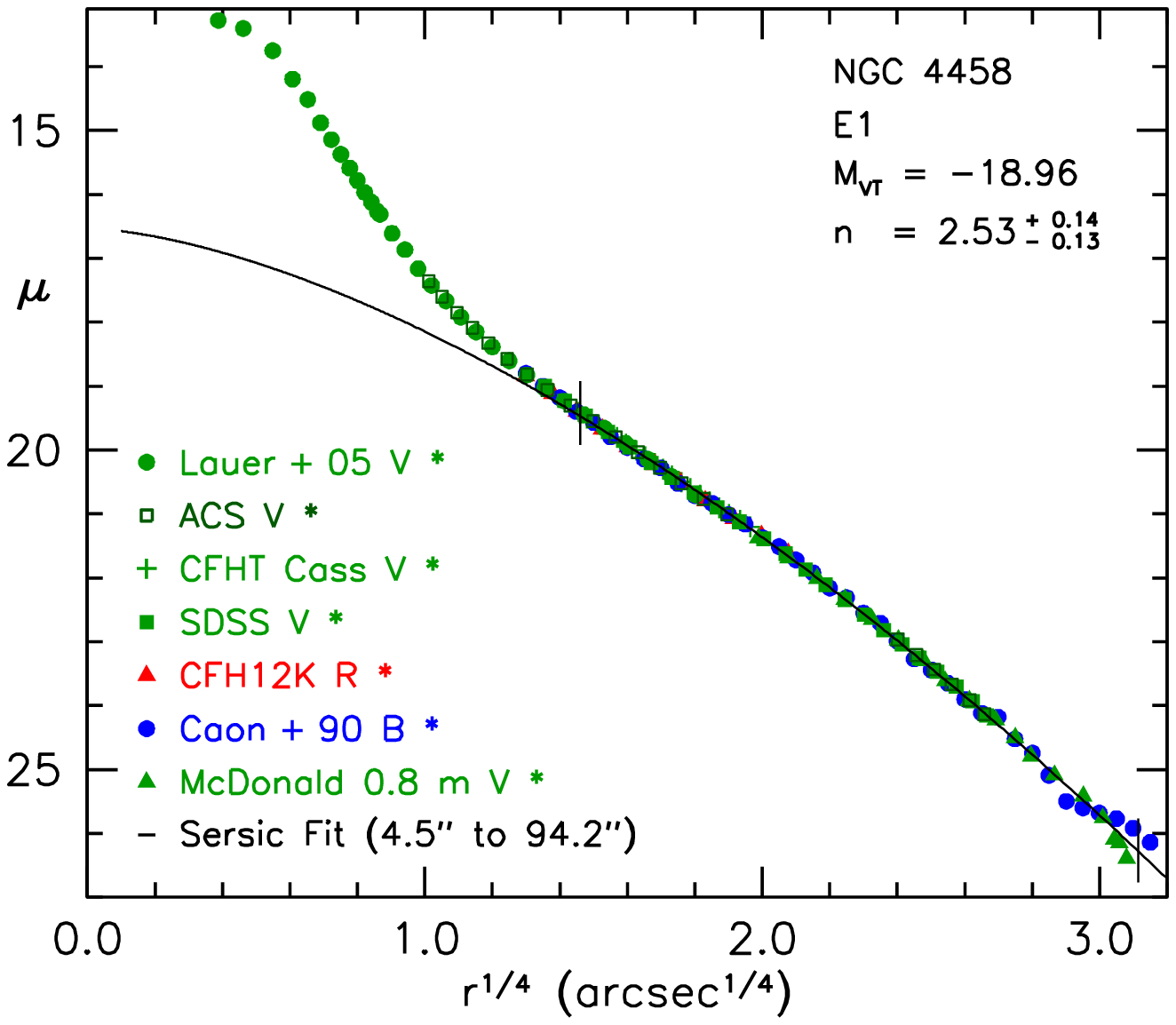}
\includegraphics{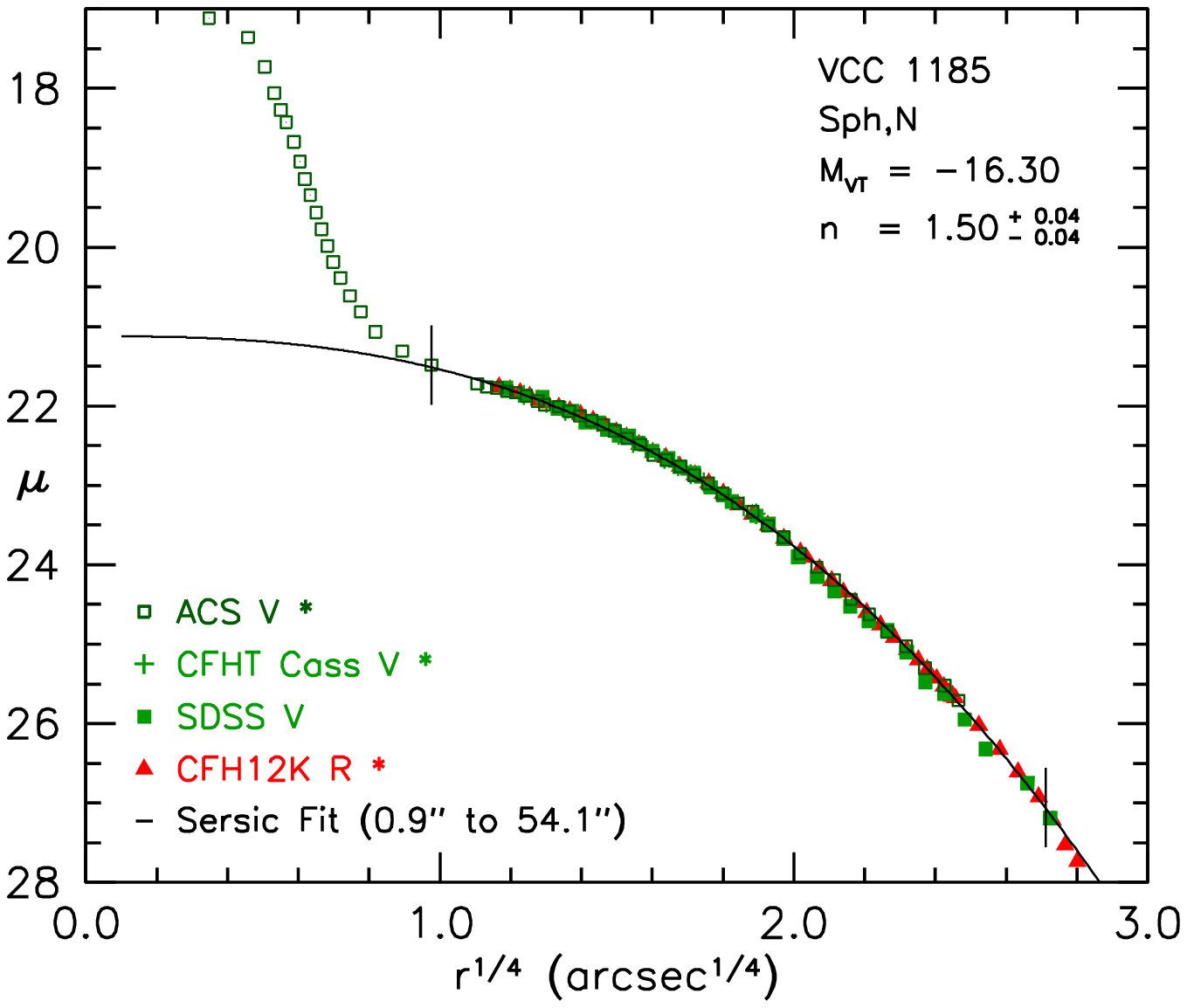}

\caption{Surface brightness profiles of 3 galaxies from KFCB.~NGC\ts4486~(M{\ts}87) is an elliptical galaxy with a central 
``core''; i.{\ts}e., central ``missing light'' with respect to the inward extrapolation of the outer S\'ersic function fit
(black curve).  NGC 4458  is an elliptical galaxy with central ``extra light'' above the inward extrapolation of the outer
S\'ersic fit.  VCC 1185 is a Sph galaxy with a nuclear star cluster (type Sph,N) in addition to its S\'ersic-function
main body.  This figure illustrates the robust profiles that are derived by using many images that provide
data in overlapping ranges of radii (e.{\ts}g.,{\it  HST} data near the center; large-field CFHT data at large $r$).
\pretolerance=15000  \tolerance=15000 
}

\end{figure}

\cl{\null}
\vskip -28pt
\eject

\vfill\eject

      Figure 1.61 strongly confirms the dichotomy between E and Sph galaxies as found in Kormendy (1985, 1987), 
Binggeli \& Cameron (1991) and 
Bender \etal (1992, 1993).
Note that the Sph sequence approaches the E sequence near its middle, not near its faint end.

\vfill

\begin{figure}[hb]

\includegraphics{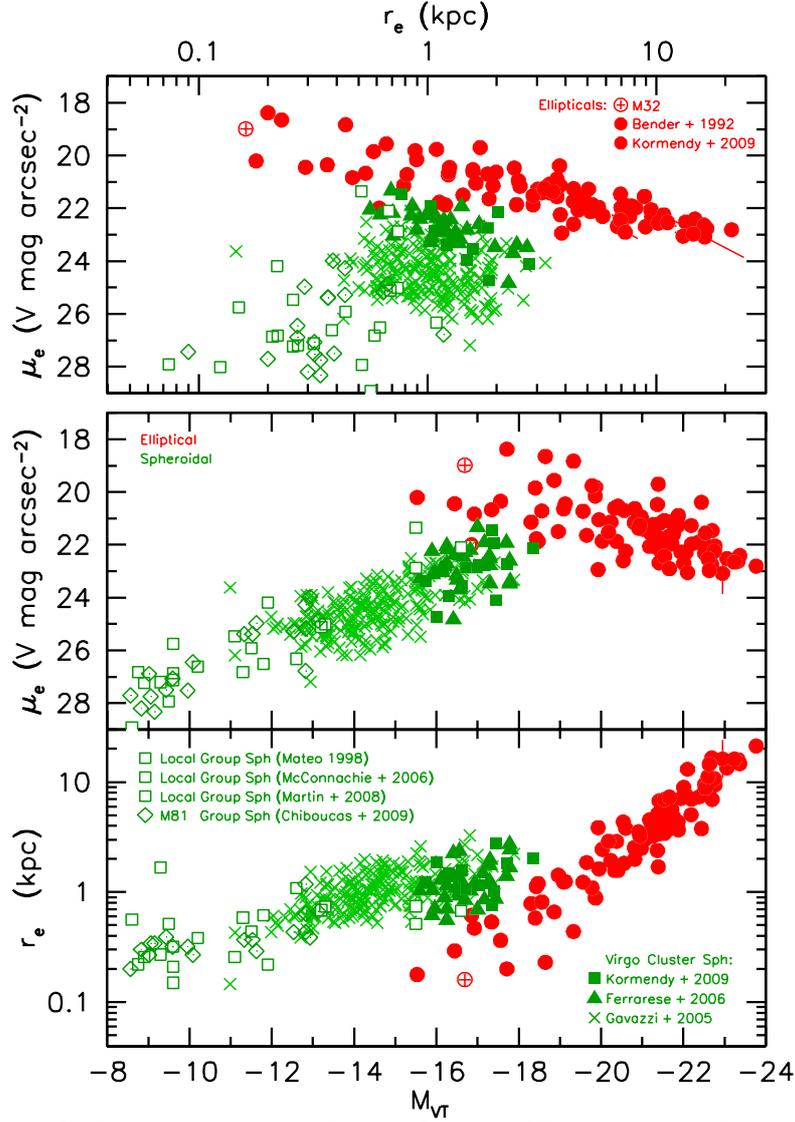}

\caption{Global parameter correlations from KFCB and Kormendy (2009) with Sph galaxies in the Local
and M{\ts}81 groups updated from Kormendy \& Bender (2012).   This figure shows 90 ellipticals and 295
spheroidals.  One elliptical plots in the Sph sequence when effective parameters are
used; difficult cases such as this one were classified in KFCB using parameters measured at the radius that 
contains 10\ts\% of the total light.  Reason: the E{\ts}--{\ts}Sph dichotomy is most pronounced when
near-central parameters are used (contrast Fig.~1.57 with this figure).
\pretolerance=15000  \tolerance=15000 
}

\end{figure}

\cl{\null}
\vskip -28pt
\eject

      The small-$r_e$, bright-$\mu_e$ end of the sequence of ellipticals is defined in part 
by galaxies like M{\ts}32 that are sometimes called ``compact ellipticals''.  As noted above, {\it compact 
ellipticals are not a special class.}  They are continuously connected to brighter ellipticals in essentially all 
parameters.  Moreover, M{\ts}32 is no longer unique, as it appeared to be in Fig.~1.57.  We now know 
of a number of M{\ts}32 analogs
(Binggeli \etal 1985;
Lauer \etal 1995;
Faber \etal 1997;
KFCB).
Figure 1.61 illustrates and KFCB reviews evidence that M{\ts}32 is normal for its low $L$.
However, it is often suggested that these galaxies are compact only because they have been tidally stripped
by much larger companions (e.{\ts}g., 
Faber 1973;
Ferrarese \etal 2006;
Bekki \etal 2001;
Chen \etal 2010).
Kormendy \& Bender (2012) review why it is not plausible that this is the explanation for why small Es are compact.
This issue is important, so I enumerate the arguments here:

\begin{enumerate}[(a)]\listsize
\renewcommand{\theenumi}{(\alph{enumi})}
\item{Compact ellipticals are not always companions of brighter galaxies (Wirth \& Gallagher 1984).  
      Some are so isolated that no tidal encounter with a big galaxy is likely ever to have happened (e.{\ts}g., 
      VCC 1871: Kormendy \& Bender 2012).}
\item{Compact Es do not have small S\'ersic indices suggestive of tidal truncation.  In fact,
      they have the same range of S\'ersic indices $n \sim 2$ to 3.5 as isolated coreless ellipticals. 
      For example, M{\ts}32 has $n \simeq 2.9$, larger than the median value for coreless ellipticals.  
      Numerical simulations show that major mergers of gas-poor galaxies like the ones in
      the nearby Universe make remnants that have exactly the above range of S\'ersic indices (Hopkins \etal 2009a).}
\item{Many Sph galaxies also are companions of bright galaxies, but we do not argue that they have been truncated amd
      thereby made abnormally compact.  An example is NGC~205, which is shown by the open square at $M_{VT} = -16.6$
      in Fig.~1.61. It is much fluffier than M{\ts}32.}
\item{Figure 1.68 below will show that the compact end of the E sequence is also defined by tiny bulges.  
      Classical bulges and ellipticals have closely similar parameter correlations.  Most classical 
      bulges that appear in our correlation diagrams do not have bright companion galaxies.}
\item{In Fig.~1.61, the ellipticals from M{\ts}32 to cD galaxies define projections of the ``fundamental plane'' correlations 
     (Djorgovski \& Davis 1987; \hbox{Faber \etal 1987;} Djorgovski \etal 1988; Bender \etal 1992).  Its interpretation is 
      well~known: galaxy parameters are controlled by the Virial theorem modified by small nonhomologies.  N-body simulations 
      of major galaxy mergers reproduce the E-galaxy fundamental plane, not the Sph parameter sequence that is almost 
      perpendicular to it (Robertson \etal 2006; Hopkins \etal 2008, 2009b).}
\end{enumerate}

\noindent Kormendy and Bender conclude: ``some compact Es may~have~been~pruned slightly, but tidal
      truncation is not the reason why the E sequence extends to the left of where it is approached by the
      Sph sequence in [Fig.\ts1.61].''

\subsubsection{Classical bulges and ellipticals satisfy the same fundamental plane parameter correlations.~I.~Bulge-disk decomposition}

      Point (d) above anticipates the result of this subsection:~classical bulges are essentially indistinguishable
from elliptical galaxies of the same luminosity.  This in turn was further anticipated when I defined classical bulges
to be elliptical galaxies that happen to live in the middle of a disk.  Here, the time has come to ante up
the evidence by adding classical bulges to Fig.~1.61.  

      Figure 1.62 emphasizes the most important requirement for this analysis.  For each disk galaxy, it is necessary 
to decompose the observed brightness distribution into (pseudo)bulge and disk parts.  This is a fundamental
part of the classification of the central component as classical or pseudo.  It provides {\it separately\/} the parameters
of the bulge and the disk, both of which we need.  For some applications, a kinematic decomposition is also needed.  

      Photometric decomposition is the crucial requirement that allows us to ask whether classical bulges satisfy the parameter
correlations for ellipticals.  Absent such a decomposition, even the distinction between ellipticals and spheroidals is 
blurred.  This is part of the reason why Ferrarese \etal (2006); Chen \etal (2010), and Glass \etal (2011) do not see the E{\ts}--{\ts}Sph 
dichotomy.   If bulges and disks are combined in various proportions and then measured as one-component galaxies,
it is inevitable that the resulting parameters will be intermediate between those of bulges and disks and that including 
them will blur the distinction between the bulge and Sph\ts$\approx${\ts}disk sequences in Fig.~1.61 (see Figs.~76 and 77 in KFCB
and Figs.~1.63 and 1.64 here). 

\vfill

\begin{figure}[hb]



 \includegraphics{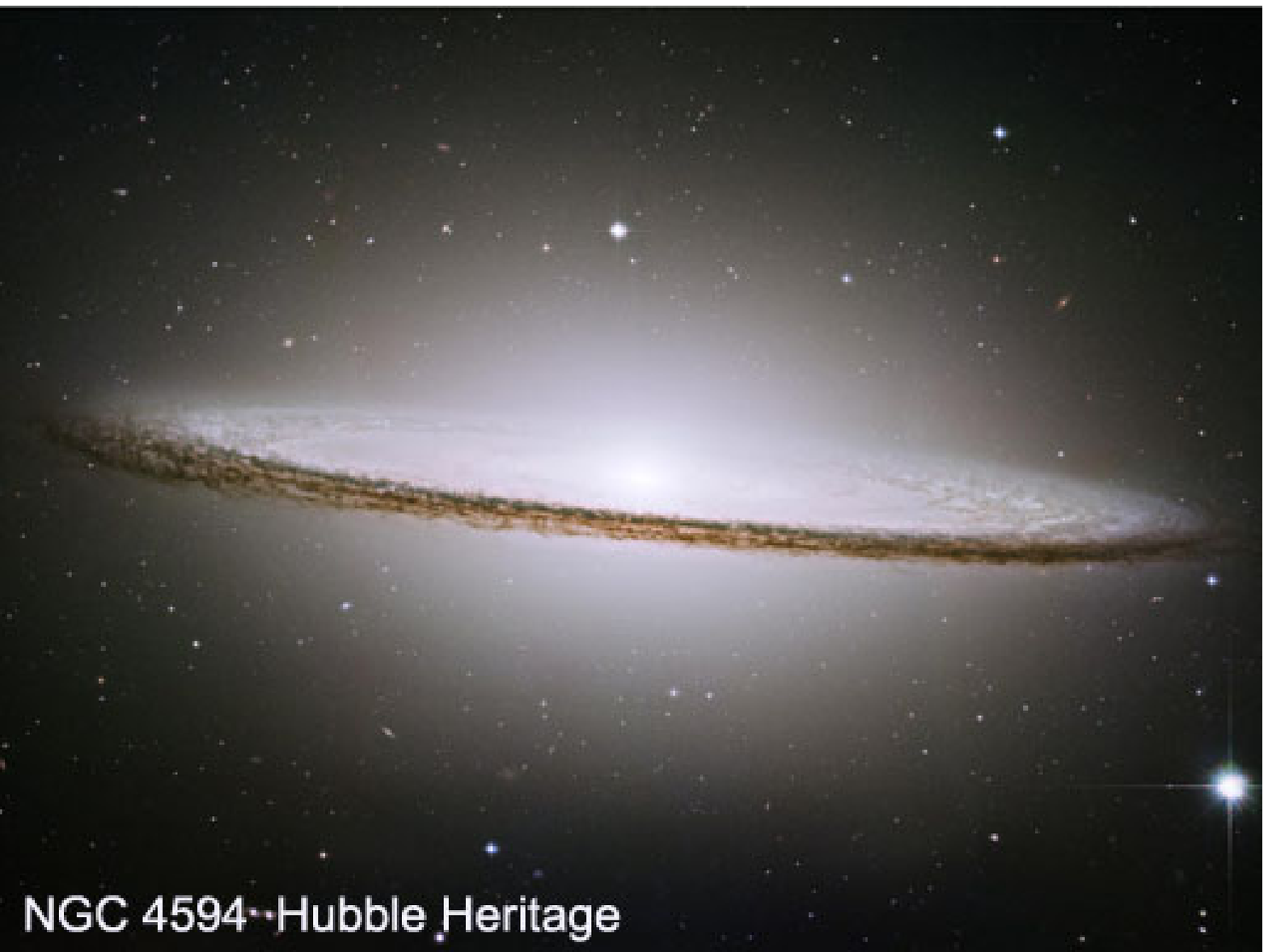}

 \includegraphics{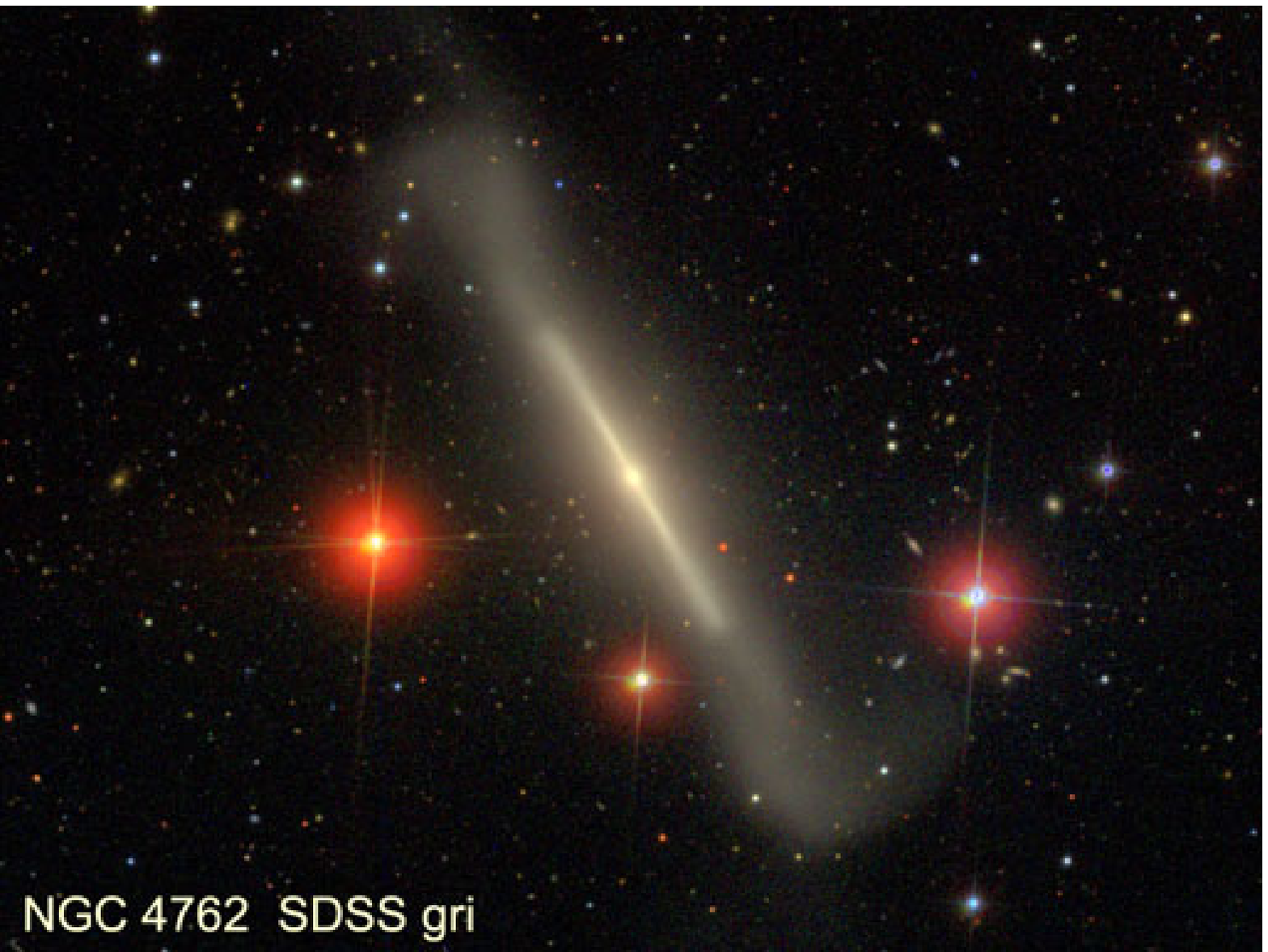}

\caption{(left) Sombrero Galaxy and (right) NGC 4762, the second-brightest S0 galaxy
in the Virgo cluster.  These galaxies illustrate why bulge-disk decomposition is necessary.
NGC 4594 is an Sa galaxy with $B/T = 0.93 \pm 0.02$ (Kormendy 2011b).  Without
photometric decomposition, we measure essentially only the bulge.  We learn nothing about the
disk.  If an S0 version of this galaxy (e.{\ts}g., NGC 3115) were viewed face-on, it would
be difficult even to discover the disk (Hamabe 1982).  In contrast, NGC 4762 is an edge-on S0
with a tiny bulge; $B/T = 0.13 \pm 0.02$ (Fig.~1.63).  Without photometric decomposition, 
we measure essentially only the disk.  We learn nothing about the bulge.   
\pretolerance=15000  \tolerance=15000 
}

\end{figure}

\cl{\null}
\vskip -28pt
\eject

      The need for bulge-disk decomposition can best be understood using an analogy. Imagine studying a population
of people, horses and people who ride on horses. Knowing nothing about them, one might measure parameters and
plot parameter correlations (linear size, mass, \dots) to look for different physical populations and regularities within each
population that might drive interpretation. We need to be careful, because some parameters (volume mass density within this 
analogy; mass-to-light ratio for galaxies) prove to be insensitive to structural differences.~Still, careful parameter~study
is promising. But the biggest people are bigger than the smallest horses. If random people are paired with random horses
and the resulting population of people+horses, together with some pure people and some pure horses, are analyzed as
one-component systems, it is inevitable that a complete continuity will be found between people and horses. But it would
be wrong to conclude that people are the same as horses. Rather, if one decomposes people and horses when they occur
together and measures their parameters separately, it will be found that some parameter correlations clearly separate
people of various sizes from horses of various sizes, even though their size distributions overlap. Further study will also
show that certain special parameters (semi-trivially:~number of arms versus number of legs in this analogy; near-central parameters in
the cases of galaxies) are especially helpful in distinguishing the physically different populations that are under study.
The one elliptical galaxy (red point) that lies within the sequence of Sphs (green points) in some panels of Fig.~1.61 was
classified using central parameters (Fig.~34 in KFCB).

      It feels strange to ``beat this dead horse'' (I'm sorry -- I could not resist): the need for component
decomposition has been understood for more than 30 years.  It quickly became standard analysis
(Kormendy 1977a; 
Burstein 1979; 
Kent~1985).
It is still so now
(Peng et al. 2002;
Knapen \etal 2003;
de Souza \etal 2004;
Laurikainen \etal 2004, 2005, 2007;
Courteau \etal 2007;
M\'endez-Abreu \etal 2008;
Weinzirl \etal 2009).
The structure (this section) and formation physics (Section 1.8) of bulges and disks are very different, and it blurs
our vision of both to analyze them as single-component systems.

\subsubsection{Small-bulge S0 galaxies and the transition from S0 to Sph galaxies}

      Kormendy \& Bender (2012:~KB2012) collect bulge and disk parameters from a variety of sources for or do photometry 
and bulge-disk decomposition of all S0 galaxies from the {\it HST} ACS Virgo Cluster survey (C\^ot\'e \etal 2004; 
Ferrarese \etal 2006).  This section reviews the results.  Classical bulges are added to the parameter correlation diagrams 
in Fig.~1.68.  But another and -- it will turn out -- especially interesting result will be to extend the Sph sequence to higher 
luminosities.  Kormendy \& Bender (2012) conclude that {\it Sph galaxies and S0 disks (but not bulges) are continuous in their 
parameter correlations.  That is, Sph galaxies are bulgeless S0s.}

      Three galaxies serve here to illustrate the transition from S0 galaxies with large classical bulges and flat disks 
to Sph galaxies with no bulges and with structure that can be vertically disky or thick.\ts~We start with NGC\ts4762.  
Figure 1.62 shows that it differs from our canonical picture of Hubble classification (Sandage 1961) in which S0 galaxies are transition
objects between elliptical and Sa galaxies.  The bulge-to-total luminosity ratio $B/T$ is a classification parameter; $B/T \equiv 1$ for 
ellipticals, and $B/T$ is intended to decrease along the sequence E\ts--{\ts}S0\ts--{\ts}Sa{\ts}--{\ts}Sb\ts--{\ts}Sc.  With some
noise, this is observed (Simien \& de Vaucouleurs 1986).  But Sidney van den Bergh (1976) already recognized 
that some S0 galaxies such as NGC\ts4762 have small bulges and, except for their cold gas content and spiral structure, are more similar 
in their overall structure to Sbc galaxies than they are to Sa galaxies.  As an alternative to the Hubble (1936) ``tuning-fork diagram'',
he proposed a ``parallel sequence classification'' in which S0 galaxies form a sequence \hbox{S0a{\ts}--{\ts}S0b{\ts}--{\ts}S0c} with decreasing
$B/T$ that parallels the sequence \hbox{Sa{\ts}--{\ts}Sb{\ts}--{\ts}Sc} of spiral galaxies with similar, decreasing $B/T$ ratios.  Van den Bergh
suggested that late-type S0 galaxies with small bulges are defunct late-type spiral galaxies that were transformed by environmental processes
such as ram-pressure stripping of cold gas by hot gas in clusters.  The KB2012 bulge-disk decompositions of NGC 4762 
and similar galaxies quantitatively confirm van den Bergh's picture, as follows.

      The brightness profile of NGC 4762 measured along the major axis of the disk is shown in Fig.~1.63 (left).
It shows a central bright and relatively round bulge and, at larger radii, three shelves in a very flat \hbox{edge-on}~disk.
The inner shelf is somewhat subtle, but the steep decrease in surface brightness between the middle and outer shelves is obvious
in Fig.~1.62.  What is this complicated structure?  This may seem like a tricky problem, but in fact, it is easy.  Relatively face-on 
galaxies that have two or three shelves in their brightness distributions are very common.  The ones with two shelves are the oval-disk
galaxies discussed in Section 1.3.3.  To get a third shelf, it is just necessary to add an early-type bar -- these have shallow
radial brightness gradients interior to a sharp outer end.  Now, the bar normally fills its attendant lens in one dimension (Section 1.4.3.4
and Fig.~1.17).  But consider a non-edge-on SB(lens)0 galaxy such as NGC 2859 (Fig.~1.9) or NGC 2950 (Fig.~1.17) in which the bar has a
skew orientation (not along either the apparent major or apparent minor axis).  If we rotated either of these galaxies about a horizontal line 
through the center in the corresponding figure until the galaxy was seen edge-on, its disk would show three shelves in its major-axis
profile.  Exterior to the bulge, the innermost shelf would be the bar, the next would be the lens, and the third would be
the outer disk.  This is how Kormendy \& Bender interpret Fig.~1.63 (left).  Thus NGC 4762 is an 
edge-on SB(lens)0 galaxy.  Bars and lenses have shallow brightness gradients at small $r$, so profile decomposition~is~easy.  
The bulge S\'ersic index $n = 2.29 \pm 0.05$ and round shape identify it as classical.  Importantly,
$B/T = 0.13 \pm 0.02$ is very small.  So Kormendy \& Bender (2012) classify NGC 4762 as SB(lens)0bc.  Note in Fig.~1.63 (right)
how measuring NGC 4762 as a single-component system (green point with brown center) mixes parameters of the classical bulge (brown point)
and disk (green cross).  Only after bulge-disk decomposition do we see that the tiny classical bulge of NGC 4762 helps to define the compact 
extension of the E{\ts}--{\ts}bulge parameter sequence.

\vfill

\begin{figure}[hb]

\includegraphics{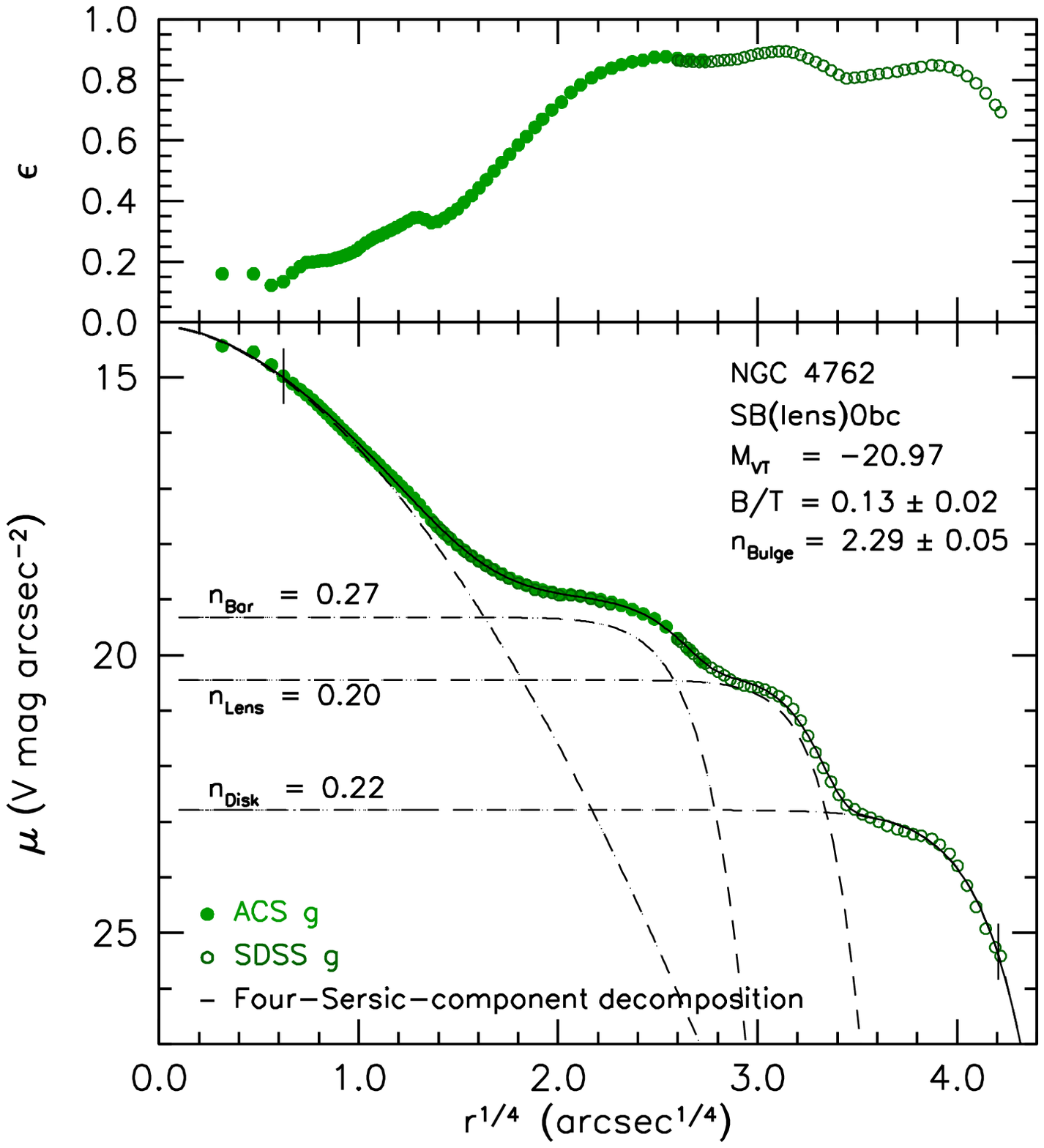}

\includegraphics{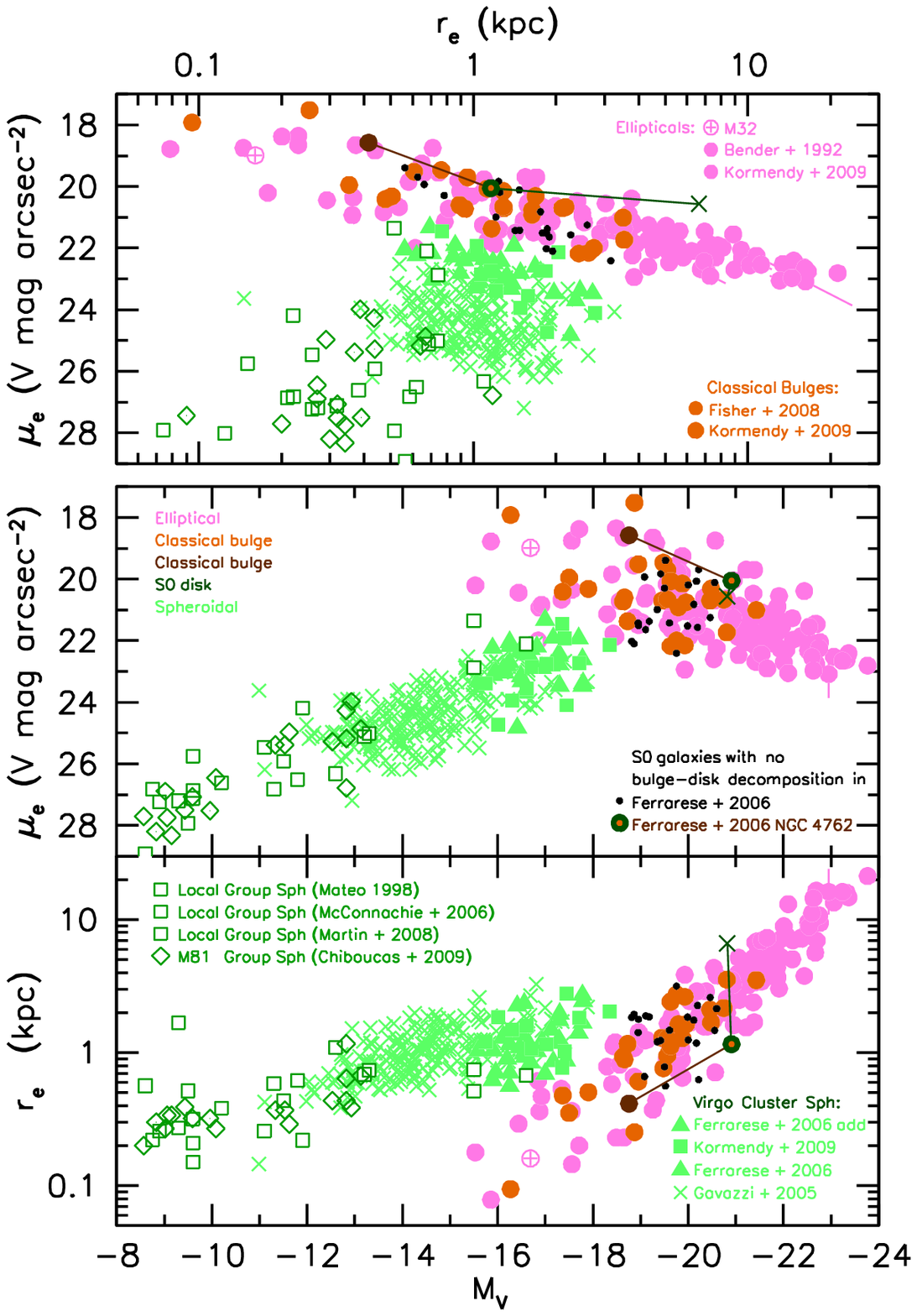}

\caption{(left) Ellipticity and surface brightness along the major axis of NGC\ts4762
measured by fitting elliptses to the isophotes in the ACS and SDSS $g$-band images.  The dashed curves 
show a decomposition of the profile inside the fit range (vertical dashes).  The bulge, bar, lens and disk 
are represented by S\'ersic functions with indices $n$ given in the figure.  Their sum 
(solid curve) fits the data with an RMS of 0.033 V mag arcsec$^{-2}$.
(right) Parameter correlations showing the results of the bulge-disk 
decomposition.  The green filled circles with the brown centers show the 
total parameters measured by Ferrarese \etal (2006) for the bulge and disk together.  They are connected
by straight lines to the parameters of the bulge (dark brown filled circles) and disk 
(dark green crosses). From KB2012.
}

\end{figure}

\eject

      NGC 4452 is closely similar to NGC 4762 but is even more extreme.  Figure 1.64 (left) shows that it, too, is an
edge-on SB(lens)0 galaxy. The decomposition robustly shows that NGC 4452 has only a very tiny pseudobulge with $n \simeq 1.06 \pm 0.14$
(recall classification criterion (6) in Section 1.5.3) and $PB/T = 0.017 \pm 0.004$.  This is an SB(lens)0c galaxy.

\vfill

\begin{figure}[hb]


\includegraphics{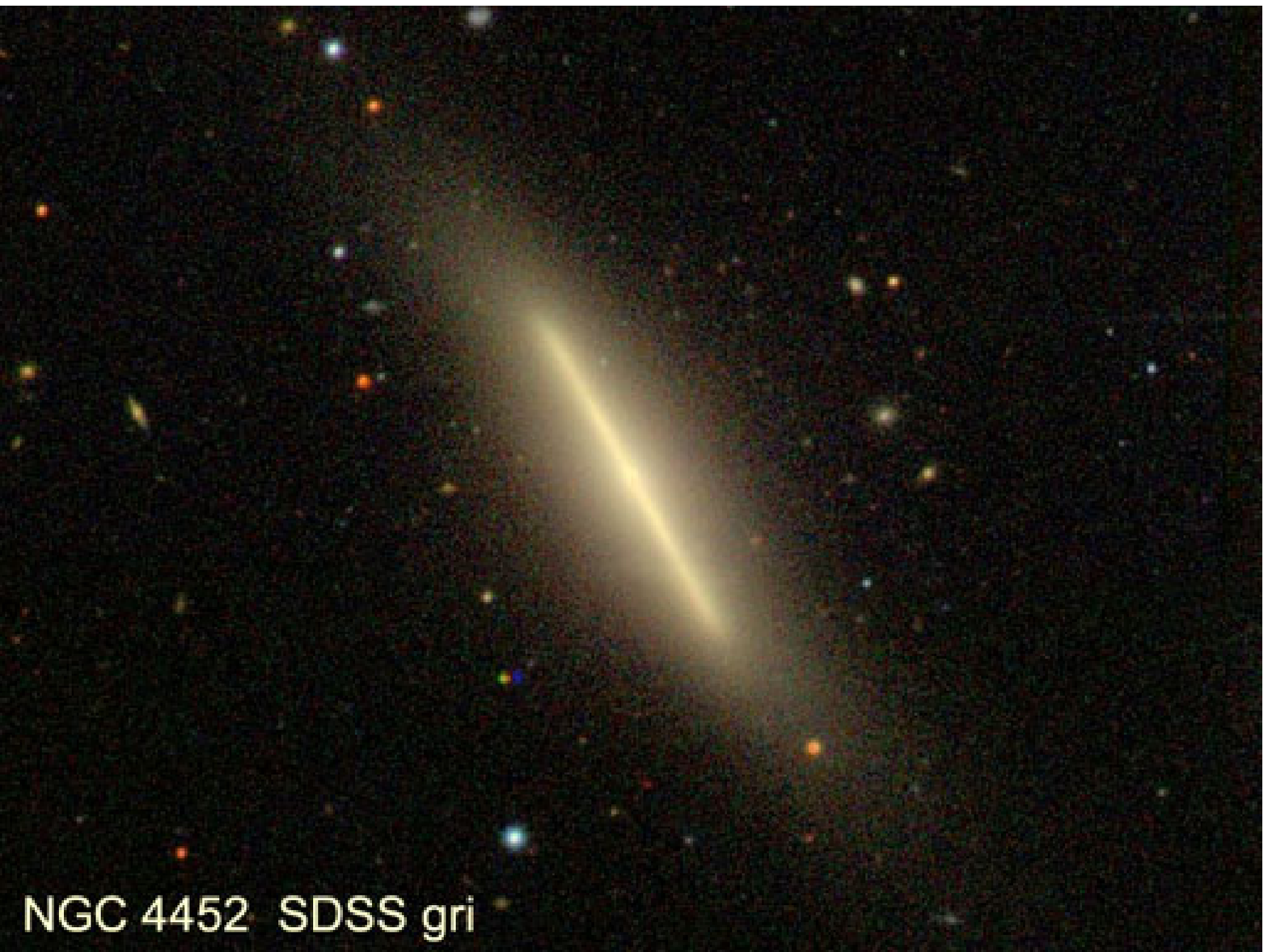}

\includegraphics{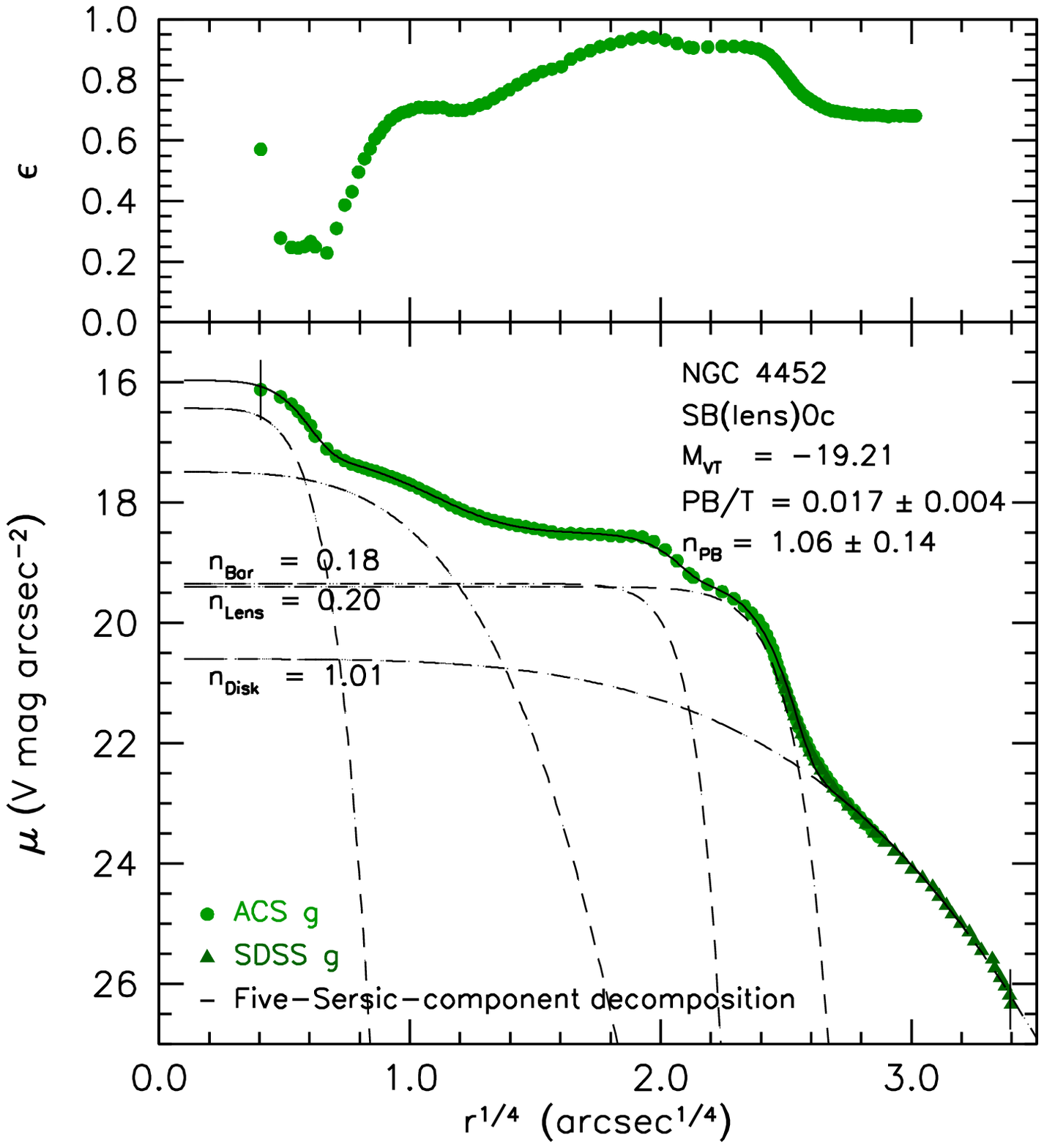}

\includegraphics{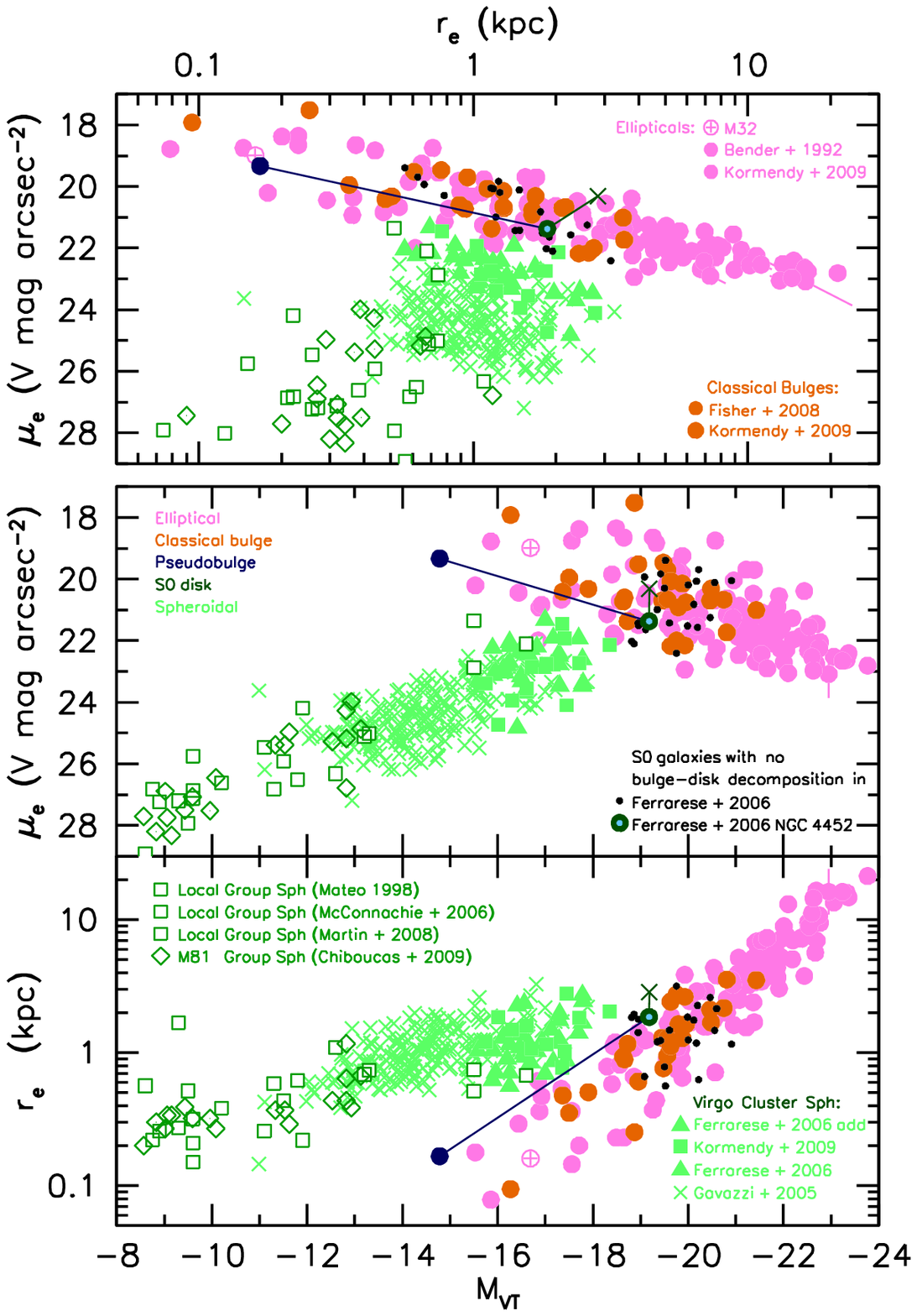}

\caption{(top left) SB(lens)0 galaxy NGC\ts4452.  The tiny pseudobulge is almost invisible.  The inner disk 
is edge-on and very flat; it again consists of two 
shelves in surface~brightness.  Including the outer, thicker disk, these three shelves are 
signatures of a bar, lens and disk.  (bottom left) Ellipticity $\epsilon$ and 
surface brightness $\mu_V$ along the major axis of NGC 4452.  The five dashed curves show a decomposition of 
the profile inside the fit range (vertical dashes).  The nucleus, 
bulge, bar, lens and disk are represented by S\'ersic functions with indices $n$ given in 
the figure.  Their sum (solid curve) fits the data with RMS = 0.044 V mag arcsec$^{-2}$. 
(bottom right) Parameter correlations showing the results of the bulge-disk 
decomposition.  The green filled circles with the blue centers show the 
total parameters measured by Ferrarese \etal (2006) for the bulge and disk together.  These points are connected
by straight lines to the parameters of the pseudobulge (blue filled circles) and disk (dark green crosses).
From KB2012.
}

\end{figure}

\cl{\null}
\vskip -28pt
\eject

      The parameter correlations in Figs.~1.63 and 1.64 serve to emphasize how bulge-disk decomposition improves our understanding 
of the E sequence.  The small black filled dots show the parameters measured by Ferrarese \etal (2006) for the ACS Virgo
cluster survey S0s.  They do not violate the E sequence.  But they do combine bulge and disk properties 
into one set of parameters, so they fail to show something that is very important.  In each of these two galaxies,
the bulge is tiny, comparable in luminosity to the smallest ellipticals.  The classical bulge of NGC 4762 helps to 
define the extension of the E sequence toward objects that are more compact than any spheroidal.  Even the tiny pseudobulge
of NGC\ts4452 lies~near~the compact end of the E$+$bulge sequence (cf.~Figs.~1.42 and 1.43, which show other, similarly
compact and tiny pseudobulges).  Figures 1.68 and 1.69 will summarize the parameter correlations for classical bulges
and S0 disks, respectively.  Here, I want to emphasize two things.  First, there exist S0 galaxies with
classical-bulge-to-total luminosity ratios $B/T$ that range from almost 1 to essentially zero.  The pseudobulge
in NGC 4452 is so small that one cannot hide a significant classical bulge in that galaxy.  Second, both NGC 4762 
and NGC 4552 have vertically thickened and warped outer disks.  Both galaxies have nearby companions.  Kormendy \&
Bender (2012) interpret these results as indicating that the outer disks are tidally warped and being heated dynamically
in the vertical direction.  They present evidence that many other S0 and Sph galaxies in the Virgo
cluster are dynamically heated, too.  Thus NGC 4762 and NGC 4552 are ``missing links'' that have some properties of
S0 galaxies and some properties of the brightest Sph galaxies.  


      NGC 4638 is even more spectacularly an S0 -- Sph transition object.  Figure 1.65 shows (bottom) the large-scale structure 
and (top) an embedded, edge-on disk and bulge in an enlargement from {\it HST\/} images.  When we wrote KB2012, this
structure was, to our knowledge, unique.  Figure 1.65 (bottom) suggests that NGC 4638 is an edge-on
S0 whose bulge happens to be very boxy.  This would be interesting but not unique; boxy bulges are discussed in Section 1.5.2.9.
But already in the bottom panel of Fig.~1.65, the structure looks suspiciously unusual: the brightness gradient in the boxy 
structure is very shallow, like that in its companion, the normal Sph,N galaxy NGC 4637.  The top panel of Fig.~1.65 shows an
almost-round, small bulge in NGC 4638 with a normal, steep brightness gradient.  To our surprise, {\it the brightness profile robustly shows\/}
(Fig.~1.66) {\it that the outer boxy structure has a S\'ersic brightness profile with $n = 1.11 \pm 0.12$ characteristic of the
main body of a Sph galaxy.  This profile is not concave-upward, as it would be if the bulge and the boxy structure where part of
the same component with $n \gg 4$.}

\vfill\eject

\cl{\null}

\vfill

\begin{figure}[hb]



 \includegraphics{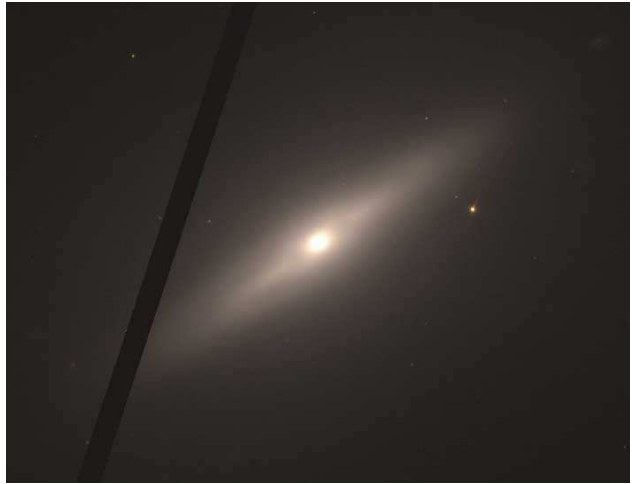}

 \includegraphics{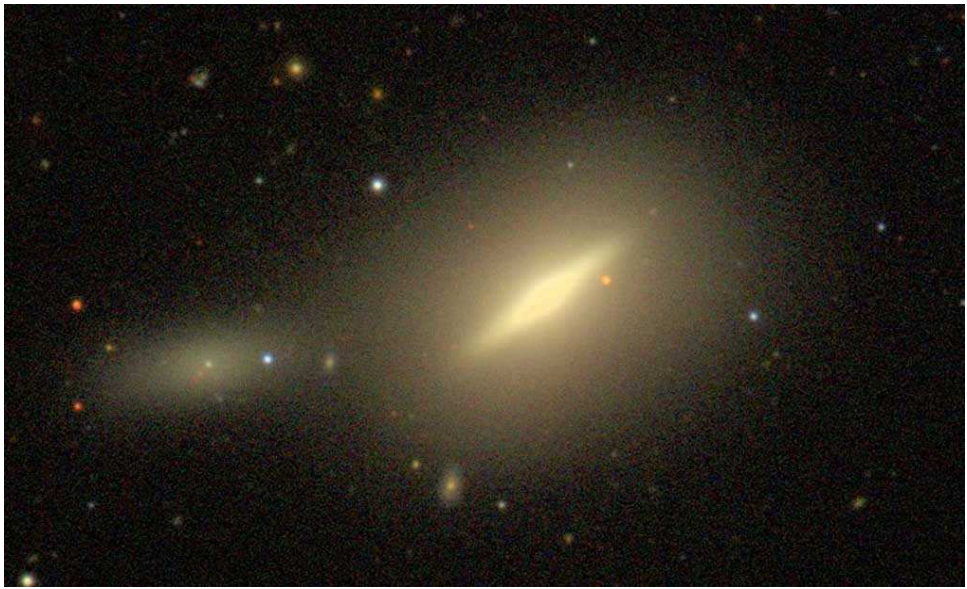}

\caption{(top) Color image of NGC 4638 = VCC 1938 made from the {\it HST} ACS $g$, mean of $g$ and $z$, and $z$ images.  
This image shows the edge-on disk and central bulge.  Brightness is proportional to the square root of intensity, 
so the brightness gradient in the bulge is much steeper than that in the boxy halo.  The very red foreground star 
near the NE side of the disk is also evident in the bottom image.
(bottom) Color image of NGC 4638 = VCC 1938 from {\tt WIKISKY}.  The brightness ``stretch'' emphasizes
faint features, i.{\ts}e., the extremely boxy, low-surface-brightness halo in which the S0 disk and bulge are
embedded.  The elongated dwarf to the west of NGC 4638 is the Sph,N galaxy NGC 4637.  Like many other spheroidals,
NGC~4637 is flatter than any elliptical.  Note also that VCC 2048 (not illustrated) is another ``missing link'' 
galaxy with both S0 and Sph properties: like NGC 4637, it is flatter than any elliptical; its main body is clearly a Sph, 
but it contains an embedded, tiny S0 disk (see KB2012, from which the above images are taken).
}

\end{figure}

\cl{\null}
\vskip -28pt
\eject

      Kormendy \& Bender (2012) therefore conclude that {\it NGC 4638 contains three structural components, and edge-on
S0 galaxy that consists of an $n = 3.6 \pm 1.4$ classical bulge plus an $n = 0.5 \pm 0.1$ Gaussian disk embedded in 
a normal Sph galaxy with $n = 1.11 \pm 0.12$.  I.{\ts}e., NGC 4638 has the properties of both an S0 and a Sph galaxy.}
VCC 2048 is similar (Fig.~1.65 caption).

      It is instructive to compare the parameters of the three components of NGC 4638 with their counterparts in pure
S0 and Sph galaxies (Fig.~1.67).  The classical bulge helps to define the compact end of the normal E{\ts}--{\ts}Sph 
parameter sequence.  It is within a factor or $\sim$\ts2 as compact as M{\ts}32.  The disk proves to have the highest
effective brightness of any S0 disk shown in Fig.~1.69.  The reasons are (1) that it is edge-on, so the path length
through it is large, and (2) that its profile is Gaussian rather than exponential; the strong outer truncation 
results in small $r_e$ and hence bright $\mu_e$.  The boxy component is consistent with the extrapolation of the
Sph sequence; it is the brightest Sph galaxy known in the Virgo cluster. 

\vfill

\begin{figure}[hb]

\includegraphics{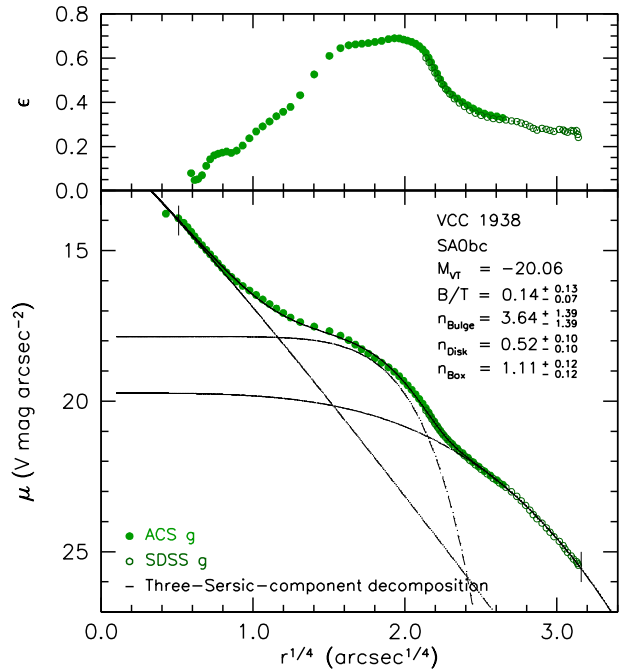}

\caption{Ellipticity $\epsilon$ and surface brightness $\mu_V$ along the major axis of NGC 4638 as measured 
on the {\it HST} ACS and SDSS $g$ images.  Dashed curves show a three-S\'ersic-function decomposition of 
the profile inside the fit range (vertical dashes).  The bulge is small, but it is classical.
The disk has a Gaussian profile, as do many other S0s discussed in KB2012.  Remarkably, the outer, boxy halo 
is clearly distinct from the bulge and disk and has a S\'ersic index $n = 1.11 \pm 0.12$.
The sum of the components (solid curve) fits the data with an RMS of 0.054 V mag arcsec$^{-2}$.  From KB2012.
}

\end{figure}

\cl{\null}
\vskip -28pt
\eject

      NGC 4638 lives in a high-density part of the Virgo cluster where strong dynamical heating is plausible.
Kormendy \& Bender interpret the boxy Sph part of the galaxy as the dynamically heated remnant of the outer disk.
Because these stars are no longer part of a disk, the disk that remains has a strongly truncated, i.{\ts}e., Gaussian 
profile.  

      KB2012 discusses additional evidence that higher-luminosity Sphs are, by and large, more disky. 
This is consistent with the suggestion that dynamical heating is one of the S\ts$\rightarrow${\ts}Sph transformation
processes and that this heating has the smallest effect on the biggest, most robust galaxies.

\vfill

\begin{figure}[hb]

\includegraphics{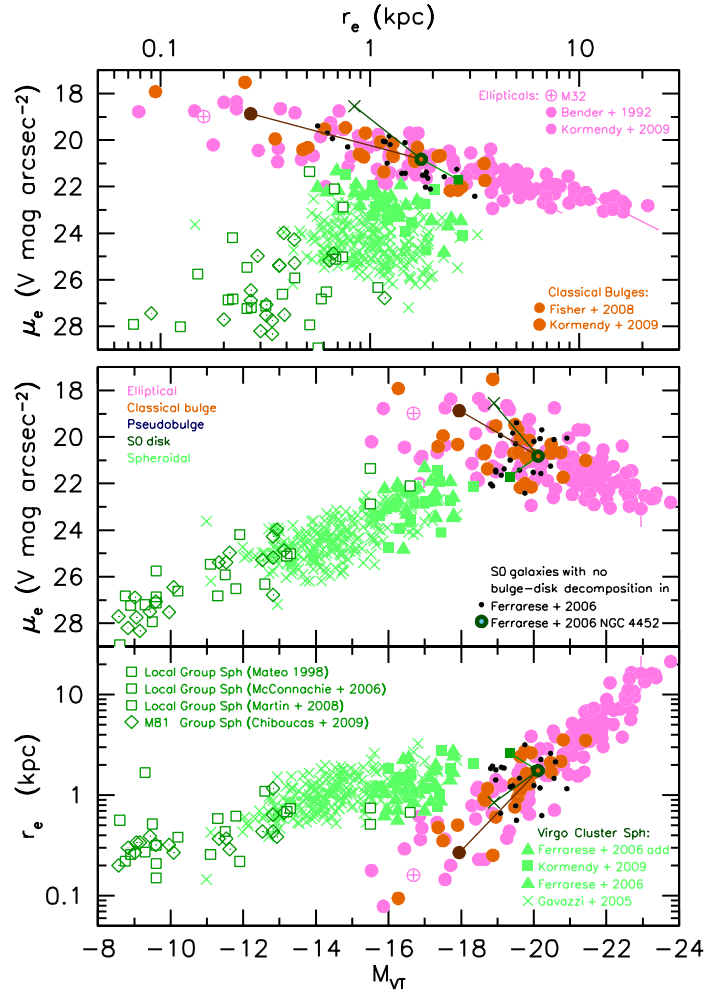}

\caption{Parameter relations showing results of the bulge-disk-Sph decomposition of NGC 4638.  The green circles 
with the brown centers show the total parameters measured by Ferrarese \etal (2006) for all components together.  These points 
are connected by lines to the parameters of the classical bulge (brown circles), the disk (green crosses),
and the Sph halo (green square).  From KB2012.
\pretolerance=15000  \tolerance=15000 
}

\end{figure}

\cl{\null}
\vskip -28pt
\eject

\subsubsection{Interim summary and road map}

      In Section 1.7.4.2, our discussion of the E{\ts}--{\ts}Sph dichotomy branched out in a new direction --
the close relationship between Sphs and S0 galaxy disks.  Section 1.7.5 pursues this.~Meanwhile, it is useful to summarize 
where~we~are.

      Section 1.7 is about environmental secular evolution.  The ``bottom line'' will be that a
variety of  environmental processes appear to have transformed some intermediate-Hubble-type spiral galaxies 
into S0s and some late-type spiral and Magellanic irregular (Im) galaxies into Sphs.  Sph galaxies will prove to be bulgeless
S0s.  ``Missing link'' galaxies that have some S0 and some Sph properties are the new subject that entered
the above discussion.  

      Recall that we were in the process of investigating the E{\ts}--{\ts}Sph dichotomy.  That is, even though they 
{\it look\/} similar, elliptical and spheroidal galaxies have quantitatively different structural parameters
and parameter correlations.  This imples that they have different formation histories -- histories that we are
in the process of deciphering.  I reviewed the history of the above discovery,
concentrating on how improved measurements and enlarged galaxy samples have strengthened the evidence for the
dichotomy.  Originally not recognized (Fig.~1.55), it was first found using small galaxy samples (Fig.~1.57) and
since has been confirmed using 90 ellipticals and 295 spheroidals (Fig.~1.61).
Our next aim has been to add classical bulges, to increase the sample size and to further show that tiny
ellipticals are not compact because they are tidally stripped.  This led us into a discussion of bulge-disk
decomposition and a description of three example galaxies, two of which have classical bulges that are substantially 
as compact as the smallest ellipticals.  In our standard picture of bulge formation by major mergers, these
bulges would have formed before their attendant disks.  It is implausible that such bulges are compact because they
were tidally stripped.

The bulge parameters measured and collected in KB2012 now allow us to
``pay the piper'' in confirming our definition of bulges as (essentially) ellipticals that live in the middle of a
disk.  This is the subject of Section 1.7.4.4.  I then return to Sph and S0 galaxies in Section 1.7.5.

\subsubsection{Classical bulges and ellipticals satisfy the same fundamental plane parameter correlations.~II.~Results}

      Figure 1.68 shows the parameter correlations from Fig.~1.61 with 57 bulges added.  Of these, 35 are known to be 
classical via their parameters and the discussion in the source papers (see the key).  I also add 22 bulges from 
Baggett \etal (1998); they are shown with open circles, because we cannot be certain that they are classical.  I examined
all of these galaxies and ensured as well as possible (using Section 1.5.3) that their bulges are classical.

\vfill\eject

      Figure 1.68 confirms the assumptions that underlie our definition of classical bulges:~they satisfy the same 
parameter correlations as do ellipticals.  Given the uncertainties in bulge-disk decomposition, there is no 
evidence that the scatter for classical bulges is different from that for ellipticals.  This is an update of a result 
that has been found previously, e.{\ts}g., by Fisher \& Drory (2008, 2010:~Fig.~1.41 here).  Pseudobulges can satisfy 
these relations, but they have much larger scatter, and they fade out by becoming low in surface brightness, not by becoming compact 
(Figs.~1.42 and 1.43).

\vfill

\begin{figure}[hb]

\includegraphics{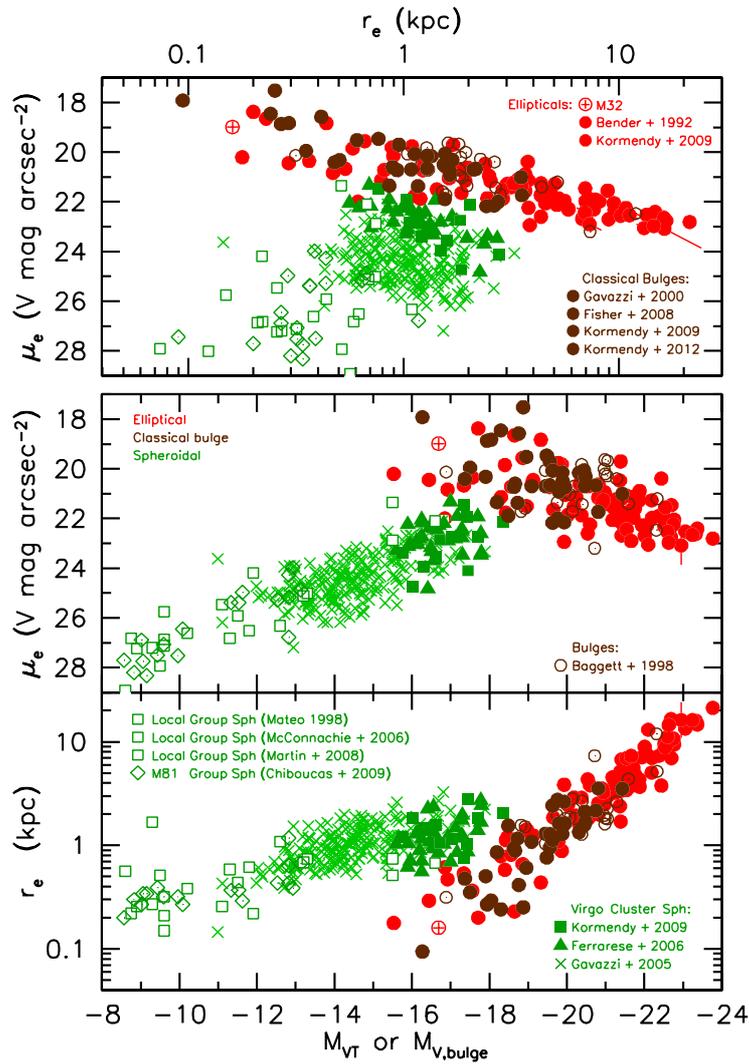}

\caption{Global parameter correlations from KFCB, from Kormendy (2009), and from Fig.~1.61 here including the sample of 
bulges from KB2012.  All ACS VCS S0s are included, three as Sphs and 23 as bulges.  For simplicity, 
points in further figures encode bulge type but not the source of the data.  
\pretolerance=15000  \tolerance=15000 
}

\end{figure}

\cl{\null}
\vskip -28pt
\eject

\subsection{Sph galaxies are bulgeless S0 galaxies}

      Figure 1.69 shows Fig.~1.68 with the disks of S0 galaxies added.  Kormendy \& Bender (2012)
conclude that spheroidals are continuous in their parameter correlations with the disks (but not the bulges) of S0
galaxies.  People~call a galaxy an S0 if it has smooth, nearly elliptical isophotes and {\it two components\/},
a bulge and a disk.  If it has no bulge and only one, shallow-surface-brightness-gradient component, 
we give it a different name -- a spheroidal.

\vfill

\begin{figure}[hb]

\includegraphics{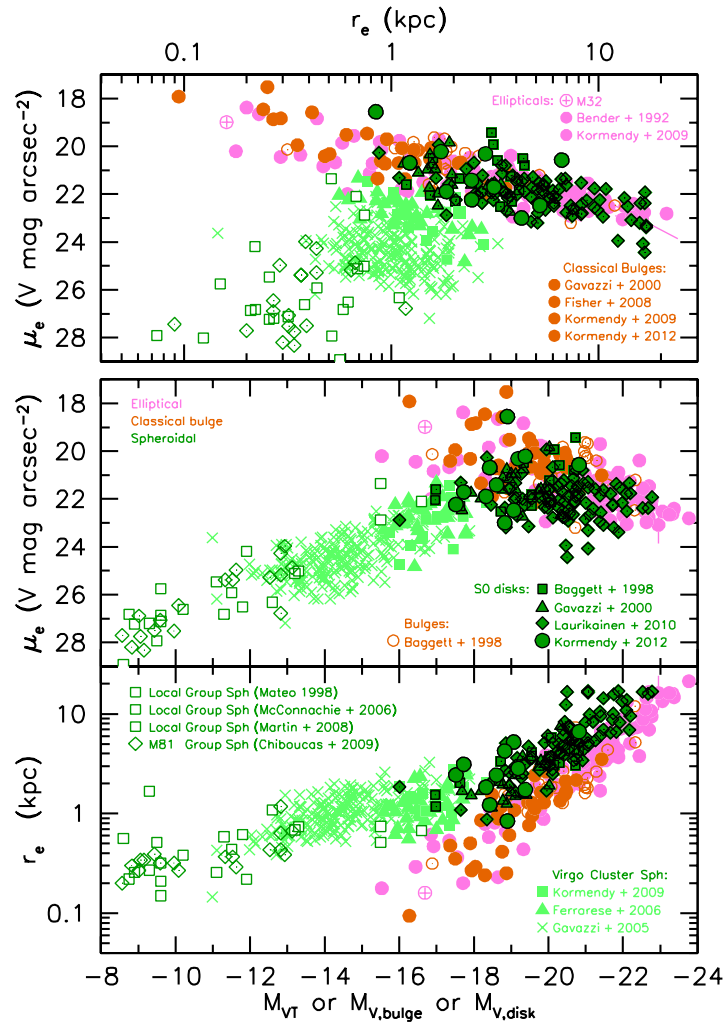}

\caption{Parameter correlations for ellipticals, bulges and Sphs with disks of 126 S0s added 
(green points outlined in black).  Bulges and disks of S0 galaxies are plotted 
separately.   The middle panel shows the Freeman (1970) result that disks
of big galaxies tend to have the same central surface brightness $\mu_0 = \mu_e - 1.822$
mag arcsec$^{-2}$ for an exponential.
\omit{Here, $\mu_e \simeq 22.0$ $V$ mag arcsec$^{-2}$ corresponds to $\mu_0 = \mu_e - 1.82 \simeq
20.2$ $V$ mag arcsec$^{-2}$ or about 21.1 $B$ mag arcsec$^{-2}$.  This is slightly brighter
than the Freeman value of 21.65 $B$ mag arcsec$^{-2}$ because many of the disk parameters are
not corrected to face-on orientation. }   We conclude that Sphs are continuous with the disks 
but not the bulges of S0 galaxies.  Updated from KB2012.
\pretolerance=15000  \tolerance=15000 
}

\end{figure}

\cl{\null}
\vskip -28pt
\eject

\cl{\null}

\begin{figure*}

\vskip 2.75truein

\includegraphics{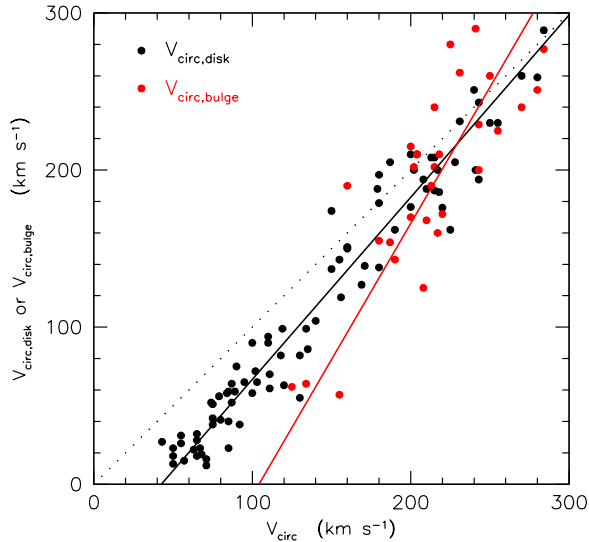}

\caption{Maximum rotation velocity of the bulge $V_{\rm circ,bulge}$ (red points)
and disk $V_{\rm circ,disk}$ (black points) derived in bulge-disk-halo decompositions of the
rotation curves of galaxies whose outer, dark matter rotation velocities are $V_{\rm circ}$.  
Equality of the visible and dark matter rotation velocities is shown by the dotted~line.
Every red point has a corresponding black point, but many galaxies are bulgeless and then
only a disk was included in the decomposition.  
This figure illustrates the well known rotation~curve~conspiracy, 
$V_{\rm circ,bulge} \simeq V_{\rm circ,disk} \simeq V_{\rm circ}$ for~the~halo
(Bahcall \& Casertano 1985; van Albada \& Sancisi 1986; Sancisi \& van Albada~1987).
It shows that the conspiracy happens mostly for galaxies with $V_{\rm circ} \sim 200$~km~s$^{-1}$.    
The lines are least-squares fits with variables symmetrized around 200 km s$^{-1}$. 
The bulge correlation is steeper than that for disks; 
bulges disappear at $V_{\rm circ} \simeq 104 \pm 16$ km s$^{-1}$. 
From Kormendy \& Bender (2011) and Kormendy \& Freeman (2013).  
\pretolerance=15000  \tolerance=15000 
}

\end{figure*}

\phantom{00000000}
\vskip -39pt

      That bulges disappear where Fig.\ts1.69 suggests is shown~in~Fig.\ts1.70.  Rotation curve 
decompositions confirm what our experience tells us: bulges disappear at $V_{\rm circ} \sim 100$
km s$^{-1}$ or $M_{\rm V,{\ts}disk} \sim -18$ (Tully \& Fisher 1977).  There is noise; e.{\ts}g.,
M{\ts}33 has $M_{\rm V,{\ts}disk} = -19.0$ and $V_{\rm circ} \simeq 135 \pm 10${\ts}km{\ts}s$^{-1}$ 
(Corbelli 2003) and no bulge (Kormendy \& McClure 1993).  But of course, we also
expect that disks fade when they are transformed from S$+$Im~to~S0.  Figure 1.70 is an important observational
``target'' for future work: the formation physics that underlies it is largely unknown.  But there
is ample evidence that bulges disappear approximately where the Sph and S0 disk sequences meet
in Fig.~1.69.  This is enough to explain the different names.

      Kormendy \& Bender (2012) suggest that the kink in the $\mu_e${\ts}--{\ts}$M_V$ correlation that
happens roughly at the transition from S0 disk to Sph tells us where the correlation turns into a sequence of
decreasing baryon retention at lower galaxy luminosity.  It is not an accident that this happens roughly where
the bulge contribution to the gravitational potential well disappears.

\vfill \eject

\subsection{Spiral and irregular galaxies have the same structural correlations as S0 galaxy disks and Sph galaxies}

      Kormendy's (1985, 1987) conclusion that Sph galaxies are defunct dS$+$Im galaxies depended critically on the
observation (Fig.\ts1.57) that they~all~have the same structural parameter correlations.~That result was based~on~a~small 
number of galaxies and has never been checked.  KB2012 updates and extends this test with 407 galaxies 
that cover the complete luminosity range from the tiniest dwarf irregulars to the brightest Sc disks.  Figure 1.71 shows 
that S$+$Im galaxies do indeed have the same parameter correlations as S0 disks and spheroidals.\ts~Therefore they
are closely related.

\vfill

\begin{figure}[hb]

\includegraphics{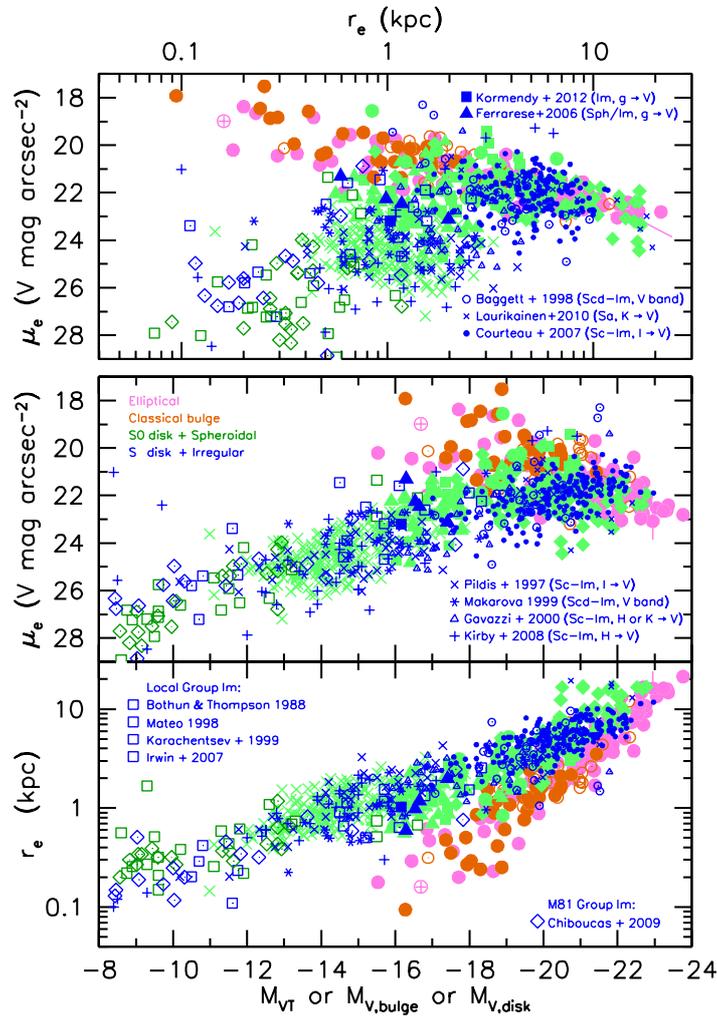}

\caption{Fig.~1.69 correlations with disks of Sa{\ts}--{\ts}Im galaxies added
(blue points for 407 galaxes from 14 sources listed in the keys).  
When bulge-disk decomposition is needed, the components are plotted separately.  
From KB2012.
\pretolerance=15000  \tolerance=15000 
}

\end{figure}

\cl{\null}
\vskip -28pt
\eject

\subsection{A revised parallel-sequence classification of galaxies}

      Figure 1.72 shows our proposed revision of Sidney van den Bergh's (1976) morphological classification scheme based on the foregoing 
observations.  

      Van den Bergh put S0 galaxies in a sequence that parallels the spirals; the classification parameter that 
determines the stage along either sequence is the (pseudo)bulge-to-total luminosity ratio, $(P)B/T$.  Pseudo~and~classical 
bulges are not distinguished; in a classification based on small-scale~images, this is the only practical strategy.  Only $(P)B/T$ 
and not parameters such as spiral arm pitch angle determine the stage, so van den Bergh's classification of spirals 
is not quite the same as Sandage's or de Vaucouleurs's.~We do not address this issue.  Figure 1.72 adopts van den Bergh's 
theme of placing S0s and spirals in parallel sequences based only on $(P)B/T$.  

      Kormendy \& Bender extend ven den Bergh's discussion in two ways. \vskip -18pt \phantom{0000000000000000} 
     \begin{enumerate}[(a)]\listsize
     \renewcommand{\theenumi}{(\alph{enumi})}
\item{They resolve the uneasy aspect of van den Bergh's paper that he listed~no~S0c or later-type S0 galaxies.  Based on a 
      comparison of $(P)B/T$ ratios of S0s with spirals of known Hubble type, they find several 
      of the ``missing'' late-type S0s; e.{\ts}g., the S0bc galaxy NGC 4762; the S0c galaxy NGC 4452 (Section 1.7.4.2).  
      NGC 4452 is also singled out as an S0c by Cappellari \etal (2011), who independently propose a parallel-sequence 
      classification based on kinematic maps.  A few other S0cs are known (Laurikainen \etal 2011; Buta 2012). }
\item{They place Sph galaxies in parallel with Im galaxies.  They note that, in a more detailed classification that includes
      Sd and Sm galaxies, some Sphs (e.{\ts}g., ones with nuclear star clusters) would be placed in parallel with late-type
      (especially Sm) spirals, and others (e.{\ts}g., ones without nuclei) would be put in parallel with Ims.
      Adding Sph galaxies at the late-type end of the S0 sequence for the first time finds a natural home for them in a
      morphological classification scheme.}
\end{enumerate}

\vfill

\begin{figure}[hb]


\includegraphics{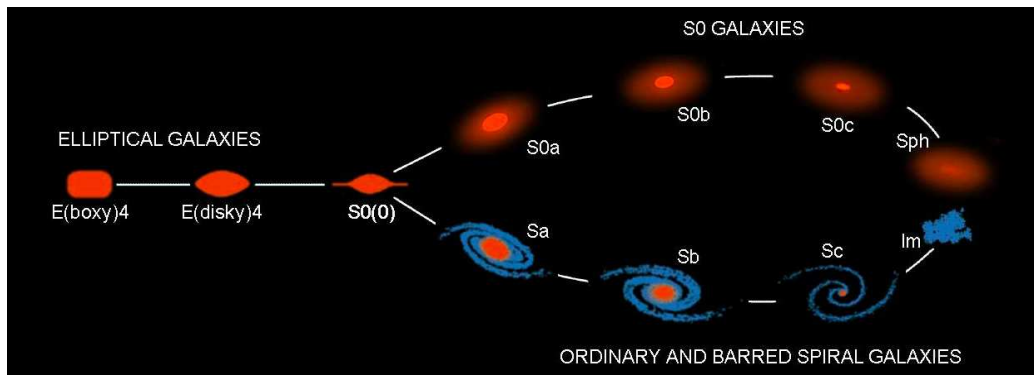}

\caption{Revised parallel-sequence morphological classification of galaxies.
The E types are from Kormendy \& Bender (1996).  Transition objects between
spirals and S0s (van den Bergh's anemic galaxies) exist but are not illustrated.
Bulge-to-total ratios decrease toward the right; Sc and S0c galaxies have tiny or no
pseudobulges. Sph and Im galaxies are bulgeless.  From KB2012.
         }

\end{figure}

\cl{\null}
\vskip -28pt
\eject

      It is important to understand which observations lead to Fig.~1.72\footnote{Allan Sandage (2004)
          accused Sidney van den Bergh of hermeneutical circularity in setting up his parallel-sequence
          classification, which  -- he thought -- involved too much interpretation.  A prosaic but sufficient
          paraphrase is ``circular reasoning''.  The basic idea is this: a morphological classification of galaxies should be
          set up based only on observations and not on interpretation (see Section 1.3.4).  The aim is that regularities revealed by the 
          classification will later aid interpretation.  However, if some interpretation is used in setting up the classification, 
          then the ``aid'' that the classification can provide is foreordained.  This is circular reasoning.  In practice, science 
          is not as ``black and white'' as Sandage suggests.  Even the greatest scientists (Sandage explicitly picked Hubble as one
          of these) set up classifications with future interpretation~in~mind.  They make decisions about which observations to 
          treat as relevant and which ones to treat as secondary.  Van den Bergh did this faithfully; Sandage was just uneasy about 
          how important those decisions were.  It should be clear from these remarks that I respect both sides of the argument.
          In the end, further advances reviewed here have, I claim, vindicated van den Bergh's decisions.  For a classification 
          to be successful, it must ultimately motivate a clearcut paradigm of interpretation.  Van den
          Bergh's parallel-sequence classification has done this.}. 
They involve quantitative parameter measurements, but they do not involve interpretation.  First, the
observations that establish E{\ts}--{\ts}S0{\ts}--{\ts}Sph continuity:
     \begin{enumerate}[(a)]\listsize
     \renewcommand{\theenumi}{(\alph{enumi})}
\item{Galaxies with smooth, nearly elliptical isophotes, little cold gas and little star formation range in bulge-to-total
      luminosity ratio from $B/T = 1$ to $B/T = 0$. Here, the existence of a bulge component and the measurement of $B/T$
      are based on quantitative surface photometry, on nonparametric measurements of structural parameters for elliptical and Sph
      galaxies by integrating the observed isophotes, on parametric (S\'ersic-function-based) bulge-disk decomposition for
      disk galaxies, and on the resulting structural parameter correlations (Figs.~1.68, 1.69).  When $B/T = 1$, we call
      the object an elliptical; when $1 > B/T > 0$, we call the central component a bulge and the outer component -- if flat\footnote{This
      is to prevent confusion with cD galaxies, which have cluster-sized debris halos, not disks.}~--
      a disk, and when $B/T = 0$, we call the galaxy a spheroidal.}
\item{In the structural correlations between effective radius $r_e$, effective brightness $\mu_e \equiv \mu(r_e)$ and total
      absolute magnitude, Sph galaxies define a sequence that is continuous with the disks but not the bulges of S0 galaxies.
      There is some overlap in luminosity between Sphs and S0 disks.}
\item{NGC\ts4762, NGC\ts4452, NGC\ts4638 and VCC\ts2048 are galaxies that have both~S0 and Sph properties.  We know this
      because all four galaxies are seen edge-on.  All contain flat disks.  Three contain a tiny (pseudo)bulge
      (VCC 2048 contains only a nuclear star cluster).  The thick outer components of all four galaxies have parameters --
      including S\'ersic indices $n \sim 1$ -- that are indistinguishable from those of Sphs.  That is, these galaxies consist of
      S0 central parts embedded in Sph or Sph-like outer halos.  This helps to establish S0{\ts}--{\ts}Sph continuity.}
\item{Bigger Sph galaxies tend to be dynamically more S0-disk-like:~they have larger ratios of rotation
      velocity to velocity dispersion (van Zee \etal 2004).  Note: at all $L$, some Sphs rotate slowly
      (see KB2012 for a review).
     }
\end{enumerate}
\noindent These observations justify our conclusion that Sph galaxies are continuous in their properties with S0 disks,
which in turn motivates our juxtaposition of Sph galaxies with S0cs.  In essence, Sph galaxies are bulgeless S0s.

\vfill\eject

\noindent ~~Observations that suggest parallel sequences of S$+$Im and S0$+$Sph galaxies:
     \begin{enumerate}[(a)]\listsize
     \renewcommand{\theenumi}{(\alph{enumi})}
\item{For every $B/T$ ratio that is observed in an S0 or Sph galaxy, there are S or Im galaxies that have
      corresponding, similar $B/T$ ratios.  We see a continuous transition from S0 to E as
      $B/T \rightarrow 0$.  We do not know whether Sas also have a continuous
      transition $B/T \rightarrow 0$.  The Sombrero galaxy (NGC 4594) has one of the largest bulge-to-total ratios known,
      $B/T = 0.93 \pm 0.01$ (Kormendy \& Bender 2013).  I know no Sa with larger $B/T$.  Thus it is prudent to retain a 
      classification S0(0) that is intermediate between elliptical and both~Sa~and~S0a.}
\item{Except for details such as spiral structure, the {\it global\/} structure of spirals and S0s is similar.
      For any generic Sa, Sb, or Sc galaxy, there are similar S0a, S0b, or S0c galaxies.  In particular,
      the bulges of spiral and S0 galaxies both satisfy the E parameter correlations.  The fractions of classical
      and pseudo bulges are similar at similar stages along the tuning fork (Kormendy \& Kennicutt 2004).  And the disks
      of S$+$Im galaxies have almost the same parameter correlations as Sph galaxies and S0 disks (Fig.~1.71).}
\item{Some Sph galaxies contain low-contrast spiral structure; therefore they contain embedded disks
      (Jerjen \etal 2000, 2001;
       Barazza \etal 2002; 
       De Rijcke \etal 2003;
       Graham \etal 2003; 
       Ferrarese \etal 2006; 
       Lisker \etal 2006, 2007, 2009).}
\item{Many dSph companions of our Galaxy contain intermediate-age stellar populations 
      (Da Costa 1994;
      Mateo 1998:~Fig.~1.73 here;
      Tolstoy \etal 2009).
      Both among the Galaxy's satellites and in the larger {\it HST} ACS Nearby Galaxy Treasury Survey
     (Weisz \etal 2011a, b), {\it dS, dIm and dSph galaxies have similar, heterogeneous star formation
      histories except that the star formation rate in dSph galaxies is currently zero.}~This is a matter of
      definition{\ts}--{\ts}if~a~dwarf contains gas and star formation, it is called dSph/dIm or
      dIm.  The Virgo cluster contains several examples (Ferrarese \etal 2006; KB2012).}

\vfill

\begin{figure}[hb]


 \includegraphics{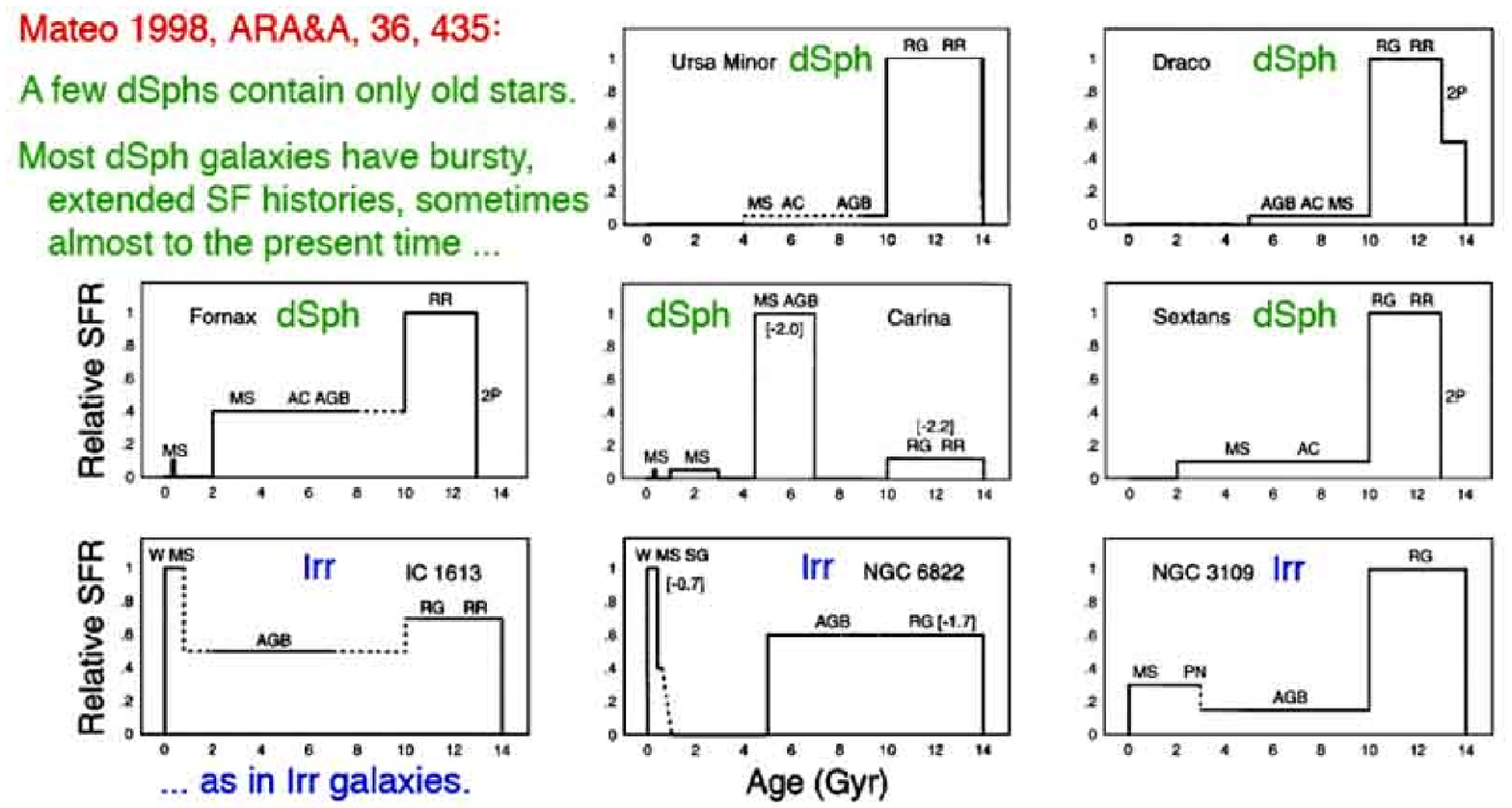}

\caption{Star formation histories of dSph and dIm galaxies from Mateo (1998).
         Relative star formation rates are shown as a function of time since the Big Bang.
         }

\end{figure}

\cl{\null}
\vskip -28pt
\eject

\item{Similarly, some spiral galaxies in clusters contain gas only near their centers, and some S0s 
      contain near-central gas and small amounts of star formation.  This is discussed in Section 1.7.9.
      Here, it again means that some S0 galaxies are less different from some spiral galaxies than optical
      images would suggest.}
\item{Van den Bergh's (1976) ``anemic galaxies'' are omitted from Fig.~1.72 for simplicity,
      but they are galaxies that are intermediate in properties between spiral and S0 galaxies.
      Their contain only low-amplitude spiral structure star formation.  The transition from
      S to anemic to S0 looks continuous.}
\end{enumerate}
\noindent Thus a substantial collection of morphological and structural parameter observations
          motivate our suggested parallel-sequence galaxy classification.  We revise it to place 
          Sph galaxies at the end of the S0 sequence, juxtaposed with the latest-type spirals and irregulars. 
          It is important to note three things.  We do not intend to imply that the luminosity function of 
          galaxies is the same at all stages of the tuning fork.  Indeed, we already know that 
          Im and Sph galaxies tend to have lower luminosities than earlier-type S and S0 galaxies.  Second,
          we do not mean to imply that galaxies are equally abundant at every stage of either the S$+$Im
          or the S0$+$Sph sequence.  Indeed, it is clear that S0c galaxies are much rarer than 
          Sphs or earlier-type S0s.  This provides a hint for interpretation.  But it is not a reason to
          change the classification.  And third, we do not intend~to~fix~what~isn't~broken.  Our suggestion
          of a parallel-sequence classification is not meant to replace Hubble classes.  We propose
          Fig.~1.72 as a complement to Hubble classification, useful because it encodes a different collection
          of observations that are relevant to a different collection of questions about formation physics.

\vskip -24pt

\phantom{000000000000000000000000}

\subsection{Parallel-sequence classification and\\bimodality in the galaxy color-magnitude relation}

      Work on galaxy formation nowadays concentrates on the history of star formation in the Universe
and on understanding stellar populations.  The iconic observation that current work tries to explain is the
 color bimodality of galaxies in the color-magnitude relation as revealed by the Sloan Digital Sky Survey (SDSS) 
at low redshifts
(Strateva \etal 2001;
Bernardi {\it et al.}~2003; 
Kauffmann \etal 2003a, b;
Hogg \etal 2002, 2004; 
Blanton \etal 2003, 2005; 
Baldry \etal 2004)
and by HST studies of galaxies at high redshifts.
Figure 1.74 shows this result and illustrates how the E{\ts}--{\ts}S0{\ts}--{\ts}Sph arm of the parallel-sequence
tuning fork relates to it.  The bright end of the prominent and thin red sequence consists of ellipticals, S0s,
and some early-type spirals.  But their luminosity functions are bounded at low $L$.  When the red
sequence is extended fainter, it must become dominated by Sphs at $M_V \ll -18$.  The deepest surveys
detect this (Blanton \etal 2005; Drory \etal 2009).

\vfill\eject

\cl{\null}

\begin{figure*}

\vskip 2.75truein


 \includegraphics{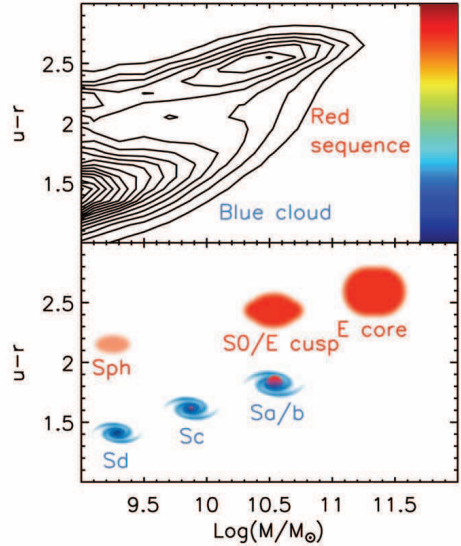}

\caption{Correspondence between our parallel-sequence classification and the color bimodality of galaxies 
in the SDSS color-magnitude relation.  The top panel shows contours of galaxy number density in the correlation 
between SDSS $u - r$ color and galaxy baryonic mass $M/M_\odot$ (Baldry \etal 2004).  The  narrow 
``red sequence''  of mostly-non-star-forming galaxies and the broader ``blue cloud'' of star-forming galaxies
are well known.  The bottom panel shows the morphological types from Fig.~1.72 that dominate in various parts of the top panel.
The rapidly rising luminosity function of spheroidals at the low-mass limit of the diagram may account for 
the contour in the top panel at (9.0, 2.2).    The ``take-home point'' is that the bright end of the red sequence 
consists of ellipticals, S0s and early-type spirals, but the faint end  is dominated by Sph galaxies.  
Adapted from KB2012.
\pretolerance=15000  \tolerance=15000 
         }

\end{figure*}

\phantom{Bloody random text to fix CUP's vertical spacing quirks.}

\vskip -70pt

\phantom{Bloody random text to fix CUP's vertical spacing quirks.}

\subsection{S+Im $\rightarrow$ S0+Sph galaxy transformation processes}

      The natural interpretation of the observations discussed in this section is that S0 and Sph galaxies
are defunct, ``red and dead'' versions of spiral and irregular galaxies that have been transformed by physical processes~to
be discovered.  Most of these turn out to be environmentally driven~and~slow.  

      The relative ordering and positioning of galaxies in the parallel-sequence classification is justified
on purely observational grounds based on choices of which results to use in the classification and which
to regard -- for present purposes -- as secondary.  However, it would be disingenuous to pretend that I and many
others have not been thinking about the underlying formation and evolution processes for a long time.  This is 
inevitable in a world where no observational curiosity goes uninterpreted for long.  In fact, there are many
candidate processes.  Astronomers frequently argue about which of many compelling theories are correct.  
My experience is that these arguments go on longest when everybody is correct.  This is one of those occasions.

      Candidate S+Im $\rightarrow$ S0+Sph galaxy transformation processes are reviewed in KB2012.
Here, I list them briefly including only the most important supporting observations:
     \begin{enumerate}[(a)]\listsize
     \renewcommand{\theenumi}{(\alph{enumi})}
\item{The main {\it internal\/} evolution process was already mentioned in Section 1.7.3.  Below a fiducial mass that
      corresponds to $V_{\rm circ} \simeq 100$ km s$^{-1}$, i.{\ts}e., just where bulges disappear (Fig.~1.70) and
      therefore galaxy names get changed from S0 to Sph (Fig.~1.69), supernova-driven winds are expected to
      expell a larger fraction of a galaxy's baryons from lower-mass objects regardless of whether they now are irregular or spheroidal
(e.{\ts}g., Larson 1974;
       Saito 1979;
       Dekel \& Silk 1986;
       Vader 1986;
       Schaeffer \& Silk 1988; see
       Hensler \etal 2004;
       Stinson \etal 2007
       for two among many more recent discussions).  This is why I suggested that the decreasing surface brightnesses of Sph and Im galaxies
       at lower luminosities (Fig.~1.71) is a baryon retention sequence.} \vskip 6pt
\item{The most thoroughly studied {\it external\/} transformation process is ram-pressure stripping of cold gas by
      hot gas in clusters and perhaps groups of galaxies.
      Suggested by Gunn \& Gott (1972), the idea has varied in popularity.  It has never gained
      widespread acceptance, perhaps in part because Dressler~(1980) argued that it was {\it not\/} the main cause of the
      morphology-density relation that spiral galaxies get less abundant whereas S0 galaxies get more abundant
      at higher galaxy densities in clusters.  Dressler argued that this result does not strongly depend on cluster richness.  
      However, examination of his \hbox{Fig.~8{\ts}--{\ts}10}~(see~Fig.\ts25
      in KB2012) shows that the ratio of S0 to S galaxies increases from low-concentration clusters to 
      high-concentration clusters to \hbox{X-ray-emitting clusters}.  An alternative hypothesis
      is that ram-pressure stripping happens more easily in clusters of all richness than simple energy arguments suggest.
      More recent results bear this out:} \vskip 6pt
\item[\phantom{(a)}]{Compelling evidence for ongoing ram-pressure stripping is provided by H$\alpha$ and H{\ts}{\sc i} observations of
      spiral galaxies in the Virgo cluster 
      (Chung \etal 2007;
      Kenney \etal 2004, 2008).
      Figure 1.75 shows some of these results.  Many spiral galaxies embedded in the X-ray gas that fills the cluster
      center show remarkable H{\ts}{\sc i} tails.  The above authors interpret them as cold gas that trails behind the galaxy
      after having been stripped from the galaxy by the hot gas in the cluster.  The spectacular H$\alpha$ filaments that point
      from the tidally disturbed NGC 4438 toward the hot-gas-rich NGC 4406 (top panel in Fig.~1.75) are similarly interpreted as
      ram-pressure stripped.  Also, many spirals near the center of the cluster are
      much smaller and more depleted in H{\ts}{\sc i} than are galaxies in the cluster outskirts 
     (Cayatte \etal 1990;
      Chung \etal 2009).
      Kormendy \& Bender (2012) note that ``the three most depleted galaxies illustrated in Fig.~8 of Chung \etal (2009) are
      NGC 4402, NGC 4405 and NGC 4064.  They have a mean absolute magnitude $M_V = -19.4 \pm 0.2$. Virtually all Sphs are fainter 
      than this.~~~If even the deep gravitational potential wells of still-spiral galaxies
      }
\end{enumerate}

\eject

\cl{\null}

\vfill

\begin{figure}[hb]


\includegraphics{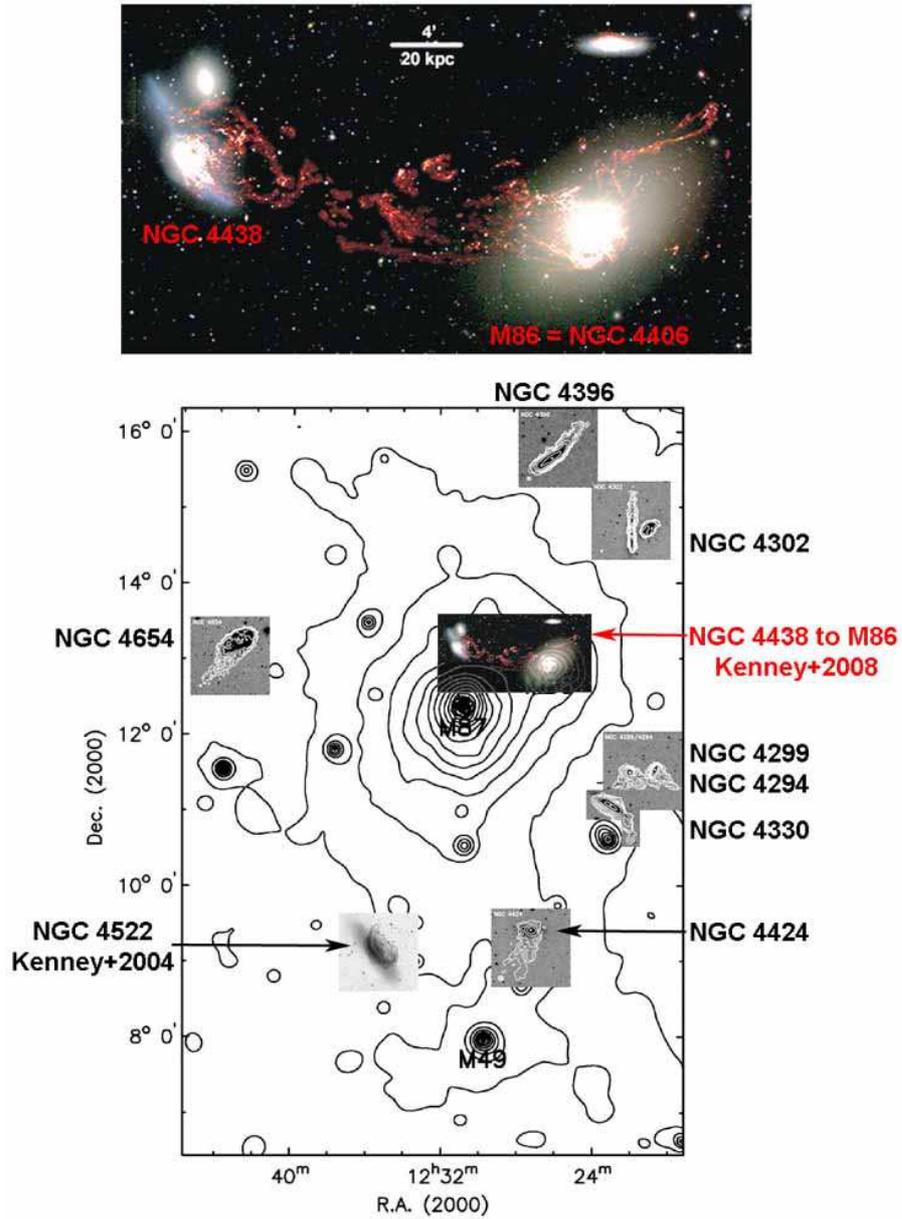}

\caption{The large panel shows 0.5 -- 2.0 keV X-ray brightness contours in the Virgo cluster as measured with {\it ROSAT}
by B\"ohringer \etal (1994).  Superposed are grayscale images of galaxies with H{\ts}{\sc i} tails indicative of
ongoing ram-pressure gas stripping (white or black contours).  The H{\ts}{\sc i} images are from Chung \etal (2007);
Kenney \etal (2004); Abramson \etal (2011).  The color inset image and large image at top show the 
H$\alpha$ emission filaments that extend from NGC 4438 to NGC 4406 (Kenney \etal 2008).   Each small inset image shows 
the galaxy centered on its position in the cluster, but the panels are magnified.  
This figure is adapted from Fig.~4 in Chung \etal (2007) and is reproduced from KB2012.
\pretolerance=15000  \tolerance=15000 
         }

\end{figure}

\cl{\null}
\vskip -28pt
\eject

     \begin{enumerate}[(a)]\listsize
     \renewcommand{\theenumi}{(\alph{enumi})}
\item[\phantom{(a)}]{suffer H{\ts}{\sc i} stripping, 
      then the shallow potential wells of dS$+$Im galaxies are more likely to be stripped.''  Substantial additional
      evidence also suggests that ram-pressure stripping is more effective than we thought (see KB2012 and
      van Gorkom \& Kenney 2013 for reviews).}\vskip 7pt
\item[\phantom{(a)}]{Can ram-pressure stripping still happen in the Local Group's much shallower gravitational
      potential well?  Compelling observations which indirectly suggest that the answer is ``yes'' are shown
      in Fig.~1.76.  Close companions of our Galaxy, of M{\ts}31, and of other nearby giant galaxies are almost all spheroidals; 
      distant companions are irregulars; Sph/Im transition galaxies tend to live at intermediate distances, and larger irregulars 
      ``survive'' at closer distances to their giant companions 
     (Einasto \etal 1974; 
      van den Bergh 1994a,{\ts}b,{\ts}2007; 
      Mateo 1998;
      Skillman \etal 2003;
      Bouchard \etal 2009;
      McConnachie 2012).
      Hints of similar effects in larger satellites are seen in the Zurich Environmental Study (ZENS:~Cibinel \etal 2012).~Like
      previous authors, Kormendy \& Bender suggest that ``ram-pressure stripping can happen even in environments that
      are gentler than cluster centers.  It may be indirect evidence for a pervasive warm-hot intergalactic medium
      (WHIM: Dav\'e \etal 2001) that is difficult to detect directly but that may be enough to convert dwarf irregulars into
      spheroidals.''  
      }\vskip 7pt
\item[\phantom{(a)}]{All this evidence suggests that ram-pressure stripping is one of the processes that transforms late-type,
      gas-rich and star-forming galaxies into red and dead S0 and Sph galaxies.}

\vfill

\begin{figure}[hb]


 \includegraphics{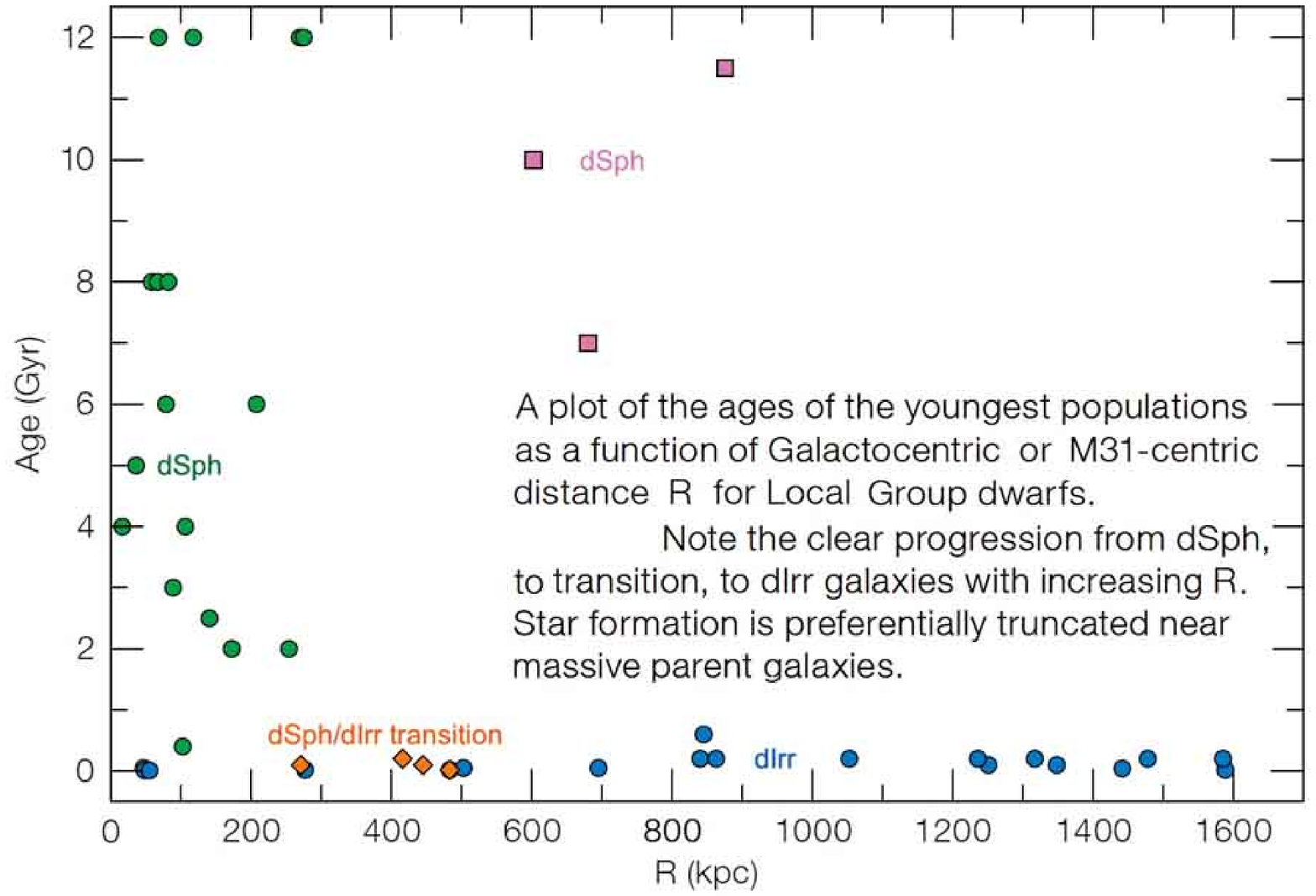}

\caption{From Mateo (2008), the ages of the youngest stellar populations in dwarf galaxy companions
versus Galactocentric or M31centric distance $R$.  Except for the Magellanic Clouds, all close companions
of our Galaxy and of M{\ts}31 are spheroidals.  Distant companions are irregulars except for three
free-flying dSphs (pink points).  The Sph/Im transition galaxies mostly lie at intermediate distances.  
         }

\end{figure}

\cl{\null}
\vskip -28pt
\eject

\item[(c)]{Galaxy harassment is a dynamical process that should operate wherever objects orbit repeatedly through
           rapidly varying gravitational potential fields, especially in virialized clusters of galaxies with velocity
           dispersions that are much larger than the internal velocities of galaxies.  It is the cumulative effect of 
           many encounters with other galaxies and with the cluster potential.  Simulations show that it
           strips outer mass, heats disks and promotes gas inflow toward the center that presumably feeds star formation
          (Moore\ts{\it et{\ts}el.}\ts1996, 1998; 
           Lake \etal 1998).
           A variant is tidal stirring of dwarf galaxies on elliptical orbits around our Galaxy or M{\ts}31 
          (Mayer \etal 2001a,{\ts}b, 2006). 
           Kormendy \& Bender concur with the above authors in suggesting that harassment converts late-type disks into spheroidals 
           and more robust, earlier-type spirals into hotter systems that resemble S0s.  A benefit of this picture is that
           inflowing gas can feed star formation; this helps to explain why S0 disks and spheroidals -- which must fade substantially
           after star formations stops -- do not have much lower surface brightnesses than current versions of S$+$Im progenitors
           (see Ferguson \& Binggeli 1994 for a review of this problem).  Gravity is not negotiable.  Its effects are clean.
           It is encouraging how many observations can be tied together into a coherent picture if harassment is one of the galaxy
           transformation processes:} \vskip 6pt
\item[\phantom{(a)}]{(1) Dynamical heating plausibly explains why faint spheroidals are not flat, why many bright
           spheroidals contain disks (either observed directly when seen edge-on or inferred from their spiral structure), and
           why the outer parts of our ``rosetta stone'' galaxies NGC 4762, NGC 4552, NGC 4638 and VCC 2048 are vertically thick
           whereas their more robust inner parts are flat.} \vskip 7pt
\item[\phantom{(a)}]{2) Sph and Im galaxies have similar distributions of axial ratios 
          (Ferguson \& Sandage 1989;
           Binggeli \& Popescu 1995).
           The latter authors conclude that ``there is no evidence for a difference between the flattening distributions of {\it nucleated\/}
           dE,N and classical (giant) Es''.~However, in my experience, many~Sph,N galaxies are flatter than any elliptical.
           NGC~4637 in Fig.~1.65 and VCC~2048 in Fig.~10 of KB2012 are examples.  Ferguson \& Sandage note that
           ``The similarity of flattenings of dE (bright, no N) and Im types removes one of the previous objections to the
           hypothesis that some dwarf ellipticals could be stripped dI's''.  The exact engineering needs further study,
           but dynamical heating added to the fact that the smallest galaxies are not flat anyway provides a promising
           way to explain the flattening observations.} \vskip 6pt
\item[\phantom{(a)}]{3) Intracluster light is believed to consist of stars that have been stripped by harassment from
           individual galaxies.  In the Virgo cluster, it is irregular~and still in the early stages of formation
          (Mihos \etal 2005, 2009;
           Arnaboldi \etal 1996, 2002, 2004;
           Castro-Rodrigu\'ez \etal 2009;
           Arnaboldi \& Gerhard 2010).
           In rich clusters, it is widely observed 
          (Thuan \& Kormendy 1977;
           Adami~{\it et~al.} 2005;
           Krick \& Bernstein 2007; 
           Gonzalez \etal 2007;
           Okamura 2011).
           When intracluster light is very bright, it is called a ``cD halo'' 
          (Morgan \& Lesh 1965;
           Oemler 1976; 
           Schombert 1988).
           These halos are robustly understood to consist of tidally liberated stars and disrupted galaxies
          (Richstone 1976;
           Dressler 1979; 
           Kelson \etal 2002).
           If gravitational harassment can produce all these effects, it is difficult to see how the mere heating of 
           smaller galaxies could be avoided.} \vskip 5pt
\item[\phantom{(a)}]{4) Kormendy \& Bender (2012) suggest that the ``new class of dwarfs that are of huge size (10000 pc in diameter 
           in the extreme) and of very low surface brightness of about 25 B mag arcsec$^{-2}$ at the center'' discovered by Sandage
           \& Binggeli (1984) are ``spheroidals that have been harassed almost to death''.} \vskip 5pt
\item[\phantom{(a)}]{5) Anisotropic dynamical heating is a natural way to try to explain triaxial and slowly rotating Sphs (e.{\ts}g.,
           Bender \& Nieto 1990).
           This idea is consistent with the observation that the brightest Sphs are in many cases the most disky and rapidly rotating
           ones.  However, unusually violent encounter histories can allow a small number of Sphs to be anisotropic even at
           the highest masses.} \vskip 5pt
\item[\phantom{(a)}]{6) Carefully engineered encounter histories can make Sph galaxies that have kinematically decoupled
           subsystems, even counterrotation of the harassed outer parts with respect to the inner galaxy 
          (De Rijcke \etal 2004; 
           Gonz\'alez-Garc\'\i a \etal 2005).~Counterrotating systems are seen in VCC\ts510 (Thomas \etal 2006).} \vskip 5pt
\item[(d)]{Starvation of continued infall of cold gas from the cosmological structure hierarchy is frequently discussed as
           an S$+$Im $\rightarrow$ S0$+$Sph transformation process (e.{\ts}g.,
           Larson \etal 1980;
           Balogh \etal 2000;
           Bekki \etal 2002;
           Boselli \etal 2009).
           Absent such infall, star formation at currently observed rates generally uses up the available gas in a few Gyr
          (Larson \etal 1980;
           Boselli \etal 2009). 
           Given the observation that the center of the Virgo cluster and {\it a fortiori\/} the centers of rich clusters
           of galaxies are dominated by hot gas (e.{\ts}g., Fig.~1.75), it is difficult to see how starvation can be avoided.
           These are not environments where low-density cold gas can survive to feed continued accretion onto galaxies for 
           billions of years after the cluster acquires a large velocity dispersion.}
\end{enumerate}

      Thus many processes 
(1) may explain the growing dark matter dominance (i.{\ts}e., baryon deficiency) of lower-mass dwarf galaxies and 
(2) can potentially transfrom S$+$Im galaxies into S0$+$Sph galaxies.  
Kormendy \& Bender (2012) emphasize that ``the relevant question is not `Which one of these mechanisms is correct?'  It is 
`How can you stop any of them from happening?'  It seems likely to us that all of the above processes matter.''

      In this regard, I conclude by emphasizing the following points.  

      Most papers (Boselli \etal 2009 is an exception) investigate one process; when they get into trouble 
explaining some particular observation, they conclude that this process is not the answer.~If all above processes 
happen, then there is more potential to understand all of the diagnostic observations.  Theorists like to ask clean 
questions, investigating one process at a time.  There are good reasons for this.  But Nature does everything~together.  
Eventually, we will have to do likewise if we expect to understand galaxy evolution.  The hellishly complicated interplay of 
different processes is a feature, not a bug.  We cannot avoid this problem forever.

      Second, one observation that is frequently cited to disfavor ram-pressure stripping and strangulation is that bulges
are systematically bigger in S0s than in spirals (e.{\ts}g., Dressler 1980).  But (i) small pseudobulges in late-type spirals 
skew the distribution of S-galaxy bulges to smaller luminosities; if such galaxies are transformed before secular evolution 
has time to make pseudobulges, the result is a Sph, not an S0.  Then it will not be counted among the S0s.  Also, (ii) the 
distribution of S0 bulges is skewed toward high luminosity by the frequent misclassification of the highest ellipticals as S0s 
(KFCB).  This happens because their high S\'ersic indices $n \gg 4$ give them a ``core-halo'' appearance that persuades 
classifiers to call~them~S0s.  An example is the elliptical galaxy NGC 4406, which is classified S0 by Sandage \& Tammann (1981).
Similarly, giant ellipticals~are~commonly~classified~S0$_3$ when they contain nuclear dust disks; e.{\ts}g., the elliptical 
NGC\ts4459 (KFCB).  (iii) When accurate bulge-disk decompositions are carried out, the folklore that S0 galaxies mostly have 
large bulges is not confirmed.  Among~S0s discussed in KFCB and in KB2012, about half have $(P)B/T < 0.5$ and 
six have $(P)B/T$ \lapprox \ts1/3, the value for the Sb~galaxy~M{\ts}31.
Finally, (iv) the distribution of $B/T$ in the progenitor galaxies of any transformation process is a strong
function of environment -- bulgeless disks are preferentially made in the field, and merger remnants are preferentially made
in clusters such as Virgo (Kormendy \etal 2010).  So field spirals do not fairly sample the potential progenitors of 
S0 galaxies in clusters.

      Another observation that is frequently cited to disfavor ram-pressure stripping and strangulation is that
S0s have higher surface brightnesses than spirals.  Disk fading that follows the shutdown of star formation might lead
us to expect the opposite effect.  But Fig.~1.71 shows little sign of such an effect.  Note that the surface brightnesses 
of both S0 and spiral disks are not corrected for inclination, so they are treated in the same way.  However, internal absorption 
is important in spirals and not in S0s.  So internal-absorption-corrected surface brightnesses of spiral galaxy disks would be brighter 
than those of S0 disks.  Also, harassment -- like any effect that rearranges angular momentum -- persuades some gas to fall 
toward the center and should increase the surface brightness there via star formation.

      Finally, Kormendy \& Bender (2012) point out that S $\rightarrow$ S0 transformation does not require the removal
of all gas nor the quenching of all star formation.  Some S0s still contain gas, especially molecular gas near their
centers (e.{\ts}g., 
Welch \& Sage 2003; 
Sage \& Wrobel 1989; 
Thronson \etal 1989; 
Devereux \& Young 1991; 
Young \etal 1995).
These S0s form stars.  The Virgo spirals whose outer H{\ts}{\sc i} distributions are truncated
have normal central molecular gas content (Kenney \& Young 1986).  We do not need to solve
the  problem of removing all gas from the deepest parts of galaxy potential wells.

      It is clear that much work -- much {\it complicated\/} work -- is still needed on the messy baryonic physics of 
galaxy evolution.  But I am encouraged to think that the still unknown details of the various transformation processes 
do not threaten our overall picture that at least some S0s and likely all Sph galaxies are defunct spiral and irregular galaxies.

\vskip -30pt

\phantom{"Gobble gobble gobble" invisibly to fix CUP section spacing}

\subsection{Environmental secular evolution --\\``An Idea Whose Time Has Come''}

      Morphological observations such as those encoded in the parallel-sequence classification of Fig.~1.72 lead to the
robust conclusion that many S0s are closely related to spiral galaxies and that essentially all Sph galaxies are closely 
related to the latest-type spirals and irregulars.  Figure 1.72 does not directly tell us what that relationship is.  
However, in recent years, rapidly improving observations of transformation processes in action, including
H{\ts}{\sc i} and H$\alpha$ tails, 
relatively recent star formation in dSph galaxies, 
vertically thick outer disks in interacting S0s and
``rosetta stone'' galaxies with both S0 and Sph properties, 
strongly imply that some or all of a collection of environmental processes transform spiral and irregular
galaxies into red and dead S0 and spheroidal galaxies.  This happens especially in rich clusters, but small
galaxies can suffer transformation even in relatively quiescent environments such as the Local Group.  
Kormendy \& Bender (2012) suggest that ``environmental secular evolution is An Idea Whose Time Has Come.''

\vskip -30pt

\phantom{"Gobble gobble gobble" invisibly to fix CUP section spacing}

\section{Toward a Comprehensive Picture of Galaxy Formation}

      This section ties together our standard picture of galaxy formation by hierarchical clustering  
(lectures by Shlosman,{\ts}Scoville,{\ts}Calzetti) 
with our School's subject of secular evolution 
(lectures by Athanassoula, Binney, Buta, Peletier, van Gorkom and me).  
For pedagogical reasons, it is useful to introduce this standard picture here.  Shlosman (2012) provides more detail.

      What is at stake for future work?  We have a formation paradigm: quantum density fluctuations 
in non-baryonic dark matter form immediately after the Big Bang and then get stretched by the expansion of the Universe; gravity drives 
hierarchical clustering that causes the fluctuations to grow, separate, collapse and form galaxy halos; the baryons cool inside
the halos to form stars and visible galaxies.  Spirals form when halos accrete gas that dissipates and forms disks.  Ellipticals 
form when galaxies collide and merge; then dynamical violence scrambles disks into ellipsoidal Es.  This picture is well supported 
by theory and observations.  What are the remaining puzzles{\ts}--{\ts}the cracks in the paradigm?~They are short-cuts to progress.

      First, what is understood?  Hierarchical clustering of dark matter initial density fluctuations is nowadays calculated in
exquisite detail (Fig.~1.77).  Results are in excellent agreement with observations of large-scale structure. 					

      Our job is illustrated in Fig.~1.78.  Hierarchical clustering of dark matter is well understood 
(background image).~More tricky is the physics of baryonic galaxy formation within dark halos.~It is possible 
that all remaining problems with our formation picture on galaxy scales are problems of baryonic physics.

\vfill

\begin{figure}[hb]


 \includegraphics{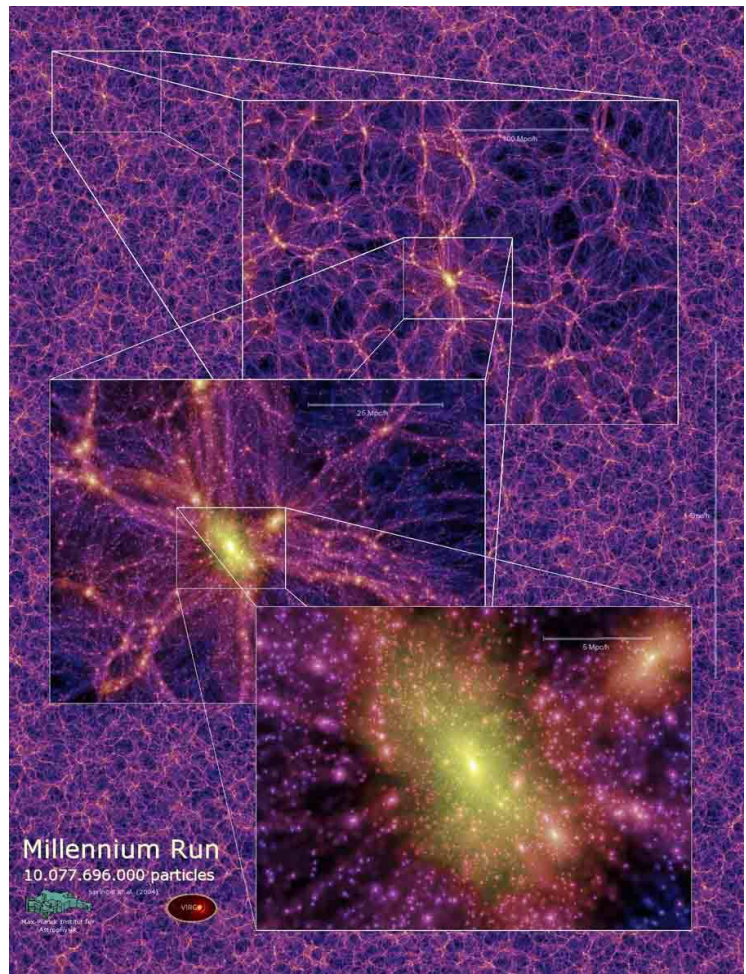}

\caption{The Millennium Simulation (Springel \etal 2005) is the iconic example of an $n$-body calculation of the formation of large-scale and
         galaxy-sized structures via hierarchical clustering of primordial quantum density fluctuations that have been
         stretched by the expansion of the Universe and increased in contrast by self-gravity. 
         }

\end{figure}

\cl{\null}
\vskip -28pt
\eject

      The good news is that ellipticals are fairly well understood.   Simulations of the hierarchical growth of galaxies 
suggest that they change back and forth between spiral and elliptical, depending on whether their recent history was dominated 
by a major merger or by cold gas dissipation (Fig.~1.79).  Today's students did not live through the revolution in our understanding
that resulted from Toomre's (1977a) introduction of mergers to our lexicon of galaxy formation.   I therefore review this subject briefly.
  
      The first bad news is that we do not know how to form bulgeless galaxies.  Continued bulge growth is inherent to the story in Fig.~1.79.  
Once you have a bulge, you cannot get rid of it.~This problem was reviewed in Section 1.6.1.

      The other bad news is that galaxy formation by hierarchical clustering of cold dark matter still has problems on the size scales 
of individual galaxies.  Reviews of the subject run the gamut from very optimistic (Primack 2004) to sober (Silk \& Mamon 2012) to very 
pessimistic (Kroupa 2012).  This is a sign of a subject in flux{\ts}--{\ts}of~cracks in the paradigm.  These are opportunities.

      The comfortable news is that we have a growing understanding of secular evolution in disk galaxies.  It happens now in low-
and intermediate-density environments.  But finding pseudobulges in Virgo S0s shows that it had time to happen
even in the progenitor environments of some present-day clusters.
\vfill

\begin{figure}[hb]


 \includegraphics{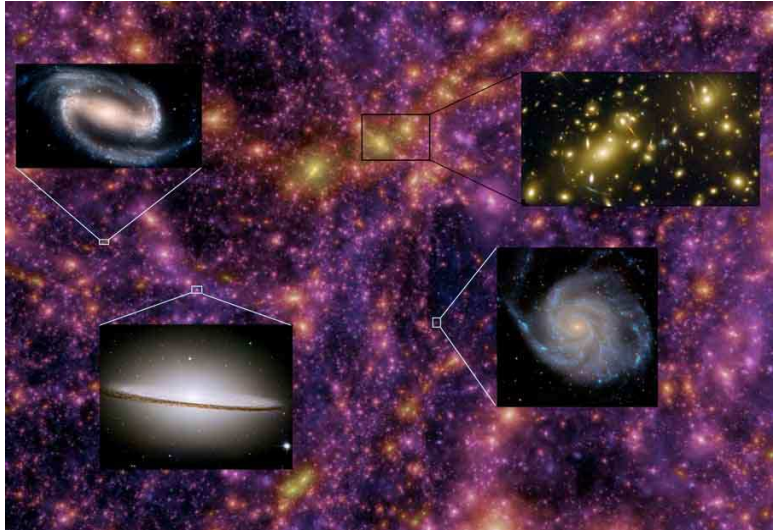}

\caption{Theme of building a comprehensive picture of galaxy formation by studying the physics of baryonic galaxies as
         embedded in the dark matter hierarchy represented here by the Millennium Simulation.
         High-density environments are dominated by merger remnants{\ts}--{\ts}giant ellipticals in rich clusters.
         We understand them fairly well.  Low-density environments are dominated by pure-disk galaxies such as M{\ts}101; 
         we do not understand how they form.~Bulge-dominated spirals~like the Sombrero live in intermediate environments.  
         Barred and other galaxies that undergo secular evolution also tend to live in intermediate-density environments.  
         }

\end{figure}

\cl{\null}
\vskip -28pt
\eject

\cl{\null}

\vfill

\begin{figure}[hb]


 \includegraphics{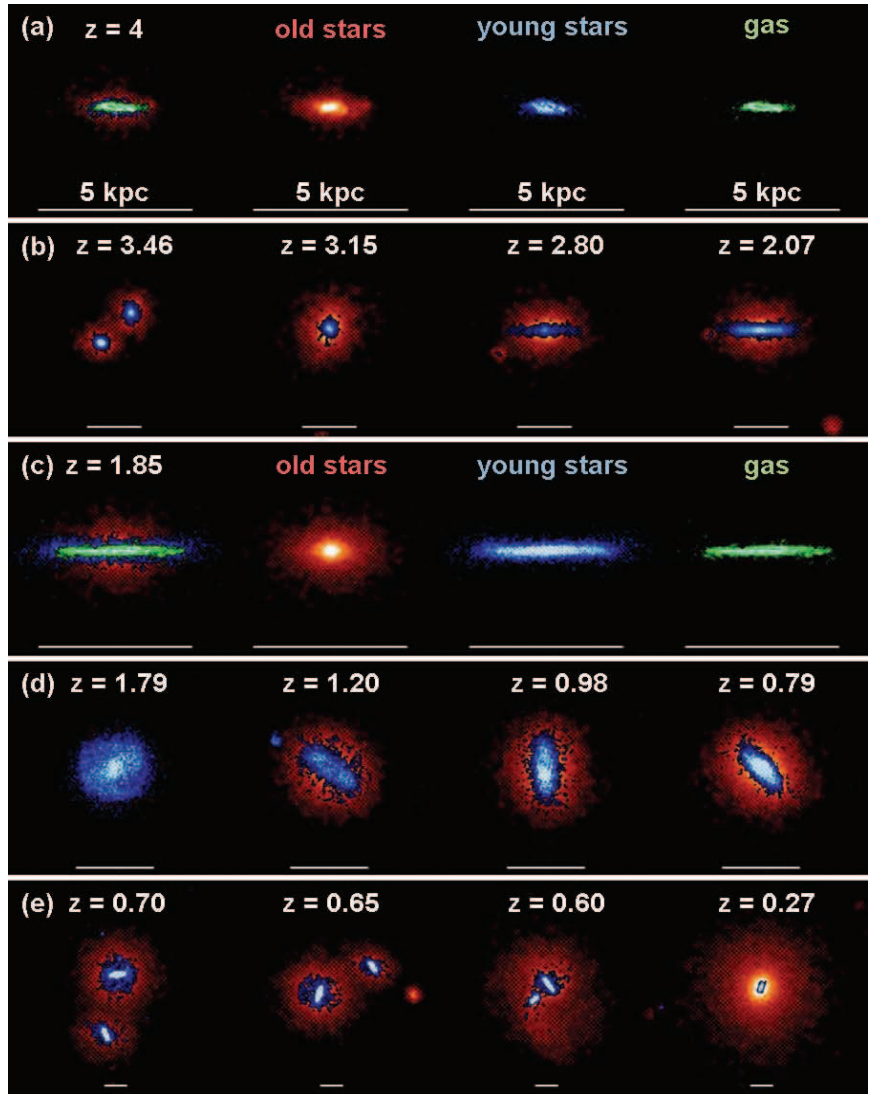}

\caption{Example of the evolution of a single galaxy by hierarchical clustering (Steinmetz \& Navarro 2002, 2003). 
Colors denote old stars, young stars $<$\ts200 Myr old and cold gas (see keys).  Scale bars are 5 kpc in all panels.
Panel (a) shows the most massive progenitor at $z = 4$; it already contains both old stars and a gas disk.  In panel (b),
a classical bulge forms in a major merger at $z \simeq 3$ and then regrows a disk by later infall of cold gas.
Panel (c): at $z = 1.8$, the galaxy looks like an early-type spiral with a dense bulge surrounded by a young disk.
Panel (d): At $z \simeq 1.6$, tidal forcing by a companion shown in the $z = 1.2$ image triggers a bar.
The satellite is accreted at $z = 1.18$, but the bar prominent in the young component survives
for several more Gyr.  Panel (e): At $z = 0.7$, the galaxy merges with another galaxy
that has about half of its mass.  The result is an elliptical galaxy at $z = 0.27$.  This could accrete more 
gas and form a Sombrero-galaxy-like system, but it cannot get rid of its large bulge.
         }

\end{figure}

\cl{\null}
\vskip -28pt
\eject

\subsection{Formation of ellipticals by major galaxy mergers}

      It is hard to describe to today's students what a revelation it was when Alar Toomre (1977a) presented his 
hypothesis that all ellipticals are created from progenitor disk galaxies by the dynamical violence of 
galaxy collisions and mergers.  The feeling had been that stars are too small to collide, so interacting
galaxies merely pass through each other.  We missed two points.  First, tidal effects are easily 
strong enough to scramble cold, rotating disks into dynamically hot ellipticals.  Doing this work takes
energy out of the orbits and soon causes the galaxies to merge.~Second, large and massive 
halos of dark matter surround visible galaxies and give them much bigger collision cross sections than we 
thought when we saw only the visible stars.  It is no accident that the merger picture became established 
soon after~we~realized that dark matter is real (Faber \& Gallagher 1979).  Mergers-in-progress turn out to
explain a whole zoo of previously mysterious peculiar galaxies~(e.{\ts}g., Figs.\ts1.80{\ts}--{\ts}1.82).
Toomre's suggestion that mergers make all ellipticals~is, as far as we know, exactly correct.~And Toomre's
(1980) additional hypothesis that mergers make classical bulges robustly looks to be exactly correct,~too.

\vfill

\begin{figure}[hb]


 \includegraphics{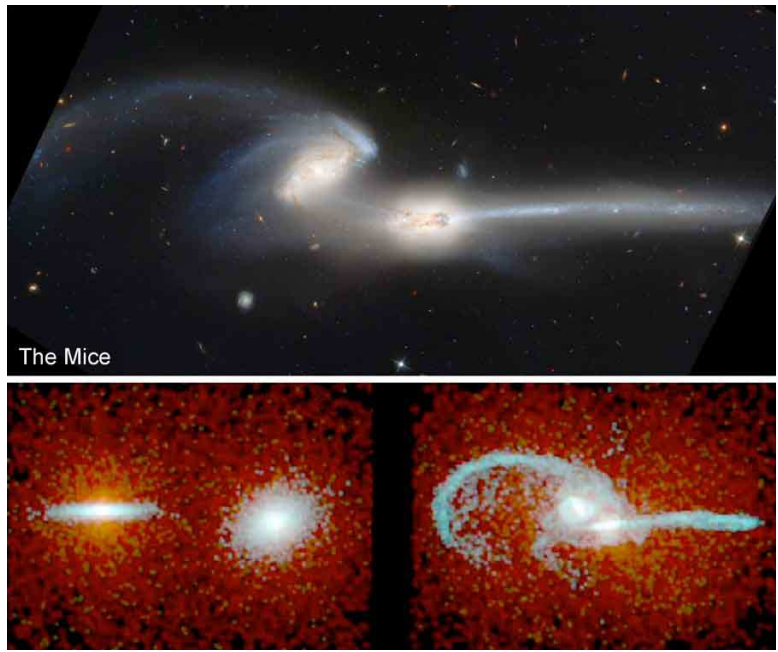}

\caption{Formation of peculiar galaxies such as NGC\ts4676 (``the mice'') by ongoing gravitational
         encounters.  Such encounters explain most objects in
         (e.{\ts}g.)~Arp's (1966) {\it Atlas of Peculiar Galaxies\/}.  (top) Hubble Heritage image. 
         (bottom left) Our view of the initial conditions and (bottom right) the moment when the
         configuration matches the galaxies for an $n$-body simulation
         (Barnes 1998, 2004) of two infinitely thin disks (blue particles) embedded in
         spherical dark halos (red particles).
         }

\end{figure}

\cl{\null}
\vskip -28pt
\eject

      I was in the audience for Toomre's (1977a) talk; we all remember it~as~a historic moment.
But not everyone was immediately captured by the new ideas, and I confess that I was a partial agnostic for longer 
than most.  Kormendy (1989) was a late paper that tried to keep merger enthusiasm from getting out of control.
Its point is still  correct~-- the sequence of increasing density in smaller ellipticals (Fig.~1.68) is a 
sequence of increasing dissipation with accompanying starbursts during formation.  As emphasized all along by Toomre,
merger progenitors generally contain gas; crunching gas likes to make stars, and so star formation is an integral
part of spiral-spiral mergers.
 
      By $\sim$\ts1990, we understood that ultraluminous {\it IRAS\/} galaxies (``ULIRGS'') are prototypical 
dissipative mergers in progress 
(Joseph \& Wright 1985; 
Sanders \etal 1988a, b;
Sanders \& Mirabel 1996; 
Rigopoulou \etal 1999; 
Dasyra \etal 2006a, b;
see Dasyra \etal 2006c and KFCB Section 12.3.2 for reviews).  Figure 1.81 shows the most famous example.  Dust-shrouded starbursts generally
dominate the far-infrared luminosity $L > 10^{12}$ $L_\odot$ (see Joseph 1999 and KFCB for reviews).  Their structural
parameters are consistent with the E fundamental plane
(Kormendy \& Sanders 1992; 
Doyon \etal 1994; 
Genzel \etal 2001; 
Tacconi \etal 2002; 
Veilleux \etal 2006; 
Dasyra \etal 2006a, b). 
Stellar velocity dispersions $\sigma \simeq 100$ to 230 km s$^{-1}$ show that local ULIRGs are making moderate-luminosity 
ellipticals; i.{\ts}e., 
the disky-coreless side of the E{\ts}--{\ts}E dichotomy in Figs.~1.4 and 1.60
(Genzel \etal 2001; 
Tacconi \etal 2002; 
Dasyra \etal 2006a, b; c). 

\vfill

\begin{figure}[hb]


 \includegraphics{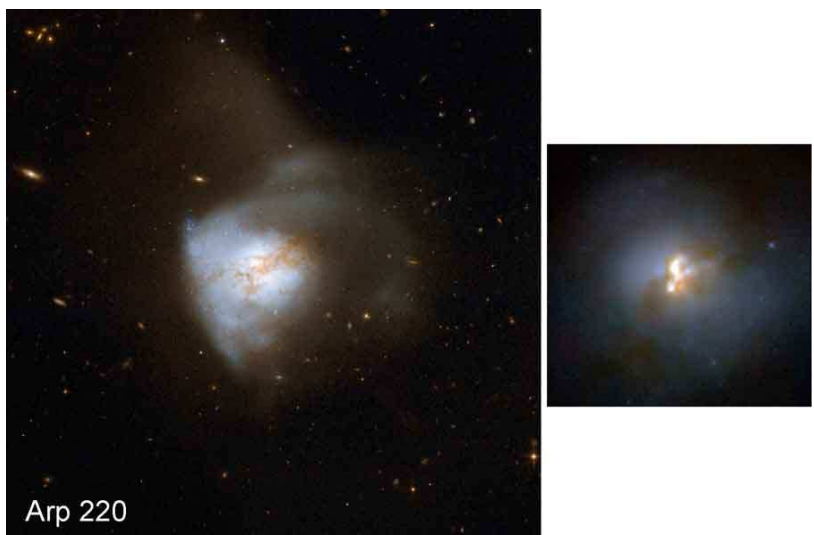}

\caption{Arp 220, the prototypical Ultraluminous Infrared Galaxy, 
         an elliptical galaxy being formed by a dissipative merger accompanied by
         a dust-shrouded starburst.  At left is a Hubble Heritage image.  The {\it HST\/} NICMOS
         $JHK$ image at right reveals two remnant nuclei separated by 0\sd98 $\simeq$ 360 pc (Scoville \etal 1998).
         }

\end{figure}

\cl{\null}
\vskip -28pt
\eject

\noindent 

      Central to the rapid acceptance of the merger picture was the extensive observational evidence for mergers in progress
that was published by Fran\c cois Schweizer (e.{\ts}g., 1978, 1980, 1982, 1987, 1996).  Figure 1.82 shows the particularly
convincing example if NGC 7252 (the ``atoms for peace'' galaxy).  It has both diagnostic tidal tails and ``ripples'' or 
``shells'' in the light distribution that trace edge-on caustics of wrapped~former~disks.  Shells, too, became standard
merger diagnostics~(Malin~\&~Carter 1980, 1983; Schweizer \& Seitzer 1988).   Shells are seen in absorption as well as in
stars; an example of the explanation of a previously mysterious peculiar galaxy as an S0 that contains an accreted and now 
phase-wrapped dust disk is NGC 4753 (Steiman-Cameron \etal 1992).  The correlation that stronger fine structure 
such as shells is seen in ellipticals with younger~stellar~populations~(bluer colors and stronger
H$\beta$ absorption lines) further~supported~the~merger picture and filled in the evolution time sequence between 
mergers in progress and old, completely relaxed and phase-mixed ellipticals 
(Carter \etal 1988; \phantom{00000000}

\vfill

\begin{figure}[hb]


 \includegraphics{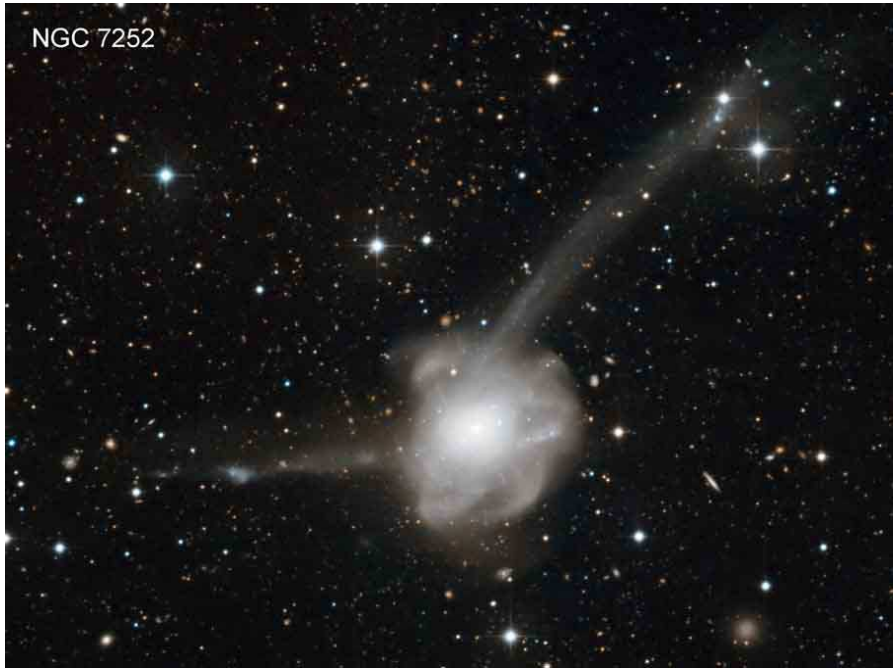}

\caption{NGC 7252 is another prototypical merger-in progress that is making an elliptical galaxy (Schweizer 1982).  
         Two tidal tails that point in roughly opposite directions and that have opposite velocity differences with 
         respect to the systemic velocity are the simplest diagnostic signature of a merger in progress (see~the~seminal 
         paper by Toomre \& Toomre 1972).  The essential point is that dynamical clocks run most slowly at the largest
         radii, so remnant tidal tails persist long after the main bodies of interacting galaxies have merged.  For H{\ts}{\sc i}
         observations and $n$-body models of NGC 7252, see Hibbard \etal (1994, 1995, 1996).
         }

\end{figure}

\cl{\null}
\vskip -28pt
\eject

\noindent Schweizer \etal 1990;
Schweizer \& Seitzer 1992).
Compelling further support was provided both by detailed H{\ts}{\sc i} observations (e.{\ts}g., 
Hibbard \etal 1994, 1995, 1996, 2001a,{\ts}b -- see {\tt http://www.nrao.edu/astrores/HIrogues}) and especially by $n$-body simulations
(the master of the art is Josh Barnes 1988, 1989, 1992).
Simulations further confirmed that mergers dump huge amounts of gas to galaxy centers, thereby feeding starbursts (e.{\ts}g., 
Barnes \& Hernquist 1991, 1992, 1996;
Mihos \& Hernquist 1994;
Hopkins \etal 2009a).
By the time of the reviews of
Schweizer (1990, 1998);
Barnes \& Hernquist (1992), 
Kennicutt (1998c) and
Barnes (1998), 
the merger revolution in our understanding of elliptical galaxies was a ``done deal''.

      A variant of the merger picture involves the observation that many~\hbox{high-$z$} galaxies are dominated by
$10^8$\ts--\ts$10^9$ $M_\odot$, kpc-size star forming clumps (e.{\ts}g.,
Elmegreen \etal 2005, 2007, 2008a, 2009a,{\ts}b;
Bournaud \etal 2007;
Genzel \etal 2008;
F\"orster Schreiber \etal 2009;
Tacconi \etal 2010).
Bournaud's galaxy UDF\ts1668 (Fig.~1.83) is remarkably similar to the initial conditions used by van Albada (1982) 
to simulate the collapse of lumpy initial conditions.  He showed (Fig.~1.83 here) that relatively gentle collapses 
produce S\'ersic-function profiles with \hbox{$n \simeq 2${\ts}--{\ts}4} like those in real classical bulges 
(e.{\ts}g., Fisher \& Drory 2008).  This is confirmed in modern $n$-body merger simulations (e.{\ts}g., Hopkins \etal 2009a).

\vfill

\begin{figure}[hb]


 \includegraphics{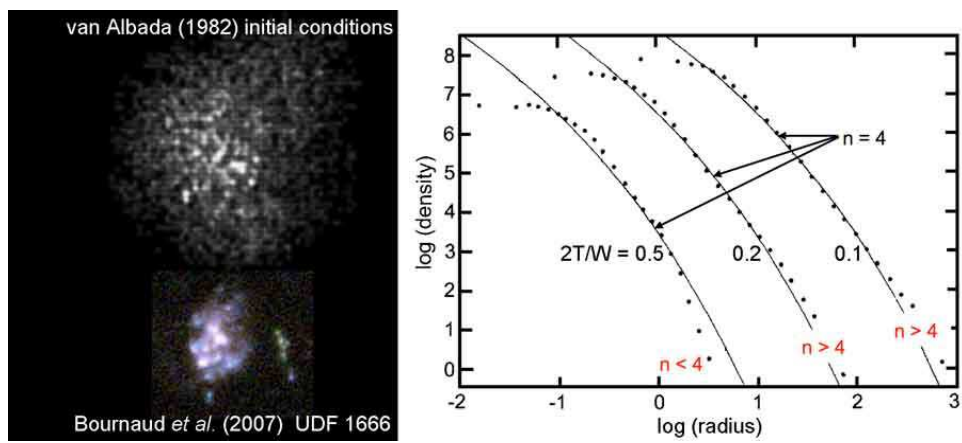}

\caption{Mergers of two galaxies that consist mostly of stars make S\'ersic (1968)
         function remnants with indices $n \sim 2${\ts}--{\ts}4.~An early illustration of this~is~van
         Albada (1982), whose initial conditions look remarkably similar to the
         clumpy high-$z$ galaxy UDF 1666 studied by Bournaud \etal (2007).  It is 
         an example~of~the clump instability picture discussed by Elmegreen \etal (2008b).
         Van Albada's initial conditions were parameterized by the ratio
         of twice the total kinetic energy~to the negative of the potential energy.~In virial
         equilibrium, $2T/W = 1$.~For~smaller values, van Albada found that
         gentle collapses ($2T/W = 0.5$) make S\'ersic profiles with~$n < 4$, whereas
         violent collapses ($2T/W$\ts\lapprox\ts0.2) make $n$\ts\gapprox 4.~This 
         is a sign that the clumps discussed by Elmegreen \etal (2008b) merge to make classical bulges.
         }

\end{figure}

\cl{\null}
\vskip -28pt
\eject

\cl{\null}

      Elmegreen \etal (2008b) model gas-rich galaxy disks in the early Universe and find that they
violently form clumps like those observed (Fig.~1.84).  The clumps quickly merge and make 
a high-S\'ersic-index bulge.  It rotates slowly.  Rotation velocities decrease with 
increasing distance from~the~disk plane.~So these are classical bulges, and this is a variant on
the merger picture.  From many colorful conversations with Allan Sandage, I suspect that he would
have welcomed this ``ELS with lumps'' picture (ELS = Eggen \etal 1962).

\vfill

\begin{figure}[hb]


 \includegraphics{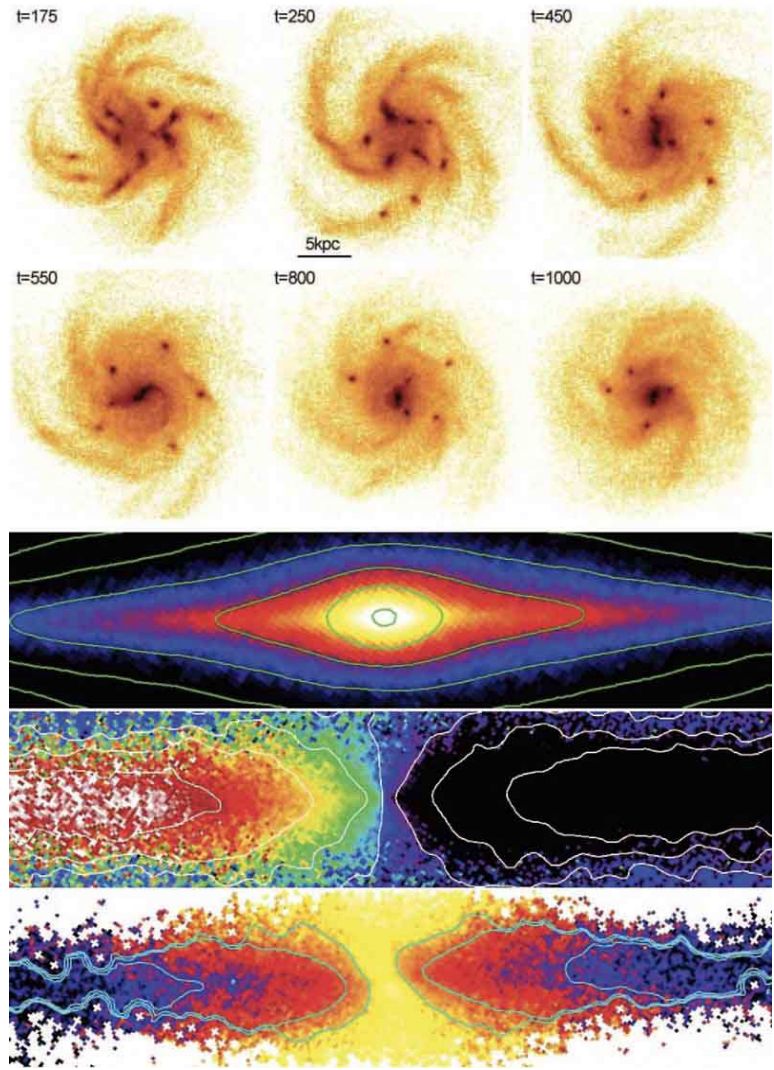}

\caption{A variant of the merger picture involves high-$z$ disks that are
         unstable to the formation of large clumps which quickly merge to form a classical bulge.
         It has ellipsoidal (not cylindrical) rotation (bottom). From Elmegreen \etal (2008b).
         }

\end{figure}

\cl{\null}
\vskip -28pt
\eject

\subsection{Mergers and secular evolution both happen in a hierarchically clustering Universe}

       ``It is impossible to remove the problem of galaxy formation from its cosmological context of 
hierarchical clustering'' (Jones 1992). Begun in papers like White \& Rees (1978), our picture of
hierarchical clustering has reached a remarkable level of sophistication.  The Millenniumm Simulation 
is one example from a vast literature.  I want to emphasize again that the merger formation of elliptical 
galaxies that was discussed in Section\ts1.8.1, the secular evolution of disk galaxies that was discussed
in Sections \hbox{1.2{\ts}--{\ts}1.6}, and the environmental secular evolution that was discussed in 
Section\ts1.7 all happen within the cosmological context of hierarchical clustering.  Much work remains to
be done in connecting the story of galaxy formation on the kpc scales of most studies of individual
galaxies with the Mpc scales where \hbox{$n$-body} dark matter simulations are at their best.
Current work is dominated by the complicated physics of baryons, including the effects of
reionization, dissipation, star formation, energy feedback and active galactic nuclei.   We like to 
think that galaxies are mature objects and that our job is to study galaxy evolution to see how they
got that way.  {\it But $\sim$\ts2/3 of the baryons in the Universe do not yet live in galaxies or have not 
yet cooled and formed stars} (e.{\ts}g.,
Fukugita \etal 1998;
Dav\'e \etal 2001;
Read \& Trentham 2005).
{\it Galaxy formation is much less ``finished'' than we like to think!}  Our job is far from finished, too.
For all of us students of galaxies, this~is~good~news.

\omit{
\cl{\null}
\vfill
\begin{figure}[hb]
\includegraphics{./DM-VM-Spectrum.eps}
\caption{From Read \& Trentham (2005): the field galaxy baryonic mass function (black
         data points and spline fits) compared with the dark matter halo masss spectrum
         from Weller \etal (2005).  The shortfall of visible galaxies at low masses is
         usually interpreted to be caused by the ejection of baryons by supernova-driven
         energy feedback.  The shortfall at high masses is conventionally interpreted as
         due to energy feedback from active galactic nuclei (AGNs) that keeps the gas
         halos of galaxies and of clusters of galaxies too hot to cool and form stars.
         Heating can also be provided by continued gas infall from the surrounding structure
         hierarchy (Dekel \& Silk 2005, 2008).  An additional important effect is that some
         large halos are occupied by groups and clusters of galaxies and not by individual
         galaxies. Finally, even at intermediate masses, there is a significant shortfall
         in baryonic galaxies compared to expectations based on the canonical cosmological
         baryon fraction.  This is interpreted to be due to the fact that many baryons 
         still reside in a ``warm-hot intergalactic medium'' (WHIM) that surrounds even
         galaxies such as the Milky Way (Dave \etal 2001).  The ``take-home message'' is
         this: We like to think that galaxies are mature objects and that our job is to 
         study galaxy evolution to see how they got that way.  But about 2/3 of the baryons
         in the Universe do not yet live in galaxies or have not yet cooled and formed
         stars.  Galaxy formation is much less ``finished'' than we like to think!
         }
\end{figure}
\cl{\null}
\vskip -28pt
\eject
}

\section*{Acknowledgments}

It is a great pleasure to thank Jes\'us F\'alcon-Barroso and Johan Knapen for organizing
this 2011 workshop, for inviting me to give the introductory and closing lectures, and
for meticulously editing this book.
I am also grateful to J\'esus for his untiring efforts to make our visit to the
Canary Islands such a pleasure.  I also especially want to thank IAC students
Judit Bakos, 
Javier Blasco Herrera, 
Santiago Erroz, 
Adriana de Lorenzo-C\'aceres,
Mireia Montes, 
Agnieszka Ry\'s, 
Jos\'e Ram\'on S\'anchez-Gallego and
Marja Seidel
for their help with the workshop and for their kindness to Mary and to me.  And finally, 
I would like to thank all of the above and all of the students for making this an 
extraordinarily pleasureable and productive experience.  It was a pleasure to meet you all, 
and I hope to see you again often. Warmest best wishes from Mary and from me for successful and 
fulfilling careers.

Mary and I very much enjoyed our interactions with the other lecturers,
Lia{\ts}Athanassoula,
James{\ts}Binney,
Albert{\ts}Bosma,
Ron{\ts}Buta,
Daniela{\ts}Calzetti,
Reynier Peletier,
Nick Scoville,
Isaac Shlosman and
Jacqueline van Gorkom.

      These lectures were prepared, delivered and in large part written up during two visits
to the Max-Planck-Institut f\"ur Extraterrestrische Physik, Garching-bei-M\"unchen, Germany and the
Observatory of the Ludwig-Maximilians-Universit\"at, Munich, Germany.  It is a great pleasure~to~thank 
Managing Director Ralf Bender and the staff of both institutes for their wonderful hospitality 
and financial support.  I also warmly thank Christa Ingram and Bettina Niebisch of the 
Max-Planck-Institut f\"ur Extraterrestrische Physik for transcribing my oral lectures.
And I sincerely thank the many people who allowed me to reproduce figures (see captions).

      As with all my papers, I owe a huge debt of gratitude to Mary Kormendy for her editorial help
and for her patience and understanding during the preparation of this review.

      This work makes extensive use of data products from the digital image database 
of the Sloan Digital Sky Survey (SDSS).  Funding for 
SDSS and SDSS-II was provided by the Alfred P.~Sloan Foundation, the Participating Institutions, 
the National Science Foundation, the US Dept.~of Energy, the National Aeronautics and Space 
Administration, the Japanese Monbukagakusho, the Max Planck Society and the Higher
Education Funding Council for England.  The SDSS is managed by the Astrophysical Research 
Consortium for the Participating Institutions.~They are the 
American Museum of Natural History, Astrophysical Institute~Potsdam, University of Basel, 
University of Cambridge, Case Western Reserve University, University of Chicago, Drexel 
University, Fermilab, the Institute for Advanced Study, the Japan Participation Group, 
Johns Hopkins University, the Joint Institute for Nuclear Astrophysics, the Kavli 
Institute for Particle Astrophysics and Cosmology, the Korean Scientist Group, the 
Chinese Academy of Sciences (LAMOST), Los Alamos National Laboratory, the 
Max-Planck-Institute for Astronomy, the Max-Planck-Institute for Astrophysics,
New Mexico State University, Ohio State University, University of Pittsburgh, 
University of Portsmouth, Princeton University, United States Naval Observatory
and the University of Washington.

     This research depended critically on extensive use of NASA's  Astrophysics Data System 
bibliographic services.
I also made extensive use of the NASA/IPAC Extragalactic~Database~(NED), 
which is operated by the Jet Propulsion Laboratory and the California Institute of Technology
under contract with NASA.  
And I used the HyperLeda electronic database (Paturel \etal 2003) at
{\tt http://leda.univ-lyon1.fr} and the image display tool SAOImage DS9 developed by the Smithsonian 
Astrophysical Observatory.  

My work on secular evolution is supported by NSF grant AST-0607490 and by the 
Curtis T.~Vaughan, Jr.~Centennial Chair in Astronomy. 

Note added November 2013:  This review was finished and submitted in mid-June 2012.  The astro{\kern 1pt}-ph
version is essentially identical to the published paper except for differences in spelling (British 
there and American here) and for improvements here in a few figures.  The present version remains
up to date except for one important new paper by Sellwood (2013).  This is a broad
review of the theory of ``Secular Evolution in Disk Galaxies'' that overlaps some of the present
subjects but that also discusses many additional fundamental topics.  
It is an important complement to the present paper and to Kormendy \& Kennicutt (2004).

\vs

\centerline{\bfseries References} \vskip 2pt

\newcommand{\nhi}{\noindent \hangindent=20pt}

\frenchspacing


\nhi Adami, C., Slezak, E., Durret, F., \etal 2005, A\&A, 429, 39

\nhi Abraham, R.~G., Merrifield, M.~R., Ellis, R.~S., Tanvir, N.~R., \& Brinchmann, J.~1999, MNRAS, 308, 569 

\nhi Abramson, A., Kenney, J.~D.~P., Crowl, H.~H., \etal 2011, AJ, 141, 164

\nhi Aguerri, J. A. L., Elias-Rosa, N., Corsini, E. M., \& Mu\~noz-Tu\~n\'on, C. 2005, A\&A, 434, 109

\nhi Andredakis, Y.~C., Peletier, R.~F., \& Balcells, M.~1995, MNRAS, 275, 874

\nhi Andredakis, Y.~C., \& Sanders, R.~H.~1994, MNRAS, 267, 283

\nhi Arnaboldi, M., Gerhard, O.~2010, Highlights of Astronomy, 15, 97

\nhi Arnaboldi, M., Gerhard, O., Aguerri, J.~A.~L., \etal 2004, ApJ, 614, L33

\nhi Arnaboldi, M., Freeman, K. C., Mendez, R. H., \etal 1996, ApJ, 472, 145  

\nhi Arnaboldi,{\ts}M., Aguerri,{\ts}J.{\ts}A.{\ts}L., Napolitano,{\ts}N.{\ts}R., \etal 2002, AJ, 123, 760   

\nhi Arp, H.~C.~1966, {\it Atlas of Peculiar Galaxies\/} (Pasadena: California Inst. of Technology)


\nhi Athanassoula, E.~1992, MNRAS, 259, 345    

\nhi Athanassoula, E.~2003, MNRAS, 341, 1179   

\nhi Athanassoula, E.~2005, Ann.~N.~Y.~Acad.~Sci., 1045, 168          

\nhi Athanassoula, E.~2012, in {\it XXIII Canary Islands Winter School of Astrophysics, Secular Evolution of Galaxies}, ed.~J.~Falc\'on-Barroso
                    \& J.~H.~Knapen (Cambridge: Cambridge University Press), in press

\nhi Athanassoula, E., Bosma, A., Cr\'ez\'e, M., \& Schwarz, M.~P.~1982, A\&A, 107, 101  

\nhi Athanassoula, E., Lambert, J.~C., \& Dehnen, W.~2005, MNRAS, 363, 496

\nhi Athanassoula, E., \& Misiriotis, A.~2002, MNRAS, 330, 35         %


\nhi Baade, W.~1944, ApJ, 100, 147

\nhi Baade, W.~1963, {\it The Evolution of Stars and Galaxies} (Cambridge: Harvard University Press)

\nhi Baes, M., Buyle, P., Hau, G.~K.~T., \& Dejonghe, H.~2003, MNRAS, 341, L44

\nhi Baggett, W.~E., Baggett, S.~M., \& Anderson, K.~S.~J.~1998, AJ, 116, 1626

\nhi Bahcall, J. N., \& Casertano, S. 1985, ApJ, 293, L7


\nhi Baldry, I. K., Glazebrook, K., Brinkmann, J., \etal 2004, ApJ, 600, 681

\nhi Balogh, M.~L., Navarro, J.~F., \& Morris, S.~L.~2000, ApJ, 540, 113  

\nhi Banerjee, A., Matthews, L.~D., \& Jog, C.~J.~2010, NewA, 15, 89      

\nhi Barazza, F.~D., Binggeli, B., \& Jerjen, H.~2002, A\&A, 391, 823     

\nhi Barentine, J.~C., \& Kormendy, J., ApJ, in press (arXiv:1205.6876)

\nhi Barnes, J.~E.~1988, ApJ, 331, 699

\nhi Barnes, J.~E.~1989, Nature, 338, 123

\nhi Barnes, J.~E.~1992, ApJ, 393, 484

\nhi Barnes, J.~E.~1998, in {\it 26$^{\rm th}$ Advanced Course of the Swiss Society of Astronomy and Astrophysics, Galaxies: 
        Interactions and Induced Star Formation}, ed. D. Friedli, L. Martinet, \& D. Pfenniger (New York: Springer-Verlag), 275 

\nhi Barnes, J.~E.~2004, {\it MNRAS}, {\bf 350}, 798

\nhi Barnes, J.~E., \& Hernquist, L.~1992, ARA\&A, 30, 705

\nhi Barnes, J.~E., \& Hernquist, L.~E.~1991, ApJ, 370, L65            

\nhi Barnes, J.~E., \& Hernquist, L..~1996, ApJ, 471, 115



\nhi Bekki, K., Couch, W.~J., Drinkwater, M.~J., \& Gregg, M.~D.~2001, ApJ, 557, L39  

\nhi Bekki, K., Couch, W.~J., \& Shioya, Y.~2002, ApJ, 577, 651


\nhi Bender, R., Burstein, D., \& Faber, S.~M.~1992, ApJ, 399, 462

\nhi Bender, R., Burstein, D., \& Faber, S.~M.~1993, ApJ, 411, 153



\nhi Bender, R., Kormendy, J., \etal 2008, in preparation                   


\nhi Bender, R., \& Nieto, J.-L.~1990, A\&A, 239, 97  






\nhi Benedict, G. F., Howell, D. A., Jorgensen, I., Kenney, J. D. P., \& Smith, B.~J. 2002, AJ, 123, 1411 

\nhi Berentzen, I., Heller, C. H., Shlosman, I., \& Fricke, K. J. 1998, MNRAS, 300, 49

\nhi Bernardi, M., Sheth, R. K., Annis, J., \etal 2003, AJ, 125, 1882


\nhi Binggeli, B., \& Cameron, L.~M.~1991, A\&A, 252, 27

\nhi Binggeli, B., \& Popescu, C.~C.~1995, A\&A, 298, 63

\nhi Binggeli, B., Sandage, A., \& Tammann, G.~A.~1985, AJ, 90, 1681

\nhi Binggeli, B., Sandage, A., \& Tammann, G.~A.~1988, ARA\&A, 26, 509

\nhi Binggeli, B., Sandage, A., \& Tarenghi, M.~1984, AJ, 89, 64

\nhi Binggeli, B., Tammann, G.~A., \& Sandage, A.~1987, AJ, 94, 251




\nhi Binney, J.~2004, in {\it IAU Symposium 220, Dark Matter in Galaxies}, ed. S. D. Ryder,
             D. J. Pisano, M. A. Walker \& K. C. Freeman (San Francisco:{\thinspace}ASP), 3



\nhi Binney, J., \& Tremaine, S.~1987, {\it Galactic Dynamics}, (Princeton: Princeton University Press)

\nhi Blanton, M.~R., Eisenstein, D., Hogg, D.~W., Schlegel, D.~J., \& Brinkmann, J.~2005, ApJ, 629, 143 

\nhi Blanton, M. R., Hogg, D. W., Bahcall, N. A., \etal 2003, ApJ, 594, 186 

\nhi Blanton, M.~R., Lupton, R.~H., Schlegel, D.~J., \etal 2005, ApJ, 631, 208

\nhi Blitz, L., \& Spergel, D. N. 1991, ApJ, 379, 631



\nhi B\"ohringer, H., Briel, U. G., Schwarz, R. A., \etal 1994, Nature, 368, 828

\nhi Boroson, T.~1981, ApJS, 46, 177

\nhi Boselli, A., Boissier, S., Cortese, L., \& Gavazzi, G.~2009, Astr.~Nachr., 330, 904

\nhi Bosma, A.~1981a, AJ, 86, 1791

\nhi Bosma, A.~1981b, AJ, 86, 1825



\nhi Bosma, A., Ekers, R.~D., \& Lequeux, J.~1977b, A\&A, 57, 97                  

\nhi Bosma, A., Gadotti, D.~A., de Blok, W.~J.~G., \& Athanassoula, E.~2010, in {\it Galaxies in Isolation: Exploring Nature Versus Nurture}, 
                ed.~L.~Verdes-Montenegro, A.~del Olmo, \& J.~Sulentic (San Francisco: ASP), 53

\nhi Bosma, A., van der Hulst, J.~M., \& Sullivan, W.~T.~1977a, A\&A, 57, 373     

\nhi Bothun, G. D., \& Thompson, I. B. 1988, AJ, 96, 877

\nhi Bouchard, A., Da Costa, G.~S., \& Jerjen, H.~2009, AJ, 137, 3038

\nhi Bournaud, F., \& Combes, F. 2002, A\&A, 392, 83                             

\nhi Bournaud, F., Combes, F., \& Semelin, B.~2005, MNRAS, 364, L18              

\nhi Bournaud, F., Elmegreen, B.~G., \& Elmegreen, D.~M.~2007, ApJ, 670, 237     


\nhi Brooks, A. M., Governato, F., Quinn, T., Brook, C. B., \& Wadsley, J. 2009, ApJ, 694, 396

\nhi Burkert, A., \& Tremaine, S.~2010, ApJ, 720, 516

\nhi Burstein, D.~1979, ApJ, 234, 435

\nhi Buta, R.~1990, in {\it Galactic Models}, Ann. N. Y. Acad. Sci., 596, 58

\nhi Buta, R.~1995, ApJS, 96, 39  

\nhi Buta, R.~2011, in {\it Planets, Stars, and Stellar Systems, Vol.~6, Extragalactic Astronomy and Cosmology}, ed.~W.~C.~Keel 
     (New York: Springer-Verlag), in press (arXiv:1102.0550)

\nhi Buta, R.~2012, in {\it XXIII Canary Islands Winter School of Astrophysics, Secular Evolution of Galaxies}, ed.~J.~Falc\'on-Barroso
                    \& J.~H.~Knapen (Cambridge: Cambridge University Press), in press


\nhi Buta, R.~J., Byrd, G.~G., \& Freeman, T.~2004, AJ, 127, 1982  

\nhi Buta, R., \& Combes, F.~1996, Fund. Cosmic Phys., 17, 95

\nhi Buta, R., \& Crocker, D.~A.~1991, AJ, 102, 1715

\nhi Buta, R.~J., Corwin, H.~G., \& Odewahn, S.~C.~2007, {\it The de Vaucouleurs Atlas of Galaxies} (Cambridge: Cambridge University Press)



\nhi Buta, R., Treuthardt, P. M., Byrd, G. G., \& Crocker, D. A. 2000, AJ, 120, 1289    

\nhi Caon, N., Capaccioli, M., \& Rampazzo, R. 1990, A\&AS, 86, 429



\nhi Cappellari,{\ts}M., Emsellem,{\ts}E., Krajnovi\'c,{\ts}D., \etal 2011, MNRAS, 416, 1680

\nhi Carollo, C.~M.~1999, ApJ, 523, 566

\nhi Carollo, C.~M., Ferguson, H.~C., \& Wyse, R.~F.~G., ed.~1999, {\it The Formation of Galactic Bulges} (Cambridge: Cambridge University Press)

\nhi Carollo, C.~M., \& Stiavelli, M.~1998, AJ, 115, 2306

\nhi Carollo, C.~M., Stiavelli, M., de Zeeuw, P.~T., \& Mack, J.~1997, AJ, 114, 2366

\nhi Carollo, C.~M., Stiavelli, M., \& Mack, J.~1998, AJ, 116, 68

\nhi Carollo, C.~M., Stiavelli, M., Seigar, M., de Zeeuw, P.~T., \& Dejonghe, H.~2002, AJ, 123, 159 

\nhi Carter, D., Prieur, J.~L., Wilkinson, A., Sparks, W.~B., \& Malin, D.~F.~1988, MNRAS, 235, 813

\nhi Castro-Rodrigu\'ez, N., Arnaboldi, M., Aguerri, J. A. L., \etal 2009, A\&A, 507, 621

\nhi Cayatte, V., van Gorkom, J. H., Balkowski, C., \& Kotanyi, C. 1990, AJ, 100, 604

\nhi Ceverino, D., \& Klypin, A.~2007, MNRAS, 379, 1155    

\nnhi Chen, C.-W., C\^ot\'e, P., West, A., Peng, E.~W., \& Ferrarese, L.~2010, ApJS, 191, 1

\nhi Chiboucas, K., Karachentsev, I.~D., \& Tully, R.~B.~2009, AJ, 137, 3009

\nhi Chung, A., van Gorkom, J.~H., Kenney, J.~D.~P., \& Vollmer, B.~2007, ApJ, 659, L115

\nhi Chung, A., van Gorkom, J.~H., Kenney, J.~D.~P., Crowl, H., \& Vollmer, B.~2009, AJ, 138, 1741  

\nhi Cibinel, A., Carollo, C.~M., Lilly, S.~J., \etal 2012, ApJ, in press (arXiv: 1206.6496)


\nhi Combes, F. 2008a, in {\it ASP Conference Series, Vol. 390, Pathways Through an Eclectic Universe}, 
     ed. J. H. Knapen, T. J. Mahoney, \& A. Vazdekis (San Francisco: ASP), 369

\nhi Combes, F.~2008b, in {\it ASP Conference Series, Vol. 396, Formation and Evolution of Galaxy Disks}, ed.~J. G.
                      Funes \& E.~M.~Corsini (San Francisco: ASP), 325

\nhi Combes, F.~2010, in {\it IAU Symposium 271, Astrophysical Dynamics: From Stars to Galaxies}, ed. N. Brummell, A.S. Brun, 
     M. S. Miesch \& Y. Ponty (Cambridge: Cambridge University Press), 119

\nhi Combes, F.~2011, Mem. Soc. Astron. Ital. Suppl., 18, 53  

\nhi Combes, F., Debbasch, F., Friedli, D. \& Pfenniger, D.~1990, A\&A, 233, 82

\nhi Combes, F., \& Sanders, R.~H.~1981, A\&A, 96, 164

\nhi Comer\'on, S., Knapen, J.~H., Beckman, J.~E., \etal 2010, MNRAS, 402, 2462

\nhi Corbelli, E. 2003, MNRAS, 342, 199

\nhi Corbelli, E., \& Salucci, P.~2000, MNRAS, 311, 441     

\nhi Corsini, E.~M., Debattista, V.~P., \& Aguerri, J.~A.~L.~2003, ApJ, 599, L29                       

\nhi C\^ot\'e, P., Blakeslee, J.~P., Ferrarese, L., \etal 2004, ApJS, 153, 223  




\nhi Courteau, S., de Jong, R.~S., \& Broeils, A.~H.~1996, ApJ, 457, L73

\nhi Courteau,{\ts}S., Dutton,{\ts}A.{\ts}A., van den Bosch,{\ts}F.{\ts}C., \etal 2007, ApJ, 671, 203  


\nhi Curran, S.~J., Polatidis, A.~G., Aalto, S., \& Booth, R.~S.~2001a, A\&A, 368, 824  

\nhi Curran, S.~J., Polatidis, A.~G., Aalto, S., \& Booth, R.~S.~2001b, A\&A, 373, 459

\nhi Da Costa, G. S. 1994, in {\it ESO/OHP Workshop on Dwarf Galaxies}, ed. G. Meylan \& Ph. Prugniel (Garching: ESO), 221

\nhi Dasyra, K. M., Tacconi, L.~J., Davies, R.~I., \etal 2006a, ApJ, 638, 745

\nhi Dasyra, K. M., Tacconi, L.~J., Davies, R.~I., \etal 2006b, ApJ, 651, 835

\nhi Dasyra, K.~M., Tacconi, L.~J., Davies, R.~I., \etal 2006c, NewAR, 50, 720

\nhi Dav\'e, R., Cen, R., Ostriker, J. P., \etal 2001, ApJ, 552, 473




\nhi Davis, L.~E., Cawson, M., Davies, R.~L., \& Illingworth, G.~1985, AJ, 90, 169 

\nhi Debattista, V.~P., \& Shen, J.~2007, ApJ, 654, L127

\nhi de Blok, W.~J.~G., Walter, F., Brinks, E., \etal 2008, AJ, 136, 2648  

\nhi Dekel, A., \& Birnboim, Y.~2006, MNRAS, 368, 2  


\nhi Dekel, A., \& Silk, J.~1986, ApJ, 303, 39

\nhi De Rijcke, S., Dejonghe, H., Zeilinger, W. W., \& Hau, G. K. T. 2003, A\&A, 400, 119

\nhi De Rijcke, S., Dejonghe, H., Zeilinger, W. W., \& Hau, G. K. T. 2004, A\&A, 426, 53

\nhi de Souza, R. E., Gadotti, D. A., \& dos Anjos, S. 2004, ApJS, 153, 411

\nhi de Vaucouleurs, G.~1959, {\it Handbuch der Physik}, 53, 275

\nhi de Vaucouleurs, G. 1963, ApJS, 8, 31  

\nhi de Vaucouleurs, G., de Vaucouleurs, A., Corwin, H.~G., \etal 1991, {\it Third Reference Catalogue of Bright Galaxies} (Berlin:~Springer-Verlag) (RC3)

\nhi Devereux, N. A., \& Young, J. S. 1991, ApJ, 371, 515

\nhi Di Matteo, T., Springel, V., \& Hernquist, L.~2005, Nature, 433, 604

\nhi Dixon, M.~E.~1971, ApJ, 164, 411

\nhi Djorgovski, S., \& Davis, M.~1987, ApJ, 313, 59

\nhi Djorgovski, S., de Carvalho, R., \& Han, M.-S.~1988, in {\it ASP Converence Series, Vol.~4, The Extragalactic 
             Distance Scale}, ed.~S.~van den Bergh \& C.~J.~Pritchet (San Francisco: ASP), 329

\nhi D'Onghia, E., Burkert, A., Murante, G., \& Khochfar, S. 2006, MNRAS, 372, 1525

\nhi Doyon, R., Wells, M., Wright, G. S., \etal 1994, ApJ, 437, L23

\nhi Dressler, A. 1979, ApJ, 231, 659

\nhi Dressler, A.~1980, ApJ, 236, 351

\nhi Dressler, A., 1989, in {\it IAU Symposium 134, Active Galactic Nuclei}, ed.~D.~E. Osterbrock
             \& J.~S.~Miller (Dordrecht: Kluwer), 217


\nhi Driver, S.~P., Popescu, C.~C., Tuffs, R.~J., \etal 2007, MNRAS, 379, 1022

\nhi Drory, N., Bundy, K., Leauthaud, A., \etal 2009, ApJ, 707, 1595

\nhi Dutton, A. A. 2009, MNRAS, 396, 121


\nhi Eggen, O.~J., Lynden-Bell, D., \& Sandage, A.~R.~1962, ApJ, 136, 748

\nhi Einasto, J., Saar, E., Kaasik, A., \& Chernin, A. D. 1974, Nature, 252, 111

\nhi Elmegreen, B.~G., Bournaud, F., \& Elmegreen, D.~M.~2008a, ApJ, 684, 829 

\nhi Elmegreen, B.~G., Bournaud, F., \& Elmegreen, D.~M.~2008b, ApJ, 688, 67  

\nhi Elmegreen, B.~G., \& Elmegreen, D.~M.~2005, ApJ, 627, 632

\nhi Elmegreen, B.~G., Elmegreen, D.~M, Fernandez, M.~X., \& Lemonias, J.~J. 2009b, ApJ, 692, 12

\nhi Elmegreen, B.~G., Galliano, E., \& Alloin, D.~2009, ApJ, 703, 1297      

\nhi Elmegreen,{\ts}D.{\ts}M.,{\ts}Elmegreen,{\ts}B.{\ts}G,{\ts}Marcus,{\ts}M.{\ts}T.,\ts\etal2009a,{\ts}ApJ,{\ts}701,{\ts}306

\nhi Elmegreen, D.~M., Elmegreen, B.~G, Ravindranath, S., \& Coe, D.~A.~2007, ApJ, 658, 763




\nhi Englmaier, P., \& Gerhard, O. 1997, MNRAS, 287, 57

\nhi Erwin, P., Beckman, J. E., \& Vega Beltran, J. C. 2004, in {\it Penetrating Bars Through Masks of Cosmic Dust: The Hubble 
              Tuning Fork Strikes a New Note}, ed.~D.~L.~Block, I.~Puerari, K.~C.~Freeman, R.~Groess, \& E.~K.~Block (Dordrecht: Kluwer), 775

\nhi Erwin, P., Vega Beltr\'an, J.~C., Graham, A.~W., \& Beckman, J.~E.~2003, ApJ, 597, 929  


\nhi Faber, S.~M.~1973, ApJ, 179, 423

\nhi Faber, S.~M., \& Lin, D.~N.~C.~1983, ApJ, 266, L17

\nhi Faber, S.~M., Dressler, A., Davies, R. L., Burstein, D., Lynden-Bell, D. 1987, in {\it Nearly Normal Galaxies: From the Planck
                                 Time to the Present}, ed. S.~M.~Faber (New York: Springer), 175 

\nhi Faber, S.~M., \& Gallagher, J.~S.~1979, ARA\&A, 17, 135

\nhi Faber, S.~M., \& Jackson, R.~E.~1976, ApJ, 204, 668

\nhi Faber, S.~M., Tremaine, S., Ajhar, E.~A., \etal 1997, AJ, 114, 1771


\nhi Falc\'on-Barroso, J., Bacon, R., Bureau, M., \etal 2006, MNRAS, 369, 529

\nhi Falc\'on-Barroso, J., Peletier, R.~F., \& Balcells, M.~2002, MNRAS, 335, 741

\nhi Ferguson, H. C., \& Sandage, A. 1989, ApJ, 346, L53

\nhi Ferguson, H.~C., \& Binggeli, B.~1994, A\&AR, 6, 67

\nhi Ferrarese, L.~2002, ApJ, 578, 90

\nhi Ferrarese L, \& Merritt D. 2000, ApJ, 539, L9

\nhi Ferrarese, L., C\^ot\'e, P., Jord\'an, A., \etal 2006, ApJS, 164, 334


\nhi Fisher, D.~B. 2006, ApJ, 642, L17

\nhi Fisher, D.~B., \& Drory, N.~2008, AJ, 136, 773

\nhi Fisher, D.~B., \& Drory, N.~2010, ApJ, 716, 942

\nhi Fisher, D.~B., \& Drory, N.~2011, ApJ, 733, L47					

\nhi Fisher, D.~B., Drory, N., \& Fabricius, M.~H.~2009, ApJ, 697, 630

\nhi F\"orster Schreiber, N.~M., Genzel, R., Bouch\'e, N., \etal 2009, ApJ, 706, 1364

\nhi Freeman, K.~C.~1970, ApJ, 160, 811

\nhi Freeman, K. C. 1975, in {\it IAU Symposium 69, Dynamics of Stellar Systems}, ed. A. Hayli (Dordrecht: Reidel), 367

\nhi Friedli, D., \& Benz, W.~1993, A\&A, 268, 65

\nhi Fukugita, M., Hogan, C.~J., \& Peebles, P.~J.~E.~1998, ApJ, 503, 518


\nhi Gadotti, D.~A.~2008, MNRAS, 384, 420

\nhi Gadotti, D.~A.~2009, MNRAS, 393, 1531                                             


\nhi Gadotti, D.~A., \& Kauffmann, G.~2009, MNRAS, 399, 621

\nhi Galliano, E., Alloin, D., Pantin, E. \etal 2008, A\&A, 492, 3                                            

\nhi Gavazzi, G., Donati, A., Cucciati, O., \etal 2005, A\&A, 430, 411

\nhi Gavazzi, G., Franzetti, P., Scodeggio, M., Boselli, A., \& Pierini, D.~2000, A\&A, 361, 863

\nhi Gebhardt, K., Bender, R., Bower, G., \etal 2000, ApJ, 539, L13


\nhi Genzel, R., Burkert, A., Bouch\'e, N., \etal 2008, ApJ, 687, 59

\nhi Genzel, R., Eisenhauer, F., \& Gillessen, S. 2010, Rev. Mod. Phys., 82, 3121 

\nhi Genzel, R., Tacconi, L.~J., Eisenhauer, F., \etal 2006, Nature, 442, 786

\nhi Genzel, R., Tacconi, L.~J., Rigopoulou, D., Lutz, D., \& Tecza, M.~2001, ApJ, 563, 527

\nhi Gerin, M., Combes, F., \& Athanassoula, E.~1990, A\&A, 230, 37                    

\nhi Glass, L., Ferrarese, L., C\^ot\'e, P., \etal 2011, ApJ, 726, 31

\nhi Gonzalez, A. H., Zaritsky, D., \& Zabludoff, A. I. 2007, ApJ, 666, 147

\nhi Gonz\'alez-Garc\'\i a, A. C., Aguerri, J. A. L., \& Balcells, M. 2005, A\&A, 444, 803

\nhi Governato, F., Brook, C., Mayer, L., \etal 2010, Nature, 463, 203

\nhi Graham, A.~W.~2004, ApJ, 613, L33

\nhi Graham, A.~W.~2008, ApJ, 680, 143

\nhi Graham, A.~W., Erwin, P., Trujillo, I., \& Asensio Ramos, A.~2003, AJ, 125, 2951  

\nhi Granato, G.~L., De Zotti, G., Silva, L., Bressan, A., \& Danese, L.~2004, ApJ, 600, 580


\nhi Greene, J.~E., Peng, C.~Y., Kim, M., \etal 2010, ApJ, 721, 26                     

\nhi G\"ultekin, K., Richstone, D. O., Gebhardt, K., \etal 2009, ApJ, 698, 198

\nhi Gunn, J.~E., \& Gott, J.~R.~1972, ApJ, 176, 1

\nhi Hamabe, M. 1982, PASJ, 34, 423

\nhi H\"aring, N., \& Rix, H.-W.~2004, ApJ, 604, L89

\nhi Harris, G.~L.~H., \& Harris, W.~E.~2011, MNRAS, 410, 2347

\nhi Harris, J., Calzetti, D., Gallagher, J.~S., Conselice, C.~J., \& Smith, D.~A.~2001, AJ, 122, 3046  

\nhi Hasan, H., \& Norman, C. 1990, ApJ, 361, 69

\nhi Hasan, H., Pfenniger, D., \& Norman, C. 1993, ApJ, 409, 91

\nhi Hensler, G., Theis, Ch., \& Gallagher, J.~S.~2004, A\&A, 426, 25

\nhi Hibbard, J.~E., Guhathakurta, P., van Gorkom, J.~H., \& Schweizer, F.~1994, AJ, 107, 67

\nhi Hibbard, J.~E., \& Mihos, J.~C.~1995, AJ, 110, 140

\nhi Hibbard, J.~E., van der Hulst, J.~M., Barnes, J.~E., \& Rich, R.~M.~2001a, AJ, 122, 2969

\nhi Hibbard, J.~E., \& van Gorkom, J.~H. 1996, AJ, 111, 655

\nhi Hibbard, J.~E., van Gorkom, J.~H., Rupen, M.~P., \& Schiminovich, D.~2001b, in {\it ASP Conference Series, Vol.~240,
       Gas and Galaxy Evolution}, ed. J. E. Hibbard, M. P. Rupen, \& J. H. van Gorkom (San Francisco: ASP), 659

\nhi Ho, L.~C.~2007, ApJ, 668, 94

\nhi Ho, L.~C.~2008, ARA\&A, 46, 475

\nhi Ho, L. C., Filippenko, A. V., Sargent, W. L. W., \& Peng, C. Y. 1997, ApJS, 112, 391

\nhi Hogg, D.~W., Blanton, M., Strateva, I., \etal 2002, AJ, 124, 646  

\nhi Hogg, D.~W., Blanton, M. R.. Brinchmann, J., \etal 2004, ApJ, 601, L29 


\nhi Hopkins, P.~F., Bundy, K., Croton, D., \etal 2010, ApJ, 715, 202      

\nhi Hopkins, P.~F., Cox, T.~J., Dutta, S.~N., Hernquist, L., Kormendy, J., \& Lauer, T.~R.                      2009a, ApJS, 181, 135  

\nhi Hopkins, P.~F., Cox, T.~J., \& Hernquist, L.                                                                2008,  ApJ, 689, 17    


\nhi Hopkins, P.~F., Hernquist, L., Cox, T.~J., di Matteo, T., Robertson, B., \& Springel, V.                     2006, ApJS, 163, 1


\nhi Hopkins, P.~F., Hernquist, L., Cox, T.~J., Kere\v s, D., \& Wuyts, S.                                           2009b, ApJ, 691, 1424  



\nhi Hopkins, P.~F., Lauer, T.~R., Cox, T.~J., Hernquist, L., \& Kormendy, J.                                     2009c, ApJS, 181, 486   

\nhi Hopkins, P.~F., Somerville, R. S., Cox, T. J., \etal 2009d, MNRAS, 397, 802   



\nhi Hozumi, S.~2012, PASJ, 64, 5

\nhi Hozumi, S., \& Hernquist, L. 1999, in {\it ASP Conference Series vol.~182, Galaxy Dynamics}, ed.~D.~R.~Merritt, 
     M.~Valluri, \& J.~A.~Sellwood (San Francisco: ASP), 259

\nhi Hozumi, S., \& Hernquist, L. 2005, PASJ, 57, 719

\nhi Hu, J. 2008, MNRAS, 386, 2242                                                        


\nhi Hubble, E.~1936, {\it The Realm of the Nebulae} (New Haven: Yale Univ.~Press)

\nhi Irwin, M. J., Belokurov, V., Evans, N. W., \etal 2007, ApJ, 656, L13

\nhi Jarrett, T.~H., Chester, T., Cutri, R., Schneider, S.~E., \& Huchra, J.~P.~2003, AJ, 125, 525

\nhi Jerjen, H., \& Binggeli, B.~1997, in {\it The Second Stromlo Symposium:
             The Nature of Elliptical Galaxies}, ed.~M. Arnaboldi \etal (San Francisco: ASP), 239


\nhi Jerjen, H., Kalnajs, A., \& Binggeli, B.~2000, A\&A, 358, 845  

\nhi Jerjen, H., Kalnajs, A., \& Binggeli, B.~2001, in {\it ASP Conf. Ser. 230, Galaxy Disks and Disk Galaxies}, ed.~J. G. Funes \& E. M. Corsini (San Francisco: ASP), 239

\nhi Jogee, S., Barazza, F.~D., Rix, H.-W., \etal 2004, ApJ, 615, L105

\nhi Jones, B.~J.~T.~1992, unpublished lectures at the Nordic Summer School, Galaxies: Structure, Formation and Evolution,
         Gr{\"a}ft$^\circ${\kern-5pt}avallen, Sweden


\nhi Joseph, R. D. 1999, Ap\&SS, 266, 321

\nhi Joseph, R. D., \& Wright, G. S. 1985, MNRAS, 214, 87

\nhi Kalnajs, A.~1973, Proc.~Astron.~Soc.~Australia, 2, 174

\nhi Karachentsev, I., Aparicio, A., \& Makarova, L. 1999, A\&A, 352, 363

\nhi Karachentsev, I.~D., Karachentseva, V.~E., \& Parnovsky, S.~L.~1993, Astr.~Nachrichten, 314, 97


\nhi Kauffmann, G., Heckman, T.~M., White, S.~D.~M., \etal 2003a, MNRAS, 341, 33 

\nhi Kauffmann, G., Heckman, T.~M., White, S.~D.~M., \etal 2003b, MNRAS, 341, 54 

\nhi Kauffmann, G., White, S.~D.~M., \& Guiderdoni, B.~1993, MNRAS, 264, 201

\nhi Kautsch, S.~J.~2009, PASP, 121, 1297

\nhi Kautsch, S.~J., Grebel, E.~K., Barazza, F.~D., \& Gallagher, J.~S.~2006, A\&A, 445, 765

\nhi Kelson, D. D., Zabludoff, A. I., Williams, K. A., \etal 2002, ApJ, 576, 720

\nhi Kenney, J.~D., \& Young, J.~S.~1986, ApJ, 301, L13


\nhi Kenney, J.~D.~P., Tal, T., Crowl, H.~H., Feldmeier, J., \& Jacoby, G.~H.~2008, ApJ, 687, L69

\nhi Kenney, J.~D.~P., van Gorkom, J.~H., \& Vollmer, B.~2004, AJ, 127, 3361  

\nhi Kennicutt, R.~C.~1989, ApJ, 344, 685     

\nhi Kennicutt, R.~C.~1998a, ApJ, 498, 541    

\nhi Kennicutt, R.~C.~1998b, ARA\&A, 36, 189  

\nhi Kennicutt, R.~C.~1998c, in {\it 26$^{\rm th}$ Advanced Course of the Swiss Society of Astronomy and Astrophysics, Galaxies: 
        Interactions and Induced Star Formation}, ed. D. Friedli, L. Martinet, \& D. Pfenniger (New York: Springer-Verlag), 1

\nhi Kent, S.~M.~1985, ApJS, 59, 115

\nhi King, I.~1962, AJ, 67, 471

\nhi King, I.~R.~1966, AJ, 71, 64

\nhi Kirby, E. M., Jerjen, H., Ryder, S. D., \& Driver, S. P. 2008, AJ, 136, 1866

\nhi Knapen, J.~H., Beckman, J.~E., Heller, C.~H., Shlosman, I., \& de Jong, R.~S. 1995a. ApJ, 454, 623                    

\nhi Knapen, J.~H., Beckman, J.~E., Shlosman, I., \etal 1995b, ApJ, 443, L73                                                     

\nhi Knapen, J.~H., de Jong, R.~S., Stedman, S., \& Bramich, D.~M.~2003, MNRAS, 344, 527  

\nhi Knapen, J. H., Mazzuca, L. M., B\"oker, T., Shlosman, I., Colina, L., Combes, F., \& Axon, D. J.~2006, A\&A. 448, 489 

\nhi Knapen, J.~H., Sharp, R.~G., Ryder, S.~D., \etal 2010, MNRAS, 408, 797  

\nhi Koda, J., Milosavljevi\'c, M., \& Shapiro, P. R. 2009, ApJ, 696, 254

\nhi Kormendy, J.~1977a, ApJ, 217, 406

\nhi Kormendy, J.~1977b, ApJ, 218, 333

\nhi Kormendy, J.~1979a, in {\it Photometry, Kinematics and Dynamics of Galaxies}, ed.~D.~S.~Evans 
                        (Austin: Dept.~of Astronomy, Univ.~of Texas at Austin), 341

\nhi Kormendy, J.~1979b, ApJ, 227, 714

\nhi Kormendy, J.~1981, in {\it The Structure and Evolution of Normal Galaxies}, ed. S. M. Fall \&
                        D. Lynden-Bell (Cambridge: Cambridge Univ.~Press), 85

\nhi Kormendy, J.~1982a, ApJ, 257, 75                                                                   

\nhi Kormendy, J. 1982b, in {\it Twelfth Advanced Course of the Swiss Society of Astronomy and Astrophysics,
                          Morphology and Dynamics of Galaxies}, ed. L. Martinet \& M. Mayor
                          (Sauverny: Geneva Observatory), 113

\nhi Kormendy, J.~1983, ApJ, 275, 529  

\nhi Kormendy, J.~1984a, ApJ, 286, 116  

\nhi Kormendy, J.~1984b, ApJ, 286, 132  

\nhi Kormendy, J.~1985, ApJ, 295, 73

\nhi Kormendy, J.~1987, in {\it Nearly Normal Galaxies: From the Planck Time to the Present}, ed.~S.~M.~Faber (New York: Springer-Verlag), 163

\nhi Kormendy, J.~1989, ApJ, 342, L63

\nhi Kormendy, J.~1993a, in {\it The Nearest Active Galaxies}, ed.~J.~Beckman,
             L.~Colina \& H.~Netzer (Madrid: Consejo Superior de Investigaciones Cient\'\i ficas), 197

\nhi Kormendy, J.~1993b, in {\it IAU Symposium 153, Galactic Bulges}, ed. H. Dejonghe \& H. J. Habing (Dordrecht: Kluwer), 209 

\nhi Kormendy, J.~2004a, in {\it Penetrating Bars Through Masks of Cosmic Dust: The Hubble 
              Tuning Fork Strikes a New Note}, ed.~D.~L.~Block, I.~Puerari, K.~C.~Freeman, R.~Groess, \& E.~K.~Block (Dordrecht: Kluwer), 816

\nhi Kormendy, J.~2004b, in {\it Penetrating Bars Through Masks of Cosmic Dust: The Hubble 
              Tuning Fork Strikes a New Note}, ed.~D.~L.~Block, I.~Puerari, K.~C.~Freeman, R.~Groess, \& E.~K.~Block (Dordrecht: Kluwer), 831

\nhi Kormendy, J.~2008a, in {\it IAU Symposium 245, Formation and Evolution of Galaxy Bulges}, ed.~M.~Bureau,~E.~Athanassoula, \&
                        B.~Barbuy (Cambridge: Cambridge University Press), 107

\nhi Kormendy, J.~2008b, StarDate, March/April issue

\nhi Kormendy, J.~2009, in {\it ASP Conference Series, Vol.~419, Galaxy Evolution: Emerging Insights and Future Challenges}, ed.~S.~Jogee, I.~Marinova,
              L.~Hao, \& G.~A.~Blanc (San Francisco: ASP), 87

\nhi Kormendy , J., \& Barentine, J.~C.~2010, ApJ, 715, L176

\nhi Kormendy, J., \& Bender, R.~1994, in {\it ESO/OHP Workshop on Dwarf Galaxies}, 
                 ed.~G.~Meylan \& P.~Prugniel (Garching: ESO), 161

\nhi Kormendy, J., \& Bender, R.~1996, ApJ, 464, L119

\nhi Kormendy, J., \& Bender, R.~2009, ApJ, 691, L142

\nhi Kormendy, J., \& Bender, R.~2011, Nature, 469, 377

\nhi Kormendy, J., \& Bender, R.~2012, ApJS, 198, 2 (KB2012)

\nhi Kormendy, J., \& Bender, R.~2013, ApJS, in preparation

\nhi Kormendy, J., Bender, R., \& Cornell, M.~E.~2011, Nature, 469, 374

\nhi Kormendy, J., \& Cornell, M.~E.~2004, in {\it Penetrating Bars Through Masks of Cosmic Dust: The Hubble Tuning Fork Strikes 
             a New Note}, ed.~D.~L.~Block, I.~Puerari, K.~C.~Freeman, R.~Groess, \& E.~K.~Block (Dordrecht: Kluwer), 261

\nhi Kormendy, J., Drory, N., Bender, R., \& Cornell, M.~E.~2010, ApJ, 723, 54


\nhi Kormendy, J., \& Fisher, D.~B.~2005, in {\it The Ninth Texas-Mexico Conference on Astrophysics}, ed. S. Torres-Peimbert \& G. MacAlpine,
     RevMexA\&A, Serie de Conferencias, 23, 101

\nhi Kormendy, J., \& Fisher, D. B.~2008, in {\it ASP Conference Series, Vol. 396, Formation and Evolution of Galaxy Disks},
      ed.~J.~G.~Funes \& E.~M.~Corsini (San Francisco: ASP), 297

\nhi Kormendy, J., Fisher, D.~B., Cornell, M.~E., \& Bender, R.~2009, ApJS, 182, 216 (KFCB)


\nhi Kormendy, J., \& Freeman, K.~C.~2013, ApJ, in preparation

\nhi Kormendy, J., \& Gebhardt, K.~2001, in {\it 20$^{\rm th}$ Texas Symposium on Relativistic Astrophysics},
      ed.~J.~C.~Wheeler \& H.~Martel (Melville: AIP), 363

\nhi Kormendy, J., \& Ho, L.~C.~2013, ARA\&A, 51, 511 

\nhi Kormendy, J., \& Illingworth, G.~1982, ApJ, 256, 460

\nhi Kormendy, J., \& Illingworth, G.~1983, ApJ, 265, 632

\nhi Kormendy, J., \& Kennicutt, R.~C.~2004, ARA\&A, 42, 603

\nhi Kormendy, J., \& McClure, R.~D.~1993, AJ, 105, 1793

\nhi Kormendy, J., \& Norman, C.~A.~1979, ApJ, 233, 539


\nhi Kormendy, J., \& Richstone, D.~1995, ARA\&A, 33, 581

\nhi Kormendy, J., \& Sanders, D.~B.~1992, ApJ, 390, L53

\nhi Krick, J. E., \& Bernstein, R. A. 2007, AJ, 134, 466

\nhi Kristen, H., J\"ors\"ater, S., Lindblad, P.~O., \& Boksenberg, A.~1997, A\&A, 328, 483  

\nhi \hbox{Kroupa, P.~2012, PAS Australia, \tt http://www.publish.csiro.au/paper/} {\tt AS12005.htm} (arXiv:1204.2546)

\nhi Kuhn, T.~S.~1970, {\it The Structure of Scientific Revolutions} (Chicago: Univ. of Chicago Press)

\nhi Kuz'min, G.~1956, Astron.~Zh, 33, 27

\nhi Lake, G., Katz, N., \& Moore, B.~1998, ApJ, 495, 152

\nhi Larson, R. B. 1974, MNRAS, 169, 229

\nhi Larson, R. B.. Tinsley, B.~M., \& Caldwell, C.~N.~1980, ApJ, 237, 692  

\nhi Lauer, T.~R., Ajhar, E.~A., Byun, Y.-I., \etal 1995, AJ, 110, 2622


\nhi Lauer, T. R., Faber, S. M., Gebhardt, K., \etal 2005, AJ, 129, 2138

\nhi Lauer, T. R., Faber, S. M., Lynds, C. R., \etal 1992, AJ, 103, 703

\nhi Lauer, T.~R., Faber, S. M., Richstone, D., \etal 2007, ApJ, 662, 808

\nhi Laurikainen, E., \& Salo, H.~2002, MNRAS, 337, 1118

\nhi Laurikainen, E., Salo, H., \& Buta, R. 2005, MNRAS, 362, 1319

\nhi Laurikainen, E., Salo, H., Buta, R., \& Knapen, J. H. 2007, MNRAS, 381, 401

\nhi Laurikainen, E., Salo, H., Buta, R., \& Knapen, J. H. 2011, MNRAS, 418, 1452

\nhi Laurikainen, E., Salo, H., Buta, R., \& Knapen, J. H., \& Comer\'on, S.~2010, MNRAS, 405, 1089

\nhi Laurikainen, E., Salo, H., Buta, R., \etal 2006, AJ, 132, 2634

\nhi Laurikainen, E., Salo, H., Buta, R., \& Vasylyev, S.~2004, MNRAS, 355, 1251


\nhi Lindblad, B.~1956, Stockholms Obs. Ann., 19, No.~7                               

\nhi Lindblad,{\ts}P.{\ts}A.{\ts}B., Lindblad,{\ts}P.{\ts}O., \& Athanassoula,{\ts}E.~1996, A\&A, 313, 65

\nhi Lindblad, P.~O.~1999, NewAR, 9, 221

\nhi Lisker, T., Brunngr\"aber, R., \& Grebel, E.~K.~2009, AN, 330, 966          

\nhi Lisker, T., Grebel, E.~K., \& Binggeli, B.~2006, AJ, 132, 497               

\nhi Lisker, T., Grebel, E.~K., Binggeli, B., \& Glatt, K.~2007, ApJ, 660, 1186  

\nhi Liu, Y., Zhou, X., Ma, J., \etal 2005, AJ, 129, 2628


\nhi Lynden-Bell, D., \& Kalnajs, A.~J.~1972, MNRAS, 157, 1

\nhi Lynden-Bell, D., \& Pringle, J.~E.~1974, MNRAS, 168, 603

\nhi Lynden-Bell, D., \& Wood, R.~1968, MNRAS, 138, 495

\nhi MacArthur L. A., Courteau, S., \& Holtzman, J. A. 2003, ApJ, 582, 689

\nhi Magorrian, J., Tremaine, S., Richstone, D., \etal 1998, AJ, 115, 2285 

\nhi Makarova, L. 1999, A\&AS, 139, 491

\nhi Malin, D.~F., \& Carter, D.~1980, Nature, 285, 643

\nhi Malin, D.~F., \& Carter, D.~1983, ApJ, 274, 534

\nhi Maller, A. H., Katz, N., Kere\v s, D., Dav\'e, R., \& Weinberg, D. H. 2006, ApJ, 647, 763

\nhi Maoz, D., Barth, A. J., Ho, L. C., Sternberg, A., \& Filippenko, A. V. 2001, AJ, 121, 3048  

\nhi Marconi, A., \& Hunt, L.~K.~2003, ApJ, 589, L21

\nhi Marinova, I., Jogee, S., Weinzirl, T., \etal 2012, ApJ, 746, 136  

\nhi Martin, N.~F., de Jong, J.~T.~A., \& Rix, H.-W. 2008, ApJ, 684, 1075

\nhi Mateo, M.~1998, ARA\&A, 36, 435

\nhi Mateo, M.~2008, The Messenger, 134, 3

\nhi Matthews, L.~D. 2000, AJ, 120, 1764                                          

\nhi Matthews, L.~D., Gallagher, J.~S., \& van Driel, W.~1999a, AJ, 118, 2751     

\nhi Matthews, L.~D., van Driel, W., \& Gallagher, J.~S.~1999b, astro-ph/9911022  

\nhi Mayer, L., Governato, F., Colpi, M., \etal 2001a, ApJ, 547, L123

\nhi Mayer, L., Governato, F., Colpi, M., \etal 2001b, ApJ, 559, 754

\nhi Mayer, L., Mastropietro, C., Wadsley, J., Stadel, J., \& Moore, B. 2006, MNRAS, 369, 1021

\nhi McConnachie, A.~W. 2012, AJ, 144, 4

\nhi McConnachie, A.~W., \& Irwin, M.~J.~2006, MNRAS, 365, 1263  



\nhi McLure, R.~J., \& Dunlop, J.~S.~2002, MNRAS, 331, 795

\nhi M\'endez-Abreu, J., Aguerri, J.~A.~L., Corsini, E.~M., \& Simonneau, E.~2008, A\&A, 478, 353  




\nhi Merritt, D.~2006, ApJ,, 648, 976

\nhi Mihalas, D., \& Routly, P.~M.~1968, {\it Galactic Astronomy} (San Francisco: Freeman)

\nhi Mihos, J. C., Harding, P., Feldmeier, J., \& Morrison, H. 2005, ApJ, 631, L41

\nhi Mihos, J. C., \& Hernquist, L. 1994, ApJ, 437, L47

\nhi Mihos, J. C., Janowiecki, S., Feldmeier, J. J., Harding, P., \& Morrison, H. 2009, ApJ, 698, 1879

\nhi Mihos, J.~C., McGaugh, S.~S., \& de Bolk, W.~J.~G.~1997, ApJ, 477, L79     


\nhi Milosavljevi\'c, M., Merritt, D., Rest, A., \& van den Bosch, F.~C.~2002, MNRAS, 331, L51


\nhi Moore, B., Katz, N., Lake, G., Dressler, A., \& Oemler, A.~1996, Nature, 379, 613  

\nhi Moore, B., Lake, G., \& Katz, N.~1998, ApJ, 495, 139  

\nhi Morgan, W.~W.~1951, Publ.~Obs.~Univ.~of Michigan, 10, 33

\nhi Morgan, W. W., \& Lesh, J. R. 1965, ApJ, 142, 1364

\nhi Mulder, P.~S.~1995, A\&A, 303, 57  




\nhi Noguchi, M.~1987, MNRAS, 228, 635                                 

\nhi Noguchi, M. 1988, A\&A, 203, 259                                  

\nhi Noguchi, M.~1996, ApJ, 469, 605                                   

\nhi Norman, C.~A., \& Hasan, H.~1990, in {\it Dynamics and Interactions of Galaxies}, ed.~R.~Wielen (New York: Springer-Verlag), 479

\nhi Norman, C.~A., May, A., \& van Albada, T.~S.~1985, ApJ, 296, 20

\nhi Norman, C.~A., Sellwood, J.~A., \& Hasan, H.~1996, ApJ, 462, 114

\nhi Nowak, N., Thomas, J., Erwin, P., \etal 2010, MNRAS, 403, 646

\nhi Oemler, A. 1976, ApJ, 209, 693

\nhi Okamura, S. 2011, Paper presented at the ESO Workshop on Fornax, Virgo, Coma et al.: Stellar Systems in High Density Environments,
     ed. M. Arnaboldi,~{\tt http://www.eso.org/sci/meetings/2011/}\hfill\par\quad {\tt fornax\_virgo2011/posters.html}

\nhi Ostriker, J.~P., \& Peebles, P.~J.~E.~1973, ApJ, 186, 467

\nhi Oswalt,{\ts}T.{\ts}D.,{\ts}Smith,{\ts}J.{\ts}A.,{\ts}Wood,{\ts}M.{\ts}A.,{\ts}\&{\ts}Hintzen,{\ts}P.{\ts}1996, Nature,{\ts}382,{\ts}692

\nhi Paturel, G., Petit, C., Prugniel, Ph., \etal 2003, A\&A, 412, 45

\nhi Peebles, P. J. E., \& Nusser, A. 2010, Nature, 465, 565



\nhi Peletier, R. F., Davies, R. L., Illingworth, G. D., Davis, L. E., \& Cawson, M. 1990, AJ, 100, 1091

\nhi Peletier,{\ts}R.{\ts}F., Falc\'on-Barroso,{\ts}J., Bacon,{\ts}R., \etal 2007a, MNRAS, 379, 445

\nhi Peletier, R.~F., Ganda, K., Falc\'on-Barroso, J., \etal 2007b, in {\it IAU Symp. 241, Stellar Populations as 
    Building Blocks of Galaxies}, ed. A. Vazdekis \& R. Peletier (Cambridge: Cambridge Univ.~Press), 485

\nhi Peletier, R.~F., Ganda, K., Falc\'on-Barroso, J., \etal 2008, in {\it IAU Symposium 245, Formation and 
     Evolution of Galaxy Bulges\/}, ed. M.~Bureau, E. Athanassoula, \& B.~Barbuy (Cambridge: Cambridge Univ.~Press), 271

\nhi Peng, C. Y., Ho, L. C., Impey, C. D., \& Rix, H.-W. 2002, AJ, 124, 266


\nhi Pfenniger, D.~1984, A\&A, 134, 373                  

\nhi Pfenniger, D.~1985, A\&A, 150, 112                  

\nhi Pfenniger, D., \& Friedli, D.~1991  A\&A, 252, 75

\nhi Pfenniger, D., \& Norman, C.~1990, ApJ, 363, 391

\nhi Pfenniger, D., \& Norman, C.~A.~1991, in {\it IAU Sumposium 146, Dynamics of Galaxies and Their Molecular Cloud Distributions}, 
             ed.~F.~Combes \& F.~Casoli (Dordrecht: Kluwer), 323

\nhi Pildis, R.~A., Schombert, J.~M., \& Eder, J.~A.~1997, ApJ, 481, 157

\nhi Plummer, H.~C.~1911, MNRAS, 71, 460

\nhi Primack, J.~R.~2004, in {\it IAU Symposium 220, Dark Matter in Galaxies}, ed. S.~D. Ryder, D.~J.~Pisano, M.~A.~Walker, \& K.~C.~Freeman
     (San Francisco, ASP), 53

\nhi Raha, N., Sellwood, J.~A., James, R.~A., \& Kahn, F.~D.~1991, Nature, 352, 411


\nhi Rautiainen, P., \& Salo, H.~2000, A\&A, 362, 465

\nhi Rautiainen, P., Salo, H., \& Laurikainen, E.~2005, ApJ, 631, L129

\nhi Ravindranath, S., Ho, L.~C., \& Filippenko, A.~V.~2002, ApJ, 566, 801

\nhi Read, J.~I., \& Trentham, N.~2005, Phil. Trans. R. Soc. London, A363, 2693

\nhi Regan, M.~W., Thornley, M. D., Helfer, T. T. \etal 2001, ApJ, 561, 218

\nhi Regan, M.~W., Vogel, S.~N., \& Teuben, P.~J.~1997, ApJ, 482, L143

\nhi Renzini, A.~1999, in {\it The Formation of Galactic Bulges}, ed.~C.~M.~Carollo, H.{\ts}C.{\ts}Ferguson \& R.{\ts}F.{\ts}G.{\ts}Wyse (Cambridge:
             Cambridge Univ. Press),\ts9

\nhi Richstone, D. O. 1976, ApJ, 204, 642

\nhi Richstone, D., Ajhar, E. A., Bender, R., \etal 1998, Nature, 395, A15

\nhi Rigopoulou, D., Spoon, H. W. W., Genzel, R., \etal 1999, AJ, 118, 2625

\nhi Roberts, W.~W.~1969, ApJ, 158, 123

\nhi Roberts, W.~W., Roberts, M.~S., \& Shu, F.~H.~1975, ApJ, 196, 381

\nhi Robertson, B., Cox, T.~J., Hernquist, L., \etal 2006, ApJ, 641, 21

\nhi Robertson,{\ts}B., Yoshida,{\ts}N., Springel,{\ts}V., \& Hernquist,{\ts}L. 2004, ApJ, 606, 32

\nhi Sage, L. J., \& Wrobel, J. M. 1989, ApJ, 344, 204

\nhi Saito, M. 1979, PASJ, 31, 193

\nhi Sakamoto, K., Okamura, S., Minezaki, T., Kobayashi, Y., \& Wada, K. 1995, AJ, 110, 2075  

\nhi Salo, H., Rautiainen, P., Buta, R., \etal 1999, AJ, 117, 792


\nhi Sancisi, R., \& van Albada, T.{\ts}S. 1987, in {\it IAU Symposium 117, Dark Matter in the Universe},
     ed.{\ts}J.{\ts}Kormendy \& G.{\ts}R.{\ts}Knapp (Dordrecht:{\ts}Reidel) 67

\nhi Sandage, A.~1961, {\it The Hubble Atlas of Galaxies}, (Washington: Carnegie Institution of Washington)

\nhi Sandage, A.~1975, in {\it Stars and Stellar Systems, Vol.~9, Galaxies~and~the Universe}, ed.~A.~Sandage, M.~Sandage,
     \& J.~Kristian (Chicago: Univ. of Chicago Press), 1

\nhi Sandage, A.~2004, in {\it Penetrating Bars Through Masks of Cosmic Dust: The Hubble Tuning Fork Strikes 
             a New Note}, ed.~D.~L.~Block, I.~Puerari, K.~C.~Freeman, R.~Groess, \& E.~K.~Block (Dordrecht: Kluwer), 39

\nhi Sandage, A., \& Bedke, J.~1994, {\it The Carnegie Atlas of Galaxies} (Washington: Carnegie Institution of Washington)

\nhi Sandage, A., \& Binggeli, B.~1984, AJ, 89, 919

\nhi Sandage, A., Binggeli, B., \& Tammann, G.{\ts}A.~1985a, in {\it ESO Workshop on the Virgo Cluster of Galaxies}, ed.{\ts}O.-G.{\ts}Richter 
     \& B.{\ts}Binggeli (Garching: ESO), 239

\nhi Sandage, A., Binggeli, B., \& Tammann, G.~A.~1985b, AJ, 90, 1759


\nhi Sandage, A., \& Tammann, G.~A.~1981, {\it Revised Shapley-Ames Catalog of Bright Galaxies} (Washingtom: Carnegie Institution of Washington)

\nhi Sanders, D. B., \& Mirabel, I. F. 1996, ARA\&A, 34, 749

\nhi Sanders, D. B., Soifer, B. T., Elias, J. H., \etal 1988a, ApJ, 325, 74

\nhi Sanders, D. B., Soifer, B. T., Elias, J. H., Neugebauer, G., \& Matthews, K. 1988b, ApJ, 328, L35

\nhi Sanders, R.~H., \& Tubbs, A.~D.~1980, ApJ, 235, 803

\nhi Schaeffer, R., \& Silk, J.~1988, A\&A, 203, 273

\nhi Schechter, P.~1976, ApJ, 203, 297

\nhi Schechter, P.~L., \& Dressler, A.~1987, AJ, 94, 563

\nhi Schlegel, D.~J., Finkbeiner, D.~P., \& Davis, M.~1998, ApJ, 500, 525 

\nhi Schmidt, M.~1959, ApJ, 129, 243

\nhi Schombert, J. M. 1988, ApJ, 328, 475

\nhi  Schweizer, F. 1978, in {\it IAU Symposium 77, Structure and Properties of Nearby Galaxies}, ed. E. M. Berkhuijsen \& R.~Wielebinski
      (Dordrecht: Reidel), 279

\nhi  Schweizer, F. 1980, ApJ, 237, 303  


\nhi  Schweizer, F. 1982, ApJ, 252, 455


\nhi  Schweizer, F. 1987, in {\it Nearly Normal Galaxies: From the Planck Time to the Present}, ed. S. M. Faber (New York: Springer-Verlag), 18

\nhi  Schweizer, F. 1990, in {\it Dynamics and Interactions of Galaxies}, ed. R. Wielen (New York: Springer-Verlag), 60

\nhi  Schweizer, F. 1996, AJ, 111, 109   

\nhi  Schweizer, F. 1998, in {\it 26$^{\rm th}$ Advanced Course of the Swiss Society of Astronomy and Astrophysics, Galaxies: 
        Interactions and Induced Star Formation}, ed. D. Friedli, L. Martinet, \& D. Pfenniger (New York: Springer-Verlag), 105

\nhi  Schweizer, F., \& Seitzer, P. 1988, ApJ, 328, 88  

\nhi  Schweizer, F., \& Seitzer, P. 1992, AJ, 104, 1039

\nhi  Schweizer, F., Seitzer, P., Faber, S.~M., \etal 1990, ApJ, 364, L33

\nhi Scoville, N. Z., Evans, A. S., Dinshaw, N., \etal 1998, ApJ, 492, L107

\nhi Seigar, M., Carollo, C.~M., Stiavelli, M., de Zeeuw, P.~T., \& Dejonghe, H.~2002, AJ, 123, 184

\nhi Sellwood, J.~A.~1980, A\&A, 89, 296  

\nhi Sellwood, J.~A.~2000, in {\it ASP Conference Series 197, Dynamics of Galaxies: from the Early Universe
     to the Present}, ed.~F.~Combes, G.~A.~Mamon, \& V.~Charmandaris (San Francisco: ASP), 3

\nhi Sellwood, J.~A.~2006, ApJ, 637, 567                    

\nhi Sellwood, J.~A.~2008, ApJ, 679, 379                    

\nhi Sellwood, J.~A.~2013, Rev. Mod. Phys., in press (arXiv:1310.0403)


\nhi Sellwood, J.~A., \& Carlberg, R.~G.~1984, ApJ, 282, 61 

\nhi Sellwood, J.~A., \& Moore, E.~M.~1999, ApJ, 510, 125

\nhi Sellwood, J.~A., \& Wilkinson, A.~1993, Rep.~Prog.~Phys., 56, 173

\nhi S\'ersic, J.~L.~1968, {\it Atlas de Galaxias Australes} (C\'ordoba: Observatorio Astron\'omico, Universidad de C\'ordoba)


\nhi Shen, J., \& Debattista, V.~P.~2009, ApJ, 690, 758

\nhi Shen, J., Rich, R.~M., Kormendy, J., \etal 2010, ApJ, 720, L72

\nhi Shen, J., \& Sellwood, J.~A.~2004, ApJ, 604, 614    

\nhi Shlosman, I.~2012,  in {\it XXIII Canary Islands Winter School of Astrophysics, Secular Evolution of Galaxies}, ed.~J.~Falc\'on-Barroso
                    \& J.~H.~Knapen (Cambridge: Cambridge University Press), in press

\nhi Shu, F., Najita, J., Ostriker, E., \etal 1994, ApJ, 429, 781

\nhi Shu, F.~H., Milione, V., \& Roberts, W.~W.~1973, ApJ, 183, 819

\nhi Shu, F.~H., Najita, J., Ostriker, E.~C., \& Shang, H.~1995, ApJ, 455, L155

\nhi Shu, F.~H., Stachnik, R.~V., \& Yost, J.~C.~1971, ApJ, 166, 465

\nhi Silk, J., \& Mamon, G.~A.~2012, Res.~A\&A, 12, 917 (arXiv:1207.3080)

\nhi Silk, J., \& Rees, M.~J.~1998, A\&A, 331, L1

\nhi Simien, F., \& de Vaucouleurs, G.~1986, ApJ, 302, 564

\nhi Simkin, S. M., Su, H. J., \& Schwarz, M. P.~1980, ApJ, 237, 404

\nhi Skillman, E.~D., C\^ot\'e, S., \& Miller, B.~W.~2003, AJ, 125, 593

\nhi Somerville, R.~S., Hopkins, P.~F., Cox, T.~J., Robertson, B.~E., \& Hernquist, L.~2008, MNRAS, 391, 481  

\nhi Sparke, L.~S., \& Sellwood, J.~A.~1987, MNRAS, 225, 653

\nhi Springel, V., White, S. D. M., Jenkins, A., \etal 2005, Nature, 435, 629  

\nhi Steiman-Cameron, T.~Y., Kormendy, J., \& Durisen, R.~H.~1992, AJ, 104, 1339

\nhi Steinmetz, M., \& Navarro, J.~F.~2002, NewA, 7, 155

\nhi Steinmetz, M., \& Navarro, J.~F:~2003, NewA, 8, 557

\nhi Stinson, G.~S., Dalcanton, J.~J., Quinn, T., Kaufmann, T., \& Wadsley, J.~2007, ApJ, 667, 170


\nhi Strateva, I., Ivezi\'c, \v Z., Knapp, G. R., \etal 2001, AJ, 122, 1861 

\nhi Tacconi, L.~J., Genzel, R., Lutz, D., \etal 2002, ApJ, 580, 73

\nhi Tacconi, L.~J., Genzel, R., Neri, R., \etal 2010, Nature, 463, 781

\nhi Tasca, L.~A.~M., \& White, S.~D.~M.~2011, A\&A, 530, A106


\nhi Thomas, D., Brimioulle, F., Bender, R., \etal 2006, A\&A, 445, L19

\nhi Thronson, H. A., Tacconi, L., Kenney, J., \etal 1989, ApJ, 344, 747

\nhi Thuan, T. X., \& Kormendy, J. 1977, PASP, 89, 466


\nhi Tolstoy, E., Hill, V., \& Tosi, M. 2009, ARA\&A, 47, 371





\nhi Toomre, A.~1963, ApJ, 138, 385   

\nhi Toomre, A.~1964, ApJ, 139, 1217  

\nhi Toomre A.:~1977a, in {\it The Evolution of Galaxies and Stellar Populations}, ed. B. M. Tinsley \& R. B. Larson (New Haven:
             Yale Univ.~Obs.), 401

\nhi Toomre A.~1977b, ARA\&A, 15, 437

\nhi Toomre, A.~1981, in {\it The Structure and Evolution of Normal Galaxies}, ed.
            S.~M.~Fall \& D.~Lynden-Bell (Cambridge: Cambridge Univ.~Press), 111

\nhi Toomre, A.~1990, in {\it Dynamics and Interactions of Galaxies}, ed.~R.~Wielen (New York:~Springer-Verlag), 292

\nhi Toomre, A., \& Toomre, J.~1972, ApJ, 178, 623

\nhi Tremaine, S.~1981, in {\it The Structure and Evolution of Normal Galaxies}, ed.
            S.~M.~Fall \& D.~Lynden-Bell (Cambridge: Cambridge Univ.~Press), 67

\nhi Tremaine S.~1989, in {\it Dynamics of Astrophysical Disks}, ed.~J.~A.~Sellwood (Cambridge: Cambridge Univ. Press), 231

\nhi Tremaine, S., \& Weinberg, M. D. 1984, ApJ, 282, L5

\nhi Tremaine, S., Gebhardt, K., Bender, R., \etal 2002, ApJ, 574, 740


\nhi Treuthardt, P., Salo, H., \& Buta, R.~2009, AJ, 137, 19


\nhi Tully, R.~B., \& Fisher, J.~R.~1977, A\&A, 54, 661

\nhi Vader, J.~P.~1986, ApJ, 305, 669

\nhi Valenzuela, O., \& Klypin, A.~2003, MNRAS, 345, 406

\nhi van Albada, T.~S.~1982, MNRAS, 201, 939

\nhi van Albada, T.~S. \& Sancisi, R. 1986, Phil.~Trans.~R.~Soc.~London, A320, 447

\nhi van den Bergh, S. 1976, ApJ, 206, 883  




\nhi van den Bergh, S.~1994a, AJ, 107, 1328

\nhi van den Bergh, S.~1994b, ApJ, 428, 617

\nhi van den Bergh, S. 2007, {\it The Galaxies of the Local Group} (Cambridge: Cambridge University Press)



\nhi van der Kruit, P.~C., Jim\'enez-Vicente, J., Kregel, M., \& Freeman, K.~C.~2001, A\&A, 379, 374  

\nhi van der Kruit, P.~C., \& Freeman, K.~C.~2011, ARA\&A, 49, 301




\nhi van Gorkom, J.~H., \& Kenney, J.~D.~P. 2013, ARA\&A, in preparation

\nhi van Zee, L., Skillman, E. D., \& Haynes, M. P. 2004, AJ, 128, 121


\nhi Visser, H.~C.~D.~1980, A\&A, 88, 149

\nhi Volonteri, M., Natarajan, P., \& G\"ultekin, K.~2011, ApJ, 737, 50



\nhi Weinzirl, T., Jogee, S., Khochfar, S., Burkert, A., \& Kormendy, J.~2009, ApJ, 696, 411

\nhi Weisz, D.~R., Dalcanton, J. J., Williams, B. F., \etal 2011a, ApJ, 739, 5 

\nhi Weisz, D.~R., Dolphin, A. E., Dalcanton, J. J., \etal 2011b, ApJ, 743,~8

\nhi Welch, G. A., \& Sage, L. J. 2003, ApJ, 584, 260

\nhi White, S.~D.~M., \& Rees, M.~J.~1978, MNRAS, 183, 341



\nhi Winget, D. E., \& Kepler, S. O. 2008, ARA\&A, 46, 157

\nhi Wirth, A., \& Gallagher, J.~S.~1984, ApJ, 282, 85

\nhi Yoshida, M., Ohyama, Y., Iye, M., \etal 2004, AJ, 127, 90  

\nhi Young, J. S., Xie, S., Tacconi, L., \etal 1995, ApJS, 98, 219


\nhi Zwicky, F.~1942, Phys.~Rev., 61, 489

\nhi Zwicky, F.~1951, PASP, 63, 61

\nhi Zwicky, F.~1957, {\it Morphological Astronomy} (Berlin: Springer-Verlag)   

\end{document}